\documentclass[aps,showpacs,preprintnumbers,nofootinbib,superscriptaddress,groupedaddress]{SPhdThesis}

\usepackage[T1]{fontenc}
\usepackage{lmodern}
\usepackage{amsmath}
\usepackage{graphicx,amssymb,amsmath,amsthm,amsfonts,epsfig,epsf}
\usepackage{epstopdf}
\definecolor{darkred}{rgb}{0.5,0,0}
\usepackage{aas_macros}
\usepackage{bm}
\usepackage{dcolumn}
\usepackage[utf8]{inputenc}
\usepackage{latexsym}
\usepackage{rotating}
\usepackage{longtable}

\usepackage{enumerate}
\usepackage{tensor,multirow}
\usepackage{mathtools}
\usepackage{url}
\usepackage{epigraph}
\usepackage{titlesec}
\usepackage[sort&compress,numbers]{natbib}
\usepackage{doi}
\usepackage{hyperref}
\hypersetup{colorlinks=false}
\usepackage{colortbl}
\usepackage[toc,page]{appendix}
\usepackage{minitoc}
\usepackage{emptypage}
\def\nn{\nonumber}

\def\be{\begin{equation}}
\def\ee{\end{equation}}
\def\ba{\begin{align}}
\def\ea{\end{align}}
\newcommand{\beq}{\begin{eqnarray}}
\newcommand{\eeq}{\end{eqnarray}}
\newcommand{\ben}{\begin{enumerate}}
\newcommand{\een}{\end{enumerate}}

\newcommand{\nocontentsline}[3]{}
\newcommand{\tocless}[2]{\bgroup\let\addcontentsline=\nocontentsline#1{#2}\egroup}

\renewcommand{\vec}[1]{\boldsymbol{#1}}

\newcommand{\mndd}{_{\mu\nu}}

\newcommand{\ijdd}{_{ij}}
\newcommand{\ijuu}{^{ij}}

\newcommand{\rgb}{\mathcal{R}_{\rm GB}}
\newcommand{\til}{~}
\newcommand{\oo}{\left(}
\newcommand{\cc}{\right)}
\newcommand{\ooq}{\left[}
\newcommand{\ccq}{\right]}

\newcommand{\warn}[1]{{\textcolor{red}{\sf{[IN PROGRESS]}} }}

\makeatletter
\titleformat{\part}[display]
  {\Huge\scshape\filright}
  {\partname~\thepart:}
  {20pt}
  {\thispagestyle{epigraph}}
\makeatother
\SgSetTitle{Universidade de Lisboa\\ Instituto Superior Técnico}
\SgSetAuthorDegrees{Lorenzo Annulli}
\SgSetAuthor{Challenging theories of gravitation:\\ dark matter, compact objects and gravitational waves}
\SgSetYear{{\bf 2020}}
\SgSetDegree{
 \begin{tabular}{rl}
    {\bf Supervisor:}& Doctor Vítor Manuel dos Santos Cardoso\\
    {\bf Co-Supervisor:}& Doctor Leonardo Gualtieri
\end{tabular}}
\SgSetDepartment{Thesis approved in public session to obtain the PhD Degree in Physics\\
Jury final classification: Pass with Distinction}
\SgSetUniversity{$\;$}
\SgSetDeclarationDate{{\bf 2020}}

\begin{document}

		\SgAddTitle%
		
	  \cleardoublepage
		
		\SgAddTitlesecond%
		
		\cleardoublepage
		
		\begin{frontmatter}
		
		\begin{resumo}

O melhor modelo para descrever a interação gravitacional é a teoria da Relatividade Geral. Observações variadas corroboram a legitimidade da teoria, por comparação com a mais antiga descrição newtoniana. Entre outros, podemos encontrar a explicação para a precessão anómala do periélio de Mercúrio, o desvio para o vermelho (gravitacional) da luz e a previsão da evolução orbital dos pulsares binários. Além disso, a Relatividade Geral prevê a existência de ondas gravitacionais, ondulações do espaço-tempo produzidas por massas aceleradas. 

Graças a uma rede conectada de interferômetros chamada LIGO / Virgo, as ondas gravitacionais produzida pela coalescência de corpos astrofísicos massivos e compactos foram medidas diretamente. Estas observações recentes abriram o caminho para uma forma completamente nova de testar a interação gravitacional. As ondas gravitacionais emitidas por buracos negros, estrelas de neutrões ou outras fontes compactas, transportam as assinaturas dos seus progenitores e do meio em que estes nasceram e evoluíram. Por isso, as ondas gravitacionais transportam informação sobre a gravidade. 

A possibilidade de usar ondas gravitacionais para obter um entendimento mais profundo de problemas em aberto dentro da Relatividade Geral motiva o trabalho desenvolvido nesta tese. Cada secção é essencialmente dedicada a desafiar o modelo atual de gravitação, às vezes incluindo novos campos de matéria ainda não descobertos, e outras vezes modificando a estrutura teórica da Relatividade Geral. 

Na primeira parte deste manuscrito, são discutidas as consequências astrofísicas da possível presença de campos escalares que permeiam galáxias. Em particular, a inclusão de um novo escalar fundamental como um dos constituintes da matéria escura desconhecida pode resolver alguns dos problemas da cosmologia moderna (por exemplo, ausência de concentrados de matéria escura ou fraqueza de atrito dinâmico em galáxias anãs). Assim, fornece-se um quadro detalhado da interação de buracos negros massivos e estruturas escalares de matéria escura. 

A segunda parte é dedicada à análise da geração e propagação das ondas gravitacionais. São analisados os efeitos da dispersão das ondas gravitacionais pela presença de binários entre a fonte e os observadores. Posteriormente, é examinada a aproximação do limite próximo como uma ferramenta interessante para investigar a colisão de objetos compactos extremos. Esses corpos podem ser vistos como estrelas compactas que imitam o espaço-tempo de um buraco negro, mas não possuem singularidades nem horizontes. As propriedades das ondas gravitacionais produzidas por tais eventos são discutidas também. Por fim, a evolução dinâmica de campos escalares em torno de binários de buracos negros é também formulada dentro da abordagem de limite próximo. 

A última parte desta tese foca-se em mecanismos de instabilidade em torno de buracos negros e estrelas, em dois modelos alternativos de gravitação. O primeiro considera espaços-tempos em teorias com acoplamentos não triviais entre um grau de liberdade escalar e a curvatura do espaço-tempo. O segundo investiga a contingência de ter novos acoplamentos com um campo vetorial. No último cenário, novas soluções de estrelas de neutrões são também discutidas.

\noindent{\bf{Palavras-chave:}} Relatividade geral; Ondas gravitacionais; Objetos compactos; Halos de matéria escura; Escalarização espontânea.
\end{resumo}%

		\begin{abstract}

The most accurate model to describe the gravitational interaction is the well-known theory of General Relativity. Several observational evidences corroborate the legitimacy of the theory compared to the older Newtonian gravity. Among others, we can find the explanation for the deviation of the precession of Mercury's perihelion, the gravitational redshift of light and the prediction of the orbital decay of binary pulsars. General Relativity furthermore predicts the existence of gravitational waves, i.e. spacetime {\it ripples} produced by accelerated masses. 

Thanks to a connected network of interferometers called LIGO/Virgo, gravitational waves from the coalescence of massive and compact astrophysical bodies have been measured directly. These recent observations paved the way to a completely new route to test the gravitational interaction. Gravitational waves emitted by black holes, neutron stars or other compact sources, carrying the signatures of their generators and of the environment in which they live, will provide crucial knowledge about the underlying theory of gravitation.

The possibility of using gravitational waves to obtain a deeper understanding of open problems within General Relativity motivates the work developed in this thesis. Each part is essentially devoted to challenging the current model of gravitation, sometimes including yet undiscovered new matter fields, and other times modifying the theoretical framework of General Relativity. 
 
In the first part of this manuscript, I discuss the astrophysical consequences of the presence of scalar fields permeating galaxies. Remarkably, including a new fundamental scalar as one of the constituent of the unknown {\it dark matter} may solve some of the problems of modern Cosmology (e.g. absence of dark matter cusps or weakness of dynamical friction in dwarf galaxies). Hence, a detailed picture of the interaction of massive black holes and scalar dark matter structures is provided.

The second part is dedicated to the analysis of the generation and propagation of gravitational waves. The effects of gravitational wave scattering by the presence of binaries between the source and the observers are analyzed. Afterwards, I examine the close limit approximation as a promising tool to investigate the collision of extreme compact objects. These bodies can be seen as compact stars that mimic the spacetime of a black hole, but neither possess singularities nor horizons. The properties of the gravitational waves produced by such events are discussed. Finally, the dynamical evolution of scalar fields around black hole binaries is also formulated within the close limit approach.

The last part of this thesis focuses on unstable mechanisms around black holes and stars, in two alternative models of gravitation. The first considers binary spacetimes in theories with non-trivial couplings between a scalar degree of freedom and the spacetime curvature. The second instead delves into the contingency of having new couplings with a vector field. In the latter scenario, I show novel neutron star solutions arising accordingly.

\noindent{\bf Key-words:} General Relativity; Gravitational waves; Compact objects; Dark matter haloes; Spontaneous scalarization.
\end{abstract}%

	    \begin{citcit}

\vspace*{\fill}
\noindent
\makebox[0.8\textwidth]{\it A mio padre}

\vfill

\end{citcit}

			\cleardoublepage
		
    	\begin{acknowledgements}

Four years over the shoulder of a {\it Maestro} are all I wish to any PhD student through their career. The luck I had in being guided by Vitor is something for which I will always be thankful. Sharing my personal and professional growth with him and Leonardo helped me in all possible ways. Many thanks to both of you!

I am grateful to all the members of CENTRA and GRIT, and especially to those who have collaborated with me during these years. They have made countless hours of seminars, group meetings and brainstorming always fruitful and interesting. 

I am also extremely thankful to all the members of the Jury and to all the colleagues that, with careful questions and thoughtful comments, helped me improving an earlier version of this manuscript.

A good work environment is mandatory to work well. I want to thank All my office-mates for providing it with their constant kindness and smile.

Rodrigo, this PhD was a roller coaster of emotions, inside and outside of the office. The continuous confrontation, our fights and amazing travels made this work possible, you made me a better physicist and colleague. Thank you.

My Greek community,  Thanasis and Kiriakos, two Friends before than Colleagues, I will always have a special space for you in my heart. $\epsilon\upsilon\chi\alpha\rho\iota\sigma\tau\omega$.

My Portuguese community, Zé, Pedro and Isma, you made me feel part of this Country, you gave me your Time and Love. Obrigado guys.

During my stay in Lisbon I have changed several houses, but I always felt Home. Inside and outside our nest, I met so many great people. Diogo, Vera, Rebecca, Chrysalena, Nanda, Bruno, Nico, Isa, Arianna, Francisco, Andrea, Gianluca and all the amazing people have been next to me in these years, thank you so much.

A special thought goes also to all my Friends in Italy and around the World. Zim, Gheb, Erni, Edo, Gio, Fra, Miki, Emy, Leo and Everyone is constantly waiting for me to come back R(h)ome, the fundamental role you have in my life has always accompanied me in these years abroad.

The same goes for my Family, people with a big heart, with enormous patience and infinite Love, you helped in shaping who I am now. I will be forever thankful. A special hug also goes to my Grandmas and to Valerio... Welcome Home!

This wonderful journey would have been impossible without You. You taught me way more than any book can do. I Love you so much Anna, I am honored to share my Life with you.

I am also indebted to the Funda\c{c}ao para a Ci\^{e}ncia e a Tecnologia for the grant awarded me in the framework of the Doctoral Programme IDPASC-Portugal, which made it possible for me to complete this thesis. I also acknowledge partial financial support provided under the European Union's H2020 ERC Consolidator Grant ``Matter and strong-field gravity: New frontiers in Einstein's theory'' and the GWverse COST Action CA16104, ``Black holes, gravitational waves and fundamental physics''.

I acknowledge the warm hospitality of the Theory Institute at CERN and Perimeter Institute where parts of this work work has being done. I am also indebted to Ana Sousa Carvalho for producing some of the figures presented in this manuscript.
\end{acknowledgements}%
    	
    		\cleardoublepage
    	
    	\dominitoc
    	\SgAddToc
    	
    	\cleardoublepage
    	
     	\SgAddLof
     	
     	\cleardoublepage
     	
		\begin{preface}
The research presented in this thesis has been carried out at the Center for Astrophysics
and Gravitation (CENTRA) in the Physics department of the Instituto Superior Técnico - Universidade de Lisboa.

I declare that this thesis is not substantially the same as any that I have submitted for a degree, diploma or other qualification at any other university and that no part of it has already been or is concurrently submitted for any such degree, diploma or other qualification.

The following thesis has been the result of several collaborations. Chapters\til\ref{chapter:Theoretical_framework},\til\ref{chapter:Newtonian_boson_stars} and\til\ref{chapter:external_perturbers_NBS} are the outcome of the synergy with Prof. Vitor Cardoso and the PhD Colleague Rodrigo Vicente. Chapters\til\ref{chapter:scattering_processes} and\til\ref{Scattering of gravitational wave} have been carried out with Dr. Laura Bernard, Prof. Diego Blas and Prof. Vitor Cardoso. Finally, chapters\til\ref{chapter:CLAP} and\til\ref{chapter:spontaneous_vectorization} are the outcome of the work with Prof. Leonardo Gualtieri and Prof. Vitor Cardoso.

A complete list of the articles included in this thesis is displayed below:

\begin{itemize}

\item[$\diamond$]  \cite{Annulli:2020ilw}: {L. Annulli, V. Cardoso and R. Vicente}, ``Stirred and shaken: Dynamical behavior of boson stars and dark matter cores'', \href{https://reader.elsevier.com/reader/sd/pii/S0370269320307474?token=F2DAC1C98ED2CAB3F387A0C370038FC1AED4A103858AD5B630695B82FF5ADC6C3D07DF5A5F4B4F3C59C04FD01393D0FD}{\it Phys.Lett.B 811 (2020) 135944 }, \href{https://arxiv.org/pdf/2007.03700.pdf}{arXiv:2007.03700v1 [astro-ph]}

\item[$\diamond$] \cite{Annulli:2020lyc}: {L. Annulli, V. Cardoso and R. Vicente}, ``Response of ultralight dark matter to supermassive black holes and binaries'', \href{https://journals.aps.org/prd/pdf/10.1103/PhysRevD.102.063022}{\it Phys.Rev D102 (2020)}, \href{https://arxiv.org/pdf/2009.00012.pdf}{arXiv:2009.00012v1 [gr-qc]}

\item[$\diamond$]\cite{Annulli:2018quj}: {L. Annulli, L. Bernard, D. Blas and V. Cardoso}, ``Scattering of scalar, electromagnetic and gravitational waves from binary systems'', \href{https://journals.aps.org/prd/pdf/10.1103/PhysRevD.98.084001}{\it Phys.Rev. D98 (2018)}, \href{https://arxiv.org/pdf/1809.05108.pdf}{arXiv:1809.05108v1 [gr-qc]}

\item[$\diamond$] 	\cite{Annulli:2021dkw}: {L. Annulli, V. Cardoso and L. Gualtieri}, ``Generalizing the close limit approximation of binary black holes'', \href{https://arxiv.org/pdf/2104.11236.pdf}{arXiv:2104.11236 [gr-qc]}

\item[$\diamond$] 	\cite{Annulli:2021lmn}: {L. Annulli}, ``CLAP for modified gravity:
scalar instabilities in binary black hole spacetimes'', 
\href{https://arxiv.org/pdf/2105.08728.pdf}{arXiv:2105.08728 [gr-qc]}

\item[$\diamond$] \cite{Annulli:2019fzq}: {L. Annulli, V. Cardoso and L. Gualtieri}, ``Electromagnetism and hidden vector fields in modified gravity theories: spontaneous and induced vectorization'', \href{https://journals.aps.org/prd/pdf/10.1103/PhysRevD.99.044038}{\it Phys.Rev. D99 (2019)}, \href{https://arxiv.org/pdf/1901.02461.pdf}{arxiv:1901.02461v1 [gr-qc]}

\end{itemize}
\end{preface}%
		
		\cleardoublepage

	    \begin{acronyms}
 \begin{tabular}{rl}
    {\bf GR}& General Relativity\\
    {\bf GW}& Gravitational wave\\
    {\bf BH}& Black hole\\
    {\bf BBH}&  Binary black hole\\
    {\bf NS}& Neutron star\\
    {\bf DM}& Dark matter\\
    {\bf NBS}&  Newtonian boson star\\
    {\bf SP}& Schr\"{o}dinger-Poisson\\
    {\bf EMRI}& Extreme mass ratio inspiral\\
    {\bf PN}& Post Newtonian\\
    {\bf CM}& Center of mass\\
    {\bf EM}& Electromagnetism\\
    {\bf LL}&  Landau Lifshitz\\
    {\bf 2p}&  Dipole\\
    {\bf SW}& Scalar wave\\
    {\bf TT}& Transverse traceless\\
    {\bf CLAP}& Close limit approximation\\
    {\bf ECO}& Extreme compact object\\
    {\bf NR}&  Numerical Relativity\\
    {\bf QNM}& Quasi normal mode\\
    {\bf BL}&  Brill Lindquist\\
    {\bf ADM}& Arnowitt Deser Misner\\
    {\bf EsGB}& Einstein scalar Gauss Bonnet\\
    {\bf HN}& Hellings Nordtvedt\\
    {\bf CD}& Constant density (star)\\
    {\bf EOS}& Equation of state\\
    {\bf Poly}& Polytropic
    \end{tabular}
\end{acronyms}%
	    
	    \cleardoublepage
	    
	    \begin{citcit}

\vspace*{\fill}
\noindent
\makebox[0.8\textwidth]{\it "Truth is the ultimate power. When the truth comes around, all the lies have to run and hide"}

\noindent
\makebox[0.8\textwidth]{- O'Shea Jackson}

\vfill

\end{citcit}

	    \cleardoublepage
	    
		\end{frontmatter}

\chapter{Introduction}

The eternal epistemological debate over how to perceive objects and phenomena in Nature began centuries ago questioning the very {\it existence} of things around us. Attempting to distinguish between what belongs to Metaphysics and what to Physics, this abiding philosophical controversy ended up postulating systematic guidelines on how to build a scientific theory, how it should relate to reality and its ultimate purposes. Among others, philosophical paradigms, such as Positivism,  Coherentism and Epistemological Anarchism, profoundly influenced our contemporary vision of the Universe\til\cite{Comte1855-COMTPP,Olsson2012-OLSCTO-2,Feyerabend1975-FEYAM}. Despite all the possible differences of such gnosiological approaches, we can safely assume today that natural entities and their behaviour can be interpreted, and comprehensively described, through the structure of scientific theories. How to specifically construct and justify a theory is then a subject on its own, that evades the scope of this thesis. However, one might argue that the evolution of a theory runs on various, but interconnected, paths.  Like parallel trains with some intersecting stations, a physical model needs to be founded both on a mathematical {\it framework} and on {\it numbers}. The former can be conceived as an ensemble of concepts, rules and formulas that allow for the logical coherence of the theory itself. The latter is the outcome of measurements taken directly from natural processes through experiments. The best theory is then the one that explains the largest possible number of phenomena, and allows for making further predictions within the given framework. 
This cycle of predictions and experiments is the basis of every modern Science: as a lichen, an organism built on the mutual relation between algae and fungi, experiments and theory feed each other in a never ending circle. 

Nowadays, after various scientific revolutions, we face a period of {\it corroboration} of ideas and models developed by the joint efforts of scientists around the world. Despite the ubiquitous presence of unexplained problems in science, the more the current theories are not {\it falsified} by subsequent experiments, the more they consolidate as common knowledge. This is the case of the best up-to-date model describing gravitation:  General Relativity (GR). Its incredible predicting power found its climax in the first detections of gravitational waves (GWs) passing through the Earth, produced by coalescing black holes (BHs)~\cite{Abbott:2016blz,Abbott:2016nmj,Abbott:2017vtc,Abbott:2017oio,Abbott:2020tfl}. These events demonstrate once again the great heritage that Albert Einstein left us since the beginning of the last century. 

In GR, GWs are perturbations of the gravitational field that propagate at the speed of light. GWs are commonly produced by accelerated masses in a system with a non-zero degree of spherical/rotational asymmetry, whose characteristics (mass, acceleration, orbit, etc.) shape the typical frequency and amplitude of the emitted waves. Among others, typical sources of GWs are binary systems in orbit, non-spherical supernovae, etc.. Thanks to the GWs capability to distort the spacetime at their passage, laser interferometers observatories, like LIGO and Virgo, were able to detect GWs. In simple words, the detection consisted in the observation of the deformed path that light experiences when a GW is passing through. In this Introduction we will not display a thorough analysis of the properties of a GW, however, we refer the reader to Chapter\til\ref{chapter:CLAP} for a brief review of the various part in which a GW can be decomposed and on some detail about how to obtain GWs from linearized Einstein's equations.

Nevertheless, one may wonder why {\it challenge} the best model we have to describe why ``things fall''. 
\begin{quote}
{\it ``Whenever a theory appears to you as the only possible one, take this as a sign that you have neither understood the theory nor the problem which it was intended to solve.''}\\
\hspace*{\fill} {``Objective Knowledge: An Evolutionary Approach''} (Oxford U. Press)
\end{quote}
This was Sir Karl Raimund Popper's view in 1972, regarding the continuous fight over the acceptance of a cemented scientific knowledge and its contrasts with potential new developments\til\cite{Popper1962-POPCAR,Popper1972-POPOKA}; it magnificently summarizes one of the aspects of a modern view of the scientific method, initially developed by Galileo Galilei at the beginning of the 17th century. In the words of the Austrian-British philosopher one may also find the deeper roots behind the work shown in this manuscript. The continuous effort in trying to falsify GR will inevitably bring more knowledge, expanding its validity and eventually leading to the comprehension of the phenomena not fully captured today within the general relativistic framework  (even at that point though, it is worth stressing that the scientific endeavour is a never-ending one!).

Shifting the focus now on {\it how to} challenge GR, it is worth to pinpoint some of its actual open problems. Among others, one may find the lack of a profound understanding of the nature of singularities~\cite{Penrose:1964wq,Penrose:1969,Cardoso:2017soq,Cardoso:2017cqb}, and the origin of dark energy or dark matter~\cite{Weinberg:1988cp,Bertone:2018krk}. These (yet) unresolved issues clearly show that there is still room for possible extensions or modifications of Einstein's theory. Furthermore, they may serve as a guide for deeper scientific investigations.

The above-mentioned direct detections of GWs act as one of the possible stations where purely theoretical studies interconnect with experiments, offering novel information from uncharted energy scales and {\it spacetime} curvatures. In other words, the LIGO/Virgo observations of GWs produced by BHs and massive stars provided the first insights on regimes where dynamical gravitational interactions dominate over the other known fundamental forces -- also known as the {\it strong gravity regime}.  Furthermore, they posed certain constraints on GR~\cite{TheLIGOScientific:2016src,Barack:2018yly,Bird:2016dcv,Cornish:2017jml,Ezquiaga:2017ekz,Creminelli:2017sry,Annala:2017llu}, and on some modified gravity models~\cite{Yunes:2016jcc,Baker:2017hug,Sakstein:2017xjx,Cardoso:2019rvt}. These exciting discoveries can be seen as the first step on a long road to a new understanding of the gravitational universe. Having the great possibility of measuring waves coming directly from such extreme environments, it is even more essential to try and challenge theories of gravitation. Like a magnifying glass, GWs may shed light on fundamental open questions to which we would otherwise be blind; boosting our understanding on the nature of the gravitational interaction will help assess the foregoing deeper problems in the theory.

Th strong gravity regime is the lowest common denominator throughout this thesis. Being far away from the borders established by the capability of any human-made laboratory, it assumes a crucial role in testing GR. The usual stage where such configuration occurs is in the proximity of {\it compact objects}, like BHs or neutron stars (NSs). Being among the sources that generate detectable GWs, these astrophysical bodies provide the best environments to test gravity, and might help sharpen our knowledge on the entire Universe.

In the next Sections, one will find the underlying motivations behind the work developed in each Chapter in more detail. The following three paragraphs highlight the state-of-the-art of gravitational physics, describing several interconnected directions to challenge theories of gravitation.

\section*{\large Precision gravitational wave astronomy}

The advent of third generation detectors~\cite{Hild:2010id,Punturo:2010zza,Maggiore:2019uih} and the space-based LISA mission~\cite{Audley:2017drz} (together with the planned Tajii program\til\cite{10.1093/nsr/nwx116}) will increase the number and accuracy of GW observations, starting a new, {\it precision} gravitational wave astronomy era. With high quality data and low instrumental noise, GWs from massive and distant compact objects will provide a statistical and systematic vision of the objects populating the cosmos, giving access to virtually all its visible parts~\cite{Maggiore:2019uih,Audley:2017drz}. This new opening on the universe motivates part of the work in this thesis: an increasing ability to probe the properties of compact objects will help us in studying and improving models that describe the gravitational interaction and matter fields. Hence, in the next Chapters we will present various toy models that try to challenge GR and that, in some cases, might be constrained by future GWs observations.

Nonetheless, this new precise GW astronomy era will already answer some of the fundamental open questions in the field. Among others, the observation of inspiralling compact objects will determine their mass and spin to ashtonishing levels of accuracy by astronomy standards~\cite{Berti:2004bd,AmaroSeoane:2007aw} and will impose strong constraints on non-trivial radiation channels~\cite{Barausse:2016eii,Cardoso:2016olt,Arvanitaki:2016qwi,Brito:2017zvb}. Precise measurements of the gravitational waveform may reveal whether the objects have non-zero tidal Love numbers (i.e. parameters that indicate the rigidity of a body),
potentially discriminating BHs from other hypothetic compact objects~\cite{Cardoso:2017cfl,Sennett:2017etc,Cardoso:2017njb,Cardoso:2017cqb}. Accurate observations will also test long-held beliefs about how matter behaves in curved spacetime. As an example, the consequences of non-trivially embedding the Maxwell field in highly curvature spacetimes are shown in Chapter\til\ref{chapter:spontaneous_vectorization}. In addition, astronomical measurements on binary pulsar systems~\cite{Lange:2001rn,Antoniadis:2013pzd}, together with observations of GWs emitted by binaries containing NSs, such as GW170817~\cite{TheLIGOScientific:2017qsa}, have improved our knowledge about compact stars. Theoretical predictions about NS spacetimes and the equation of state of matter at such high densities will be compared with
observational data, improving our knowledge of non-vacuum extreme geometries. 

This enormous potential for new science requires the careful control of systematic factors. Environmental effects, such as accretion disks, nearby stars, electric or magnetic fields, a cosmological constant or even dark matter, can possibly blur what is otherwise a clear picture of compact binaries~\cite{Barausse:2007dy,Barausse:2007ph,Kocsis:2011dr,Yunes:2011ws,Macedo:2013qea,Barausse:2014tra,Barausse:2014pra}. Along these lines, in Chapter\til\ref{Scattering of gravitational wave} one can find a specific example of the effect of compact binaries on the propagation of GWs. Precision GW astronomy can also inform us on the nature and distribution of dark matter, providing information on the local dark matter density where the process is taking place~\cite{Eda:2013gg,Barausse:2014tra,Barausse:2014pra}. In fact, a non-trivial dark matter environment may change the inspiral of a compact binary, via accretion or dynamical friction. In addition, if dark matter consists of new fundamental light fields, then rotating BHs can become lighthouses of GWs~\cite{Brito:2015oca,Hui:2016ltb,Bertone:2018krk,Baibhav:2019rsa}. The main astrophysical consequences of such scenarios are extensively discussed in Part {\bf I}.

This new experimental window will also influence our perception of the most intriguing and simple astrophysical bodies of the Universe: BHs. A fundamental result of vacuum GR is that all isolated, stationary and
asymptotically flat BHs belong to the same family of solutions\til\cite{Israel:1967wq,Hawking:1971vc} -- the Kerr family~\cite{Kerr:1963ud} -- fully described by just two parameters, mass and angular momentum~\cite{Chrusciel:2012jk,Robinson:2004zz,Cardoso:2016ryw}. These instrinsic characteristics determine the relaxation mechanisms of BHs formed after the merger of two compact bodies. Hence, an accurate analysis of the final, ringdown phase of the GW signal will allow us to perform tests of the ``BH'' nature of the newly formed object~\cite{Cardoso:2016rao,Cardoso:2017njb,Cardoso:2017cqb}. Thus, to some extent, testing the Kerr nature of BHs means testing GR. Furthermore, foundational questions regarding these fascinating objects are associated with the presence of horizons. Particularly, some of these issues concern the breakdown of determinism associated with Cauchy horizons or the fate of singularities of the classical equations~\cite{Penrose:1964wq,Penrose:1969,Cardoso:2017soq,Cardoso:2017cqb,Cardoso:2019rvt}. While possible pathological behavior is conjectured to be hidden behind horizons, questions remain concerning the effect of quantum gravity on the near-horizon structure or even on horizons themselves: do horizons exist?  Are the objects we observe really BHs, or are they extreme (and exotic) compact objects (ECOs) which mimic the BH behaviour? GW astronomy can have an important role in this matter, by constraining the existence of ``echoes'' produced in the last stage of the formation of a BH mimicker, or assessing the tidal properties of the coalescing compact objects~\cite{Cardoso:2016rao,Cardoso:2016oxy,Abedi:2016hgu,Nielsen:2018lkf,Abedi:2018pst,Lo:2018sep,Tsang:2018uie,Uchikata:2019frs,Abbott:2020jks,Wang:2020ayy,Maselli:2017cmm,Agullo:2020hxe,Cardoso:2019rvt}. Chapter\til\ref{chapter:CLAP} is entirely dedicated to the generation of GWs from these compact sources.

\section*{\large Alternative theories}

The collection of unexplained problems in the universe, as the nature of dark energy and dark matter (DM) (see also below), makes theories which modify or expand GR important on their own\til\cite{Sotiriou:2007yd,Sotiriou:2008rp}. Furthermore, the lack of a consistent theory of quantum gravity, that might account, for example, for the aforementioned issues with the existence of singularities, makes the search for alternatives both timely and interesting. Conversely, the astonishing agreement between GR predictions and experimental observations has strongly constrained the plausible alternative frameworks. Hence, in the next paragraphs of this Section we focus on {\it which} might be suitable strategies to study alternative theories, that is, challenging GR. In view of this, one may distinguish two ways to tackle the study of alternative theories.

On one side, tests of GR and its alternatives are based on the capability to constrain the parameters of each theory with the highest precision. In order to be sensitive to such small deviations, one needs to compute the gravitational waveforms using the full framework of an alternative theory. To study GW generation in modified gravity, a thorough study of the properties of the theory is needed, namely, carrying out a spacetime decomposition (e.g. 3+1)\til\cite{Arnowitt:1962hi,Gourgoulhon:2007ue,alcubierre2008introduction,baumgarte2010numerical,shibata2015numerical}, understanding if the theory is well-posed and constructing physically motivated initial data. Notably, this program has been carried out for only a few theories~\cite{Salgado:2005hx,Salgado:2008xh,Berti:2013gfa,Shibata:2013pra,Torii:2008ru,Yoshino:2011qp,Witek:2020uzz,Julie:2020vov,East:2020hgw}. Then, one may perform the time evolution of relevant physical systems and obtain GWs solving the modified gravity evolution equations. Thus, having an expanded GWs catalogue, the network of GWs interferometers will allow for precision tests of the promising alternative, making the search for new physics also possible~\cite{Berti:2015itd,Barack:2018yly,Cardoso:2019rvt,Berti:2005ys,Berti:2016lat,Yang:2017zxs}. Additionally, some modified theory allows also for the presence of ECOs. As already mentioned, having accurate waveforms from such bodies would prove to be useful in the procedure of match filtering between the GW observations and a future, extended, GWs catalogue (see Chapter\til\ref{chapter:CLAP}). 

On the other side, tests of gravity comprise also {\it smoking guns} for new physics: unique predictions of an alternative theory. Thanks to such peculiar phenomena, one may therefore discriminate between GR and its competitors. In view of this, it is crucial to search for such distinctive mechanisms in the GW signals produced by compact bodies\til\cite{Damour:1993hw,Cardoso:2011xi,Berti:2015itd,Barausse:2020rsu,Brito:2015oca}. If a compact object in an alternative theory differs from its GR counterpart, it might give rise to valuable smoking guns for the modified model. Hence, given the important role that BHs possess in different theories of gravitation, it might be useful to introduce the no-hair conjecture and no-hair theorems now. With no-hair conjecture it is usually meant that BHs cannot be described by any number other than their mass, electric charge and angular momentum. This conjecture is directly motivated by the above-mentioned {\it uniqueness} theorems by Israel and Hawking\til\cite{Hawking:1971vc,Israel:1967wq} (see the Kerr hypothesis above). Any other charge that might describe a BH spacetime is called {\it hair}, and the no-hair conjecture can be summarized with the sentence: ``BHs have no-hair''\til\cite{Misner:1974qy,Ruffini:1971bza}. 

In alternative theories possessing extra scalar degree-of-freedom, as Brans-Dicke or Bergmann-Wagoner scalar-tensor gravity\til\cite{Brans:1961sx,Berti:2015itd}, the no-hair conjecture has been proved, giving rise to the so-called no-scalar hair theorems\til\cite{Hawking:1972qk,Bekenstein:1971hc,cmp/1103842741,PhysRevD.51.R6608,Sotiriou:2011dz,Sotiriou:2013qea,Sotiriou:2014pfa,PhysRev.164.1776}. These theorems state that, if the scalar is time-independent, BH solutions in scalar-tensor theories are the same as those in GR: the scalar must be trivial and the spacetime it is described by known BH solutions in GR (i.e. the Kerr metric). Overcoming these theorems is possible, if one relaxes one or more of its assumptions. Considering for instance an oscillating time-dependent complex scalar field in a Kerr spacetime may lead to BHs with non-trivial scalar hair\til\cite{Herdeiro:2014goa}. Furthermore, it is also possible to have BH solutions that differ from GR ones, but that are still described only by their mass and angular momentum. In this case, one commonly refers to hair of the {\it second kind}, that are not associated with extra charges, but are non-trivial functions of the BH's general relativistic parameters (e.g. the BH mass). In the context of more complicated alternative theories, with extra couplings between the scalar and the gravitational sector for example, one or more of the hypothesis of no-scalar hair theorems might fall. Once again, evading such no-go results might produce BH solutions with non-trivial hair. One example of those occurs in theories that allow BHs with non-trivial scalar charges through a mechanism called {\it spontaneous scalarization}\til\cite{Silva:2017uqg,Doneva:2017bvd,Witek:2018dmd,Silva:2018qhn,Minamitsuji:2018xde,Doneva:2019vuh,Fernandes:2019rez,Minamitsuji:2019iwp,Cunha:2019dwb,Andreou:2019ikc,Ikeda:2019okp}. One interesting aspect of such solutions, that motivates part of this thesis, resides in the possibility of having new and interesting dynamics in binary systems, when compact objects encompass new charges (or hair).

Historically, spontaneous scalarization arose in the framework of scalar-tensor theories, when a new fundamental scalar degree-of-freedom couples with gravity with strength parametrized by a coupling constant $\beta$ (the nature of the coupling itself might differ between alternative theories). For certain coupling strengths $\beta$ one finds static solutions in scalar-tensor theory with a trivial scalar, equivalent to those of GR, which are also stable. However, there are couplings for which a GR solution is unstable and triggers a ``tachyonic'' instability, leading to stars with non-trivial charges. These compact stars are said to be {\it scalarized}~\cite{Damour:1992we}. The possibility to ``awake'' a new fundamental field is a valuable smoking gun for these alternative theories, as it leads, for example, to dipolar emission of radiation. Due to its non-perturbative nature, spontaneous scalarization of NSs avoids the strong constraints set by solar system experiments, established in the regime where the gravitational forces are relatively close to the Newtonian ones\til\cite{Will:2014kxa,Ramazanoglu:2016kul}. Stringent constraints on spontaneous scalarization of NSs in massless scalar-tensor theories arise from pulsar timing~\cite{Antoniadis:2012vy,Berti:2015itd}. For massive scalars instead the constraints become weaker. In fact, in this case NS binary systems radiate only when each component of the binary is close enough to interact with the scalar field of the companion\til\cite{Ramazanoglu:2016kul}, becoming more important closer to the merger. If the field is too massive the scalar is never excited. Let us stress that the above constraints uses NSs systems. Binaries containing only BHs cannot be used to test dipolar scalar radiation in scalar-tensor gravity because of the no-hair theorem applying in those theories (see discussion above). However, a similar scalarization mechanism might also happen in vacuum BH spacetimes, when non-trivial coupling between the scalar and the curvature are present\til\cite{Silva:2017uqg,Doneva:2017bvd}. Hence, a search for a non-trivial scalar charges in BH binaries might also be performed in the near future. In this case, constraints on BH charges in modified gravity might arise directly from GW observations, when there will be the proper numerics to model GWs generated by BHs in alternative theories (see\til\cite{East:2020hgw} for instance).

Scalarization phenomena may occur also for vector, tensor and spinor fields\til\cite{Ramazanoglu:2017xbl,Doneva:2017duq,Annulli:2019fzq,Kase:2020yhw,Ramazanoglu:2019gbz,Ramazanoglu:2017yun,Ramazanoglu:2018hwk,Minamitsuji:2020hpl}. In theories including a vector field, like the Einstein-Maxwell theory,
a massless vector field is embedded in curved spacetime through the standard ``comma-goes-to-semicolon''
rule~\cite{1975pbrg.book.....L}, but there are endless other possibilities. Ultimately, it is up to the observations to
determine the appropriate description. For instance, in a simple and elegant extension proposed by Hellings-Nordtvedt~\cite{Hellings:1973zz}, a non-minimal coupling between the curvature and the vector field is introduced. The consequences on the structure of compact stars of such new coupling are discussed in Chapter\til\ref{chapter:spontaneous_vectorization} where spontaneously vectorized stars are shown in detail.

Furthermore, scalar instabilities may also happen in compact binary spacetimes. Hence, these mechanisms may provide distinctive observables during the inspiral of compact objects. As an example, Chapter\til\ref{chapter:EsGB} describes scalar perturbations of binaries in theories of gravity with a new fundamental scalar degree-of-freedom. The chosen model is Einstein-scalar-Gauss-Bonnet gravity (EsGB), which admits the Schwarzschild geometry as well as BHs with scalar hair as solutions. Accordingly, scalar perturbations grow unbounded around binary systems. This ``dynamical scalarization'' process is easier to trigger: it occurs at lower values of the coupling constant of the theory, compared to the corresponding process for isolated BHs. These results emphasize the importance of having waveforms for BH binaries in alternative theories, in order to consistently perform tests beyond GR.

\section*{\large Dark matter}

One of the greatest open problem in Physics regards the nature and properties of DM. The term DM commonly refers to an hypothetical form of matter that should account for roughly $27\%$ of the total mass–energy density present in the universe. The extraordinary role that DM played in shaping the cosmos, providing the proper condition for the formation of structures for instance, makes the search for its unknown character of rare importance. Furthermore, since this unknown form of matter interacts only gravitationally with the environment, gaining insight into DM physics offers a unique opportunity to test the fundamental laws of gravitation.

The theoretical existence of DM has a long history, that notably started with Lord Kelvin and Henri Poincaré\til\cite{Bertone:2016nfn}. However, the first and most important evidences for the existence of DM date back to the works of Zwicky in 1933\til\cite{Zwicky:1933gu} and Rubin and Ford in 1970\til\cite{Rubin:1970zza}. The former studied the dispersion velocities of the galaxies forming the Coma cluster. Assuming an average mass of $\sim 10^9 M_\odot$ per galaxy, Zwicky computed their average kinetic energy and their typical velocity dispersion. Then, estimating the cluster mass based only on the visible matter content, he concluded that the total mass was not enough to keep the cluster bound together; the presence of extra {\it dark} mass was necessary to match the observed data. The latter instead used the the rotation curve of galaxies to highlight the ubiquitous presence of DM. Specifically, computing the circular velocity profile of stars and gas as a function of their distance from the galactic center, they found a discrepancy between expected values (computed again only through visible masses) and the observed ones. Despite the lack of a model describing the DM composition, these seminal works firmly consolidated its existence as a necessary aspect of the universe.

Nowadays, the most accredited model for DM is the cold DM model (CDM), that requires non-relativistic velocities for its constituents. Through the last century, a number of candidates were proposed to explain the nature of CDM. Among others, one may find weakly interacting massive particles (WIMPs), massive compact halo objects (MACHOs), axions etc.\til\cite{Essig:2013lka}. Each of those successfully described at least some of the observational evidences required to be a DM candidate, as, for instance, accounting for the observed power spectrum of the cosmic microwave background (CMB)\til\cite{Planck:2015fie}, or the large-scale structure of the universe. Conversely, none of the current model containing massive particles, of both baryonic or non-baryonic origin, successfully describes what happens at the scale of a galaxy or less $\sim 1 $kpc, being in most cases inconsistent
with observations. For instance, the discrepancy between the number density of galaxies and the the predicted number density of DM haloes, the expected DM density cusps in the centers of galaxies, or the weakness of dynamical friction in dwarf galaxies are example of such small-scale issues\til\cite{Weinberg:2013aya}.

Supported by the observational evidence of the existence of the Higgs boson\til\cite{Aad:2012tfa} (the first scalar particle ever detected) and inspired by axion-like particles, in which massive scalars were introduced to solve the strong CP violation present in quantum chromodynamics (QCD)~\cite{Peccei:1977hh}, models of DM comprising an ultralight scalar field increased their popularity in the last decades~\cite{Robles:2012uy,Hui:2016ltb,Bar:2019bqz,Bar:2018acw,Desjacques:2019zhf,Davoudiasl:2019nlo}. The scalar mass in these models is of the order of $10^{-22} {\rm eV}$. The fundamental underlying reason for the choice of such small mass (compare to the usual scale of particle masses in the Standard Model) resides in having a typical de Broglie wavelength comparable with (sub-)galactic scales
\begin{equation}
\frac{\lambda_{\rm de Broglie}}{2\pi}\sim 1.9 \,{\rm kpc} \left(\frac{10^{-22}{\rm eV}}{\mu}\right)\left(\frac{10\, {\rm km\, s^{-1}}}{v}\right)\,.
\end{equation}
Remarkably, thanks to this typical length-scale, that gives them the name {\it fuzzy} DM models\til\cite{Hu:2000ke}, such DM configurations can explain the large-scale structure of the universe, as well as account for some of the above-mentioned small-scale open issues of particle-like CDM models~\cite{Matos:1999et,Suarez:2013iw,Li:2013nal,Hui:2016ltb}.

In Part {\bf I} of this thesis the possibility of having scalar particles as one of the constituent of DM is thoroughly analyzed. Using a theory of a scalar field in flat space as a starting point, it can be shown that localized time-independent solutions cannot exist~\cite{Derrick:1964ww}. This powerful result limits the ability of fundamental scalars to describe possible novel objects where the scalar is confined (see also the no-hair paragraph above). A promising way to circumvent such no-go result is to consider time-dependent fields.
Within this more general framework, it can be shown that BHs can stimulate the growth of structures in their vicinities~\cite{Herdeiro:2014goa,Brito:2015oca},
and that new self-gravitating solutions are possible. Such objects can describe dark stars which have so far gone undetected~\cite{Barack:2018yly,Cardoso:2019rvt,Giudice:2016zpa,Ellis:2017jgp}. Surprisingly, the simplest solution of this kind also seems to be a good description of structures we know to exist: dark matter cores in haloes. As argued above, these {\it fuzzy} DM models require ultralight bosonic fields, of which the axion is a prototypical example, or generalizations thereof, such as
axion-like particles~\cite{Jaeckel:2010ni,Essig:2013lka}, ubiquitous in string-inspired scenarios~\cite{Arvanitaki:2009fg,Acharya:2015zfk}. Remarkably, these boson condesates provide a natural alternative to the standard structure formation through DM seeds and to the cold DM paradigm. A similar, albeit much wider, phenomenology arises in models of ultralight vector fields, such as dark photons, that are also a generic prediction of string theory~\cite{Goodsell:2009xc}. 

The core of these scalar DM haloes is also called Newtonian boson star (NBS). The study of the dynamics of such objects is interesting for a number of reasons. As DM candidates, it is important to understand the stability of such configurations, and the way they interact with surrounding bodies (stars, BHs, etc)~\cite{Macedo:2013qea,Khlopov:1985}. For example, the mere {\it presence} of a star or planet might change the local DM density. The motion of a compact binary can, in principle, stir the surrounding DM to such an extent that a substantial emission of scalars takes place. When a star crosses one of these extended bosonic configurations, it may change its properties to the extent that the configuration simply collapses or disperses; in the eventuality that it settles down to a new configuration, it is important to understand the timescales involved. 

Understanding the behavior of DM when moving perturbers drift by, or when a binary inspirals within a DM medium, is fundamental for attempting to detect DM via GWs. In the presence of a non-trivial environment accretion, gravitational drag and the self-gravity of the medium all contribute to a small, but potentially observable, change of the GW phase~\cite{Eda:2013gg,Macedo:2013qea,Barausse:2014tra,Hannuksela:2018izj,Cardoso:2019rou,Baumann:2019ztm,Kavanagh:2020cfn}. Understanding the backreaction on the environment seems to be one crucial ingredient in this endeavour~\cite{Kavanagh:2020cfn}.

\section*{\large Plan of the thesis}

For the sake of clarity, the structure of this manuscript is summarized as follows.

Part {\bf I} is entirely dedicated to the study of a promising DM candidate: ultralight scalar fields. Chapters\til\ref{chapter:Theoretical_framework} and\til\ref{chapter:Newtonian_boson_stars} describe the theoretical framework needed to introduce and study NBSs and their excitations. The perturbative scheme to compute sourceless and sourced perturbations is outlined. The Quasi Normal Modes (QNMs) of such structures are also shown. Chapter\til\ref{chapter:external_perturbers_NBS} examines a number of cases in which perturbers are placed statically or in motion inside NBS. Scalar fluxes due to BHs oscillating or binaries orbiting inside NBSs are computed. Effects on the GWs generation are also discussed.

In Part {\bf II} various aspects of the GWs generation and propagation are considered. Chapter\til\ref{Scattering of gravitational wave} shows the effects of the scattering between a GW and an intersecting binary system. The cross section for such events is also computed for the first time. Results from this Chapter show that, given the current values of the population of compact objects in the universe, the GWs emitted from distant sources will not suffer modifications due to scattering processes with binaries during their propagation. In Chapter\til\ref{chapter:CLAP}, a generalization of the Close Limit Approximation (CLAP) is developed. Notably, approximate waveforms from head-on collision of ECOs are shown for the first time. An analysis of the QNMs of binary BHs is also provided.

Finally, Part {\bf III} investigates instability mechanisms in alternative theories. These processes can be treated both at a linear and a non-linear level. Chapter\til\ref{chapter:EsGB} focuses on linear scalar instabilities in EsGB, using the generalized CLAP formalism developed in Chapter\til\ref{chapter:CLAP}. Results therein highlight the fundamental role of the non-trivial interactions during the collision of compact objects. Chapter\til\ref{chapter:spontaneous_vectorization} deals instead with instabilities arising from a simple non-minimally coupled vector-tensor theory. A linear analysis shows the onset of such unstable processes, while a non-linear analysis delineates the outcome of the instability: vectorized NSs.

Unless otherwise stated, geometrized units, where $G=c=1$, are used (i.e. energy
and time have units of length). A $\left(-,+,+,+\right)$ convention for the spacetime metric is also employed.

	    \cleardoublepage
\epigraphhead[450]{{\it This part is based on Refs.\til\cite{Annulli:2020lyc,Annulli:2020ilw}}}
\part{Ultralight fields and dark matter}

  \cleardoublepage

\chapter{Theoretical framework}\label{chapter:Theoretical_framework}
\minitoc
In this first Chapter of Part {\bf I} we start by describing the framework of theories that lead to scalar-made objects.

We consider two different theories of scalar fields, yielding localized objects with a static energy-density profile, but with a time-periodic scalar. The first theory describes a self-gravitating massive scalar, and the resulting objects are known as boson stars~\cite{Kaup:1968zz,Ruffini:1969qy,Liebling:2012fv}. In the limit of weak gravitational fields, these bodies reduce to the so-called Newtonian boson stars. These are objects made of very light fields (in particular, bosons with a mass~$\sim 10^{-22}\,{\rm eV}$), and describe very well most cores of DM haloes\til\cite{Hui:2016ltb,Hu:2000ke,Amendola:2005ad,Hlozek:2014lca}.

The second theory describes a non-linearly-interacting scalar in flat space, yielding solutions known as Q-balls: non-topological solitons which arise in a large family of field theories admitting a conserved charge $Q$, associated with some continuous internal symmetry~\cite{Coleman:1985ki}. Q-balls are not particularly well motivated as a DM candidate, but serve as an additional example of a scalar configuration to which our formalism can be directly applied. For this reason, a detailed treatment of such objects is given in Appendix\til\ref{ScalarQ}.

After describing how to build scalar field structures, we focus on the computation of the physical quantities that describe the dynamics of small perturbations evolving on a scalar field configurations. For instance, being interested in evaluating the effects that a scalar environment might cause on binaries orbiting inside them, in Sec.\til\ref{section:The_fluxes} we illustrate how to compute the energy emitted at spatial infinity during such processes. This total flux of energy, in general, contains contribution both from the background and from the particles in motion into it. Hence, in Sec.\til\ref{sec:The perturber's fluxes} we show how to extract solely the energy lost by the perturbers during such interaction. This might be especially interesting in light of the computation of the dynamical friction effects that occurs in these environments\til\cite{Hui:2016ltb}.

\section{The theory}\label{section:The_theory}
We start considering a general $U(1)$-invariant, self-interacting, complex scalar field $\Phi(x^\mu)$ minimally coupled to gravity described by the action
\begin{equation}
\mathcal{S}\equiv \int d^4x \sqrt{-g}\left(\frac{R}{16 \pi}-\frac{1}{2} g^{\mu\nu}\partial_{\mu}\Phi\partial_{\nu}\Phi^*-\mathcal{U}\right)\,,\label{action}
\end{equation}
where $R$ is the Ricci scalar of the spacetime metric $g_{\mu \nu}$, $g\equiv \det(g_{\mu \nu})$ is the metric determinant, and $\mathcal{U}(|\Phi|^2)$ is a real-valued, $U(1)$-invariant, self-interaction potential. 
For a weak scalar field $|\Phi| \ll 1$, the self-interaction potential is $\mathcal{U}\sim \mu^2|\Phi|^2/2+{\cal O} (|\Phi|^4)$, where $\mu$ is the scalar field mass. 
By virtue of Noether's theorem, this theory admits the conserved current
\begin{equation}
j_\mu =- \frac{i}{2}\left(\Phi^*\partial_\mu \Phi-\Phi\partial_\mu \Phi^*\right)\,, \label{NoetherCurrent}
\end{equation}
and the associated conserved charge
\begin{equation} 
Q = -\int d^3x \sqrt{h} \,j_t\,,\label{NoetherCharge}
\end{equation}
where the last integration is performed over a spacelike hypersurface of constant time coordinate $t$, with $h\equiv \det(g_{\mu \nu})$ the determinant of the induced metric $h_{\mu \nu}=g_{\mu \nu}-\delta_\mu^0 \delta_\nu^0$. We shall interpret this charge as the number of bosonic particles in the system. 

The scalar field stress-energy tensor is given by
\begin{equation}\label{StressEnergy}
T^S_{\mu \nu }=\partial_{(\mu}\Phi^* \partial_{\nu)}\Phi-\frac{1}{2}g_{\mu \nu}\left(\partial_\alpha \Phi^* \partial^\alpha \Phi +2\mathcal{U}(|\Phi|^2) \right) \,,
\end{equation}
and its energy within some spatial region at an instant $t$ is
\begin{equation} \label{Energy}
E= \int d^3x \sqrt{h}\, T^S_{t t}\,.
\end{equation}
%

\section{The objects}\label{section:The_objects}
We are interested in spherically symmetric, time-periodic, localized solutions of the field equations. These will be describing, for example, new DM stars or the core of DM halos.
We take the following ansatz for the scalar in such a configuration,
\begin{equation} 
\Phi_0=\Psi_0(r)e^{-i\Omega t}\,,\label{BKG_ansatz}
\end{equation}
where $\Psi_0$ is a real-function satisfying 
\begin{equation}
\partial_r \Psi_0(0)=0 \text{ and } \lim_{r\to \infty} \Psi_0=0\,. 
\end{equation}

Our primary target are self-gravitating solutions. When gravity is included, a simple minimally coupled massive field is able to self-gravitate. Thus, we consider minimal boson stars -- self-gravitating configurations of scalar field in curved spacetime with a simple mass term potential
\begin{equation} 
\mathcal{U}_{\rm NBS}=\frac{\mu^2}{2}|\Phi|^2\,.\label{Potential_BS}
\end{equation}
In this manuscript we restrict to the Newtonian limit of these objects, where gravity is not very strong. However, many of the technical issues of dealing with NBS are present as well in a simple theory in Minkowski background.
Thus, we will also consider Q-balls~\cite{Coleman:1985ki}: objects made of a non-linearly-interacting scalar field in {\it flat space}. For these objects, we use the Minkowski spacetime metric $\eta_{\mu \nu}$ and restrict to the class of non-linear potentials 
\begin{equation} 
\mathcal{U}_{\rm Q}=\frac{\mu^2}{2}|\Phi|^2\left(1-\frac{|\Phi|^2}{\Phi_c^2}\right)^2\,,\label{Potential_Qball}
\end{equation}
where $\Phi_c$ is a real free parameter of the theory. See Appendix\til\ref{ScalarQ} for details about these objects.


What we are ultimately interested to is not in the objects {\it per se}, but rather on their dynamical response to external agents.
The response to external perturbers is taken into account, by linearizing against the spherically symmetric, stationary background,
\begin{equation} 
\Phi=\left[\Psi_0(r)+\delta \Psi(t,r,\theta, \varphi)\right] e^{-i \Omega t}\,,\label{Perturbation}
\end{equation}
with the assumption $|\delta \Psi|\ll 1$, where $\Psi_0$ is the radial profile of the unperturbed object, and $\theta$ and $\varphi$ are coordinates used to parametrize the 2-sphere. Then, the perturbation $\delta \Psi$ allows us to obtain all the physical quantities of interest, like the modes of vibration of the object, and the energy, linear and angular momenta radiated in a given process. This approach has a range of validity, $|\delta \Psi|\ll 1$, which can be controlled by selecting the perturber. As we show below, $\delta \Psi \propto m_p \mu$, where $m_p$ is the rest mass or a mass-related parameter of the external perturber. Since our results scale simply with $m_p$, it is always possible to find an external source whose induced dynamics always fall in our perturbative scheme.

For a generic point-like perturber, the stress-energy tensor is given by
\begin{equation}
T_p^{\mu \nu}=m_p \frac{u^\mu u^\nu}{u^0} \frac{\delta\left(r-r_p(t)\right)}{r^2}\frac{\delta\left(\theta-\theta_p(t)\right)}{\sin\theta} \delta\left(\varphi-\varphi_p(t)\right) \,,\label{Stress_energy_particle}
\end{equation}
where $u^\mu\equiv dx_p^\mu/d\tau$ is the perturber's 4-velocity and $x_p^\mu(t)=(t,r_p(t),\theta_p(t),\varphi_p(t))$ a parametrization of its worldline in spherical coordinates. 
\section{The gross fluxes}\label{section:The_fluxes}
The energy, linear and angular momenta contained in the radiated scalar can be obtained by computing the flux of certain currents through a 2-sphere at infinity. These currents are derived from the stress-energy tensor of the scalar (Eq.\til\eqref{StressEnergy}). A detailed computation of the above quantities is crucial to understand how small perturbations propagate inside and outside scalar structures. For instance, a massive BH moving inside a scalar environment might emit radiation in the form of scalar waves, because its mass can source scalar perturbations on the background (see Sec.\til\ref{subsection:External_perturbers}). To quantify such effects, we evaluate the flux of energy emitted during the BH motion, that reaches an observer at spatial infinity.

First, we decompose the fluctuations as
\begin{equation} 
\delta \Psi=\sum_{l,m}\int \frac{d \omega}{\sqrt{2 \pi} r} \left[Z_1^{\omega l m }Y_l^m e^{-i\omega t}+\left(Z_2^{\omega l m }\right)^*\left(Y_l^m\right)^* e^{i\omega t}\right]\,,\label{Decomposition}
\end{equation}
where $Y_l^m(\theta,\varphi)$ is the spherical harmonic function of degree $l$ and order $m$, and $Z_1(r)$ and $Z_2(r)$ are radial complex-functions. It should be noted that~$Z_1$ and~$Z_2$ are not linearly independent. In particular, for the setups considered in this work, we find~$Z_1(\omega,l,m;r)=(-1)^mZ_2(-\omega,l,-m;r)^*$. However, for generality, we do not impose any constraint on the relation between these functions at this stage. The decomposition in Eq.\til\ref{Decomposition} can be also written in the equivalent form
\begin{equation} 
\delta \Psi=\sum_{l,m}\int \frac{d \omega}{\sqrt{2 \pi} r}Y_l^m e^{-i \omega t} \left[Z_1(\omega,l,m;r)+(-1)^m Z_2(-\omega,l,-m;r)^*\right]\,.\label{Decompositionv2}
\end{equation}
Unless strictly needed, we omit the labels $\omega$, $l$ and $m$ in the functions $Z_1^{\omega l m}(r)$ and $Z_2^{\omega l m}(r)$ to simplify the notation.
For a source vanishing at spatial infinity, we will see that one has the asymptotic fields
\begin{align} 
Z_1(r \to \infty) &\sim Z_1^\infty e^{i \epsilon_1 \left(\sqrt{\left(\omega+\Omega\right)^2-\mu^2}\right) r}\nonumber\,,\\
Z_2(r \to \infty) &\sim Z_2^\infty e^{i \epsilon_2 \left(\sqrt{\left(\omega-\Omega\right)^2-\mu^2}\right)^* r}\,,\label{Asymptotics}
\end{align}
where $\epsilon_1\equiv \text{sign}(\omega+\Omega+ \mu)$ and $\epsilon_2\equiv \text{sign}(\omega-\Omega-\mu)$, and $Z_1^\infty$ and $Z_2^\infty$ are complex amplitudes which depend on the source. We choose the signs~$\epsilon_1$ and~$\epsilon_2$ to enforce the Sommerfeld radiation condition at large distances. By Sommerfeld condition we mean either: (i) outgoing group velocity for propagating frequencies; or, (ii) regularity for bounded frequencies.

Scalar field fluctuations cause a perturbation to its stress-energy tensor, which, at leading order and asymptotically, is given by
\begin{equation}
\delta T_{\mu \nu}^S(r\to \infty)\sim \partial_{(\mu}\delta \Phi^* \partial_{\nu)}\delta \Phi-\frac{1}{2}\eta_{\mu \nu}\left[\partial_\alpha \delta \Phi^* \partial^\alpha \delta \Phi +\mu^2|\delta \Phi|^2 \right] \,,
\end{equation}
with $\delta \Phi \equiv e^{-i \Omega t} \delta \Psi$.
Then, the outgoing flux of energy at an instant $t$ through a 2-sphere at infinity is given by
\be
\dot{E}^{\rm rad}= \lim_{r\to \infty} r^2 \int d\theta d\varphi \sin \theta\, \delta T_{r \mu}^S \xi_t^\mu  \,,\label{energy}
\end{equation}
with the timelike Killing vector field $\boldsymbol{\xi}_t= -\partial_t$.
Plugging the asymptotic fields~\eqref{Asymptotics} in the last expression, it is straightforward to show that the total energy radiated with frequency in the range between $\omega$ and $\omega+d\omega$ is
\begin{equation}
\frac{dE^{\rm rad}}{d\omega}=\left|\omega+\Omega\right| {\rm Re}\left[\sqrt{(\omega+\Omega)^2-\mu^2}\right]\sum_{l,m}\left|Z_1^{\infty}(\omega,l,m)+(-1)^m Z_2^{\infty}\left(-\omega,l,-m\right)^*\right|^2\,.\label{Energy_flux}
\end{equation}
In deriving the last expression we considered a process in which the small perturber interacts with the background configuration during a finite amount of time. In the case of a (eternal) periodic interaction (e.g., small particle orbiting the scalar configuration) the energy radiated is not finite. However, we can compute the average rate of energy emission in such processes, obtaining
\begin{equation} 
	\dot{E}^{\rm rad}=\int \frac{d\omega}{2\pi}\left|\omega+\Omega\right| {\rm Re}\left[\sqrt{(\omega+\Omega)^2-\mu^2}\right]\sum_{l,m}\left|Z_1^{\infty}(\omega,l,m)+(-1)^m Z_2^{\infty}\left(-\omega,l,-m\right)^*\right|^2\,.\label{Energy_flux_rate}
\end{equation}
The last expression must be used in a formal way, because, as we will see, the amplitudes~$Z_1^\infty$ and~$Z_2^\infty$ contain Dirac delta functions in frequency $\omega$. The correct way to proceed is to substitute the product of compatible delta functions by just one of them, and the incompatible by zero.~\footnote{It is easy to do a more rigorous derivation applying the formalism directly to a specific process. For generality, we let~\eqref{Energy_flux_rate} as it is.} 

The (outgoing) flux of linear momentum at instant $t$ is
\begin{equation} 
\dot{P}^{\rm rad}_i= \lim_{r\to \infty} r^2 \int  d\theta d\varphi \sin \theta\, \delta T_{r \mu}^S e_i^\mu\,,\label{momentum}
\end{equation}
with $i=\{x,y,z\}$ and where $\boldsymbol{e}_x$, $\boldsymbol{e}_y$, $\boldsymbol{e}_z$ are unit spacelike vectors in the $x$, $y$, $z$ directions, respectively. These are given by
\begin{align}
\boldsymbol{e}_x&= \sin \theta \cos \varphi \, \boldsymbol{e}_r+ \frac{\cos \theta \cos\varphi}{r}\, \boldsymbol{e}_\theta -\frac{\sin \varphi}{r \sin \theta}\, \boldsymbol{e}_\varphi\,, \nn \\
\boldsymbol{e}_y&= \sin \theta \sin \varphi \, \boldsymbol{e}_r+ \frac{\cos \theta \sin\varphi}{r}\, \boldsymbol{e}_\theta +\frac{\cos \varphi}{r \sin \theta}\, \boldsymbol{e}_\varphi\,, \nn \\
\boldsymbol{e}_z&= \cos \theta \, \boldsymbol{e}_r- \frac{\sin \theta}{r} \, \boldsymbol{e}_\theta \nn \,,
\end{align}
with $e_r^\mu=\delta_r^\mu$, $e_\theta^\mu=\delta_\theta^\mu$ and $e_\varphi^\mu=\delta_\varphi^\mu$ in spherical coordinates.
For an axially symmetric process there are only modes with azimuthal number $m=0$ composing the scalar field fluctuation~\eqref{Decomposition}. In that case, using the asymptotic fields~\eqref{Asymptotics}, we can show that the total linear momentum radiated along $z$ with frequency in the range between $\omega$ and $\omega+d\omega$ is
\begin{equation}
\frac{d P_z^{\rm rad}}{d \omega}=\sum_l\frac{2(l+1)\Theta\left[\left(\omega+\Omega\right)^2-\mu^2\right]\left|(\omega+\Omega)^2-\mu^2\right|}{\sqrt{(2l+1)(2l+3)}}\left[\Lambda_{11}(\omega,l)+2\Lambda_{12}(\omega,l)+\Lambda_{22}(\omega,l)\right]\,,\label{Momentum_flux}
\end{equation}
where $\Theta(x)$ is the Heaviside step function and we defined the functions
\begin{align*}
	\Lambda_{11}(\omega,l)&\equiv{\rm Re}\Big[Z_1^\infty(\omega,l,0)Z_1^\infty(\omega,l+1,0)^*\Big]\,,\\
	\Lambda_{12}(\omega,l)&\equiv{\rm Re}\Big[Z_1^\infty(\omega,l,0)Z_2^\infty(-\omega,l+1,0)\Big]\,,\\
	\Lambda_{22}(\omega,l)&\equiv{\rm Re}\Big[Z_2^\infty(-\omega,l+1,0)Z_2^\infty(-\omega,l,0)^*\Big]\,.
\end{align*}
Additionally, it is possible to show that no linear momentum is radiated along $x$ and $y$ in an axially symmetric process.

Finally, the outgoing flux of angular momentum along $z$ at instant $t$ is
\begin{equation}
\dot{L}^{\rm rad}_z = \lim_{r\to \infty} r^2 \int d\theta d\varphi \sin \theta\, \delta T_{r \mu}^S e_\varphi^\mu  \,,\label{angularm}
\end{equation}
with the spacelike Killing vector $\boldsymbol{e}_\varphi$. Plugging the asymptotic fields~\eqref{Asymptotics} in the last expression, it can be shown that the total angular momentum along $z$ radiated with frequency in the range between $\omega$ and $\omega+d\omega$ is 
\begin{equation}
\frac{d L_z^{\rm rad}}{d\omega}={\rm Re}\left[\sqrt{(\omega+\Omega)^2-\mu^2}\right] \sum_{l,m}m\left|Z_1^{\infty}(\omega,l,m)+(-1)^m Z_2^{\infty}\left(-\omega,l,-m\right)^*\right|^2\,.\label{AngularMomentum_flux}
\end{equation}
In the case of a periodic interaction, the angular momentum along~$z$ is radiated at a rate given by
\begin{equation}
	\dot{L}_z^{\rm rad}=\int \frac{d \omega}{2\pi}\,{\rm Re}\left[\sqrt{(\omega+\Omega)^2-\mu^2}\right] \sum_{l,m}m\left|Z_1^{\infty}(\omega,l,m)+(-1)^m Z_2^{\infty}\left(-\omega,l,-m\right)^*\right|^2\,.
\label{AngularMomentum_flux_rate}
\end{equation}

We can also compute how many scalar particles cross the 2-sphere at infinity per unit of time. This is obtained by
\begin{equation}
	\dot{Q}^{\rm rad}= \lim_{r\to \infty} r^2 \int d\theta d\varphi \sin \theta\, \delta j_{r}  \,,\label{numberf}
\end{equation}
with
\begin{equation}
	\delta j_r(r \to \infty)\sim {\rm Im}\left(\delta \Phi^*\partial_r \delta \Phi\right)\,,
\end{equation}
at leading order. Using the asymptotic fields~\eqref{Asymptotics}, we can show that the number of particles radiated in the range between~$\omega$ and~$\omega+d\omega$ is
\begin{equation}
\frac{d Q^{\rm rad}}{d \omega}=\epsilon_1 {\rm Re}\left[\sqrt{(\omega+\Omega)^2-\mu^2}\right]\sum_{l,m}\left|Z_1^{\infty}(\omega,l,m)+(-1)^m Z_2^{\infty}\left(-\omega,l,-m\right)^*\right|^2\,.\label{Particles_flux}
\end{equation}
This gives us a simple interpretation for expressions~\eqref{Energy_flux} and~\eqref{AngularMomentum_flux}. The spectral flux of energy is just the product between the spectral flux of particles and their individual energy $\Omega+\omega$; similarly, the spectral flux of angular momentum matches the number of particles radiated with azimuthal number~$m$ times their individual angular momentum -- which is also~$m$. For a periodic interaction, scalar particles are radiated at an average rate
\begin{equation}
\dot{Q}^{\rm rad}=\int \frac{d \omega}{2\pi}\,{\rm Re}\left[\sqrt{(\omega+\Omega)^2-\mu^2}\right] \sum_{l,m}\left|Z_1^{\infty}(\omega,l,m)+(-1)^m Z_2^{\infty}\left(-\omega,l,-m\right)^*\right|^2\,.
	\label{Particles_flux_rate}
\end{equation}
%
\section{The perturber's fluxes}
\label{sec:The perturber's fluxes}
One may wonder what is the relation between the radiated fluxes and the energy and momenta lost by the massive perturber ($E^{\rm lost}$,~$P_z^{\rm lost}$,~$L_z^{\rm lost}$). 
Noting that both the energy and momenta of the scalar configuration may change due to the interaction, by conservation of the total energy and momenta we know that
\begin{equation} \label{LossRad}
	E^{\rm lost}=\Delta E+E^{\rm rad},\,	P_z^{\rm lost}=\Delta P_z+P_z^{\rm rad},\,	L_z^{\rm lost}=\Delta L_z+L_z^{\rm rad}\,,	
\end{equation}
where $\Delta E$,~$\Delta P_z$ and~$\Delta L_z$ are the changes in the energy and momenta of the configuration.
So, if we have the radiated fluxes, determining the energy and momenta loss reduces to computing the change in the respective quantities of the scalar configuration.

In a perturbation scheme it is hard to aim at a direct calculation of these changes, because in general they include second order fluctuations of the scalar -- terms mixing~$\Phi_0$ with~$\delta^2 \Phi$; this does not concern the radiated fluxes, since~$\Phi_0$ is suppressed at infinity.
However, for certain setups we can compute indirectly the change in the configuration's energy~$\Delta E$. Let us see an example.
An object interacting with the scalar only through gravitation is described by a~$U(1)$-invariant action; so, Noether's theorem implies that
\begin{align}
	\nabla_ \mu \,\delta j^\mu=0\,, 
\end{align}
with
\begin{equation}\label{NoetherPert}
	\delta j^\mu= {\rm Im}\left(\delta \Phi^*\partial^\mu \delta \Phi+\Phi_0^* \partial^\mu \delta^2 \Phi+\delta^2 \Phi^* \partial^\mu \Phi_ 0  \right)\,.
\end{equation}
One may have noticed that we are neglecting the lower order perturbation
\begin{equation}
\delta j^\mu=\,{\rm Im}\left(\Phi_ 0^*\partial^\mu \delta \Phi+\delta \Phi^* \partial^\mu \Phi_0\right)\,,
\end{equation}
however, this current does not contribute to a change in the number of particles in the configuration~$\Delta Q$, because it is suppressed at large distances by the factor~$\Phi_0$ (and its derivatives). In~\eqref{NoetherPert} we are also omitting the terms involving only~$\Phi_0$, since it is easy to show that they are static and, so, do not contribute to~$\Delta Q$. Using the divergence theorem, we obtain that the number of particles is conserved,
\begin{equation} \label{DeltaQ}
	\Delta Q=-\int_{t=+\infty}d^3x \sqrt{h} \,\delta j_t+\int_{t=-\infty}d^3x \sqrt{h} \,\delta j_t=-Q^{\rm rad}\,, 
\end{equation}
which means that the number of particles lost by the configuration matches the number of radiated particles -- no scalar particles are created.
If, additionally, we can express the change in the configuration's mass in terms of the change in the number of particles -- as (we will show) it happens for NBS -- we are able to compute~$\Delta M$ from the number of radiated particles~$Q^{\rm rad}$; so, we obtain the energy loss of the perturber~$E^{\rm lost}$ using only radiated fluxes. The loss of momenta~$P_z^{\rm lost}$ and~$L_z^{\rm lost}$ can, then, be obtained through the energy-momenta relations; for example, a non-relativistic perturber moving along~$z$ satisfies
\begin{equation} 
E^{\rm lost}= \frac{\left(m_p v_{\rm i}\right)^2-\left(m_p v_{\rm i}-P_z^{\rm lost}\right)^2}{2 m_p}=P_z^{\rm lost} v_{\rm i}-\frac{(P_z^{\rm lost})^2}{2 m_p}\,, \label{EvsPloss}
\end{equation}
where~$v_{\rm i}$ is the initial velocity along~$z$.
Finally, we can compute the change in the scalar configuration momenta~$\Delta P_z$ and~$\Delta L_z$ using~\eqref{LossRad}. 

The conservation of the number of particles (\textit{i.e,} Noether's theorem) plays a key role in our scheme; it allows us to compute the change in the number of particles -- a quantity that involves the second order fluctuation~$\delta^2\Phi$ -- using only the first order fluctuation~$\delta \Phi$. When the perturber couples directly with the scalar via a scalar interaction that breaks the~$U(1)$ symmetry -- like the coupling in~\eqref{coupling_Qball} -- the number of scalar particles is not conserved; the perturber can create and absorb particles. In that case, our scheme fails and it is not obvious how to circumvent this issue to calculate of~$\Delta M$. 
In Section~\ref{section:Small_perturbations} we apply explicitly the scheme described above to compute the energy and momentum loss of an object perturbing an NBS (e.g. a plunging BH) from the radiation that reaches infinity.

\chapter{Newtonian boson stars}\label{chapter:Newtonian_boson_stars}
\minitoc

In the first part of this Chapter we show how to get self-gravitating objects in theories with a minimally coupled massive field (Eq.\til\eqref{action}), or even with higher order interactions, but taken at the Newtonian level. In the second part, perturbation theory techniques are applied to obtain, and solve, the dynamical perturbation equations induced by external perturbers in such scalar environments (e.g. orbiting binaries or plunging BHs).

The resulting NBSs have been studied for decades, either as BH mimickers, as toy models for more complicated exotica that could exist, or as realistic configurations that can describe DM~\cite{Kaup:1968zz,Ruffini:1969qy,Liebling:2012fv}. In the context of {\it fuzzy} DM models, structure formation is enhanced by haloes of condensed ultra scalar field
 ($\mu \sim 10^{-22}{\rm eV}$)\til\cite{Matos:2000ss}, whose core can be model with an NBS\til\cite{Guzman:2004wj}. Notably, the smallness of the scalar mass helps overcoming some of the relevant open problems appearing within massive particle-like cold DM models, as the absence of a DM cusp at the center of galaxies, or the unexplained small number of small mass DM haloes ($<10^{7}M_\odot$)\til\cite{Hu:2000ke}. Additionally, current cosmological observations regarding the large-scale structure of the universe remain unchanged within fuzzy DM models\til\cite{Matos:2008ag,Harko:2011jy}.

DM haloes are large objects whose compactness are typically orders of magnitude smaller than the ones of BHs or NSs. As an example, NBSs, that should account only for the core of such haloes, have compactness $M/R\sim 10^{-5}$ (in geometrized units). This compactnesses already indicate that these objects might not need a full GR formalism to be described. Furthermore, given their typical large mass ($M_{\rm halo}\sim 10^{10}M_\odot$), the motion of BHs or stars inside DM haloes can be easily treated through perturbation theory, since their mass ratio can be considered of the order $M_{\rm BH}/M_{\rm halo}\lesssim 10^{-4}$, assuming known supermassive BH masses\til\cite{Gillessen:2008qv}. Moreover, as it will be shown quantitatively in the next Section, the equations describing such scalar configurations possess a scale invariance that simplifies enormously the computation. In fact, once obtained the results for one specific configuration, we can rescale them to obtain results for the entire space of NBS configurations.

\section{Background configurations}\label{section:Background_configurations}
%
\begin{figure}	
\centering
\includegraphics[width=0.55\textwidth,keepaspectratio]{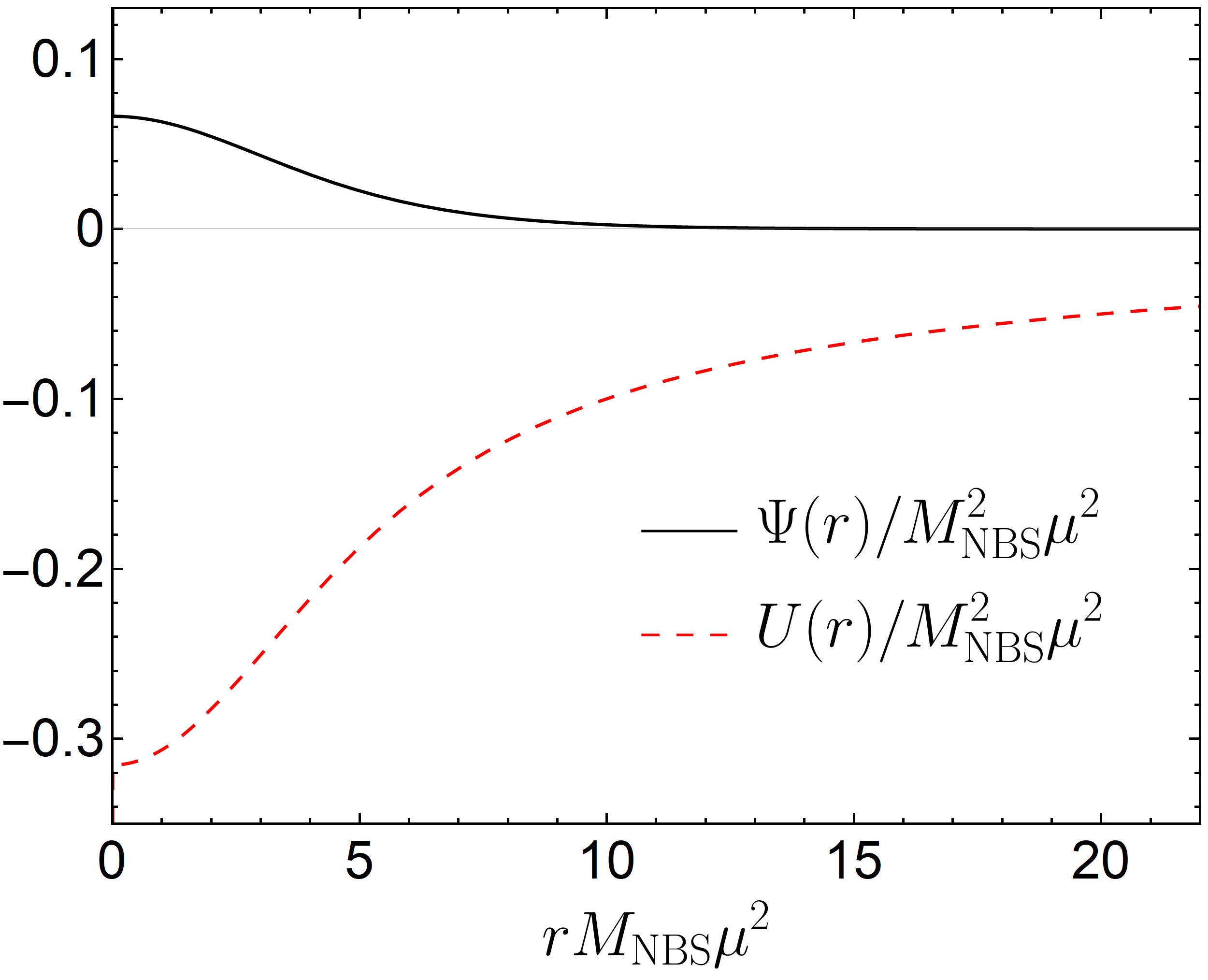} 
\caption[Universal radial profiles of an NBS.]{Universal radial profiles $\Psi(r)$ and $U(r)$ of the numerical solution of system~\eqref{EOM_BS_radial} with appropriate boundary conditions. 
Due to the scaling \eqref{eq:scaling}, this profile describes all the fundamental NBSs. They are characterized by the re-scaling invariant quantity
$\gamma/(M_{\rm NBS}^2\mu^3)\simeq0.162712$ and the mass-radius relation \eqref{eq:mass_radius_BS}.
}\label{fig:BS}
\end{figure}
The field equations for $\Phi$ and $g_{\mu \nu}$ are obtained through the variation of action~\eqref{action} with respect to $\Phi^*$ and $g_{\mu \nu}$, resulting in
\begin{align} 
\frac{1}{\sqrt{-g}}\partial_{\mu}\left(\sqrt{-g}g^{\mu\nu}\partial_{\nu}\Phi\right)&= \mu^2\Phi\,, \nn \\
R_{\mu\nu}- \frac{1}{2}R g_{\mu\nu}&= 8\pi T_{\mu\nu}^S\,.\label{EOM_BosonS}
\end{align}
Here, we are already using that $\mathcal{U}\sim \mu^2 \left| \Phi\right|^2/2$, since we want to consider a (Newtonian) weak scalar field $\left|\Phi\right| \ll 1$. The stress-energy tensor of the scalar $T_{\mu \nu}^S$ is given in Eq.~\eqref{StressEnergy}. We are interested in localized solutions of this model with a scalar field of the form~\eqref{BKG_ansatz}, with frequency 
\be
\Omega = \mu-\gamma\,.
\end{equation}
in the limit $0<\gamma\ll\mu$. In this case, the energy $\Omega$ of the individual scalar \textit{particles} forming the NBS is approximately given by their rest-mass energy $\mu$. In Appendix~\ref{app:Newtonian} we show that, using the Newtonian spacetime metric
\begin{equation}
ds^2=-\left(1+2U\right)dt^2+dr^2+r^2\left(d\theta^2+\sin^2\theta d\varphi^2\right)\,,
\end{equation}
with a weak gravitational potential $|U(r)|\ll1$, and retaining only the leading order terms, the system~\eqref{EOM_BosonS} reduces to the simpler system
\begin{align}
i \partial_t\widetilde{\Phi}&=-\frac{1}{2 \mu} \nabla^2 \widetilde{\Phi}+ \mu U \widetilde{\Phi}\,, \nn \\
\nabla^2 U&= 4\pi \mu |\widetilde{\Phi}|^2\,, 
\end{align}
where the Schrödinger field~$\widetilde{\Phi}$ is related with the Klein-Gordon field~$\Phi$ through 
\begin{equation}
\widetilde{\Phi}\equiv \sqrt{\mu} \, e^{i \mu t} \Phi\,.
\end{equation}
This is known as Schr\"{o}dinger-Poisson (SP) system (see, e.g., Ref.~\cite{Chavanis1}). 
To arrive at this description, we assume that the scalar field $\Phi$ is non-relativistic, which implies $|\partial_t\widetilde{\Phi}| \ll \mu |\widetilde{\Phi}|$.
Furthermore, using the ansatz~\eqref{BKG_ansatz} for the scalar field $\Phi$, we find
\begin{align}
\partial_r^2 \Psi+\frac{2}{r}\partial_r \Psi - 2 \mu \left(\mu U+\gamma\right) \Psi&=0\,, \nn \\
\partial_r^2 U +\frac{2}{r} \partial_r U- 4\pi \mu^2 \Psi^2&=0\,,\label{EOM_BS_radial}
\end{align}
with the constraints~$0<\gamma \ll \mu$,~$|U|\ll 1$ and~$|\Psi|\ll 1$. Remarkably, this system is left invariant under the transformation
\be
(\Psi,U,\gamma) \to \lambda^2 (\Psi, U, \gamma),\,  r \to r/\lambda\,.\label{eq:scaling}
\end{equation}
These relations imply that the NBS mass scales as $M_{\rm NBS} \to \lambda M_{\rm NBS}$ (see Eq.~\eqref{M_NBS}).
This scale invariance is extremely useful, because it allows us to effectively ignore the constraints on $\gamma$, $U$ and $\Psi$ when solving Eq.~\eqref{EOM_BS_radial}; one can always rescale the obtained solution with a sufficiently small $\lambda$, such that the constraints (i.e., the Newtonian approximation) are satisfied for the rescaled solution. Even more importantly is the fact that once a fundamental ground state NBS solution is found, all other fundamental stars can be obtained through a rescaling of that solution; obviously, the same applies to any other particular excited state.

A numerical solution of system~\eqref{EOM_BS_radial} describing all fundamental NBSs, is summarized in Fig.~\ref{fig:BS}. The appropriate boundary conditions on $\Phi$ are stated in Sec.~\ref{section:The_objects}, while here we also impose 
\begin{equation}
\partial_r U(0)=0 \text{ and } \lim_{r\to \infty} U=0\,.
\end{equation}
It is possible to show that, at large distances, the scalar decays exponentially as $\Psi \sim e^{-\sqrt{2\mu\gamma}r}/r$, whereas the Newtonian potential falls off as $-M_{\rm NBS}/r$.
Noting that the mass of an NBS is given by
\begin{equation} \label{M_NBS}
M_{\rm NBS}=4 \pi \mu^2 \int_{0}^{\infty}dr\, r^2 \left|\Psi \right|^2\,,
\end{equation}
a fundamental NBS satisfies,
\be
\frac{M_{\rm NBS}}{M_\odot} \simeq 3 \times 10^{12}\, \lambda\left(\frac{10^{-22}\, {\rm eV} }{ \mu}\right)\,,
\end{equation}
with a scaling parameter $\lambda$, such that $\{\Psi,U,\gamma/\mu\} \sim \mathcal{O}(\lambda^2)$. If one is interested in describing a DM core of mass $M \sim 10^{10} M_\odot$, this can be achieved then via a fundamental NBS made of self-gravitating scalar particles of mass $\mu \sim 10^{-22} \,{\rm eV}$, with a scaling parameter $\lambda \sim  10^{-2}$, which satisfies the Newtonian constraints. 

All the fundamental NBSs satisfy the scaling-invariant mass-radius relation
\be
M_{\rm NBS}\mu=\frac{9.1}{R\mu}\,,\label{eq:mass_radius_BS}
\end{equation}
where the NBS radius ($R$) is defined as the radius of the sphere enclosing $98\%$ of its mass. This result agrees well with previous
results in the literature~\cite{Liebling:2012fv,Boskovic:2018rub,Bar:2018acw,Membrado,Chavanis1,Chavanis2}. Comparing with some relevant scales, it can be written as
\be
\frac{M_{\rm NBS}}{M_{\odot}}=9\times 10^9\,\frac{100\, {\rm pc}}{R}\,\left(\frac{10^{-22}\,{\rm eV}}{\mu}\right)^2\,.\label{eq:mass_radius_BS2}
\end{equation}

Accurate fits for the profile of the scalar wavefunction are provided in Ref.~\cite{Kling:2017mif}. Unfortunately, these fits are defined by branches, and similar results for the gravitational potential are not discussed at length. We find that a good description of the gravitational potential of NBSs, accurate to within 1\% everywhere is the following:
\begin{align}
U&=\mu^2M_{\rm NBS}^2f\,,\\
f&=\frac{a_0+11\frac{a_0}{r_1}x+\sum_{i=2}^{9}a_ix^i-x^{10}}{(x+r_1)^{11}}\,,\\
x&=\mu^2M_{\rm NBS}r\,,r_1=1.288\,,\nonumber\\
a_0&=-5.132\,,a_2=-143.279\,,a_3=-645.326\,,\nonumber\\
a_4&=277.921\,, a_5=-2024.838\,,a_6=476.702\,,\nonumber\\
a_7&=-549.051,\, a_8=-90.244\,,a_9=-13.734\,.
\end{align}
The (cumbersome) functional form was chosen such that it yields the correct large-$r$ behavior,
and the correct regular behavior at the NBS center. For the scalar field, we find the following $1\%$-accurate expression inside the star,
\begin{align}
\Psi&=\mu^2 M_{\rm NBS}^2g\,,\\
g&=e^{-0.570459 x}\frac{\sum_{i=0}^{8}b_ix^i+b_fx^{9.6}}{(x+r_2)^{9}}\,,\\
x&=\mu^2M_{\rm NBS}r\,,r_2=1.182\,,\nonumber\\
b_0&=0.298\,,b_1=2.368\,,b_2=10.095\,,\nonumber\\
b_3&=12.552\,, b_4=51.469\,,b_5=-8.416\,,\nonumber\\
b_6&=54.141,\, b_7=-6.167\,,b_8=8.089\,,\nonumber\\
b_f&=0.310\,.
\end{align}

Finally, for future reference, the number of particles contained in an NBS is
\begin{equation}
	Q_{\rm NBS}=4 \pi \mu \int_0^\infty dr\, r^2\left|\Psi\right|^2\,,
\end{equation}
and, then, we can write the mass as~$M_ {\rm NBS}=\mu Q_{\rm NBS}$.

\section{Small perturbations} \label{section:Small_perturbations}
As shown in detail in Appendix~\ref{app:Newtonian}, small perturbations of the form~\eqref{Perturbation} to the scalar field, together with the NBS perturbed gravitational potential
\begin{equation}
U=U_0(r)+\delta U (t,r,\theta, \varphi)\,,
\end{equation}
satisfy the linearized system of equations
\begin{align}
&i \partial_{t} \delta \Psi=-\frac{1}{2 \mu}\nabla^2 \delta \Psi+ \left(\mu U_0 +\gamma\right) \delta \Psi+ \mu \Psi_0 \delta U\,,\label{Sourced_SP_System1}\\
&\nabla^2 \delta U=4 \pi\left[ \mu^2 \Psi_0 \left(\delta \Psi+\delta \Psi^*\right)+P\right]\,,\label{Sourced_SP_System2}
\end{align}
where $U_0$ is the gravitational potential of the unperturbed star, and we have included an external point-like perturber
\be
P\equiv m_p \frac{\delta \left(r-r_p(t)\right)}{r^2} \frac{\delta\left(\theta-\theta_p(t)\right)}{\sin \theta} \delta\left(\varphi-\varphi_p(t)\right)\,.\label{source_BS}
\end{equation}
The system\til\eqref{Sourced_SP_System1}-\eqref{Sourced_SP_System2} was obtained considering a non-relativistic external perturber. Note in fact that $P$ is just the non-relativistic limit of $T^p_{t t}$ given in~\eqref{Stress_energy_particle}.

This system of equations was derived for non-relativistic fluctuations, which satisfy $|\partial_t \delta \Psi|\ll \mu |\delta \Psi|$, and are sourced by a non-relativistic, Newtonian perturber. 
To study the sourceless case, we can simply set $m_p=0$.
As shown in detail in Appendix~\ref{app:Newtonian}, the perturber couples to the NBS through the total stress energy tensor entering Einstein's equation in~\eqref{EOM_BosonS}, which is taken to be the sum of the stress energy tensor of the scalar $T_{\mu \nu}^S$ (given in~Eq.\eqref{StressEnergy}) and of the perturber $T^p_{\mu \nu}$ (given in~Eq.\eqref{Stress_energy_particle}). We neglect the backreaction on the perturber's motion and treat its worldline as given.

Let us decompose the fluctuations of the scalar field as in~\eqref{Decomposition}, and the gravitational potential and the source, respectively, as~\footnote{Note that the perturbation $\delta U$ must be real-valued. Again, we will omit the labels $\omega$, $l$ and $m$ in the functions $u^{\omega l m }(r)$ and~$p^{\omega l m }(r)$ to simplify the notation.}
\begin{align}
\delta U&=\sum_{l,m}\int \frac{d \omega}{\sqrt{2 \pi} r} \left[u^{\omega l m }Y_l^m e^{-i\omega t}+\left(u^{\omega l m }\right)^*\left(Y_l^m\right)^* e^{i\omega t}\right]\,,\nn\\
\label{eq:dU_P_decompositions}
P&= \sum_{l,m} \int \frac{d \omega}{\sqrt{2 \pi} r} \left[p^{\omega l m} Y_l^m e^{-i \omega t}+\left(p^{\omega l m}\right)^* \left(Y_l^m\right)^* e^{i \omega t}\right]\,,
\end{align}
where~$p^{\omega l m}$ are radial complex-functions defined by
\begin{equation} 
	p^{\omega l m} \equiv \frac{r}{2 \sqrt{2 \pi}}\int dt d \theta d \varphi \sin \theta \,P  \left(Y_l^m\right)^* e^{i \omega t}\,.\label{p_def}
\end{equation}
Hence, using Eqs.\til\eqref{Decomposition}-\eqref{eq:dU_P_decompositions} in the system~\eqref{Sourced_SP_System1}-\eqref{Sourced_SP_System2}, we obtain the matrix equation
\begin{equation} 
	\partial_r \boldsymbol{X} -V_{\rm B}(r) \boldsymbol{X}= \boldsymbol{P}\,,\label{BS_Perturbation_Matrix_Sourced}
\end{equation}
with the vector $\boldsymbol{X}\equiv (Z_1, Z_2, u, \partial_r Z_1, \partial_r Z_2, \partial_r u)^T$, the matrix $V_B$ given by 
\be
\begin{pmatrix} 
0 & 0& 0 & 1 & 0 & 0 \\
0 & 0 & 0 & 0 & 1 & 0 \\
0 & 0 & 0 & 0 & 0 & 1 \\
V-2 \mu (\omega- \gamma) & 0 & 2 \mu^2\Psi_0 & 0 &0 & 0  \\
0 & V+2 \mu (\omega+ \gamma) & 2 \mu^2\Psi_0 & 0 & 0 & 0 \\
4\pi \mu^2 \Psi_0 & 4\pi \mu^2 \Psi_0 & V-2 \mu^2 U_0 & 0 & 0 & 0  
\end{pmatrix}\,.\nonumber
\end{equation}
with the radial potential defined as
\be
V(r)\equiv \frac{l(l+1)}{r^2}+2 \mu^2 U_0\,,
\end{equation}
and the source term
\begin{equation}
\boldsymbol{P}(r)\equiv \left(0,0,0,0,0,4\pi p\right)^T\,.
\end{equation}
Note that the condition of non-relativistic fluctuations translates, here, into the simple inequality $|\omega| \ll \mu$.

As suitable boundary conditions to solve for the fluctuations, we require both regularity at the origin,
\begin{equation}
\boldsymbol{X}(r \to 0)\sim\left(a r^{l+1},b r^{l+1},c r^{l+1},a (l+1)r^l,b (l+1)r^l,c (l+1)r^l\right)^T\,,
\end{equation}
with complex constants $a$, $b$ and $c$, and the Sommerfeld radiation condition at infinity,
\begin{equation}
\boldsymbol{X}(r \to \infty)
\sim\left(Z_1^\infty e^{i k_1 r} ,Z_2^\infty e^{i k_2 r},u^\infty, i k_1 Z_1^\infty e^{i k_1r},i k_2 Z_2^\infty e^{i k_2r},0\right)^T\,, \label{BC_sommerfeld_infinity}
\end{equation}
with 
\begin{align} \label{Wave_number_1}
k_1&\equiv\sqrt{2 \mu \left(\omega-\gamma\right)}\,, \\ \label{Wave_number_2}
k_2&\equiv-\left(\sqrt{-2 \mu \left(\omega+\gamma\right)}\right)^*\,.
\end{align}
In the last expression we are using the principal complex square root.

To calculate the fluctuations we will make use of the set of independent homogeneous solutions $\{\boldsymbol{Z_{(1)}},\boldsymbol{Z_{(2)}},\boldsymbol{Z_{(3)}},\boldsymbol{Z_{(4)}},\boldsymbol{Z_{(5)}},\boldsymbol{Z_{(6)}}\}$, uniquely determined by
\begin{align}
&\boldsymbol{Z_{(1)}}(r \to 0)\sim \Big(r^{l+1},0,0,(l+1)r^l,0,0\Big)^T\,,\nonumber \\
&\boldsymbol{Z_{(2)}}(r \to 0)\sim \Big(0,r^{l+1},0,0,(l+1)r^l,0\Big)^T\,,\nonumber \\
&\boldsymbol{Z_{(3)}}(r \to 0)\sim \Big(0,0,r^{l+1},0,0,(l+1)r^l\Big)^T\,,\nonumber \\
&\boldsymbol{Z_{(4)}}(r \to \infty)\sim \Big(e^{i k_1 r},0,0,i k_1 e^{i k_1 r},0,0\Big)^T\,,\nonumber \\
&\boldsymbol{Z_{(5)}}(r \to \infty)\sim \Big(0,e^{i k_2 r},0,0,i k_2 e^{i k_2 r},0\Big)^T\,, \nonumber \\
&\boldsymbol{Z_{(6)}}(r \to \infty)\sim \Big(0,0,u^\infty,0,0,0\Big)^T\,.
\label{BC_BS}
\end{align}
Then, the matrix
\be
F(r)\equiv\big(\boldsymbol{Z_{(1)}},\boldsymbol{Z_{(2)}},\boldsymbol{Z_{(3)}},\boldsymbol{Z_{(4)}},\boldsymbol{Z_{(5)}},\boldsymbol{Z_{(6)}}\big)
\label{eq:fundamental_matrix}
\end{equation}
is known as the fundamental matrix of the system~\eqref{BS_Perturbation_Matrix_Sourced}. Notably, the determinant of $F$ is independent of $r$.
\subsection*{The constancy of the fundamental matrix determinant}

Consider a first-order matrix ordinary differential equation
\begin{equation}\label{matrixsystem}
	\frac{d \boldsymbol{X}(r)}{dr} -V(r)\boldsymbol{X}(r)=0\,,
\end{equation}
with $\boldsymbol{X}$ a $N$-dimensional vector and $V$ a $N\times N$ matrix.
A fundamental matrix of this system is a matrix of the form $F(r)\equiv \big(\boldsymbol{X_{(1)}},...,\boldsymbol{X_{(N)}}\big)$, where $\{\boldsymbol{X_{(1)}},...,\boldsymbol{X_{(N)}}\}$ is a set of $N$ independent solutions of Eq.~\eqref{matrixsystem}. The determinant of this $N\times N$ matrix can be written as
\begin{equation*}
	\det F(r)=\epsilon^{i_1\,...\,i_N} X_{(i_1)}^1\,...\,X_{(i_N)}^N\,,
\end{equation*}
where $\epsilon$ is the Levi-Civita symbol, and $X_{(k)}^j$ is the $j$-th component of the vector $\boldsymbol{X_{(k)}}$. Using Eq.~\eqref{matrixsystem} it is easy to see that
\begin{equation}
	\frac{d }{dr}\det F=\sum_{k=1}^N \epsilon^{i_1\,...\,i_N}V^k_{\;\;\; j} \,X_{(i_1)}^1\,...\,X_{(i_k)}^j\,...\,X_{(i_N)}^N\,.
\end{equation}
Using the relation 
\begin{equation}
	\epsilon^{i_1\,...\,i_N} \,X_{(i_1)}^1\,...\,X_{(i_k)}^j\,...\,X_{(i_N)}^N=\delta_k^j \det F\,,
\end{equation}
we get
\begin{equation}
	\frac{d }{dr}\det F= {\rm Tr}(V) \det F\,.
\end{equation}
If the trace ${\rm Tr}(V)\equiv V^k_{\;\;\; k}$ is identically zero (which is always the case in this work), the determinant of the fundamental matrix is constant.

Finally, note that system~\eqref{BS_Perturbation_Matrix_Sourced} is invariant under the re-scaling
\begin{equation}
(U_0, \Psi_0,\gamma, \omega) \to \lambda^2 (U_0, \Psi_0, \gamma, \omega)\,,\quad r \to r/\lambda\,, \label{eq:scaling2}
\end{equation}
and, so, it can always be pushed into obeying the non-relativistic constraint. Additionally, for convenience, we impose that~$\delta \Psi$ and~$\delta U$ are left invariant by the re-scaling, by performing the extra transformation
\begin{align}
(Z_{1,2}, u)\to \lambda^{-3}(Z_{1,2}, u)\,, \quad m_p\to\lambda^{-1} m_p \,.
\end{align}

For a process happening during a finite amount of time the change in the NBS energy is, at leading order,
\begin{align} \label{DeltaE}
	\Delta E_{\rm NBS}&=-\int_{t=+\infty}d^3 x \sqrt{h}\, \delta T^S_{tt}+\int_{t=-\infty}d^3 x \sqrt{h}\, \delta T^S_{tt} \nn \\
	 &=\mu \Delta Q_{\rm NBS}\,,
\end{align}
since, at leading order,
\begin{align}
	\delta T^S_{tt}=\mu^2 \left(\left|\delta \Psi\right|^2+2\Psi_0{\rm Re}(\delta^2\Psi)\right)=\mu \,\delta j_t \,,
\end{align}
where~$\delta^2\Psi$ is a second order fluctuation of the scalar and we used~\eqref{NoetherPert} for the second equality.

\subsection{Validity of perturbation scheme}\label{subsection:Validity_of perturbation_scheme}
The perturbative scheme requires that $|\delta \Psi|\ll 1$, which can always be enforced by making $m_p$ as small as necessary.
On the other hand, the background construction neglects higher-order post-Newtonian (PN) contributions. A self-consistent perturbative expansion requires that such neglected terms (of order $\sim U_0^2$) do not affect the dynamics of small fluctuations (of order $\sim \delta U$). This imposes 
$m_p\gtrsim 10^4 M_{\odot}\,\left(\frac{M_{\rm NBS}}{10^{10}M_\odot}\right)^3\left(\frac{\mu}{10^{-22}\,{\rm eV}}\right)^2$,
which holds true for many systems of astrophysical interest. As shown in Appendix~\ref{app:Newtonian}, the scalar evolution equation \eqref{KG_all} is sourced by higher PN-order terms. However,
these are nearly static, or very low frequency terms, hence will make a negligible contribution for high-energy binaries or plunges. In other words, the previous constraint can be substantially relaxed in dynamical situations, such as the ones we focus on.
Finally, the Newtonian, non-relativistic approximation requires the source to have a small frequency $\lesssim 2\times 10^{-8}\,\left(\mu/10^{-22}{\rm eV}\right)\,{\rm Hz}$, in the case of a periodic motion. In Appendix~\ref{app:Newtonian} we show how to extend the formalism to include Newtonian but high frequency sources, and use it to calculate emission by a high frequency binary in Section~\ref{section:High-energy_binaries_within_boson_stars}. For plunges of nearly constant velocity $v$ piercing through an NBS, the Newtonian and non-relativistic approximation requires that $v\lesssim R\mu$. Fortunately, any NBS has $R \mu \gg 1$ and the latter condition is trivially verified.

\subsection{Sourceless perturbations}\label{subsection:Sourceless_perturbations}

Free oscillations of NBSs are fluctuations of the form
\begin{align}
\delta \Psi&= \frac{1}{\sqrt{2 \pi} r} \left[Z_1 Y_l^m e^{-i \omega t}+Z_2^*\left(Y_l^m\right)^* e^{i \omega^* t}\right]\,, \nonumber \\
\delta U&= \frac{1}{\sqrt{2 \pi} r} \left[u Y_l^m e^{-i \omega t}+u^*\left(Y_l^m\right)^* e^{i \omega^* t}\right]\,,
\end{align}
where $Z_1$, $Z_2$ and $u$ are regular solutions of system~\eqref{BS_Perturbation_Matrix_Sourced} with $P=0$, satisfying the Sommerfeld condition at infinity. These are also known as QNM solutions, and the corresponding frequency~$\omega$ is the QNM frequency.
Noting that the condition
\be
{\rm det}(F)=0\,,\label{eq:fundamental_matrix_condition}
\end{equation}
holds if and only if $\omega$ is a QNM frequency, we are able to find the NBS proper oscillation modes by solving the sourceless system~\eqref{BS_Perturbation_Matrix_Sourced}, and requiring at the same time that~\eqref{eq:fundamental_matrix_condition} is verified. These frequencies are shown in Table~\ref{table:QNM_BS_invariant}.

Additionally, notice that the sourceless system~\eqref{BS_Perturbation_Matrix_Sourced} admits also the trivial solution
\begin{align} \label{HS_trivial}
\delta \Psi_\epsilon&= \epsilon\, \Psi_0 (1+i \gamma t)\,, \nonumber \\
\delta U_\epsilon&=\epsilon\, U_0\,,	
\end{align}
with a constant $\epsilon\ll1$. This solution is valid only for a certain amount of time (while the perturbation scheme holds) and it corresponds just to an infinitesimal change of the background NBS (\textit{i.e,} an infinitesimal re-scaling of the original star) by a $\lambda=1+\epsilon/2$. This perturbation causes a static change in the number of particles in the star
\begin{equation}
	\delta Q_\epsilon= \frac{\epsilon}{2} Q_{\rm NBS}\,,
\end{equation}
and in its mass
\begin{equation}
	\delta M_\epsilon= \mu\, \delta Q_\epsilon=\frac{\epsilon}{2}  M_{\rm NBS}\,.
\end{equation}

\subsection{External perturbers}\label{subsection:External_perturbers}
In the presence of an external perturber, one needs to prescribe its motion through the source term~\eqref{source_BS}. A solution of the system~\eqref{BS_Perturbation_Matrix_Sourced} which is regular at the origin and satisfies the Sommerfeld condition at infinity can be obtained through the method of variation of parameters, and it reads
\begin{align}\label{Z1_of_r}
Z_1(r)&= 4\pi\Bigg[\sum_{n=1}^3 F_{1,n}(r) \int_\infty^r dr' F^{-1}_{n,6}(r') p(r') +\sum_{n=4}^6 F_{1,n}(r) \int_0^r dr' F^{-1}_{n,6}(r') p(r') \Bigg]\,,\\\label{Z2_of_r}
Z_2(r)&= 4\pi\Bigg[\sum_{n=1}^3 F_{2,n}(r) \int_\infty^r dr' F^{-1}_{n,6}(r') p(r') +\sum_{n=4}^6 F_{2,n}(r) \int_0^r dr' F^{-1}_{n,6}(r') p(r') \Bigg]\,,\\\label{u_of_r}
u(r)&= 4\pi\Bigg[\sum_{n=1}^3 F_{3,n}(r) \int_\infty^r dr' F^{-1}_{n,6}(r') p(r') +\sum_{n=4}^6 F_{3,n}(r) \int_0^r dr' F^{-1}_{n,6}(r') p(r') \Bigg]\,,
\end{align}
where $F_{i,j}$ is the $(i,j)$-component of the fundamental matrix defined in Eq.~\eqref{eq:fundamental_matrix}. To obtain the total energy, linear and angular momenta radiated during a given process, all we need are the amplitudes $Z_1^\infty$ and $Z_2^\infty$. These are given by
\begin{align}
Z_1^\infty&=4\pi \int_{0}^{\infty} dr' F^{-1}_{4,6}(r')p(r') \,, \label{Z1inf_BS}\\
Z_2^\infty&=4 \pi \int_{0}^{\infty} dr' F^{-1}_{5,6}(r')p(r') \,. \label{Z2inf_BS}
\end{align}
Let us now apply our framework to a few physically interesting external perturbers.
\subsection*{Plunging particle}
Consider a pointlike perturber plunging into an NBS. Without loss of generality, we can assume its motion to take place in the~$z$-axis, being described by the worldline~$x^\mu(t)=(t,0,0,z_p(t))$ in Cartesian coordinates.
Neglecting the backreaction of the fluctuations on the perturber's motion,
\begin{equation}
	\ddot{z}_p(t)=-\partial_z U_0(z_p)\,.
\end{equation}
We consider that the perturber crosses the NBS center at~$t=0$ (i.e.~$z_p(0)=0$) with velocity
\begin{equation}
\dot{z}_p(0)=-\sqrt{2\left(U_0(R)-U_0(0)\right)+v_R^2}\,,
\end{equation}
where~$v_R$ is the velocity with which the massive object enters the NBS; in other words, it is the velocity at~$r=R$.
In spherical coordinates the source reads
\begin{equation}
P=m_p \frac{\delta(\varphi)}{r^2 \sin \theta}\left[\delta\left(r-z_p(t)\right)\delta\left(\theta\right)+ \delta\left(r+z_p (t)\right)\delta\left(\theta-\pi\right)\right].\label{P_plunging}
\end{equation}

Here we do not want to be restricted to massive objects describing unbounded motions and, so, we consider also perturbers with small~$v_R$. These may not have sufficient energy to escape the NBS gravity, being doomed to remain in a bounded oscillatory motion (see Section~\ref{section:A_perturber_oscillating_at_the_center}). In these cases, we want to find the energy and momentum loss in one full crossing of the NBS and, so, we shall take the above source as "active" just during that time interval, vanishing whenever else.

Using Eq.~\eqref{p_def} the function~$p$ is
\begin{align}
	p=-\frac{m_p}{\sqrt{2 \pi}}Y_l^0(0) \delta_m^0 \frac{|t'_p(r)|}{r} \left(e^{-i \omega t_p(r)}+(-1)^l e^{i \omega t_p(r)}\right) \nn\,,
\end{align}
with~$t_p(r)\geq0$ defined by~$z_p\left[t_p(r)\right]=-r$.
This can be rewritten in the form
\begin{equation}
p=\frac{m_p}{\sqrt{2\pi}}Y_l^0(0)\, \delta_m^0\frac{|t_p'(r)|}{r}\left(\cos\left[\omega t_p(r)\right] \delta_l^\text{even}-i \sin\left[\omega t_p(r)\right] \delta_l^\text{odd}\right).
\end{equation}
The property 
\begin{equation}
p(\omega,l,0;r)=p(-\omega,l,0;r)^*\,,
\end{equation}
together with the form of system~\eqref{BS_Perturbation_Matrix_Sourced}, implies that
\begin{align}
Z_2(\omega,l,0;r)&=Z_1(-\omega,l,0;r)^*\,,\\
Z_2^\infty(\omega,l,0)&=Z_1^\infty(-\omega,l,0)^*\,.\label{property_Z1Z2_plunge_BS}
\end{align}
So, the spectral fluxes~\eqref{Particles_flux},~\eqref{Energy_flux},~\eqref{Momentum_flux} and~\eqref{AngularMomentum_flux} become, respectively,
\begin{align}
\frac{d Q^{\rm rad}}{d \omega}&=4\, {\rm Re}\left[\sqrt{2 \mu  (\omega-\gamma)}\right] \sum_l \left|Z_1^\infty(\omega,l,0)\right|^2\,,
\label{Particles_flux_BS}\\
\frac{d E^{\rm rad}}{d \omega}&= \left(\mu-\gamma+\omega\right) \frac{d Q^{\rm rad}}{d \omega}\simeq \mu \frac{d Q^{\rm rad}}{d \omega} \,,
\label{Energy_flux_BS}\\
\frac{d P_z^{\rm rad}}{d \omega}&=\sum_{l}\frac{16\mu(l+1)}{\sqrt{(2l+1)(2l+3)}}\, \Theta\left( \omega-\gamma\right)\left|\omega-\gamma\right|\text{Re}\left[Z_1^\infty(\omega,l,0) Z_1^\infty(\omega,l+1,0)^*\right]\,,\label{Momentum_flux_BS}
\end{align}
and
\be
\frac{d L_z^{\rm rad}}{d \omega}=0\,.
\end{equation}
These expressions were derived assuming a perturber in an unbounded motion. However, these are also good estimates to the energy and momenta radiated during one full crossing of the NBS by a bounded perturber, as long as its half-period is much larger than the NBS crossing time.

To compute how much energy is lost by the perturber, we need to know the change in the NBS energy~$\Delta E_{\rm NBS}$. At leading order, this is given by
\begin{equation}
	\Delta E_{\rm NBS}=\mu\, \Delta Q_{\rm NBS}=- \mu\, Q^{\rm rad}\,, 
\end{equation}
using Eq.~\eqref{DeltaE} in the first equality and~\eqref{DeltaQ} in the second.
Conservation of total energy-momenta, expressed through Eq.~\eqref{LossRad}, implies that the perturber loses the energy
\begin{align}
	E^{\rm lost}&= \Delta E_{\rm NBS}+E^{\rm rad}=\int d\omega\,(\omega- \gamma) \frac{d Q^{\rm rad}}{d \omega}\nn\\ 
	&=4\sqrt{2 \mu}\int d\omega\, {\rm Re}\left[(\omega-\gamma)^{\frac{3}{2}}\right] \sum_l \left|Z_1^\infty(\omega,l,0)\right|^2\,.
	\label{Energy_loss_BS}
\end{align}
The last expression should be understood as an order of magnitude estimate. If we had considered only the leading order contribution to~$\Delta E_{\rm NBS}$ and~$E^{\rm rad}$, we would have obtained~$E^{\rm lost}=0$. In the second equality we used higher order corrections to $E^{\rm rad}$ -- the factor~$(\omega-\gamma) \ll \mu$; but not to~$\Delta E_{\rm NBS}$. The corrections to~$\Delta E_{\rm NBS}$ may be of the same order of the corrections to~$E^{\rm rad}$ and should be included in a rigorous calculation of~$E^{\rm lost}$. We do not attempt that in this work. Interestingly, in our approximation the energy loss of the perturber matches the kinetic energy of the radiated scalar particles at infinity, as can be readily verified. The terms neglected should contain information about, for instance, the gravitational and kinetic energy of the radiated particles when they were in the unperturbed NBS. Still, we believe that Eq.~\eqref{Energy_loss_BS} is a good estimate of the order of magnitude of~$E^{\rm lost}$ and that it scales correctly with the boson star and perturber's mass,~$M_{\rm NBS}$ and~$m_p$, respectively.

For a small perturber~$m_p \mu \ll v_R$, its momentum and energy loss are related through (see Eq.~\eqref{EvsPloss})
\begin{align}
	P_z^{\rm lost}\simeq-\frac{E^{\rm lost}}{v_R}\,.
\end{align}
Using the full expression~\eqref{EvsPloss}, it is possible to see that if~$E^{\rm lost}\propto m_p^2$, then~$P^{\rm lost} \propto m_p^2$ in the limit $m_p \mu\ll v_R$. The~$E^{\rm lost}\propto m_p^2$ follows from~$Z_1^\infty \propto m_p$ (see Eq.~\eqref{Z1inf_BS}).

Conservation of total momentum, as expressed in~\eqref{LossRad}, implies that the NBS acquires a momentum 
\begin{align}
P_{\rm NBS}=P_z^{\rm lost}-P_z^{\rm rad}=-\frac{E^{\rm lost}}{v_R}-P_z^{\rm rad}\,.\label{eq:NBS_momentum}
\end{align}
In the last passage we did not included the kinetic energy associated with the momentum acquired by the boson star~$\Delta P_z$ inside~$\Delta E_{\rm NBS}$. The reason is that it is easy to check that it is subleading comparing with the correction of~$E^{\rm rad}$ considered.

\subsection*{Orbiting particles}
Consider an equal-mass binary, with each component having mass $m_p$, and describing a circular orbit of radius $r_{\rm orb}$ and angular frequency $\omega_{\rm orb}$ in the equatorial plane of an NBS. The source is modelled as
\begin{equation}
P=\frac{m_p}{r_{\rm orb}^2} \delta(r-r_{\rm orb})\delta\left(\theta-\frac{\pi}{2}\right)\left[\delta(\varphi-\omega_{\rm orb} t)+\delta(\varphi+\pi-\omega_{\rm orb} t)\right]\,. \label{P_orbiting}
\end{equation}
We are assuming that the center of mass of the binary is at the center of the NBS, but in principle our results extend to all binaries sufficiently deep inside the NBS.
Also, our methods can be applied to any binary as long as a suitable source~$P$ is given.

Using Eq.~\eqref{p_def} the source above yields
\begin{equation}
p=m_p\sqrt{\frac{\pi}{2}} \frac{Y_l^m\left(\pi/2,0\right)}{r_{\rm orb}}(1+(-1)^m)  \delta\left(r-r_{\rm orb}\right)\delta\left(\omega -m \omega_{\rm orb}\right)\,.\label{p_orbiting}\end{equation}
The perturber's motion is fully specified by a prescription relating $r_{\rm orb}$ and $\omega_{\rm orb}$; we consider Keplerian orbits
$r_{\rm orb}^3=M/\omega_{\rm orb}^2$, where $M=2m_p$ is the total mass. This setup describes either stellar-mass or supermassive BH binaries orbiting inside an NBS.
Alternatively, applying the transformation $m_p(1+(-1)^m)\to m_p$, we obtain a source that describes an extreme mass ratio inspiral (EMRI). This could be, for instance, a star of mass $m_p$ on a circular orbit around a central massive BH of mass $M_{\rm BH}$. In such case we consider the Keplerian prescription $r_{\rm orb}^3=M_{\rm BH}/\omega_{\rm orb}^2$. 

The symmetry
\be
p(\omega,l,m;r)=(-1)^m p(-\omega,l,m;r)^*\,,
\end{equation}
together with the form of system~\eqref{BS_Perturbation_Matrix_Sourced} implies
\begin{align}
Z_2(\omega,l,m;r)&=(-1)^m Z_1(-\omega,l,-m;r)^*\,,\\
Z_2^\infty(\omega,l,m)&=(-1)^m Z_1^\infty(-\omega,l,-m)^*\,.
\end{align}
These simplify the emission rate expressions~\eqref{Particles_flux_rate},~\eqref{Energy_flux_rate} and~\eqref{AngularMomentum_flux_rate}, yielding
\begin{align}
\dot{Q}^{\rm rad}&=\frac{2}{\pi} \int d\omega\, {\rm Re}\left[\sqrt{2 \mu \left(\omega-\gamma\right)}\right]\sum_{l,m} \left|Z_1^\infty(\omega,l,m)\right|^2  \,, \nn\\
\dot{E}^{\rm rad}&=\frac{2}{\pi} \int d\omega(\mu-\gamma+\omega) {\rm Re}\left[\sqrt{2 \mu \left(\omega-\gamma\right)}\right]\sum_{l,m} \left|Z_1^\infty(\omega,l,m)\right|^2  \,, \nn\\
\dot{L}_z^{\rm rad}&=\frac{2}{\pi} \int d\omega \, {\rm Re}\left[\sqrt{2 \mu \left(\omega-\gamma\right)}\right]\sum_{l,m} m \left|Z_1^\infty(\omega,l,m)\right|^2\,.
\end{align}
These can be written explicitly as
\begin{align}
\dot{Q}^{\rm rad}&=32 \pi \, \widetilde{p}^2 \sum_{l,m}{\rm Re}\left(\sqrt{2 \mu \left(m \omega_{\rm orb}-\gamma\right)}\right) \left|F_{4,6}^{-1}\left(m\omega_{\rm orb};\,r_{\rm orb}\right)\right|^2\,,\label{Particles_flux_orbitingBS2}\\
\dot{E}^{\rm rad}&=32 \pi \, \widetilde{p}^2 \sum_{l,m}{\rm Re}\left(\sqrt{2 \mu \left(m \omega_{\rm orb}-\gamma\right)}\right)(\mu-\gamma+m \omega_{\rm orb}) \left|F_{4,6}^{-1}\left(m\omega_{\rm orb};\,r_{\rm orb}\right)\right|^2\,,\label{Energy_flux_orbitingBS2}\\
\dot{L}_z^{\rm rad}&=32 \pi \, \widetilde{p}^2  \sum_{l,m}m {\rm Re}\left(\sqrt{2 \mu \left(m \omega_{\rm orb}-\gamma\right)}\right) \left|F_{4,6}^{-1}\left(m\omega_{\rm orb};\,r_{\rm orb}\right)\right|^2\,,\label{AngularMomentum_flux_orbitingBS}
\end{align}
where we defined
\begin{equation*}
	\widetilde{p}\equiv m_p \sqrt{\frac{\pi}{2}}\frac{Y_l^m(\pi/2,0)}{r_{\rm orb}}\left(1+(-1)^m\right)\,.
\end{equation*}
Equation~\eqref{Energy_flux_orbitingBS2} can be further simplified using
\begin{equation*}
	\mu-\gamma+m\omega_{orb}\simeq \mu\,,
\end{equation*}
since we are treating the scalar fluctuations as non-relativistic; that is only valid if~$\gamma \ll \mu$ and~$\omega_{\rm orb}\ll \mu$. Large azimuthal numbers~$m$ do not spoil the approximation, because the emission is strongly suppressed by~$F_{4,6}^{-1}$ in that limit.

Now we follow the same procedure that we applied in the previous section to a \textit{plunging particle}, to estimate the rate of energy loss of the binary. We start by computing, at leading order, the change in the NBS energy per unit of time:
\begin{equation}\label{eq:circular_flux_Erad}
	\dot{E}_{\rm NBS}=\mu \dot{Q}_{\rm NBS}=-\mu \dot{Q}^{\rm rad}\,,
\end{equation}
where we used Eq.~\eqref{DeltaE} in the first equality and~\eqref{DeltaQ} in the second.~\footnote{Equations~\eqref{DeltaQ} and~\eqref{DeltaE} are easy to adapt to changes happening during a finite amount of time~$\Delta t$. To get the rates of change we just need to divide these expressions by~$\Delta t$ and take the limit~$\Delta t\to 0$.}
Conservation of the total energy implies that the binary energy loss per unit of time is 
\begin{equation}\label{eq:E_loss_circular}
	\dot{E}^{\rm lost}=\dot{E}^{\rm rad}+ \dot{E}_ {\rm NBS}= 32 \pi \widetilde{p}^2 \sum_{l,m}\left(m \omega_{\rm orb}-\gamma\right){\rm Re}\left(\sqrt{2 \mu \left(m \omega_{\rm orb}-\gamma\right)}\right) \left|F_{4,6}^{-1}\left(m\omega_{\rm orb};\,r_{\rm orb}\right)\right|^2\,.
\end{equation}
Again, the last expression should be understood as an order of magnitude estimate~(the reason is discussed in the previous section where we considered a \textit{plunging particle}).

For a small perturber~$m_p\ll |\omega_{\rm orb}| r_{\rm orb}^2$, its angular momentum and energy loss are related through
\begin{equation}
	\dot{L}_z^{\rm lost}\simeq \frac{\dot{E}^{\rm lost}}{\omega_ {\rm orb}}\,.
\end{equation}
Conservation of total angular momentum, expressed through Eq.~\eqref{LossRad}, implies that per unit of time the NBS acquires the angular momentum
\begin{align}
	 \dot{L}_ {\rm NBS}=\dot{L}_z^{\rm lost}-\dot{L}_z^{\rm rad}=\frac{\dot{E}^{\rm lost}}{\omega_ {\rm orb}}-\dot{L}_z^{\rm rad}\,.
\end{align}
%

\section{Free oscillations}\label{section:Free_oscillations}
%
\begin{table}[th] 
\centering
	\begin{tabular}{c||c}
		\hline
		\hline
		$l$ &  \multicolumn{1}{c}{$\omega^{(n)}_{\rm QNM}/ \left(M_{\rm NBS}^2\mu^3\right)$} \\ 
		\hline
		\hline
		0 & $0.0682\;\,\,\,    0.121\;\,\,     0.138\;\,\,    0.146\;\,\,    0.151\;\,\,  0.154\;\,\, 0.159$\\
		1 & $\,\,0.111\,\,\;\,\,     0.134\;\,\,     0.144\;\,\,    0.149\;\,\,    0.153\;\,\,  0.157\;\,\, 0.162$\\
		2 & $\,\,0.106\,\,\;\,\,     0.131\;\,\,     0.143\;\,\,    0.149\;\,\,    0.153\;\,\,  0.156\;\,\, 0.161$\\
		\hline
		\hline
	\end{tabular} 
	\caption[Normal frequencies of an NBS.]{Normal frequencies of an NBS of mass $M_{\rm NBS}$ for the three lowest multipoles. For each multipole $l$ we show the fundamental mode ($n=0$) and the first five overtones.
	At large overtone number the modes cluster around $\gamma\simeq0.162712 M_{\rm NBS}^2\mu^3$. The first mode for $l=0$ agrees with that of Ref.~\cite{Guzman:2004wj} when properly normalized and with an ongoing fully relativistic analysis~\cite{Caio:2020comment}. The two lowest $l=0,\,1,\,2$ modes are in good agreement with a recent time-domain analysis~\cite{Guzman:2018bmo}.}
	\label{table:QNM_BS_invariant}
\end{table}
The characteristic, non-relativistic oscillations of NBSs are regular solutions of the system~\eqref{Sourced_SP_System1}-\eqref{Sourced_SP_System2}
satisfying Sommerfeld conditions~\eqref{BC_sommerfeld_infinity} at large distances. For each angular number $l$, there seems to be an infinite, discrete set of solutions which we label with an overtone index $n$, $\omega^n_{\rm QNM}$. The first few characteristic frequencies, normalized to the NBS mass, are shown in Table~\ref{table:QNM_BS_invariant}. 
They turn out to be all {\it normal mode} solutions, confined within the NBS. The characteristic frequencies are all purely real and cluster around $\gamma$.
We highlight the fact that the numbers in Table~\ref{table:QNM_BS_invariant} are universal, they hold for any NBS. The fundamental $l=0$ mode (the first entry in the Table)
had been computed previously~\cite{Guzman:2004wj}, and agrees with our calculation to excellent precision (after proper normalization). Our results are also in very good agreement with the
frequencies of the first two modes, obtained in a recent time-domain analysis~\cite{Guzman:2018bmo}.
Modes of relativistic stars have been considered in the literature~\cite{Yoshida:1994xi,Kojima:1991np,Macedo:2013jja,Macedo:2016wgh,GRITJHU}
and should smoothly go over to the numbers in Table~\ref{table:QNM_BS_invariant}. Note that modes of relativistic BSs are damped, due to couplings between the scalar and the metric and the possibility to lose energy via gravitational waves. Such damping -- which is small for the relevant polar fluctuations~\cite{Macedo:2013jja,Macedo:2016wgh,GRITJHU} -- should get smaller as one approaches the Newtonian regime, but a full characterization of the modes of boson stars is missing.
Our results show that NBSs are linearly mode stable; it would be interesting to have a formal proof,
perhaps following the methods of Ref.~\cite{Kimura:2018eiv,Kimura:2017uor}.
We point out that the stabilization of a perturbed boson star through the emission of scalar field -- known as \textit{gravitational cooling} -- has been studied previously~\cite{Seidel1994,Balakrishna:2006ru,Guzman:2006yc}.

  \cleardoublepage

\chapter{External bodies and dark matter haloes}
\label{chapter:external_perturbers_NBS}

\minitoc

\begin{figure}	
\centering
\includegraphics[width=0.5\textwidth,keepaspectratio]{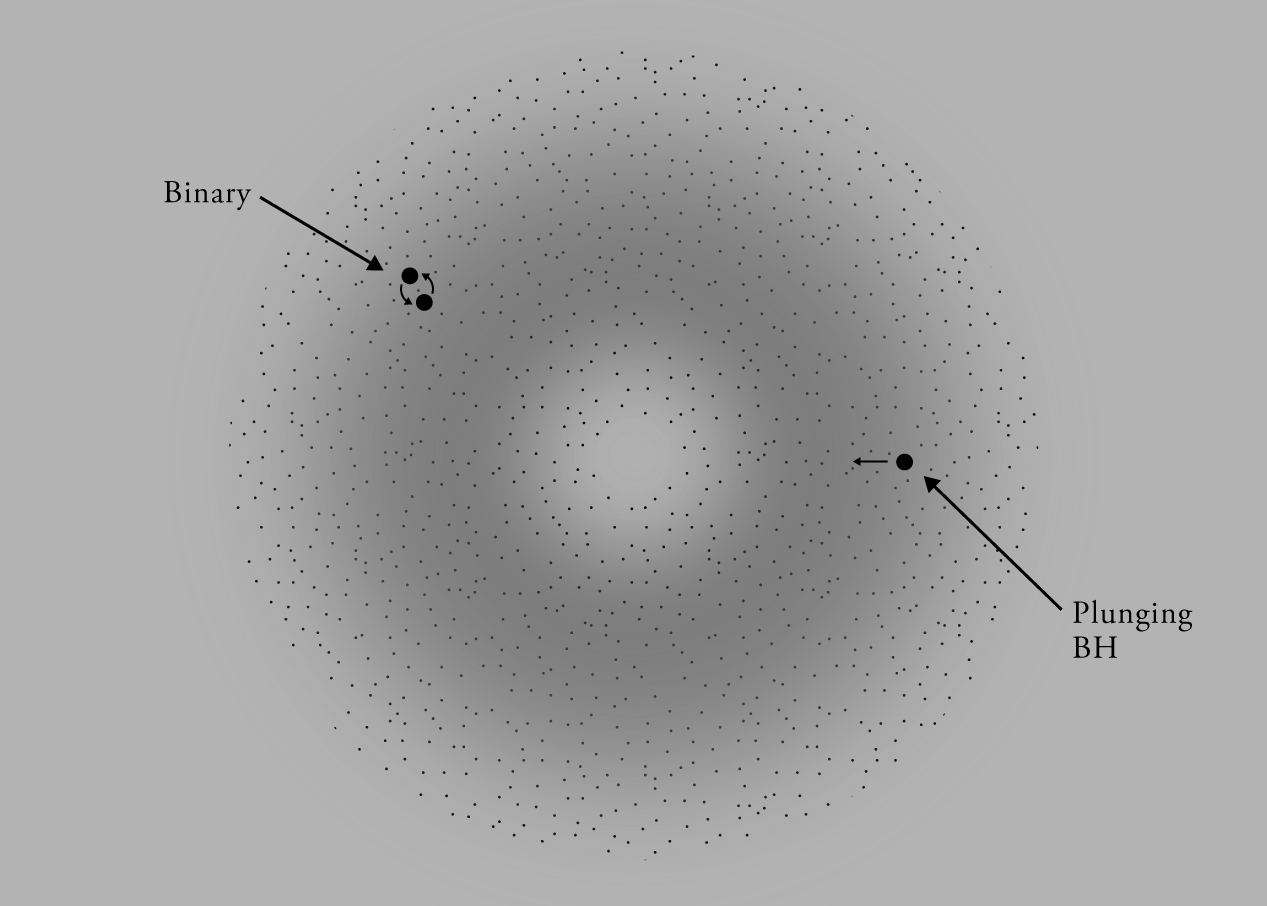} 
\caption[NBS and perturber setup.]{An equatorial slice of our setup, where a binary of two BHs or stars is orbiting inside an NBS, and a single BH is plunging through it.
Our formalism is able to accommodate both scenarios, and others.
The NBS scalar field is pictured in gray dots, and forms a large spherical configuration. The motion of the binary or of the plunging BH or star stirs the scalar profile, excites the NBS modes and may
eject some scalar field.}\label{fig:anatomy}
\end{figure}
Despite the intense study and the recent activity at the numerical relativity level~\cite{Cardoso:2016oxy,Helfer:2018vtq,Palenzuela:2017kcg,Sanchis-Gual:2019ljs,Bezares:2018qwa,Sanchis-Gual:2018oui,Widdicombe:2019woy}, the interaction of NBSs with smaller objects has hardly been studied. Notably, the variety and disparity of scales in the problem makes it ill-suited for full-blown numerical techniques, but ideal for perturbation theory. Hence, in this Chapter we investigate the response of localized scalar configurations to bodies moving in their vicinities. As an example, we will focus on stars and BHs piercing through or orbiting such DM environments.

The setup is depicted in Fig.~\ref{fig:anatomy}. The moving external bodies are modelled as point-like.
Such approximation is a standard and successful tool in BH perturbation theory~\cite{Zerilli:1971wd,Davis:1971gg,Barack:2018yvs}, in seismology~\cite{Ari} or in calculations of gravitational drag by fluids~\cite{Ostriker:1998fa,Vicente:2019ilr}. In this approximation we lose small-scale information. For light fields, the de Broglie wavelength is much larger than the size of stars, planets or BHs. Therefore we do not expect to lose important details of the physics at play. The extrapolation of our results to moving BHs or BH binaries should yield sensible answers.

A fundamental motivation towards the study of BHs in the galactic center lies in the fact that baryonic matter tends to slowly accumulate near the center of a DM structure, where it may eventually collapse, forming a massive BH. Then, gravitational collapse can impart a recoil velocity $v_{\rm recoil}$ to the BH of the order of $300\,{\rm km/s}$~\cite{1973ApJ...183..657B}, leaving the BH in an damped oscillatory motion through the DM halo, with respect to its center, with a crossing timescale
%
\begin{equation}
\tau_{\rm cross}=\sqrt{\frac{3\pi}{G\rho}}\sim 1.4\times 10^6 \,{\rm yr}\sqrt{\frac{10^3M_{\odot}\,{\rm pc}^{-3}}{\rho}}\,,
\end{equation}
and an amplitude
\begin{equation}
{\cal A}\sim 69\,{\rm pc} \sqrt{\frac{10^3M_{\odot}{\rm pc}^{-3}}{\rho}}\frac{v_{\rm recoil}}{300 \, {\rm km/s}}\,.
\end{equation}
The damping mechanism is due to dynamical friction caused by stars and DM. An intriguing result, shown later in Section\til\ref{section:A_perturber_oscillating_at_the_center}, suggests that the DM effects may be comparable to the one of stars in galactic cores. In fact, we will show how massive objects traveling through scalar media can deposit energy and momentum in the surrounding scalar field due to gravitational interaction~\cite{Hui:2016ltb,Bernard:2019nkv,Cardoso:2019dte}.

As already argued in the Introduction, another motivation to study such scalar configurations comes from the possibility of measuring GWs generated by distance sources. In fact, data from binaries in the central part of galaxies give us a unique opportunity to search for DM imprints on GW signals, through, for instance, a dephasing mechanism due to the non-vacuum environment. Hence, in the following we focus also on the scalar radiation emitted by equal mass and extreme-mass-ratio binaries and how this compares to the usual GW emitted. Quantifying the energy flowing at infinity through the scalar channel might help in understanding the detectability of such DM configurations.

The results obtained in this Chapter comes directly from the solution of the Einstein Klein-Gordon system, taken at the Newtonian level. Therefore, all the quantities presented here are evaluated considering scalars as (at least) one component of DM haloes. However, since BHs and stars interact with the DM field only gravitationally, some of the results presented in the following might be qualitatively applied also to other DM configurations, as the ones coming from a different choice of the scalar-self potential or if self-gravitating vectors are considered~\cite{Essig:2013lka}. To assess the above statement, one should perform a similar analysis, solving for the specific dynamical equations of the chosen model. This task goes beyond the scope of this thesis.
\section{A perturber sitting at the center}\label{section:A perturber_sitting_at_the_center}
%
\begin{figure}
\centering	\includegraphics[width=0.45\textwidth,keepaspectratio]{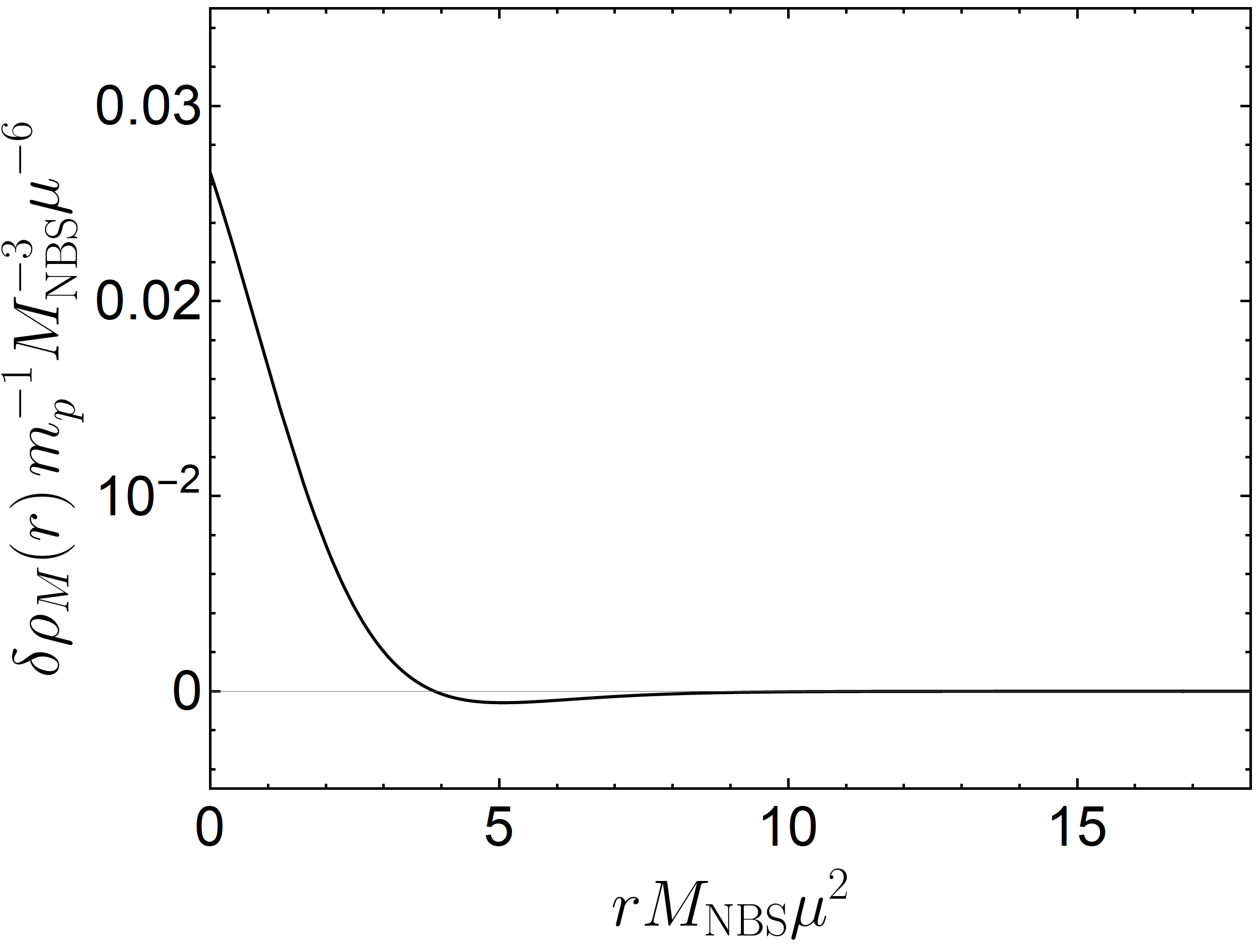}\,\,\,\includegraphics[width=0.45\textwidth,keepaspectratio]{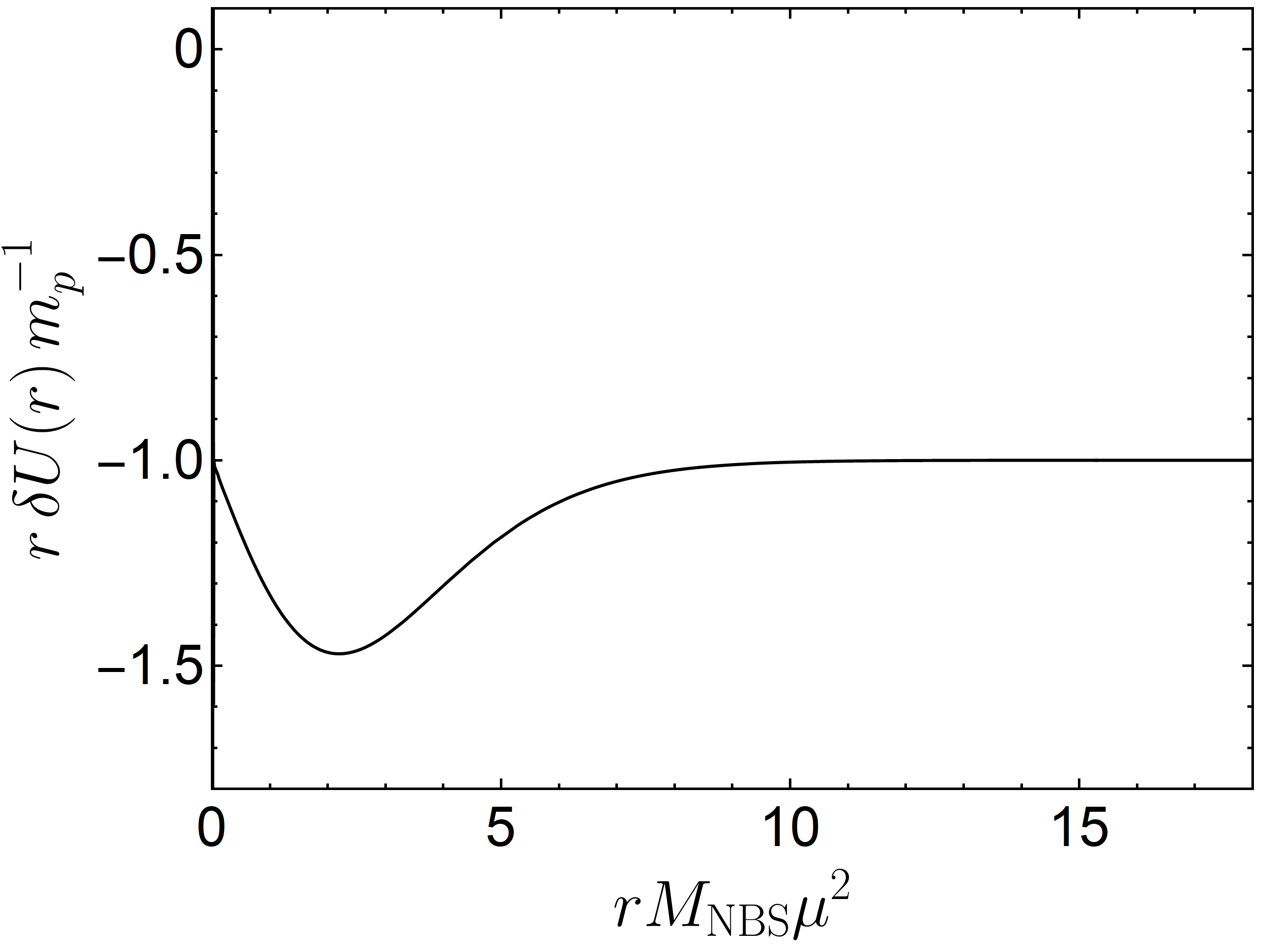}
	\caption[BH at NBS center: density and potential profiles.]{Universal perturbations induced a massive object, of mass $m_p$, sitting at the center of the scalar configuration. We assume that the perturber was brought adiabatically so that $\delta Q_{\rm NBS}=\delta M_{\rm NBS}=0$. {\bf Left panel:} perturbation in the mass density of the NBS obtained using Eq.~\eqref{deltarho}. 
{\bf Right panel:} perturbation in the gravitational potential $r \delta U= r \left(\delta U_p+\delta U_\epsilon\right)$. As expected, for large $r$, we recover the Coulombian potential $U=-m_p/r$.
}\label{fig:Particlesittingcenter}
\end{figure}
Perturbations induced by static objects located inside an NBSs, or inside a solitonic DM cores of light fields are interesting
in their own right. For perturbers localized far away, the induced tidal effects can dissipate energy
and lead to distinct signatures, both in GW signals and in the dynamics of objects close to such configurations~\cite{Mendes:2016vdr,Cardoso:2017cfl,Sennett:2017etc}.
We will not perform a general analysis of static tidal effects and will instead focus on perturbations due to a massive object at the center of an NBS.
Such object can be taken to be a supermassive BH or a neutron star, and the induced changes are important to understand how DM distribution
is affected by baryonic ``impurities.'' 

Consider then a BH or star, described by the source \eqref{source_BS}, and inducing static, spherically symmetric, real perturbations on the scalar field and gravitational potential, respectively, $\delta \Psi_p(r)$ and $\delta U_p(r)$. Then, Eqs.~\eqref{Sourced_SP_System1} and~\eqref{Sourced_SP_System2} become
\begin{align}
	\nabla^2\delta \Psi_p&= 2\mu \left(\mu U_0+\gamma\right)\delta \Psi_p+2\mu^2\Psi_0 \delta U_p\,,  \nonumber\\
	\nabla^2\delta U_p&= 4\pi \left(2\mu^2 \Psi_0\, \delta \Psi_p+P\right) \,. 
\end{align}
In the static source limit, it is straightforward to show that the matter moments are given by
\begin{equation}
p=\lim_{r_p \to 0}\,\frac{1}{2 \sqrt{2}} \frac{m_p}{r_p} \,\delta_l^0 \delta_m^0 \delta(\omega) \delta(r-r_p)  \,,
\end{equation}
for which the variation of parameters method described in Section\til\ref{subsection:External_perturbers}, implies that
\begin{align}
	\delta \Psi_p &=m_p \sum_{n=4}^{6} \frac{F_{1,n}(r)}{r}  \lim_{r_p \to 0} \left(\frac{F^{-1}_{n,6}(r_p)}{r_p}\right)\,, \nonumber \\
	\delta U_p &= m_p \sum_{n=4}^{6} \frac{F_{3,n}(r)}{r}  \lim_{r_p \to 0} \left(\frac{F^{-1}_{n,6}(r_p)}{r_p}\right)\,,
\end{align}
where the components of the fundamental matrix and its inverse are evaluated at $l=m=\omega=0$. Note that the change in the number of particles and mass of the NBS, respectively, $\delta Q_p$ and $\delta M_p$, is static, but non-zero in general. This is a consequence of the source being treated as if it was eternal. However, we know that if the perturber is brought in an adiabatic way to the center of the NBS there is no scalar radiation emitted, and, so, no change in the number of particles and mass of the star, $\delta Q_{\rm NBS}=\delta M_{\rm NBS}=0$. Fortunately, we are free to sum a trivial homogeneous solution~\eqref{HS_trivial} to enforce $\delta Q_{\rm NBS}=\delta M_{\rm NBS}=0$, while keeping $\delta \Psi=\delta \Psi_p+ \delta \Psi_\epsilon$ and $\delta U=\delta U_p+\delta U_\epsilon$ a solution of the inhomogeneous system. 
The perturbation induced in the density of particles is given by
\begin{equation}
	\delta \rho_Q=-\delta j_t=2\mu \Psi_0 \,{\rm Re}\left( \delta \Psi\right)=2\mu \Psi_0\left(\delta \Psi_p+\frac{\epsilon}{2} \Psi_0\right)\,, \nonumber
\end{equation}
and the one induced in the mass density by
\begin{equation} \label{deltarho}
\delta \rho_M=\delta T_{00}^S= \mu\, \delta \rho_Q= 2\mu^2 \Psi_0\left(\delta \Psi_p+\frac{\epsilon}{2} \Psi_0\right)\,, 
\end{equation}
where $j_t$ is the $t$-component of the Noether's current. The parameter $\epsilon$ associated with the trivial homogeneous solution must be chosen appropriately, so that
\begin{equation}
4\pi \int_{0}^{\infty} dr \,r^2 \delta \rho_Q= 4\pi \int_{0}^{\infty} dr\, r^2 \delta \rho_M=0\,.
\end{equation}

The perturbations in the mass density and gravitational potential of an NBS induced by a massive object sitting at its center are shown in Fig.~\ref{fig:Particlesittingcenter}.
Our results indicate that the particle attracts scalar field towards the center, where the gravitational potential corresponds solely to that of the point-like mass.
These results are consistent with those in Ref.~\cite{Bar:2018acw}. 
We find an insignificant change in the local DM mass density, when placing a point-like perturber at the center of an NBS; notice that $\delta \rho_M(0)/\rho_M(0)\sim 10\, m_p/M_{\rm NBS}$. Thus, a massive perturber will not enhance greatly the local DM density, which is smooth and flat for light scalars.  

On the other hand, studies with particle-like DM models find that its density close to supermassive BHs increases significantly~\cite{Gondolo:1999ef,Sadeghian:2013laa}.
This is in clear contrast to our results for light fields: a perturber does not significantly alter the local ambient density, since its size is much smaller than the scalar de Broglie wavelength.
Parenthetically, large overdensities seem to be in some tension with observations~\cite{Robles:2012uy}. Possible ways to ease the tension rely on scattering of DM by stars or BHs, or accretion by the central BH, induced by heating in its vicinities~\cite{Merritt:2002vj,Bertone:2005hw,Merritt:2003qk}.
These outcomes cannot possibly generalize to light scalars, at least not when the configuration is spherically symmetric, since there are no stationary 
BH configurations with scalar ``hair''~\cite{Hawking:1972qk,Bekenstein:1971hc,cmp/1103842741,PhysRevD.51.R6608,Sotiriou:2011dz,Herdeiro:2015waa,Cardoso:2016ryw}. But these results do prompt the questions: what happens to an NBS when a BH is placed at its center? What happens to the local scalar amplitude of an NBS when a binary is orbiting? We now turn to these issues.

\section{A black hole eating its host boson star}\label{section:A_black_hole_eating_its_host_boson_star}
As we noted, there are no stationary, spherically symmetric configurations when a non-spinning BH is placed at the center.
On long timescales, the entire NBS will be accreted into the BH, with a fraction dissipating to infinity.
This means, in particular, that our results cannot be extrapolated to when the point-like particle is a BH, and describe the system only at intermediate times. Hence, what {\it is} the lifetime of such a system, composed of a small BH sitting at the center of an NBS?
Unfortunately, most of the studies on BH growth and accretion assume a fluid-like environment~\cite{Giddings:2008gr}, an assumption that breaks down completely
here, since the de Broglie wavelength of the scalar is much larger than that of the BH. Exceptions to this rule exist~\cite{Clough:2019jpm,Hui:2019aqm}, but focus
on different aspects, and do not consider setups with the necessary difference in lengthscales.

The precise answer to this question requires full non-linear simulations in a challenging regime, with proper initial conditions.
However, in the limit we are interested in, where the BH, of mass $M_{\rm BH}\ll M_{\rm NBS}$, is orders of magnitude smaller and lighter than the NBS, 
a perturbative calculation is appropriate. Consider a sphere of radius $r_+$ centred at the origin of the NBS. The NBS is stationary, and there is a flux of energy crossing such
a sphere inwards (detailed in Appendix~\ref{app:incoming_flux}), given by
\be
\dot{E}_{\rm in}\approx 10^{-3} \mu^7 r_+^2 M_{\rm NBS}^5\,,
\end{equation}
and the same amount crossing it outwards. If such a sphere defines the BH boundary $r_+=2M_{\rm BH}$~\footnote{Actually, such a sphere should be placed outside the effective potential for wave propagation around BHs, but the difference is not relevant here.}, a fraction will be absorbed by the BH. Because of relativistic effects, 
low-frequency waves (the scalar field frequency is $\mu$ and we are in the low frequency regime with $\mu M_{\rm BH}\ll 1$) are poorly absorbed, and we find that the flux into the BH is~\cite{Unruh:1976fm} 
\begin{align} \nonumber
\dot{E}_{\rm abs}&=32\pi\left(M_{\rm BH}\mu \right)^3\dot{E}_{\rm in}=\frac{16\pi}{125}\frac{M_{\rm BH}^5}{M_{\rm NBS}^5}\left(M_{\rm NBS}\mu \right)^{10}\,,
\end{align}
where we are taking the limit $\omega \to \mu$ in the expression for the transmission. Strictly speaking, since we are in the $\omega<\mu$ regime, the limit performed above will only provide an approximate result, and a full computation of the transmission coefficient in a BH spacetime for such frequency limit should be needed. Hence, our calculation should be understood as an order of magnitude estimation. However, we have tested the above physics with a series of toy models, including the study of accretion of a massive, non-self-gravitating scalar confined in a spherical cavity with a small BH at the center (see Appendix~\ref{app:bh_bomb}). This toy model conforms to the physics just outlined. Another example in Appendix~\ref{app:string_toy} suggests that all modes
of the NBS are excited during such an accretion process, but made quasinormal (i.e., damped) by the presence of the absorption. These are all low-frequency modes, and our argument should be valid even in such circumstance.

With $\dot{E}_{\rm abs}=\dot{M}_{\rm BH}$ and fixed NBS mass, we find the timescale
\begin{equation}
\tau\sim \frac{1}{M_{\rm BH}^4M_{\rm NBS}^5\mu^{10}}=10^{24}\,{\rm yr}\,\frac{M_{\rm NBS}}{10^{10}M_{\odot}}\left(\frac{\chi}{10^4}\right)^4\left(\frac{0.1}{M_{\rm NBS}\mu}\right)^{10}\,,
\end{equation}
where $\chi\equiv M_{\rm NBS}/M_{\rm BH}$. 
In other words, the timescale for the BH to increase substantially its mass -- which we take as a conservative indicative of the lifetime of the entire NBS -- is
larger than a Hubble timescale for realistic parameters. The above timescale shows that nearly monochromatic scalar waves (the scalar background configuration) are poorly absorbed by static BH. Additionally, this results corroborate the assumptions of our perturbative scheme: the background configuration will not change sensibly during the typical astrophysical scale that we are interested in. When the material of the star is nearly exhausted, a new timescale is relevant, that of the quasinormal modes of the BH surrounded by a massive scalar.
This timescale is $\tau_{\rm QNM}\sim M_{\rm BH}(M_{\rm BH}\mu)^{-6}<\tau$~\cite{Detweiler:1980uk,Brito:2015oca}, but still typically larger than a Hubble time.

The results above showed that the system composed by a static BH and the scalar environment might considered stable on some relevant timescale. Furthermore, if we consider rotating BHs rather than static ones, under suitable circumstances (i.e. complex massive scalars with frequency above a certain threshold\til\cite{Brito:2015oca}), the entire setup may become even more stable. In fact, the BH rotation is able to provide energy, via superradiance, to the surrounding field, and sustain nearly stationary, but non-spherically-symmetric, configurations~\cite{Herdeiro:2014goa}. We will not discuss these effects here.

\section{Massive objects plunging into boson stars}\label{section:Massive_objects_plunging_into_boson_stars}
%
\begin{figure}[ht]
\centering
\includegraphics[height=5.5cm,keepaspectratio]{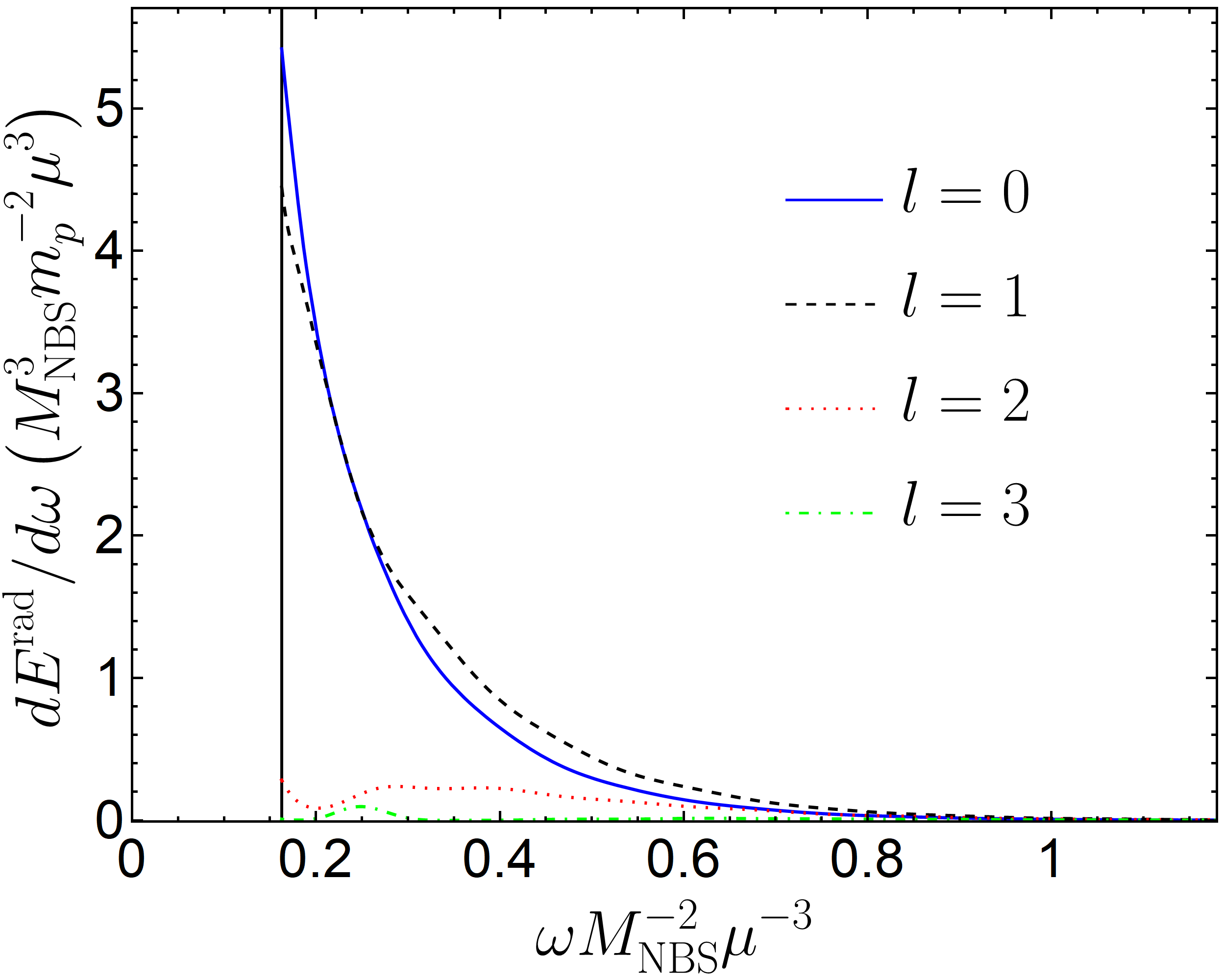}
\includegraphics[height=5.5cm,keepaspectratio]{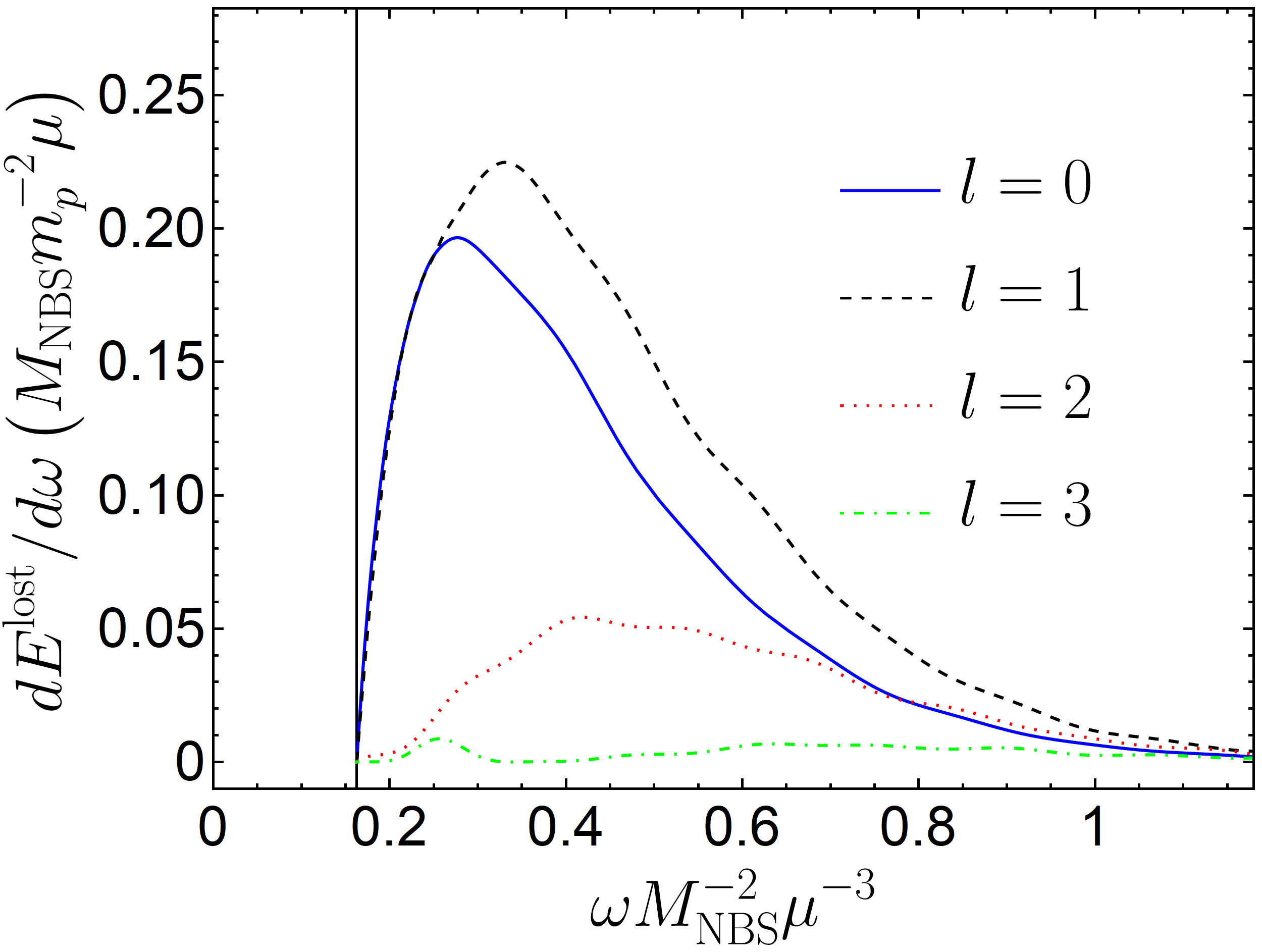}  
\includegraphics[height=5.5cm,keepaspectratio]{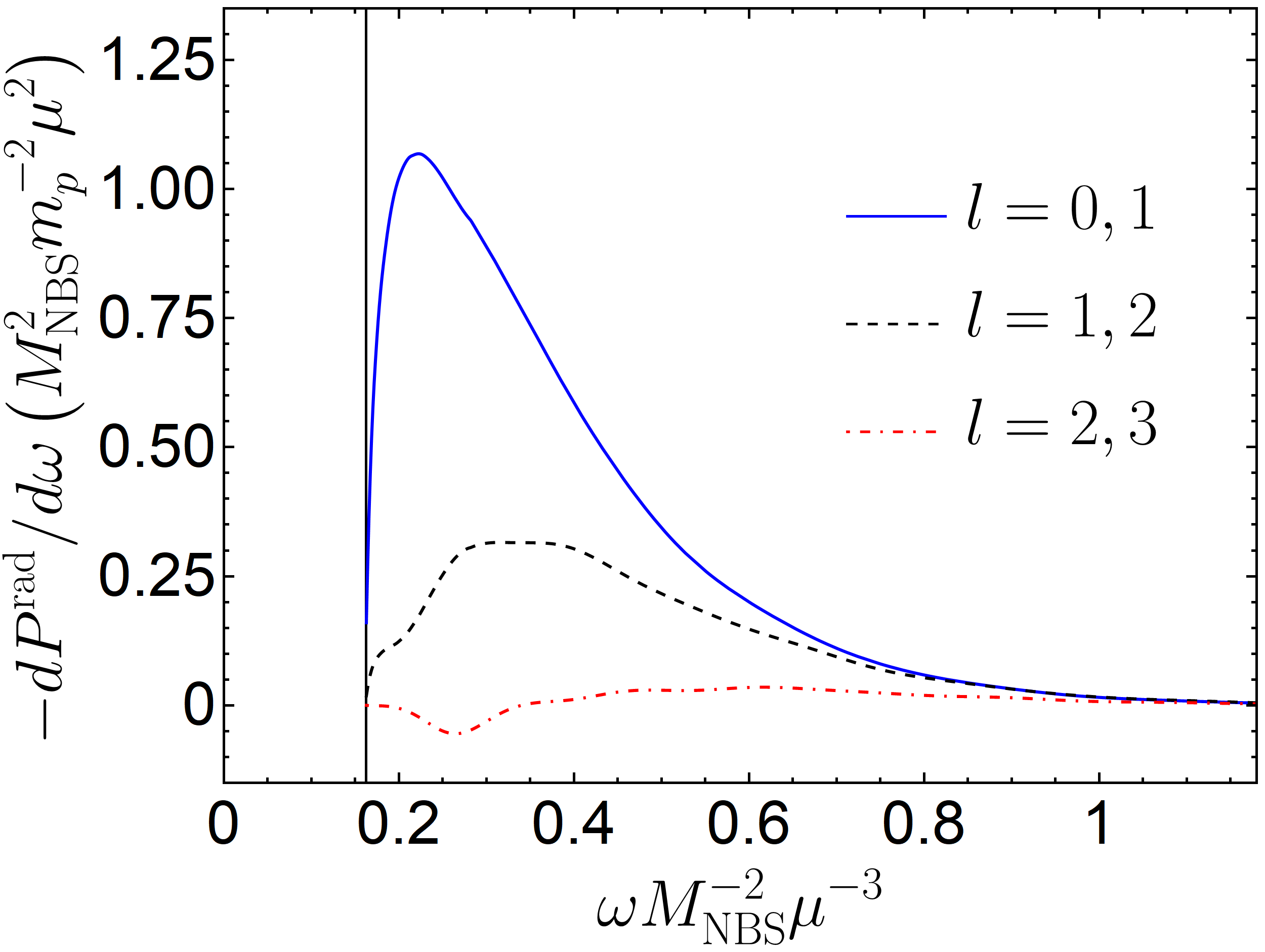}
\caption[Spectra of particle plunging in an NBS.]{Spectrum of radiation released when an object of mass $m_p$ plunges into an NBS with initial velocity $v_R\approx 0 $. Emission takes place for frequencies $\omega>\gamma$ (see Eqs~\eqref{Energy_flux_BS}-\eqref{Momentum_flux_BS}). {\bf (Upper) Left panel:} lowest multipole contribution $l=0,1,2,3$ to the total spectral flux of energy. {\bf (Upper) Right panel:} multipole contributions to the radiated kinetic energy of the scalar field. {\bf Lower panel:} spectral fluxes of linear momentum along $z$ associated with the lowest multipoles. The results obtained for other plunging velocities are summarized in Eqs~\eqref{eq:fit_BS_Erad_plunge_gravityon}-\eqref{eq:fit_BS_Prad_plunge_gravityon}.}
\label{fig:PlungingSpectrabosonstar_gravityon}
\end{figure}
Consider now a massive perturber plunging, head-on, into an NBS. The perturber is assumed to have traveled from far away, but for our purposes the only relevant quantity is the perturber velocity
when it reaches the NBS surface, $\boldsymbol{v}=-v_{R}\boldsymbol{e}_z$, with $v_R\geq 0$. This setup is described in detail in Sec.~\ref{subsection:External_perturbers}. As we argued before (and also below), this situation could describe a massive BH ``kicked'' at formation, via GW emission, in a DM core of light fields, or, simply, stars crossing an NBS. Our framework allow us to do the first self-consistent computation of the gravitational drag acting on perturbers in such systems. Including the effect of the NBS gravitational potential on the perturber motion sets a natural critical velocity in the problem, the escape velocity $v_{\rm esc}$. For the fundamental NBS described in Fig.\til\ref{fig:BS}, the velocity needed to escape from the surface of the NBS is $v_{\rm esc}\sim 0.47M_{\rm NBS}\mu$.
When the velocity is smaller than this, the crossing object should be confined in the NBS with an oscillatory motion. 
For now, we study a simple one-way motion, and assume that when the particle crosses the NBS once, it simply ``disappears''. This will allow us to estimate the dynamical friction on the perturber.
This assumption is formally correct and accurate for unbound motion. For bound oscillatory motion it is not, and we work out the full case below, in Section~\ref{section:A_perturber_oscillating_at_the_center}.

Some quantities of interest are the spectral fluxes of energy and linear momentum radiated in these processes, as well as the energy lost by the perturber. These are given, respectively, by Eqs.~\eqref{Energy_flux_BS}-\eqref{Momentum_flux_BS} and \eqref{Energy_loss_BS}. The upper left panel of Fig.~\ref{fig:PlungingSpectrabosonstar_gravityon} shows the contribution of the lowest multipoles to the total energy spectrum $d E^{\rm rad}/d \omega$ ($d E^{\rm lost}/d \omega$ upper right panel). These results were obtained through the numerical evaluation of expressions~\eqref{Energy_flux_BS}-\eqref{Energy_loss_BS} for a perturber plunging into an NBS, starting the fall from rest at~$R$. The fluxes converge exponentially with increasing values of $l$, after a sufficiently large~$l$. Our results are compatible with~$E_l^{\rm rad}\propto e^{-l}$, where~$E_l^{\rm rad}$ is the $l$-mode contribution to the energy radiated. Once the behavior of $E_l^{\rm rad}$ for large $l$ is known, we can find the total energy radiated.
For a particle plunging with zero initial velocity into an NBS we obtain $E^{\rm rad}\sim 1.28 \, m_p^2/M_{\rm NBS}$ and $E^{\rm lost}\sim 0.18 \, m_p^2 M_{\rm NBS}\mu^2$. 
Applying this procedure to other velocities, we find that the following is a good description of our results,
\begin{align}
E^{\rm rad}&=29\frac{m_p^2}{M_{\rm NBS}}\frac{e^{-3.25/X}}{X^{17/4}}\,\label{eq:fit_BS_Erad_plunge_gravityon}
\\
E^{\rm lost}&=7\,m_p^2 M_{\rm NBS}\mu^2\frac{e^{-3.54\,\left(X-0.05\right)^{-1}}}{\left(X-0.05\right)^{17/4}}\,\label{eq:fit_BS_Ekin_plunge_gravityon}
\end{align}
accurate to within $5\%$ of error for~$0\lesssim v_R\lesssim 2.5M_{\rm NBS}\mu$. This interval spans over non-relativistic astrophysical relevant velocities ({\it e.g.}, $0\lesssim v_R[{\rm km/s}]\lesssim 6000$ for the DM core of the Milky Way). 
Here, 
\be
X\equiv \frac{v_R}{M_{\rm NBS}\mu}+0.68\,.
\end{equation}

The lower central panel of Fig.~\ref{fig:PlungingSpectrabosonstar_gravityon} shows the multipolar contribution to the spectral flux of linear momentum along $z$. The linear momentum radiated also converges exponentially in $l$, after a sufficiently large $l$. For a perturber starting at rest, the total linear momentum radiated along $z$ in the whole process is $P^{\rm rad}\sim -0.43 \, m_p^2\mu$.
The fitting expression
\begin{equation}
P^{\rm rad}=-2.4\,m_p^2\mu\frac{e^{-2.26\,\left(X-0.27\right)^{-1}}}{\left(X-0.27\right)^{17/4}}\,,\label{eq:fit_BS_Prad_plunge_gravityon}
\end{equation}
is a good approximation to our results (within $5\%$ of error for~$0 \lesssim v_R\lesssim 2.5M_{\rm NBS}\mu$). 
\begin{figure}
\centering
\includegraphics[width=0.55\textwidth,keepaspectratio]{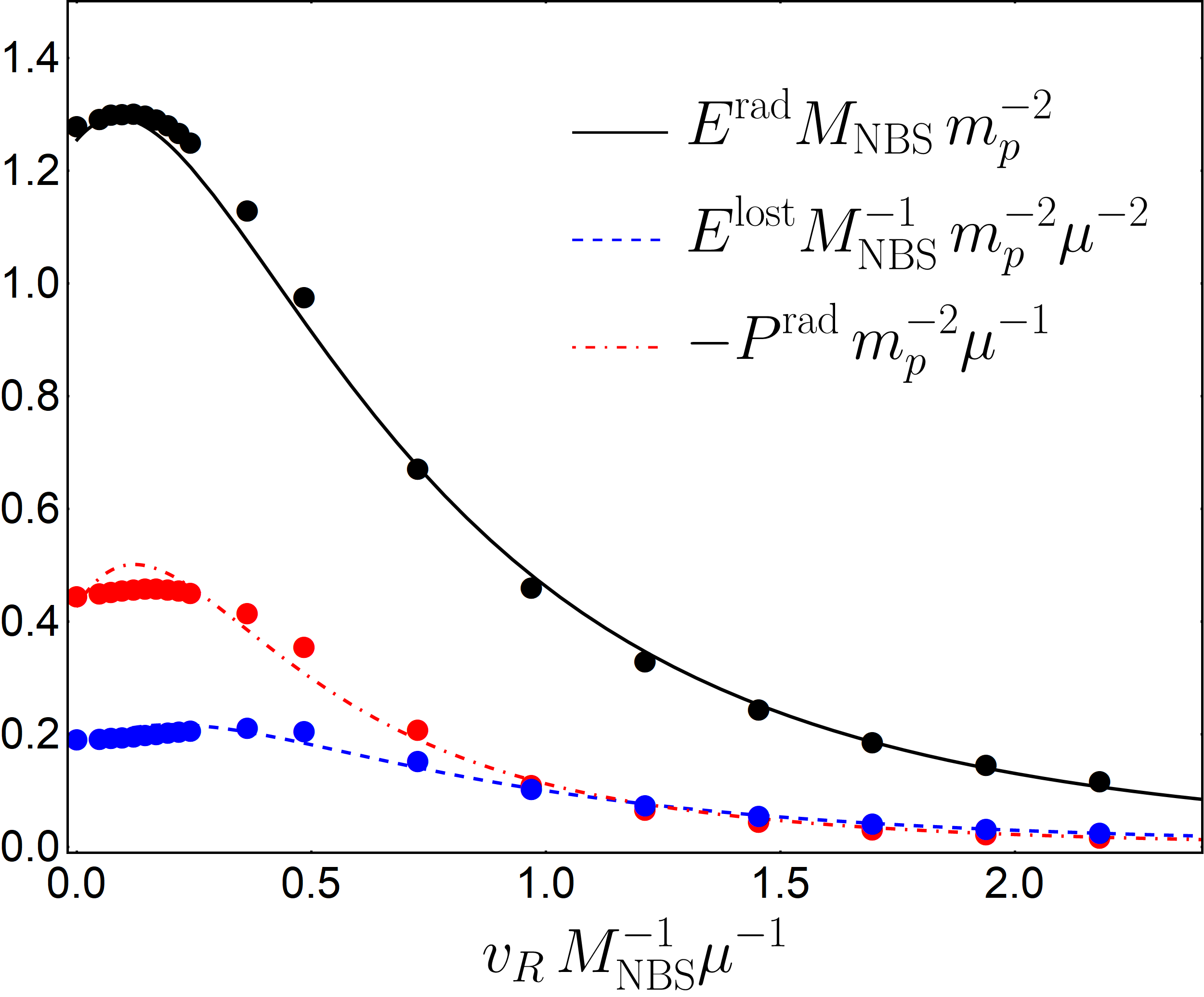} 
\caption[Energy and momentum particle plunging in an NBS.]{Total and kinetic energy, and linear momentum emitted when an object of mass $m_p$ plunges through an NBS, as a function of the initial perturber velocity. The dots correspond to the numerical data used to obtain Eqs.~\eqref{eq:fit_BS_Erad_plunge_gravityon}-\eqref{eq:fit_BS_Ekin_plunge_gravityon}-\eqref{eq:fit_BS_Prad_plunge_gravityon}.}
	\label{fig:fitsBS}
\end{figure}
Figure~\ref{fig:fitsBS} shows how the total radiated energy $E^{\rm rad}$, the total energy lost by the moving perturber $E^{\rm lost}$, and the linear momentum radiated $P^{\rm rad}$ vary with the change of initial velocity.

The momentum lost by a small plunging object ($m_p\mu \ll v_R$) is given by $P^{\rm lost}=-E^{\rm lost}/v_R$, as shown in Eq.~\eqref{EvsPloss}.
We have thus computed, in a self-consistent way, the dynamical friction acting upon a body moving within an NBS. The quantity $E^{\rm lost}$ is the actual kinetic energy lost by the perturber as it crosses the NBS. Note that, in accordance with the results for the energy lost -- in particular, its sign -- this is indeed a friction; the body will slow down. On the other hand, the results for the energy lost together with the radiated momentum show that the NBS will acquire a small momentum in the direction of the moving perturber, described by Eq.~\eqref{eq:NBS_momentum}; note the two lines crossing each other close to~$v_R= M_{\rm NBS} \mu$ in Fig.~\ref{fig:fitsBS}.

Our computations should be compared and contrasted with those of Ref.~\cite{Hui:2016ltb,Lancaster:2019mde}, where dynamical friction in these structures was estimated
without including self-gravity (therefore not accounting for the size of the scalar structure either). In contrast to those of Ref.~\cite{Hui:2016ltb}, our results are self-consistent, regular and finite at all velocities. In Appendix~\ref{app:drag}, we look at a simple toy model which indicates that the discrepancy between these results may be partially related with the trivial gravitational potential of the background medium. A non-trivial gravitational potential can confine small-frequency scalars, suppressing efficiently scalar emission. Nevertheless, the self-gravity of the scalar seems to help suppressing scalar emission for small velocities. Thus, the above results are the first self-consistent and accurate calculation of dynamical friction caused by a self-gravitating scalar on passing objects.

\section{A perturber oscillating at the center}\label{section:A_perturber_oscillating_at_the_center}
%
As a black hole forms through gravitational collapse in a DM core it can be ``kicked'', via GW emission, and left in an oscillatory motion around the center of the core. 
The reason for the kick is that collapse is, in general, an asymmetric process, and leads to emission of GWs which carry some momentum. This process is known
to lead to velocities of at most a few hundred kilometers per second~\cite{1973ApJ...183..657B}, generically smaller than the galactic escape velocity. Thus, the remnant BH
is bound to the galaxy and, in absence of dissipation, performs an oscillatory motion.

It is crucial to understand how the DM core reacts to this motion and to quantify the energy and momentum radiated and deposited in the scalar field. Similar issues were addressed in Ref.~\cite{Gualandris:2007nm}, in the context of the interaction between a kicked supermassive black hole and stars in galaxy cores.

At the center of an NBS the energy density is approximately constant~$\rho_E\simeq 4\times 10^{-3} M_{\rm NBS}^4\mu^6$. So, the motion of the perturber is
\begin{equation}
z_p(t)=	-\mathcal{A} \sin\left(\omega_{\rm osc} t\right),\,
\mathcal{A} \equiv \sqrt{\frac{3}{4 \pi} \frac{v_0^2}{\rho_ E}},\, \omega_{\rm osc}\equiv \sqrt{\frac{4 \pi \rho_E}{3}}\,,
\end{equation}
where~$v_0$ is the velocity of the perturber at the center of the core.
The source is described by
\begin{equation}
P=m_p \frac{\delta(\varphi)}{r^2 \sin \theta}\left[\delta\left(r-z_p(t)\right)\delta\left(\theta\right)+ \delta\left(r+z_p (t)\right)\delta\left(\theta-\pi\right)\right].
\end{equation}
Using Eq.~\eqref{p_def} the function~$p$ reads
\begin{equation}
p=\frac{m_p}{2\sqrt{2 \pi}}\frac{|\tau'_{1,n}(r)|}{r}Y_l^0(0) \delta_m^0 
\sum_{n \in \mathbb{Z}}\Big[e^{-i \omega \tau_{1,n}}+e^{-i \omega \tau_{2,n}}+(-1)^n \left(e^{i \omega \tau_{1,n}}+e^{i \omega \tau_{2,n}}\right)\Big]\,,
\end{equation}
where we defined
\begin{equation}
\tau_{1,n}\equiv \frac{1}{\omega_{\rm osc}}\left[\arcsin\left(\frac{r}{\mathcal{A}}\right)+2n \pi\right],\, \tau_{2,n}\equiv \frac{1}{\omega_{\rm osc}}\left[(2n+1) \pi-\arcsin\left(\frac{r}{\mathcal{A}}\right)\right] \,,
\end{equation}
as are the roots of~$r+z_p(\tau)=0$; the symmetric functions~$-\tau_{1,n}(r)$ and~$-\tau_{2,n}(r)$ are the roots of~$r-z_p(\tau)=0$. In the last expressions we are using the principal branch of the inverse sine function. It is possible to show that the function~$p$ can be expressed in the form
\begin{align}
p&=\frac{m_p}{\sqrt{2 \pi}}\frac{Y_l^0(0)}{\sqrt{\mathcal{A}^2-r^2}}\frac{\delta_m^0}{\omega_{\rm osc}}\,\Theta\left(\mathcal{A}-r\right)\nn\\
\label{p_osc_start}
&\times \sum_{n \in \mathbb{Z}}\bigg[\delta_l^{\rm even} \left(\cos\left[\omega \tau_{1,n}(r)\right]+\cos\left[\omega \tau_{2,n}(r)\right]\right)-i\, \delta_l^{\rm odd} \left(\sin\left[\omega \tau_{1,n}(r)\right]+\sin\left[\omega \tau_{2,n}(r)\right]\right)\bigg]\,,
\end{align}
Using the mathematical identities
\begin{equation}
\sum_{n \in \mathbb{Z}} \sin \left(2 n \pi \frac{\omega}{\omega_{\rm osc}}\right)=0,\,\sum_{n \in \mathbb{Z}} \cos \left(2 n \pi \frac{\omega}{\omega_{\rm osc}}\right)=\omega_{\rm osc} \sum_{n \in \mathbb{Z}} \delta(\omega-n \omega_{\rm osc})\,,
\end{equation}
together with some trivial trigonometric identities, we can rewrite~\eqref{p_osc_start} as
\begin{align} 
p&=m_p\sqrt{\frac{2}{\pi}}\frac{Y_l^0(0)}{\sqrt{\mathcal{A}^2-r^2}}\delta_m^0 \,\Theta\left(\mathcal{A}-r\right) \sum_{n \in \mathbb{Z}} \delta(\omega-2n\omega_{\rm osc}) \nonumber \\
&\times \bigg[\delta_l^{\rm even} \cos\left(2 n \arcsin\frac{r}{\mathcal{A}}\right)-i\, \delta_l^{\rm odd} \sin\left(2 n \arcsin\frac{r}{\mathcal{A}}\right)\bigg]\,.
\end{align}
With the help of the trigonometric identities
\begin{align}
\cos(2n x)&=\sum_{k=0}^{n}(-1)^k \binom{2n}{2k}\sin^{2k} x \cos^{2(n-k)}x\,, \\
\sin(2n x)&=\sum_{k=0}^{n-1}(-1)^k \binom{2n}{2k+1}\sin^{2k+1} x \cos^{2(n-k)-1}x\,,
\end{align}
the last expression can be written in the alternative form
\begin{align} 
p&=m_p\sqrt{\frac{2}{\pi}}Y_l^0(0)\delta_m^0 \,\Theta\left(\mathcal{A}-r\right)\bigg[-i\, \delta_l^{\rm odd} \sum_{k=0}^{n-1}(-1)^k \binom{2n}{2k+1}r^{2k+1}\left(\mathcal{A}^2-r^2\right)^{n-k-1}\nn\\
&+\delta_l^{\rm even}\sum_{k=0}^{n}(-1)^k \binom{2n}{2k} r^{2k}\left(\mathcal{A}^2-r^2\right)^{n-k-\frac{1}{2}}\bigg]  \sum_{n \in \mathbb{Z}} \frac{1}{\mathcal{A}^{2n}}\delta(\omega-2n\omega_{\rm osc}) \,.
\end{align}

We want to calculate the energy radiated through scalar waves due to the oscillatory motion of the massive object. First, note that the oscillation frequency is $\omega_{\rm osc}\sim 0.135M^2_{\rm NBS}\mu^3\lesssim \gamma$. Only the modes with $n\geq 1$ arrive at infinity; so, only these contribute to the energy radiated.
Applying the formalism described in Section~\ref{section:Small_perturbations}, we obtain
\begin{equation}
Z_1^\infty=4 \pi \int_0^\mathcal{A}dr'F_{4,6}^{-1}(r')p(r'),\,Z_2^\infty(\omega,l,0)=Z_1^\infty(-\omega,l,0)^*\,.
\end{equation}
The energy radiated per unit of time is (see Eq.~\eqref{Energy_flux_rate})
\begin{align}
\dot{E}^{\rm rad}&=\frac{2}{\pi} \sum_{l, n}\left(\mu-\gamma+2n\omega_{\rm osc}\right) {\rm Re}\left[\sqrt{2 \mu (2 n \omega_{\rm osc}-\gamma)}\right]|\widetilde{Z}_1^\infty(2n \omega_{\rm osc},l,0)|^2  \nn \\
&\simeq\frac{2}{\pi}\mu \sum_{l, n}  {\rm Re}\left[\sqrt{2 \mu (2 n \omega_{\rm osc}-\gamma)}\right]|\widetilde{Z}_1^\infty(2n \omega_{\rm osc},l,0)|^2 \,,
\end{align}
where we used the low-energy limit~$\gamma \ll \mu$ and~$\omega_{\rm osc}\ll \mu$, and defined
\begin{align}
\widetilde{Z}_1^\infty &\equiv 4 \pi \int_0^\mathcal{A}dr'F_{4,6}^{-1}(r')\widetilde{p}(r')\,, \\
\widetilde{p}&\equiv m_p\sqrt{\frac{2}{\pi}}\frac{Y_l^0(0)}{\sqrt{\mathcal{A}^2-r^2}} \sum_{n \in \mathbb{Z}}  \bigg[\delta_l^{\rm even} \cos\left(2 n \arcsin\frac{r}{\mathcal{A}}\right)-i\, \delta_l^{\rm odd} \sin\left(2 n \arcsin\frac{r}{\mathcal{A}}\right)\bigg]\,.
\end{align}

One can anticipate that the dominant contribution to the radiation is given by the~$n=1$ mode, which has a frequency~$\omega=2\omega_{\rm osc}$. This is the lowest frequency radiated by the perturber and, thus, we expect it to be the one carrying more energy, because the coupling between the perturber and the scalar is stronger for lower frequencies -- as will become evident in the following sections. Indeed, this is in accordance with our numerics. So, we focus on the single~$n=1$ mode. For oscillations deep inside the NBS with an amplitude~$\mathcal{A}\ll R$ -- which is where our constant density approximation holds -- we find that the following semi-analytic expression is a good description of our numerical results: 
\begin{align}
\dot{E}^{\rm rad}&=\frac{2 \sqrt{2}}{\pi} (m_p\mu)^2 \sqrt{ \frac{2\omega_{\rm osc}-\gamma}{\mu}}\sum_l c_l\, \left(\frac{\mathcal{A}}{R}\right)^{2(l+1)},
\end{align}
with the numerical constants~$c_l$. For the first multipoles we find
\begin{equation}
c_0\simeq 0.852\,, c_1 \simeq 67.7 \,,  c_2 \simeq 30.4\,, c_3 \simeq 438\,, c_4 \simeq 13.6\,, c_5\simeq 3.85\,.
\end{equation}
The above expression describes our numerics with less than~$1\%$ of error for $\mathcal{A}/R \lesssim 0.09$. These amplitudes correspond to kicks of~$v_0\lesssim 0.1 M_{\rm NBS}\mu$, which contains astrophysical relevant velocities; for the Milky Way DM core our expression covers~$v\lesssim 300\, {\rm km/s}$, which contains typical recoil velocities imparted by GW emission in gravitational collapse. Larger kicks, like the ones delivered in a merger of two supermassive BHs, have larger amplitudes and are out of our approximation. However, the framework outlined in Section~\ref{section:Small_perturbations} (without the constant density approximation) can still be applied to those cases.

Using the same reasoning that we applied to the \textit{orbiting particles} to deduce Eq.~\eqref{eq:E_loss_circular}, we can estimate the perturber's energy loss per unit of time to be
\begin{equation}
\dot{E}^{\rm lost}=\frac{2}{\pi} \sum_{l, n}\left(2n\omega_{\rm osc}-\gamma\right) {\rm Re}\left[\sqrt{2 \mu (2 n \omega_{\rm osc}-\gamma)}\right]|\widetilde{Z}_1^\infty(2n \omega_{\rm osc},l,0)|^2\,.
\end{equation}
Considering the single (dominant)~$n=1$ mode, the numerical evaluation of the last expression is well described by the semi-analytic formula
\begin{align}
\dot{E}^{\rm lost}&=\frac{2 \sqrt{2}}{\pi} (m_p\mu)^2 \left( \frac{2\omega_{\rm osc}-\gamma}{\mu}\right)^{\frac{3}{2}}\sum_l c_l\, \left(\frac{\mathcal{A}}{R}\right)^{2(l+1)}.
\end{align}
Again, this describes our numerics with less than~$1\%$ of error for small amplitude oscillations~$\mathcal{A}/R\leq 0.09$. 

One may wonder how long it takes for a kicked BH (or star) to settle down at the center of an halo, purely due to the dynamical friction caused by dark matter. When the condition
\begin{equation}\label{AdiabatCond}
\frac{\dot{E}^{\rm lost} \left(\frac{2\pi}{\omega_{\rm osc}}\right)}{\frac{1}{2}m_p \omega_{\rm osc}^2 \mathcal{A}^2} \ll 1\,,
\end{equation}
is verified, the system is suited to an adiabatic approximation, and we can compute how the amplitude changes with time by solving
\begin{align}
m_p \omega_{\rm osc}^2 \mathcal{A} \,\dot{\mathcal{A}}=-\dot{E}^{\rm lost}\,.
\end{align}
Several astrophysical systems fall within this approximation. For example, the Milky Way dark matter core has a mass~$M_{\rm NBS} \mu \sim 10^{-2}$; so, for an object forming through gravitational collapse and receiving a kick of~$300\,{\rm km/s}$, via GW emission, the adiabatic approximation is suitable if~$m_p/M_{\rm NBS} \ll 0.1$ -- which is verified by all known objects. Using only the dominant multipole~$l=0$ (which accounts for more than~$61\%$ of the total energy loss for~$\mathcal{A}/R\leq 0.09$, and more than~$89\%$ for~$\mathcal{A}/R\leq0.04 $) we obtain
\begin{align}
\mathcal{A}=\mathcal{A}_0\, e^{-t/\tau_{\rm s}}\,,
\end{align}
with the timescale
\begin{equation}
\tau_{s}\simeq \frac{56}{m_p M_ {\rm NBS} \mu^3}\sim 10^{10} {\rm yr} \left(\frac{10^{-22}\, {\rm eV}}{\mu}\right)^2\left(\frac{10^5 M_ \odot}{m_p}\right)\left(\frac{0.01}{M_ {\rm NBS}\mu}\right)\,. 
\end{equation}
So, an object kicked at the center of an NBS, interacting solely with the scalar, settles down in a timescale smaller than the Hubble time if it has a mass~$m_p\gtrsim 10^5 M_\odot$; in other words, if it is a supermassive BH. 

The above timescale is in general much larger than the period of oscillation,
\begin{equation}
	 \tau_s  \sim \frac{M_{\rm NBS}}{m_p} \left(\frac{2 \pi}{\omega_{\rm osc}}\right)\,.
\end{equation}
This suggest that treating the source as eternal is indeed a good approximation to study this process.
It is interesting to compare this result with the timescale of damping due to dynamical friction caused by stars in the galactic core. In Ref.~\cite{Gualandris:2007nm} the authors estimate that timescale to be
\begin{equation}
	\tau^*\sim 0.1\, \frac{M_{\rm c}}{m_p} \left(\frac{2 \pi}{\omega_{\rm osc}}\right) \,,
\end{equation}
where~$M_{\rm c}$ is the galactic core mass.
Using~$M_{\rm c}=M_{\rm NBS}$ we see that~$\tau^* \sim 0.1\,\tau_s$, which is smaller but still comparable to~$\tau_s$. Both ours and Ref.~\cite{Gualandris:2007nm} calculations are order of magnitude estimates, but our result further suggests that dark matter may exert a dynamical friction comparable to the one caused by stars, for processes happening in galactic cores.

\section{Low-energy binaries within boson stars}\label{section:Low-energy_binaries_within_boson_stars}
%
\begin{figure}
\centering
\includegraphics[width=0.6\textwidth,keepaspectratio]{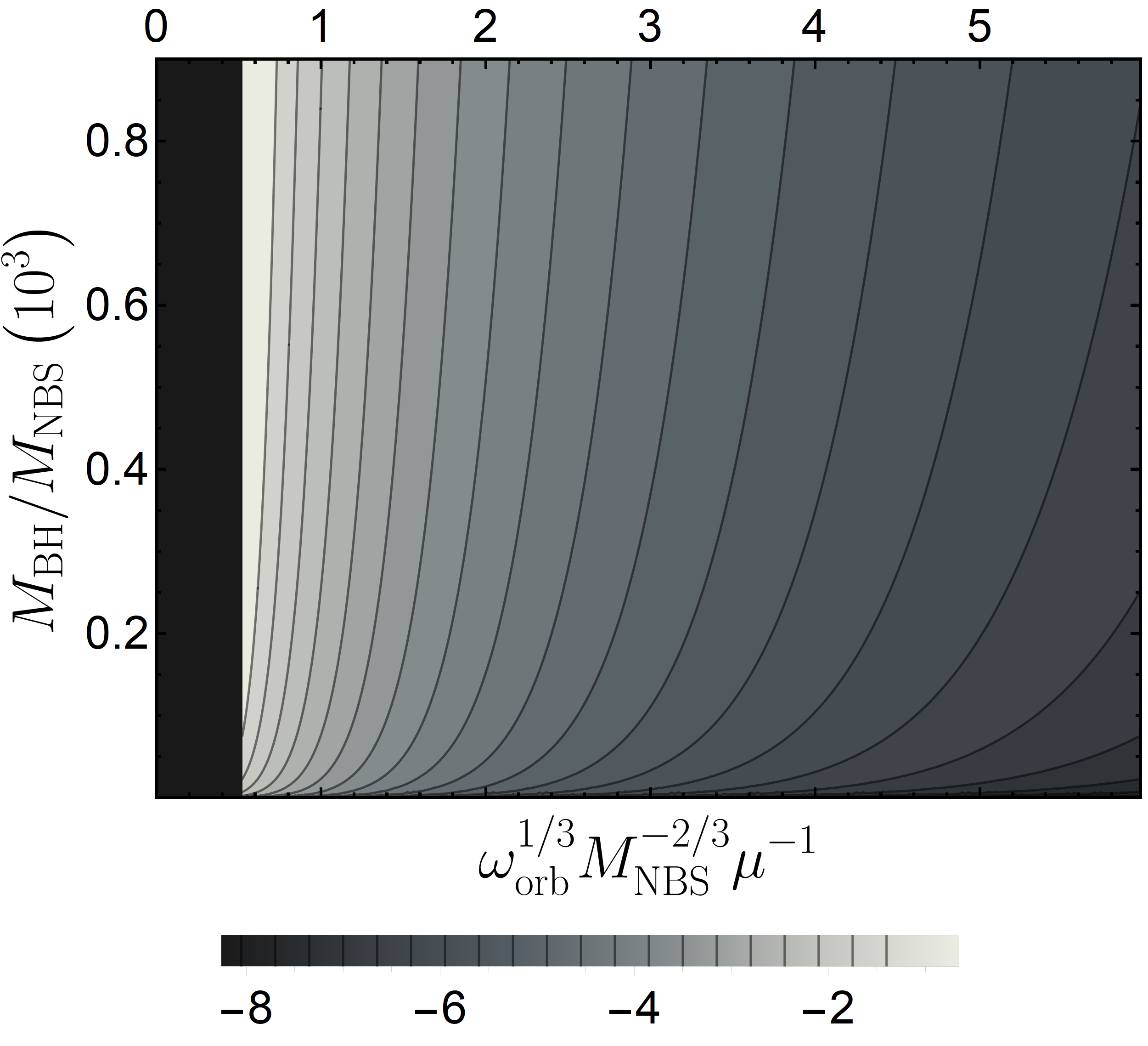} 
\caption[Scalar energy radiated by an EMRI in an NBS]{Logarithm of the universal rate of scalar energy radiated by an EMRI orbiting inside an NBS: $\log_{10}\left[\dot{E}_{\rm EMRI}^{\rm rad}\left(m_p^2 M_{\rm NBS}\mu^3\right)^{-1}\right]$.
The EMRI is described by a supermassive BH of mass $M_{\rm BH}$ sitting at the NBS center, and a star or stellar-mass BH in a circular orbit around it. 
Note that the maximum energy emitted is associated with the smallest frequency (largest distance). This is due to oscillating background field which imparts an energy $\mu$ to any wave. 
For a DM core with $M_{\rm NBS}\sim 10^{10} M_{\odot}$ and mass ratio $m_p/M_{\rm BH}\sim 10^{-4}$, the orbital distances corresponding to non-zero fluxes are in the range $r_{\rm orb}\lesssim 10^6 \,M_{\rm BH}$. For larger radii, the fluctuation has too low an energy and is confined to the structure. This explains the zero-flux (black) region on the left of the panel, corresponding to the suppression of perturbations with frequency $\omega \leq\gamma$.}  	
\label{fig:circularEMRI}
\end{figure}
We now focus on orbiting objects within such an NBS. These will describe binaries, either at an early or late stage in their life,
stirring the field and producing disturbances in the local DM profile. For example, looking at the matter moments in Eq.~\eqref{p_orbiting}, such systems can describe stars orbiting around the SgrA$^*$ BH at the center of the Milky Way. The supermassive BH has a mass $\sim 4\times 10^6 M_{\odot}$ with known companions. The closest known star, S2, has a pericenter distance of $\sim 2800M_{\rm BH}$ and a mass $m_p\sim 20 M_{\odot}$ with a large uncertainty~\cite{Abuter:2018drb,Abuter:2020dou}. Its orbit is, however, highly eccentric. Given the mass and sizes of the NBSs discussed here (i.e. which described the core of DM haloes) all these systems can be handled via perturbation techniques.
In addition, binaries close to supermassive BHs, and therefore to galactic centers, have been observed recently via electromagnetic counterparts to GWs~\cite{Graham:2020gwr}.
\subsection{Scalar emission}\label{subsection:Scalar_emission}
Let us consider first an EMRI: a perturber of mass $m_p$ orbiting a supermassive BH, of mass $M_{\rm BH}\gg m_p$ placed at the center of an NBS. Solving the perturbation equations~\eqref{BS_Perturbation_Matrix_Sourced}, with the source defined in Eq.~\eqref{p_orbiting}, with $m_p(1+(-1)^m)\to m_p$, we find that, up to $3\%$ accuracy, the fluxes of energy (Eqs.~\eqref{eq:circular_flux_Erad}-\eqref{eq:E_loss_circular}) are described by
\begin{align} 
\dot{E}^{\rm rad}_{\rm EMRI}&= 10^{-2}\, m_p^2 M_{\rm BH}^{2/3}M_{\rm NBS}^{4} \mu^{17/2}\omega_{\rm orb}^{-11/6}\Theta\left[\omega_{\rm orb}-\gamma\right]\nn\\
&\times\Big[2.66- 0.49 \, M_{\rm NBS}^{4/3}\mu^2\omega_{\rm orb}^{-2/3} +0.054 \, M_{\rm NBS}^{8/3}\mu^4\omega_{\rm orb}^{-4/3} \Big],\label{Erad_circular_BS}\\
\dot{E}^{\rm lost}_{\rm EMRI}&= 10^{-2}\, m_p^2 M_{\rm BH}^{2/3}M_{\rm NBS}^{4} \mu^{15/2}\omega_{\rm orb}^{-5/6}\Theta\left[\omega_{\rm orb}-\gamma\right]\nn\\
&\times\Big[2.70- 0.96 \, M_{\rm NBS}^{4/3}\mu^2\omega_{\rm orb}^{-2/3} +0.043 \, M_{\rm NBS}^{8/3}\mu^4\omega_{\rm orb}^{-4/3} \Big].\label{Elost_circular_BS}
\end{align}
Notice that in principle, the emission would starts for frequency larger than $\left(\gamma m^{-1}\right)$. However, since the emission in multipoles higher than the dipole is suppressed by roughly a factor $10^3$, we consider only $l=1$ in \eqref{Erad_circular_BS}.
Equations \eqref{Erad_circular_BS}-\eqref{Elost_circular_BS} were evaluated assuming a non-relativistic perturbation, therefore are valid for orbital periods 
$T=2\pi/\omega_{\rm orb}\gg 2\pi/\mu\sim 10^{-22}{\rm eV}/\mu\, {\rm yr}$. 
We show in Fig.~\ref{fig:circularEMRI} the flux of energy ($\dot{E}^{\rm rad}$) as a function of the orbital period and of the BH-NBS mass ratio. Once the orbital frequency is fixed, our results are consistent with an exponential convergence in $l$ for the flux.

The calculation above is easy to adapt to other systems. Consider an equal mass binary system ($M=2m_p$). Looking at the matter moments in Eq.~\eqref{p_orbiting}, it is clear that the first multipole moment that is going to be emitted is the quadrupole $m=2$. As a result of solving the perturbation equations, we find the following expression for the energy emitted in scalar waves, and the one lost by the orbiting particle (up to $3\%$ of accuracy)
\begin{align}
\dot{E}^{\rm rad}&= 10^{-2}\, M^{4/3} m_p^{2}M_{\rm NBS}^{4} \mu^{19/2}\omega_{\rm orb}^{-13/6}\Theta\left[2\omega_{\rm orb}-\gamma\right]\nn\\
&\times\left[1.45- 0.16 \, M_{\rm NBS}^{4/3}\mu^2\omega_{\rm orb}^{-2/3} +0.015 \, M_{\rm NBS}^{8/3}\mu^4\omega_{\rm orb}^{-4/3} \right],\label{Erad_circular_BS_equalmass}\\
\dot{E}^{\rm lost}&=10^{-2}\, M^{4/3} m_p^{2}M_{\rm NBS}^{4} \mu^{17/2}\omega_{\rm orb}^{-7/6}\Theta\left[2\omega_{\rm orb}-\gamma\right]\nn\\
&\times\left[2.97- 0.58 \, M_{\rm NBS}^{4/3}\mu^2\omega_{\rm orb}^{-2/3} +0.0051 \, M_{\rm NBS}^{8/3}\mu^4\omega_{\rm orb}^{-4/3} \right].\label{Elost_circular_BS_equalmass}
\end{align}
The expression above is valid both for solar mass BHs as well as for BH masses of the order $\sim 10^4 M_{\odot}$.

In the limit of an high-frequency ($\omega_{\rm orb} \gg \gamma, \mu U_0$), but still non-relativistic ($\omega_{\rm orb} \ll \mu$) excitation, 
the relevant equations~\eqref{Sourced_SP_System1}-\eqref{Sourced_SP_System2} can be solved analytically in closed form, noticing that $|\Psi_0 \delta \Psi|\ll |\delta U|$. 
Equation~\eqref{Sourced_SP_System2} therefore reduces simply to 
\begin{align}
\nabla^2 \delta U=4 \pi P\,,\label{Sourced_SP_System2HF}
\end{align}
which has the solution
\begin{align}
&\delta U=\frac{2}{\sqrt{2 \pi}}\sum_{l,m} \frac{u(r)}{r} Y_l^m(\theta, 0) e^{-i m \left(\omega_{\rm orb} t -\varphi\right)} \,,
\end{align}
with
\begin{equation}
u=-\left(2 \pi\right)^{3/2} m_p \left[1+(-1)^m\right]\frac{Y_l^m\left(\frac{\pi}{2},0\right)}{2l+1}\left[\left(\frac{r}{r_{\rm orb}}\right)^{-l} \Theta(r-r_{\rm orb})+\left(\frac{r}{r_{\rm orb}}\right)^{l+1} \Theta(r_{\rm orb}-r)\right]\,.\nonumber
\end{equation}
Then, using the decomposition
\begin{align}
\delta \Psi = \frac{2}{\sqrt{2 \pi}}\sum_{l,m} \frac{Z(r)}{r} Y_l^m(\theta, 0)e^{-i m\left(\omega_{\rm orb} t-\varphi\right)}\,,
\end{align}
equation~\eqref{Sourced_SP_System1} becomes
\begin{align}
\partial_r^2 Z+\left(2 \mu m\omega_{\rm orb}-\frac{l(l+1)}{r^2}\right)Z=2 \mu^2 \Psi_0  u\,.
\end{align}
Using the method of variation of parameters, we can solve the last equation imposing the Sommerfeld radiation condition at large distances and regularity at the origin. The obtained solution is, at large distances, 
\begin{align} \label{vop_hf}
Z(r \to \infty)= i \pi\mu^2Z_\infty(r\to \infty) \int_{0}^{\infty} dr' Z_0\Psi_0 u\,,
\end{align}
where $Z_0$ and $Z_\infty$ are homogeneous solutions satisfying, respectively, regularity at the origin and the Sommerfeld radiation condition at large distances, and are given by
\begin{align}
Z_0&=\sqrt{r}\, J_{l+1/2}\left(\sqrt{2 \mu m \omega_{\rm orb}}r\right)\,, \\
Z_\infty&= \sqrt{r} H^{(1)}_{l+1/2}\,\left(\sqrt{2\mu m \omega_{\rm orb}}r\right)\,,
\end{align}
with $J_\nu(x), H^{(1)}_{\nu}(x)$ Bessel and Hankel functions~\cite{Abramowitz:1970as}.
Using the asymptotic form 
\begin{equation}
Z_\infty (r\to \infty)\simeq (-i)^{l+1}\sqrt{\frac{2}{\pi}} \frac{e^{i \sqrt{2\mu m \omega_{\rm orb}}\,r }}{\left(2\mu m \omega_{\rm orb}\right)^{1/4}}\,,
\end{equation} 
and assuming that~$r_{\rm orb}\ll R$, and~$\omega_{\rm orb}/\mu\gg \left(r_{\rm orb} \mu\right)^{-2}$, the integration in~\eqref{vop_hf} converges a few wavelengths from the binary and gives
\begin{equation}
Z(r \to \infty) \simeq -(-i)^l \left(2 \pi\right)^2 \mu^{2} m_p \Psi_0(0)r_{\rm orb}^l\left[1+(-1)^m\right]  \frac{2^{-\frac{l}{2}-\frac{3}{2}}\,e^{i \sqrt{2\mu m \omega_{\rm orb}}\,r }}{\left(\mu m \omega_{\rm orb}\right)^{1-\frac{l}{2}}} \frac{Y_l^m\left(\frac{\pi}{2},0\right)}{\Gamma\left(l+\frac{3}{2}\right)}\,.
\end{equation}
So, the dominant $l=m$ modes give the scalar perturbation
\begin{align}
\delta \Psi (r\to \infty) &\simeq -8 \pi^{\frac{3}{2}}\mu^2 m_p \Psi_0(0) \sum_{m=1}^{+\infty}(-i)^m\left[1+(-1)^m\right] \nonumber \\
&\times  \frac{Y_l^m\left(\frac{\pi}{2},0\right)}{\Gamma\left(m+\frac{3}{2}\right)} \frac{(\mu m)^{\frac{m}{2}-1}(M \omega_{\rm orb})^{\frac{m}{3}}}{2^{2+\frac{m}{2}}\omega_{\rm orb}^{\left(1+\frac{m}{2}\right)}}e^{i \sqrt{2\mu m \omega_{\rm orb}}\,r }\,,
\end{align}
where we have used Kepler's law $r_{\rm orb}^3=M/\omega_{\rm orb}^2$.
Then, the flux of energy is given by
\begin{align}
\dot{E}^{\rm rad}&=- r^2 \lim_{r \to \infty}\int d\theta d\varphi \sin \theta \,T^S_{t r}= 0.28\, \pi^{3} \left(\mu m_p\right)^2 \left(\mu M_{\rm NBS}\right)^4 \nonumber \\
& \sum_{m=1}^{+\infty}\left[1+(-1)^m\right]^2\left(1+\frac{m \omega_ {\rm orb}}{\mu}\right) \left(\frac{Y_m^m\left(\frac{\pi}{2},0\right)}{\Gamma\left(m+\frac{3}{2}\right)} \frac{m^{\left(\frac{m}{2}-\frac{3}{4}\right)}(M\omega_{\rm orb})^{\frac{m}{3}}}{2^{\left(\frac{7}{4}+\frac{m}{2}\right)}(\omega_{\rm orb}/\mu)^{\left(\frac{3}{4}+\frac{m}{2}\right)}}\right)^2\,.
\end{align}
The last expression can be further simplified using~$\left(1+m\omega_ {\rm orb}/\mu\right)\simeq 1$, since we are considering low-energy excitations of the scalar field.
The same reasoning that we used to derive~\eqref{eq:E_loss_circular} can be applied here to find that the binary loses energy at a rate
\begin{equation}
\dot{E}^{\rm lost}\simeq 0.28\, \pi^{3} \left(\mu m_p\right)^2 \left(\mu M_{\rm NBS}\right)^4 \sum_{m=1}^{+\infty}\left[1+(-1)^m\right]^2 \left(\frac{Y_m^m\left(\frac{\pi}{2},0\right)}{\Gamma\left(m+\frac{3}{2}\right)} \frac{m^{\left(\frac{m}{2}-\frac{1}{4}\right)}(M\omega_{\rm orb})^{\frac{m}{3}}}{2^{\left(\frac{7}{4}+\frac{m}{2}\right)}(\omega_{\rm orb}/\mu)^{\left(\frac{1}{4}+\frac{m}{2}\right)}}\right)^2\,.
\end{equation}

These analytic results are in excellent agreement with our numerics for both EMRIs (Eq.~\eqref{Erad_circular_BS}) and equal mass binaries (Eqs.~\eqref{Erad_circular_BS_equalmass}): the leading terms agrees with the numerical within $4\%$. We remind the reader that the numerical calculations used to obtained the results in Eq.\til\eqref{Elost_circular_BS_equalmass}, has been obtained employing the variation of parameters method, described in Sec.\til\ref{subsection:External_perturbers} in detail. The numerical integration has been performed with the built-in integrator function in Wolfram Mathematica. Such agreement is a cross-check both on our numerical routine and our simple analytical description.

%
\subsection{Comparison with gravitational wave emission}\label{subsection:Comparison_with_gravitational_wave_emission}
In vacuum, the orbit of a binary system shrinks in time, due to the emission of GWs. At leading order, the loss via GWs is described by the quadrupole formula~\cite{Peters:1963ux,Poisson:1993vp},
\be
\dot{E}^{\rm GW}=\frac{32}{5}\eta^2\left(M\omega_{\rm orb}\right)^{10/3}\,,
\label{eq:quadrupole}
\end{equation}
where $\eta=m_1m_2/(m_1+m_2)^2$ is the symmetric mass ratio of a binary of masses $m_1, m_2$ and total mass $M=m_1+m_2$. The NBS instead provides an extra scalar channel for the energy loss. To estimate the flux of energy emitted in scalar waves, we consider the orbit to be circular, with the radius equal to the semi-major axis ($\sim 970$ au) of the S2 star.  For EMRIs ($m_p=\eta M$ and $M_{\rm BH}=M$), combining together Eqs.~\eqref{Elost_circular_BS}-\eqref{eq:quadrupole} we get
\begin{equation}
\frac{\dot{E}^{\rm lost}}{\dot{E}^{\rm GW}}\simeq   10^{-3}\, \left[\frac{M_{\rm NBS}}{10^{10}M_{\odot}}\right]^4\left[\frac{10^{6}M_{\odot}}{M}\right]^{2/3}\left[\frac{T}{16 {\rm yr}}\right]^{31/6}\left[\frac{\mu}{10^{-22}{\rm eV}}\right]^{17/2},
\end{equation}
where we normalized to the typical values for the EMRI composed by Sagittarius $\text{\rm A}^*$ and S2 star, surrounded by a DM halo. Since the total scalar field mass contained in a sphere of radius $r_{\rm orb}\ll R$ is negligible with respect to the mass of the central BH $M_{\rm NBS}(r_{\rm orb})/M\sim 10^{-10}$, we can consider that the entire GW flux emitted is due to the quadrupole moment of the binary alone, neglecting the gravitational field of the DM halo.

The energy balance equation imposes that the loss in the orbital energy of the binary is due to the energy carried away by scalar and gravitational waves~\cite{1989ApJ...345..434T,Stairs:2003eg} 
\be
\frac{d E^{\rm orb}}{dt}=-\left(\dot{E}^{\rm lost}+\dot{E}^{\rm GW}\right)\,.\label{eq:energy_balance}
\end{equation}

Thus, the energy loss leads to a secular change in the orbital period
\be
\dot{T}\simeq-\frac{192\pi\left(2\pi\right)^{5/3}\eta M^{5/3}}{5T^{5/3}}-\frac{5 \eta M M_{\rm NBS}^{4} T^{5/2}}{ 10^{3}\mu^{-15/2}}\,.\nonumber
\end{equation}
%
%
%
It is amusing to estimate such secular change for astrophysical parameters similar to those of S2 star orbiting around SgrA$^*$, 
\begin{align}
\dot{T}&\simeq \, -\frac{2.42}{10^{15}}  \left[\frac{M}{10^{6}M_{\odot}}\right]^{2/3}\left[\frac{T}{16 {\rm yr}}\right]^{-5/3}\left[\frac{m_p}{20 M_{\odot}}\right]\nn\\
&-\frac{4}{10^{17}}\left[\frac{M_{\rm NBS}\mu}{0.01}\right]^4\left[\frac{\mu}{10^{-22}{\rm eV}}\right]^{7/2}\left[\frac{T}{16 {\rm yr}}\right]^{5	/2}\left[\frac{m_p}{20 M_{\odot}}\right]\,,
\end{align}
which seems hopelessly small.

The period change for equal-mass binary systems follows through, and is
\be
\dot{T}=-\frac{192\pi\left(2\pi\right)^{5/3}M^{5/3}}{20T^{5/3}}-\frac{3.1 M_{\rm NBS}^{4} m_p M^{2/3}T^{17/6}}{10^{3}\mu^{-17/2}}\nonumber\,.
\end{equation}
%

\subsection*{Backreaction and scalar depletion}
One cause for concern is that our calculation assumes a fixed scalar field background $\Psi_0$, but as the binary evolves
scalar radiation is depleting the NBS of scalar surrounding the binary. Assume, conservatively, that the flux above is only removing scalar field within a sphere of radius $\sim 10 \,\ell$
centred at the binary, with the radiation wavelength $\ell=2\pi/\omega_{\rm orb}$. Then the timescale for total depletion of the scalar in the sphere is
\begin{equation}
 \tau \sim \frac{\rho R^3}{\dot{E}^{\rm rad}}\sim 10^{24}\,{\rm yr}\, \left[\frac{10^{-2}}{\mu M_{\rm NBS}}\right]^{2/3}\left[\frac{10^4}{\chi}\right]^{2/3} \left[\frac{20\,M_{\odot}}{m_p}\right]^{2}\left[\frac{10^{-22}{\rm eV}}{\mu}\right]^{11/6}\left[\frac{T}{16 {\rm yr}}\right]^{7/6}\,,
\end{equation}
that is much larger than the Hubble timescale. A similar value can be found for equal mass binary systems. Thus, our results seem to indicate that the background configuration remains unaffected by the emission of scalars by low frequency binaries.
\section{High-energy binaries within boson stars}\label{section:High-energy_binaries_within_boson_stars}
In this final Section, we now wish to focus on rapidly moving binaries, such as those suitable for LIGO or LISA sources. In such a situation, the non-relativistic regime is not appropriate. Instead, one can show that the relevant description of these systems, for which the frequencies involved $\omega_{\rm orb}\gg \mu$, is accounted for by a slight modification of the previous equations(see Appendix~\ref{app:Newtonian} for details),
\begin{align}
\nabla^2\delta U &= 4\pi P\,,\nonumber\\
-\partial^2_t \delta \Phi +\nabla^2 \delta \Phi &=2 \mu^2 \Phi\, \delta U\,.\label{eq:high_binary}
\end{align}
%

\subsection{Scalar emission close to coalescence}\label{subsection:Scalar_emission_close_to_coalescence}
%
\begin{figure}[ht]
\centering
\includegraphics[width=0.6\textwidth,keepaspectratio]{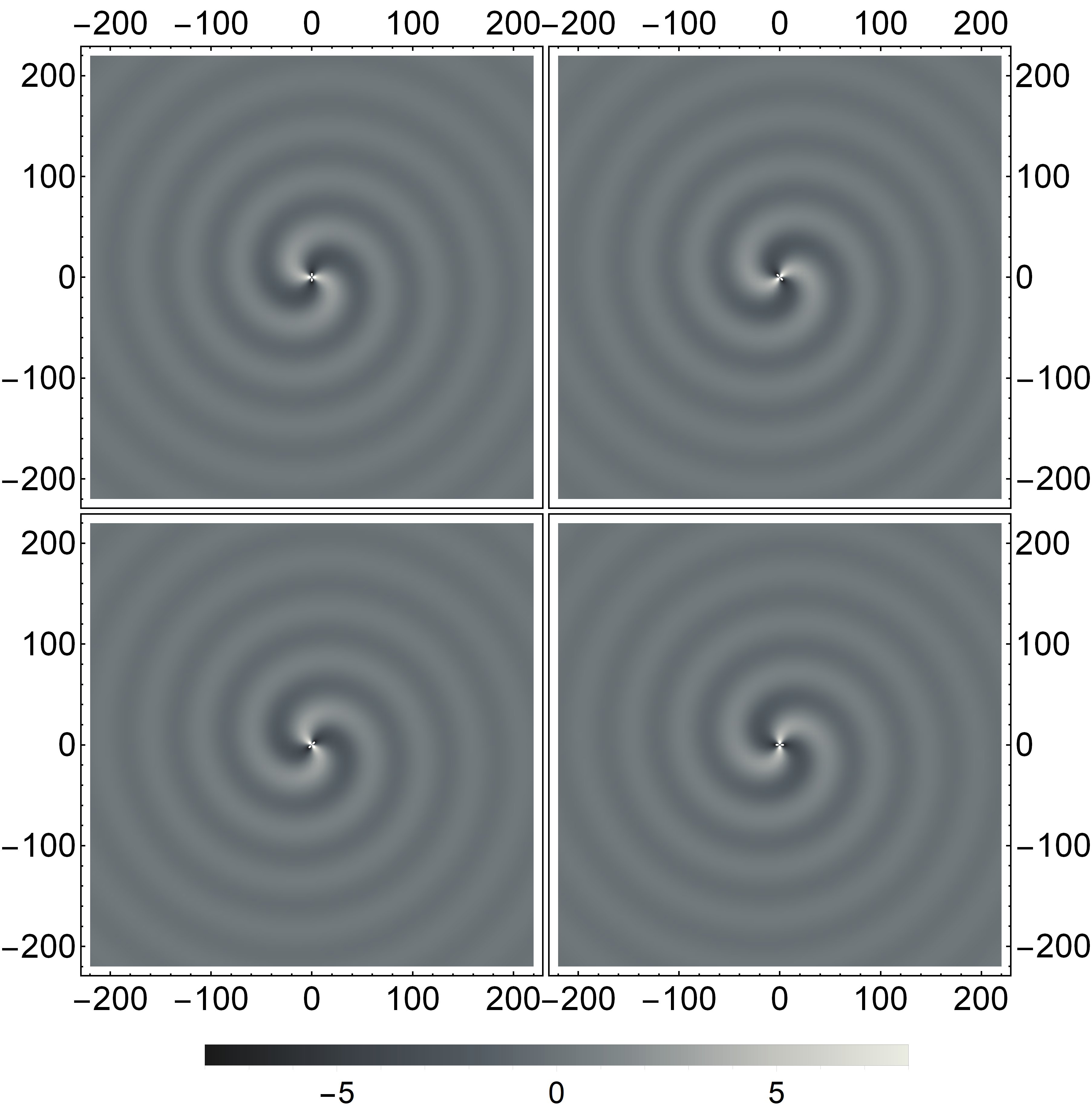}
\caption[Scalar emitted by an equal mass binary in an NBS.]{Scalar field emission from a high energy, equal-mass binary describing a circular orbit of radius $r_{\rm orb}$, evolving inside an NBS. The axis are the normalized $x/r_{\rm orb}, y/r_{\rm orb}$ respectively, and each frame represents an equatorial slice of the scalar field perturbation $10^{17} {\rm Re}\left[\delta\Phi\right]$, induced by a binary orbiting in the equatorial plane. In the upper-left panel, particles are at $(x_1,y_1)=(r_{\rm orb},0)$, $(x_2,y_2)=(-r_{\rm orb},0)$. Moving clockwise in the panels, the system evolves for an eighth of a period between each, the binary moving anti-clockwise. The binary components have the same mass, ($m_p\sim 10^6 M_\odot$) and they are orbiting inside an NBS of mass $M_{\rm NBS}\mu\sim 0.01$ with a period of $\sim 1$ day.
}
\label{fig:highfreq_timedomain}
\end{figure}

We consider two equal-mass point particles, each of mass $m_p$, on a circular motion of orbital frequency $\omega_{\rm orb}$
and radius $r_{\rm orb}$. We can solve the Poisson equation first, using a multipolar decomposition. We find
\begin{align}
U&=\sum_{lm}\frac{u_{lm}}{r}Y_{l}^m (\theta,0)e^{im\left(\phi-\phi_0\right)}\,,\\
u_{lm}&=-\frac{4\pi m_p\left(1+(-1)^m\right)\,Y_{l}^m(\pi/2,0)}{2l+1} r_{\rm orb}^{-l-1}\left[r_{\rm orb}^{2l+1}r^{-l}\Theta(r-r_{\rm orb})+r^{l+1}\Theta(r_{\rm orb}-r)\right]\,.
\end{align}
Here $\phi_0=\omega_{\rm orb}t$ is the azimuthal location of one particle; the other is at $\phi_0+\pi$. Again, if the factor $m_p\left(1+(-1)^m\right)$ is replaced by $m_p$ this same source describes a single point particle of mass $m_p$.
We now perform a Fourier transform and a multipolar decomposition of the scalar to solve Eq.~\eqref{eq:high_binary}:
\be
\delta \Phi=\frac{1}{\sqrt{2\pi}}\sum_{l,m}\int d\omega\, \frac{\delta\psi (\omega, r)}{r}e^{-i(\omega+\Omega) t} Y_{l}^m\,.
\label{eq:fourier_deltaphi}
\end{equation}
Hence, we find the following ODE for $\delta\psi$:
\be
\delta\psi''+\left((\Omega+\omega)^2-\frac{l(l+1)}{r^2}\right)\delta\psi=\sqrt{8\pi}\mu^2\Psi_{0}\,u_{lm}\delta\left(\omega-m \omega_{\rm orb}\right) \,,\nonumber
\end{equation}
Here primes stand for radial derivatives. We can now solve this using variation of constants, requiring outgoing waves at large distances and regularity at
the origin. The solution is
\begin{align}
\delta\psi&=\delta\psi_\infty \int_0^r \frac{2\sqrt{2\pi}\mu^2\Psi_{0}\,\delta\psi_H\,u_{lm}\delta\left(\omega-m \omega_{\rm orb}\right)}{i\omega}\nn\\
&+\delta\psi_H \int_r^\infty \frac{2\sqrt{2\pi}\mu^2\Psi_{0}\,\delta\psi_\infty\,u_{lm}\delta\left(\omega-m \omega_{\rm orb}\right)}{i\omega}\,,
\label{eq:totaldeltapsi_highfreq}
\end{align}
where $\omega=m\omega_{\rm orb}$ and $\delta\psi_{H,\infty}$ are homogeneous solutions,
\begin{align}
\delta\psi_{H}&=\sqrt{\frac{\pi\omega r}{2}}J_{l+1/2}(\omega r)\,,\\
\delta\psi_{\infty}&=\sqrt{\frac{\pi\omega r}{2}}\left(J_{l+1/2}(\omega r)+iY_{l+1/2}(\omega r)\right)\,.
\end{align}
 The time domain response of the NBS to the perturbations induced by a binary BH system is found solving Eq.~\eqref{eq:totaldeltapsi_highfreq} and \eqref{eq:fourier_deltaphi}. Four snapshots of one period, for two equal mass BHs are shown in Fig.~\ref{fig:highfreq_timedomain}.

A binary deep inside the NBS ($r_{\rm orb}\ll R$) and with large orbital frequency~($\omega_{\rm orb}\ll 1/r_{\rm orb}$) generates a field at large distances that is independent on the size of the NBS: the integration converges a few wavelengths
away from the binary. We find the following simple result for the dominant $l=m$ modes:
\begin{equation}
\delta\psi(r\to \infty)=i\sqrt{2\pi} m_p\left(1+(-1)^m\right)\Psi_0 \pi^{3/2}\,2^{2-m}m^{m-2} \frac{Y_{m}^m(\pi/2,0)}{\Gamma[m+3/2]}\frac{\mu^2}{\omega_{\rm orb}^2}(M\omega_{\rm orb})^{m/3}\,e^{i\omega r}\,.
\end{equation}
Here, $M=2m_p$ for the equal-mass binary. If we substitute $m_p\left(1+(-1)^m\right)\to m_p$, these results also describe an EMRI, where a single particle of mass $m_p$ is revolving around a massive BH of mass $M$ (note the crucial difference that $l=m=1$ modes are radiated for EMRIs, whereas  only even modes are emitted for equal-mass binaries).
The flux is given by
\begin{align}
\dot{E}^{\rm rad}&=- r^2 \lim_{r \to \infty}\int d\theta d\varphi \sin \theta \,\delta T^S_{t r}\nonumber \\
&=128 \pi^{3}(\mu^2 m_p \Psi_0(0))^2\left(1+(-1)^m\right)^2 \sum_{m=1}^{+\infty}\left(\frac{Y_m^m(\pi/2,0)}{\Gamma(m+3/2)} \frac{m^{m-1}(M\omega_{\rm orb})^{m/3}}{2^{m+1}\,\omega_{\rm orb}}\right)^2\,.\label{eq:energy_loss_high_binaries}
\end{align}
Since we are considering high-energy excitations of the scalar ($\omega_{\rm orb} \gg \mu$) it is easy to see that the rate of change of the NBS energy~$\dot{E}_{\rm NBS}$ is much smaller than~$\dot{E}^{\rm rad}$ (note that, at leading order, $\dot{E}_ {\rm NBS}=\mu\, \dot{Q}_ {\rm NBS}=-r^2 \lim_{r \to \infty}\int d\theta d\varphi \sin \theta \,\delta j_{r}\,.$)
so, conservation of energy (as expressed in Eq.~\eqref{LossRad}) implies that~$\dot{E}^{\rm lost}\simeq \dot{E}^{\rm rad}$.

\subsection{The phase dependence in vacuum and beyond}\label{subsection:The_phase_dependence_in_vacuum_and_beyond}
In vacuum GR, the dynamics of a binary is governed by the energy balance equation \eqref{eq:energy_balance},
together with the quadrupole formula \eqref{eq:quadrupole}.
This implies that the orbital energy of the system $E_{\rm orb}=-M^2\eta/(2r_{\rm orb})$ must 
decrease at a rate fixed by such loss. This defines immediately the time-dependence of the GW frequency to be~$f^{-8/3}=(8\pi)^{8/3}{\cal M}^{5/3}(t_0-t)/5$, where ${\cal M}$ is the chirp mass and $f=\omega_{\rm orb}/\pi$. Once the frequency evolution is known, the GW phase simply reads
\begin{equation}
\varphi(t)=2\int^t\Omega(t')dt' \,.\label{GWphase}
\end{equation}

To take into account dissipative losses via the scalar channel, we add to the quadrupole formula the energy flux~\eqref{eq:energy_loss_high_binaries}.
In the Fourier domain we can write the gauge-invariant metric fluctuations as
\begin{align}
 h_+(t)&=A_+(t_{\rm ret})\cos\varphi(t_{\rm ret}) \,,\\
 h_\times(t)&=A_\times(t_{\rm ret}) \sin\varphi(t_{\rm ret})\,,
\end{align}
where $t_{\rm ret}$ is the retarded time. The Fourier-transformed quantities are
\begin{equation}
 \tilde{h}_+= {\cal A}_+e^{i\Upsilon_+},\,\,\,  \tilde{h}_\times={\cal A}_\times e^{i\Upsilon_\times}\,.
\end{equation}
Dissipative effects are included within the stationary phase approximation, where the secular time evolution is governed by 
the GW emission~\cite{Flanagan:1997sx}. In Fourier space, we decompose the phase of the GW signal 
$\tilde{h}(f)={\cal A}e^{i\Upsilon(f)}$ as:  
\begin{equation}
 \Upsilon(f) =\Upsilon_{\rm GR}^{(0)}[1+{\rm (PN\ corrections)}+\delta_{\Upsilon}]\,,
\end{equation}
where $\Upsilon_{\rm GR}^{(0)}=3/128 ({\cal M}\pi f)^{-5/3}$ represents the leading term of the 
phase's post-Newtonian expansion, and $f=\omega_{\rm orb}/\pi$. We find the following dominant correction due to the background scalar,
\be
\delta_{\Upsilon}=\frac{16\mu^4\Psi_0^2}{51\pi^3f^4}\sim 10^{-24}\left[\frac{\mu}{10^{-22}\,{\rm eV}}\right]^4\left[\frac{10^{-4}{\rm Hz}}{f}\right]^4\left[\frac{M_{\rm NBS}\mu}{0.01}\right]^4\,,
\end{equation}
for equal-mass binaries. Such a correction corresponds to a $-6$ PN order correction~\cite{Yunes:2016jcc}. The smallness of the coefficient makes it hopeless to detect with space-based detectors. Note that pulsar timing arrays operate at lower frequencies~\cite{Barack:2018yly}, and the previous Newtonian non-relativistic analysis is necessary. However, on a positive note, the range of ultralight scalar masses may be lower than the specific value used through this Chapter ($\sim 10^{-22} {\rm eV}$). For instance, for larger haloes and scalar masses, we can find  
\be
\delta_{\Upsilon}=\frac{16\mu^4\Psi_0^2}{51\pi^3f^4}\sim 10^{-8}\left[\frac{\mu}{10^{-19}\,{\rm eV}}\right]^4\left[\frac{10^{-4}{\rm Hz}}{f}\right]^4\left[\frac{M_{\rm NBS}\mu}{0.1}\right]^4\,,
\end{equation}
that is more likely to be in the sensitivity interval of future space-based detectors\til\cite{Audley:2017drz}.
\subsection*{Backreaction and scalar depletion}
During the evolution, the binary emits scalar radiation away from the NBS. Assuming, again, that the above flux is only removing scalar field within a sphere of radius $\sim 10 \,\ell$
centred at the binary, with the radiation wavelength $\ell=2\pi/\omega_{\rm orb}$. Then the timescale for total depletion of the scalar is
\be
\tau\sim 2\times 10^{11}\,{\rm yr}\,\left(\frac{0.1}{m_p\omega_{\rm orb}}\right)^{7/3} \left(\frac{10^{-2}}{\mu M_{\rm NBS}}\right)^{2}\left(\frac{\chi}{10^4}\right)^2\nn\\
 \frac{m_p}{10^6M_{\odot}}\,,\nonumber
\end{equation}
larger than a Hubble timescale, even for binaries close to coalescence. Thus, our results seem to describe emission of scalars during the entire lifetime of a compact binary.

\section{Conclusions}\label{section:Outlook_of_Part_ I}
We showed how self-gravitating NBSs respond to time-varying, localized matter fluctuations.
These are structures that behave classically: they are composed of $N ~\sim 10^{100}\left(10^{-22}{\rm eV}/\mu\right)^2$ particles.  A binary of two supermassive BHs in the late stages of coalescence emits more than $10^{60}$ particles. Our results show some of the main astrophysical features associated with bosonic ultralight structures. For example, they are not easily depleted by binaries. Even a supermassive BH binary close to coalescence would need a Hubble time or more to completely deplete the scalar in a sphere of ten-wavelength radius around the binary. We have also shown how a consistent, self-gravitating NBS background leads to regular, finite dynamical friction acting on passing bodies, contrasting with previous calculations using 
infinite non-self-gravitating distributions~\cite{Hui:2016ltb}.

Clearly, our results can and should be extended to eccentric motion, or to self-gravitating vectorial configurations or even other non-linearly interacting scalars~\cite{Coleman:1985ki}. Our computations should also be a useful benchmark for numerical relativity simulations involving boson stars in the extreme mass ratio regime, when and if the field is able to accommodate such challenging setups. We have considered Newtonian boson stars. Extension of our results to relativistic boson stars is non-trivial, but would provide a full knowledge of the spectrum of boson stars and of their response to eternal agents. Our methods can also be extended to clouds arising from superradiant instabilities of spinning BHs~\cite{Brito:2015oca}. We don't expect qualitatively new aspects when the spatial extent of those clouds is large.

Some of the ideas exhibited in this Chapter can be of direct interest also for theories with screening mechanisms, where new degrees of freedom -- usually scalars --
are screened, via non-linearities, on some scales~\cite{Babichev:2013usa}. Such mechanisms do give rise to non-trivial profiles for the new degrees of freedom, for which many of the tools we use here should apply (see also Ref.~\cite{Brito:2014ifa}).



  \cleardoublepage
\epigraphhead[450]{{\it This part is based on Refs.\til\cite{Annulli:2018quj,Annulli:2021dkw}}}
\part{Aspects of gravitational-wave generation and propagation}

  \cleardoublepage

	\chapter{Scattering processes}
\label{chapter:scattering_processes}

\minitoc

\begin{figure}
\begin{center}
\includegraphics[width=0.6\textwidth]{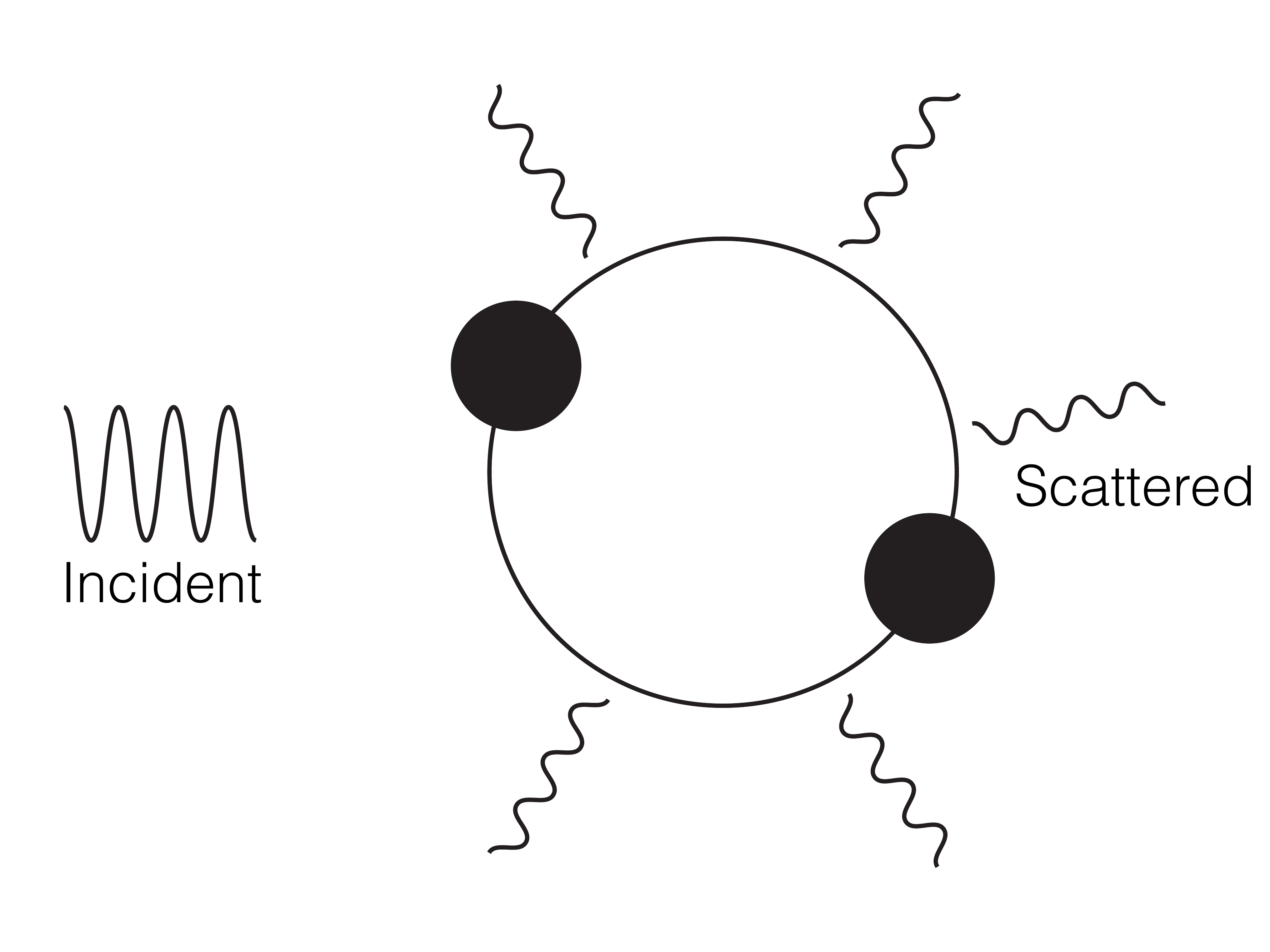}
\caption[Scattering setup.]{Scattering of an incoming wave by a binary. The wave affects the motion of the binary, which in turn re-radiates and contributes to a non-trivial scattered wave.
\label{fig:scatter}}
\end{center}
\end{figure}

The effects of the environment on the {\it propagation} of GWs are usually believed to be negligible. If the medium filling the cosmos is modelled as a perfect fluid, then GWs do not couple to it and are neither absorbed nor dispersed by such surroundings~\cite{Grishchuk:1981fp,Deruelle:1984hq}. 
These calculations have been redone for viscous fluids, for some particle dark matter models~\cite{Baym:2017xvh,Flauger:2017ged} and for more promising results for dark matter models beyond the Weakly Interacting Massive Particle paradigm\til\cite{Dev:2016hxv,Cai:2017buj}. However, the scattering of radiation by gravitationally-bound binaries, such as the one depicted in the figure\til\ref{fig:scatter}, is an unexplored subject that could furnish, in principle, measurable observables. This scattering phenomenon is the gravitational counterpart of well known and observed
electromagnetic (EM) scattering phenomena, such as the Rayleigh scattering of light, responsible for blue skies. Since the ability to do precision measurements of incoming GWs has increased to unforeseen levels, in the following sections we wish to investigate further along these lines, explicitly evaluating the cross sections for these events.

In the first Chapter of this Part, we set the stage for the GW scattering process, defining the geometrical setup and, as leading examples, we illustrate the EM scattering on a dipole (2p), a system composed by two electric charges, and the case of the scattering of scalar waves.

\section{Geometrical conventions}\label{Geometrical conventions}

\begin{figure}
\begin{center}
\includegraphics[scale=0.22]{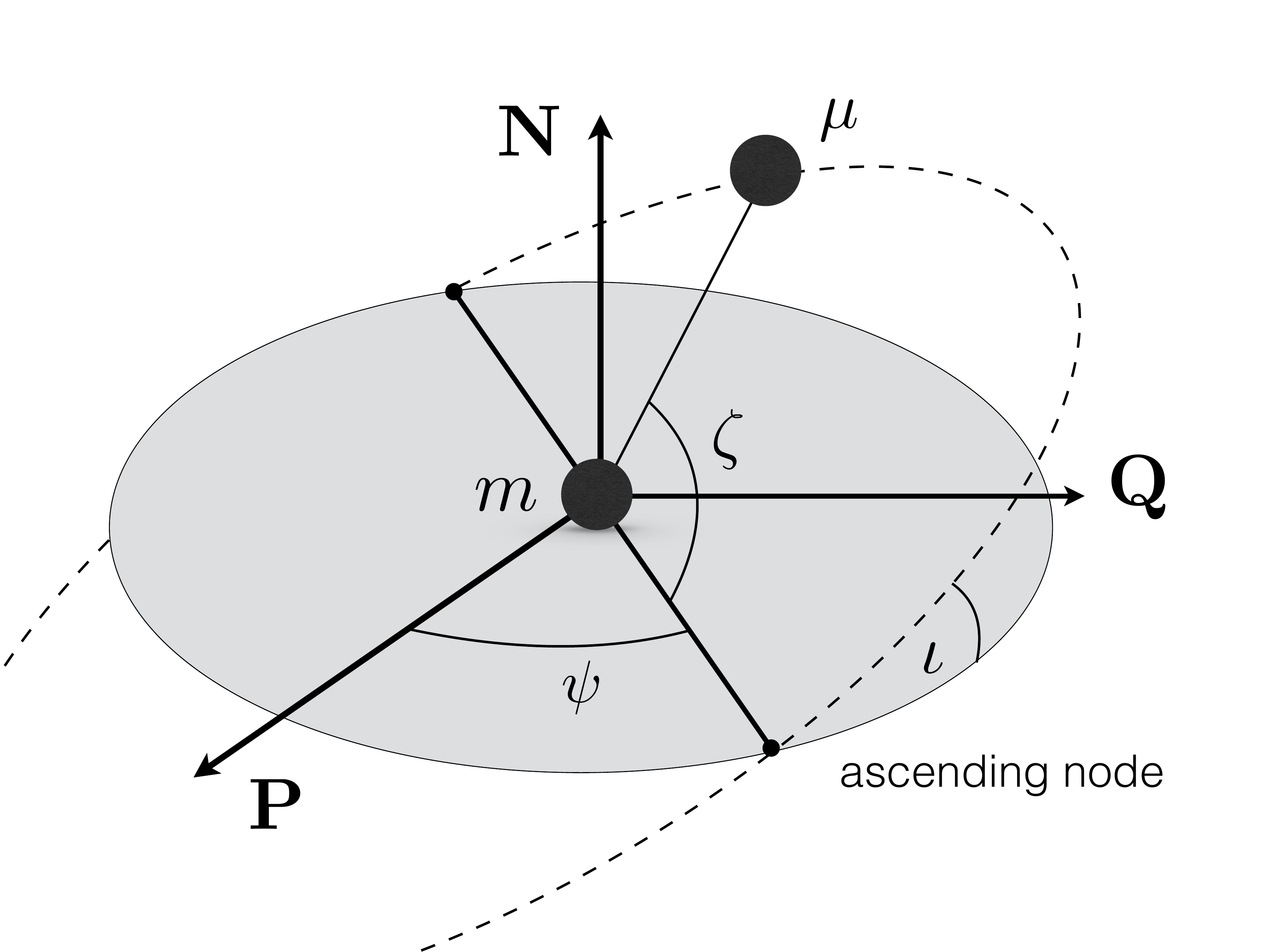}
\caption[Geometrical conventions.]{Plane of the orbit with respect to the fixed observer basis $(\mathbf{P},\mathbf{Q},\mathbf{N})$. The angle $\zeta$ is the polar angle describing the motion of the reduced mass $\mu$ in the orbital plane, while $\mathbf{L}$ and $\iota$ are respectively the total angular momentum and the angle between this vector and the direction $\mathbf{N}$. The total mass is denoted $m$.}
\label{PlaneOfTheOrbit}
\end{center}
\end{figure}

Our calculations and description of the problem involve specific but different frames. We will consider binaries, in which the motion of the individual bodies under central forces (EM or Newtonian) are described by ellipses. For a visual representation, we refer the reader to Fig.~\ref{PlaneOfTheOrbit}.

Consider an observer located along a direction $\mathbf{N}$, whose basis is $\left(\mathbf{P},\,\mathbf{Q},\,\mathbf{N}\right)$. This will be called the {\it frame of the observer} and it is fixed with respect to the observer itself. We choose as unit vector $\mathbf{P}$ the one that points toward the direction of the ascending node $\mathcal{N}$. For the Keplerian motion considered in the following, the ascending node is where the orbiting object moves north through the plane of reference. In general, in the presence of a perturbation, this freedom to choose the ascending node direction no longer exists.  Being interested on such perturbed setups, we choose to keep the basis $\left(\mathbf{P},\,\mathbf{Q},\,\mathbf{N}\right)$ in its unperturbed configuration. Additionally, we define $\psi$ as the angle between $\mathbf{P}$ and the ascending node $\mathcal{N}$; $\zeta$ the angle between the ascending node and the direction $\mathbf{n}$ and $\iota$ the angle between $\mathbf{N}$ and $\mathbf{L}$, where $\mathbf{L}$ is the angular momentum vector of the binary. 

The second frame will be the one that describes the motion of the reduced mass with respect to the center of mass (CM). This frame is defined with respect to the following directions: $\mathbf{n}$ is the radial direction with respect to the orbital motion, $\vec{\lambda}$ is the tangent one, while $\mathbf{l}$ is directed along the angular momentum direction $\mathbf{L}$. 

From classical mechanics, the following relations between the binary CM basis and the observer basis hold~\cite{PoissonWill2014}
\begin{align}\label{basis}
\mathbf{n}  = & \left(\cos\psi\,\cos\zeta-\sin\psi\,\cos\iota\,\sin\zeta\right)\mathbf{P}+\left(\sin\psi\,\cos\zeta+\cos\psi\,\cos\iota\,\sin\zeta\right)\mathbf{Q} +\sin\iota\,\sin\zeta\,\mathbf{N}, \nn \\
\boldsymbol{\lambda}  = & -\left(\cos\psi\,\sin\zeta+\sin\psi\,\cos\iota\,\cos\zeta\right)\mathbf{P} +\left(\cos\psi\,\cos\iota\,\cos\zeta-\sin\psi\,\sin\zeta\right)\mathbf{Q} +\sin\iota\,\cos\zeta\,\mathbf{N}, \nn\\
\mathbf{l} = & \sin\psi\,\sin\iota\,\mathbf{P}-\cos\psi\,\sin\iota\,\mathbf{Q}  +\cos\iota\,\mathbf{N}. 
\end{align}
Note that the unperturbed case corresponds to the configuration $\psi=0$ and $\iota=\text{const}$. In this configuration, the velocity in the CM frame is:
\begin{equation}\label{vitesse}
\mathbf{v} = \dot{r}\mathbf{n} +r\left(\dot{\zeta}+\dot{\psi}\cos\iota\right)\boldsymbol{\lambda} +r\left(\dot{\iota}\sin\zeta-\dot{\psi}\sin\iota\cos\zeta\right)\mathbf{l}\,,
\end{equation}
where $r$ is the relative position and the dot operator corresponds to a first time derivative. The frame $(\mathbf{n},\vec{\lambda},\mathbf{l})$, called CM frame in the rest of the Chapter, has time-varying basis with respect to the fixed observer frame. Lastly, we also introduce the proper frame of the wave $(\vec{e}_x,\vec{e}_y,\vec{e}_z)$, useful for the definition of the polarizations in both the EM and in the GR case. We denote $\alpha$ the angle between the $P$-axis and the ascending node $\mathcal{N}'$, $\beta$ the angle between the ascending node and $\mathbf{e}_{x}$ and $\kappa$ the angle between $\mathbf{e}_{z}$ and $\mathbf{N}$. We then have the following relations between the observer basis and the incoming GW basis:
\begin{align}\label{GWbasis}
\vec{e}_{x}  = & \left(\cos\alpha\,\cos\beta-\sin\alpha\,\cos\kappa\,\sin\beta\right)\mathbf{P}\left(\sin\alpha\,\cos\beta+\cos\alpha\,\cos\kappa\,\sin\beta\right)\mathbf{Q} +\sin\kappa\,\sin\beta\,\mathbf{N}, \nn \\
\vec{e}_{y}  = & \left(\cos\alpha\,\sin\beta+\sin\alpha\,\cos\kappa\,\cos\beta\right)\mathbf{P} \left(\sin\alpha\,\sin\beta-\cos\alpha\,\cos\kappa\,\cos\beta\right)\mathbf{Q} -\sin\kappa\,\cos\beta\,\mathbf{N}, \nn\\
\vec{e}_{z}= & -\sin\alpha\,\sin\kappa\,\mathbf{P} +\cos\alpha\,\sin\kappa\,\mathbf{Q} -\cos\kappa\,\mathbf{N}.
\end{align}
Figure \ref{PlaneOfTheOrbit} sketches the frame of the observer and of the CM.  

Finally, we will use the Keplerian parametrization of the orbit, and we perform an expansion for small eccentricities. Despite this, we will mostly concentrate on the zeroth order. Here is the parametrization we will use:
\begin{align}
\label{eqEllipserho}
r &= a\left(1-e\cos(u)\right)\,, \\
\label{eqEllipsephi}
\zeta &= v\equiv 2\arctan\left[\left(\frac{1+e}{1-e}\right)^{1/2}\tan\left(\frac{u}{2}\right)\right]\,, \\
\label{eqEllipsel}
l &\equiv n\left(t-t_{0}\right) = u-e\,\sin u \,,
\end{align}
where $a$ is the semi-major axis, $e$ is the eccentricity, $u$ and $v$ are respectively the eccentric and true anomaly, $l$ is the mean anomaly, $n$ is the mean motion and $t_{0}$ is the instant of passage at the perihelion. At Newtonian order we have that $n=\omega_0$, where $\omega_0$ is the orbital frequency of the binary system.

To avoid lengthy expressions, we will also often abbreviate the following trigonometric functions:
\begin{equation}\notag
\cos(\alpha+\beta)\equiv c_{\alpha+\beta} \text{ and }\sin(\alpha+\beta)\equiv s_{\alpha+\beta}.
\end{equation}
Furthermore, variables in bold are to be intended as vectors, while the corresponding normal ones are their corresponding magnitude. We use Greek letters to represent space-time indices and latin letters for 3-dimensional spatial indices. As the spatial indices are moved with the delta metric $\delta_{ij}$, we indifferently write them in a lower or upper position.

\section{Scattering of electromagnetic waves}\label{Scattering_EM}

We will start with an old and venerable problem, the scattering of EM waves off obstacles~\cite{Landau:1982dva}. This incursion will set the stage for both the scalar and gravitational case, while sharing some (many) features in common. We start by working out how an incoming EM wave affects a rotating dipole. This is a classical treatment that only requires linear perturbation theory. We use the change in the dipole moment induced by the incoming EM wave to compute the scattered radiation and the total scattering cross section. All these quantities are evaluated for an EM wave propagating along the direction of the observer and with the electric field oscillating in the plane of the orbit. In the high frequency limit, we recover classical results concerning scattering off oscillators. 

Specifically, we want to evaluate the effect of an incoming EM wave on a binary system of two electric charges orbiting at a frequency $\omega_0$. The monochromatic EM wave propagates along the $z$ direction and has a frequency $\Omega$.

\subsection{Unperturbed dipole physics}\label{unpertsdipolefield}

Consider a system of two charged particles, of mass $m_1$ and $m_2$, that interact through the product between the EM potential $A^{\mu}=(\Phi/c,\vec{A})$ and four-current $J_{\mu}=(c	\rho,\vec{j})$, where $\rho$ is the charge density ($\rho=q_i \delta^3(\vec{x}-\vec{x}_i)$) and $\vec{j}=\rho \vec{v}$ is the current density.
We take these charges to interact only through the Coulomb force. Consider a system of two charged particles in Minkowski flat spacetime with metric $\eta_{\mu\nu}=\mathrm{diag}(-1,1,1,1)$. Using $x^{\mu}=(x^0,x^1,x^2,x^3)=(c t,x,y,z)$ as coordinates, where $c$ is the speed of light in vacuum, the action that describes this system is
\begin{equation}
S =\int \mathrm{d}^3x \mathrm{d}t \left[-\frac{F_{\mu\nu}F^{\mu\nu}}{4\mu_0}-A_{\mu}^1 J_{\mu}^2-A_{\mu}^2 J_{\mu}^1\right]-c^2\int \mathrm{d}\tau \left(m_1+m_2\right)\,,\label{action1}
\end{equation}
in which $\mu_0$ is the magnetic vacuum permeability, $F_{\mu\nu}$ the antisymmetric EM tensor defined as $F_{\mu\nu}=\partial_{\mu}A_{\nu}-\partial_{\nu}A_{\mu}$ and $\mathrm{d}\tau=\mathrm{d}t\sqrt{1-v^2/c^2}$, where $v^2$ is the square of the three-velocity $v^i=\mathrm{d}x^i/\mathrm{d}t$. From now on, we restrict ourselves to the small velocities case, dropping all the special-relativistic terms. With all these assumptions, the $i$-th component of the equations of motion for each particles is
\begin{equation}\label{eoMunperturbed}
m_1\ddot{\vec{r}}_1^i=\frac{q_1 q_2 (\vec{r}_1-\vec{r}_2)^i}{ \lvert \vec{r}_1-\vec{r}_2 \lvert^3}, \;\;
m_2\ddot{\vec{r}}_2^i=\frac{q_1 q_2 (\vec{r}_2-\vec{r}_1)^i}{ \lvert  \vec{r}_2-\vec{r}_1 \lvert^3} \,,
\end{equation}
where $i=(1,2,3)$, $\vec{r}_{1(2)}$ represents the position vector of particle $1(2)$, $q_1,q_2$ are the electric charges and the double dot sign means a second derivative with respect to time $t$. In the CM frame, the CM vector position has zero second time derivative ($\ddot{\vec{R}}_{\rm CM}=0$), while, defining the relative position vector with respect to the radial direction defined in Sec.~\ref{Geometrical conventions} as $ \vec{r}\equiv \vec{r}_1-\vec{r}_2 =r \vec{n}$, the equations for the relative motion become
\begin{equation}
\ddot{\vec{r}}=\frac{1}{\mu}\frac{q_1 q_2}{ \mid \vec{r}\lvert^2}\vec{n},
\end{equation}
where $\mu$ is the reduced mass of the system,
\be
\mu=\frac{m_1 m_2}{m_1+m_2}\,.
\ee
We define the total mass as $m=m_1+m_2$.
Since the Coulomb force is central, the total angular momentum of the system is conserved and the motion happens on a fixed plane. The solution to the equations of motion, in analogy with the Newtonian ones, have the characteristic shape of a conic section, depending on the energy of the particles. Since we are interested in bound systems, we assume that the energy will be the one associated with bound orbits. 

We focus on the case in which the dipole is composed of two particles with equal and opposite charge and equal mass:
\begin{align}
-q_2&=q_1=q\,,\nonumber\\
m_1&=m_2=M\,.\label{charge_def}
\end{align}
From Eqs.\til\eqref{eoMunperturbed}, we find that the CM is fixed, the angular momentum of the system is constant and the motion lies in the orbital plane. The orbit of the binary can be directly obtained from
\begin{equation}
 \ddot{r}-\frac{4 L^2}{M^2 r^3}=-\frac{2q^2}{M r^2}\,,
 \end{equation}
where $L=M r^2\dot{\phi}/2$ is the magnitude of the angular momentum vector of the system and $\phi$ is the angle describing the motion of the reduced mass in the plane of the orbit (polar angle). Defining the dipole vector ($\vec{d}$) as
\begin{equation}
\vec{d}=q_1\vec{r}_1+q_2\vec{r}_2=\mu \left(\frac{q_1}{m_1}-\frac{q_2}{m_2}\right)\vec{r}\,,
\end{equation}
where $\vec{r}$ is the proper radius of the system (relative position vector in the dipole case). Introducing the vector between the CM and the observer, of magnitude $R_0$ and unit direction $\hat{\vec{R}}_0$, the generated EM wave has vector potential, electric field and magnetic one given by
\begin{equation}\label{EandMfields}
\vec{A}=\frac{1}{c R_0}\dot{\vec{d}},\,\vec{H}=\frac{1}{c^2 R_0}\ddot{\vec{d}}\times\hat{\vec{R}}_0,\, \vec{E}=\frac{1}{c^2 R_0}(\ddot{\vec{d}}\times\hat{\vec{R}}_0)\times\hat{\vec{R}}_0.
\end{equation}
This is a well-known result, a dipole emits only if it is accelerated. Finally, the expression for the intensity of the emitted energy is given by \cite{Landau:1982dva}:
\begin{equation}\label{genEmittDipole}
\mathrm{d}I=c\frac{H^2}{4\pi}R_0^2 \mathrm{d}o\rightarrow I=\frac{2}{3c^3}\ddot{d}^2,
\end{equation}
where we averaged over one period of the orbit and $\mathrm{d}o$ is the solid angle in the $\hat{\vec{R}}_0$ direction.

\subsection{Scattering from a rotating dipole}
The binary above is now hit by an EM wave described by a vector potential $A^{\mu}_{\Omega}$.
For definiteness, the wave propagates along the $\vec{e}_z$ axis, parallel to the direction $\mathbf{N}$ and to the angular momentum of the system $\vec{L}$. In this way, the $x$-$y$ plane of the orbital frame, of the observer and also of the wave are all parallel between each other and perpendicular to the $z$ direction in the observer frame. 

The action~\eqref{action1} needs to be complemented by adding both the scalar and the vector potentials of the perturbation,
\begin{align}
A^{\mu}_1 &\rightarrow A^{\mu}_1+A^{\mu}_{\Omega}=\left(\frac{\Phi_1}{c}+\frac{\Phi_{\Omega}}{c},\vec{A}_1+\vec{A}_{\Omega}\right),\\
A^{\mu}_2 &\rightarrow A^{\mu}_2+A^{\mu}_{\Omega}=\left(\frac{\Phi_2}{c}+\frac{\Phi_{\Omega}}{c},\vec{A}_2+\vec{A}_{\Omega}\right)\,.
\end{align}
Using the definitions of EM fields~\footnote{In order to pass to the old vectorial picture, in this section $\vec{B}=\mu_0 \vec{H}$ is the magnetic field in vacuum.} and potentials,
\begin{equation}
\vec{E}=-\vec{\nabla}\Phi -\frac{\partial}{\partial t}\vec{A}\;\;\; \text{ and }\;\;\; \vec{B}=\vec{\nabla}\times \vec{A}\,,
\end{equation}
one finds,
\begin{align}
m_1 \vec{a}_1&=q_1 \vec{E}_2+q_1 \vec{E}_{\Omega}+q_1 \vec{v}_1\times \vec{B}_2+q_1 \vec{v}_1\times \vec{B}_{\Omega}\,,\nn\\
\label{AppForEoMinEBecc}
m_2 \vec{a}_2&=q_2 \vec{E}_1+q_2 \vec{E}_{\Omega}+q_2 \vec{v}_2\times \vec{B}_1+q_2 \vec{v}_2\times \vec{B}_{\Omega}.
\end{align}
Dropping the last two terms of Eqs.~\eqref{AppForEoMinEBecc} by assumptions of small internal velocities compared to the speed of light, we get
\begin{align}
m_1 \vec{a}_1&=q_1 \vec{E}_2+q_1 \vec{E}_{\Omega},\nn\\
\label{EoMinEBecc}
m_2 \vec{a}_2&=q_2 \vec{E}_1+q_2 \vec{E}_{\Omega}.
\end{align} 
Finally, in the CM frame we have
\begin{align}
\ddot{\vec{R}}_{\rm CM}&=\frac{q_1+q_2}{m_1+m_2}(\vec{E}_{\Omega})_{\rm CM},\\
\label{CMacc}
\ddot{\vec{r}}&=\frac{1}{\mu}\frac{q_1 q_2}{ \lvert r\lvert^2}\vec{n}+\left(\frac{q_1}{m_1}-\frac{q_2}{m_2}\right)\left(\vec{E}_{\Omega}\right)_{\rm CM}.
\end{align}
where $\left(\vec{E}_{\Omega}\right)_{\rm CM}$, means that the quantity under consideration has to be properly expressed in the CM frame. Using the equations of motion~\eqref{EoMinEBecc} and transforming all the quantities in the CM frame, we find the total angular momentum variation in time,
\begin{equation}\label{L_TimeDerivative}
\frac{\mathrm{d} \vec{L}}{\mathrm{d} t}=
\mu \frac{2q}{M} \vec{r}\times\left (\vec{E}_{\Omega}\right)_{\rm CM}\,.
\end{equation}
Here, we used already the specific setup described by Eq.~\eqref{charge_def}.

\subsection*{Equations of motion}
As we have shown in Eq.~\eqref{L_TimeDerivative}, the time variation of the angular momentum is given by the cross product of the relative position vector and the external perturbing force $\vec{F}_{\Omega}$,
\begin{equation}
\dot{\vec{L}}\sim\vec{r}_{12}\times \vec{F}_{\Omega}\,.
\end{equation}
An electric field on the plane of the orbit changes the magnitude of the angular momentum, but not its direction. We should highlight that this simplification still captures the dynamics of the scattering, allowing us to give an analytic treatment of the process. In order to further simplify our calculations, we consider the unperturbed motion happening in circular orbits. Therefore, the equations that describe the perturbation of such kind of trajectory are given by
\begin{align}
 \ddot{r}-r\dot{\phi}^2&=-\frac{2 q^2}{M r^2}+\frac{ q E_{\Omega}}{ M}\left(c_{\gamma-\Omega t-\phi(t)}+c_{\gamma+\Omega t-\phi(t)}\right),\\
\label{system1tan}
 2 \dot{r}\dot{\phi}+r \ddot{\phi}&=\frac{q E_{\Omega}}{ M} \left(s_{\gamma-\Omega t-\phi(t)}+s_{\gamma+\Omega t-\phi(t)}\right),
\end{align}
where $\gamma$ is the angle between the direction of polarization of the electric field and the $P$ direction, in the plane of the orbit. 
Here, and in this section only, $\phi$ is the polar angle describing the orbital motion in the $x$-$y$ plane. The constant $E_\Omega$ is the amplitude of the electric field.
The presence of the perturbation in the right hand side of the second equation spoils the constancy of the angular momentum but, at first order in $E_{\Omega}$, one can find the relation linking $\dot{\phi}$ and $L$. Let us write,
\begin{equation}
2 \dot{r}\dot{\phi}+r \ddot{\phi}\equiv\frac{2}{M}\frac{1}{r}\frac{\mathrm{d}}{\mathrm{d} t}\left(\frac{M}{2}r^2\dot{\phi}\right)=\frac{2}{M}\frac{1}{r}\frac{\mathrm{d}}{\mathrm{d} t}L(t),\label{Kepler_system}
\end{equation}
where $L(t)$ is the angular momentum magnitude. Since without any external perturbation the angular momentum is conserved (and equal to a constant $L_{\circ}$), we can expand $L(t)$ in powers of the electric field
\begin{equation}
L(t)=L_{\circ}+E_{\Omega} L_1(t)+{\cal O}\left(E_{\Omega}^2\right)\,.
\end{equation} 
Making use of this, a similar expansion for $r(t)$ and for $\phi(t)$ can be found,
\begin{align}
\label{power in r}
r(t)&=r_{\circ}+E_{\Omega} g(t)+{\cal O}\left(E_{\Omega}^2\right)\,,\\
\label{power in phi}
\phi(t)&=\phi(0)+t \, \dot{\phi}=\phi(0)+t\left(\omega_0+E_{\Omega} Z_p+{\cal O}\left(E_{\Omega}^2\right)\right)\,,
\end{align}
where $r_{\circ}$ is the orbital radius of the unperturbed motion, $\phi(0)=\phi_0$ is the initial angular position of the reduced mass in the $x$-$y$ plane and $Z_p$ is the first order correction in the orbital frequency due to the external perturbation. Using Eqs.~\eqref{system1tan} we find
\begin{equation}\label{Ldot}
\dot{L}_1(t)=\frac{q\, \, r_{\circ}}{2} \left(s_{\gamma_0-\phi_0-t \Omega - t\omega_0}+s_{\gamma_0-\phi_0+t \Omega - t\omega_0}\right)\,.
\end{equation}
where we kept only the zero order in the $\phi(t)$ expansion because $\dot{L}_1$ is already a first order quantity. For the unperturbed circular motion, $\dot{\phi}=\omega_0$ is constant. Thus, integrating Eq.~\eqref{Ldot} with $\phi(t)=\omega_0 t$, one finds,
\begin{align}
L_1(t)&=\int_0^t \mathrm{d}t'[\dot{L}_1(t')]\nonumber\\
&=q r_{\circ} \left(\frac{\omega_0 c_{\gamma-\phi_0}}{\Omega ^2-\omega_0^2}+\frac{c_{\gamma-\phi_0+t \Omega - t\omega_0}}{2 (\omega_0-\Omega )}+\frac{c_{\gamma-\phi_0-t \Omega - t\omega_0}}{2 (\omega_0+\Omega)}\right)\,.\nonumber
\end{align}
Finally, the total angular momentum to first order in the external field is
\begin{align}
L(t)&=L_{\circ}+E_{\Omega} L_1(t)\nonumber\\
&= L_{\circ}+\frac{E_{\Omega} q r_{\circ}\omega_0 c_{\gamma-\phi_0}}{\Omega^2-\omega_0^2}+\frac{E_{\Omega} q r_{\circ}}{2}\bigg[\frac{c_{\gamma-\phi_0+t \Omega - t\omega_0}}{\omega_0-\Omega}+\frac{c_{\gamma-\phi_0-t \Omega - t\omega_0}}{\omega_0+\Omega}\bigg].\label{L_dot}
\end{align}
From the definition of angular momentum, from Eqs.~\eqref{power in r} and \eqref{power in phi} and  up to ${\cal O}\left(E_{\Omega}^2\right)$,
\begin{align}
L(t)&=\frac{1}{2}M r(t)^2 \dot{\phi}(t)\nonumber\\
&=\frac{M}{2} r_{\circ}^2\omega_0+\left(\frac{M}{2}r_{\circ}(r_{\circ}Z_p+2\omega_0 g(t))\right)E_\Omega\,.
\end{align}
We can compare with Eq.~\eqref{L_dot} order by order, to get
\begin{align}
L_{\circ}=\frac{M}{2} r_{\circ}^2\omega_0\,,
\end{align}	
at order zero, and
\begin{equation}
Z_p=\frac{q}{M r_{\circ}}\left(\frac{2\omega_0 c_{\gamma-\phi_0}}{\Omega ^2-\omega_0^2}+\frac{c_{\gamma-\phi_0+t \Omega - t\omega_0}}{(\omega_0-\Omega )}+\frac{c_{\gamma-\phi_0-t \Omega - t\omega_0}}{(\omega_0+\Omega)}\right)-\frac{2 \omega_0 g(t)}{r_{\circ}}\,.
\end{equation}
Now that we have used the Keplerian polar equation to get the angular perturbation due to the incoming wave, we substitute this result in $\dot{\phi}^2$ in the radial equation \eqref{system1tan} in order to find the equation governing $g(t)$. Then substituting the expansions  given by Eqs.~\eqref{power in r}-~\eqref{power in phi},
\begin{equation}
\ddot{g}(t)-r_\circ(\omega_0+E_\Omega Z_p)^2=-\frac{2q^2}{M(r_\circ+E_\Omega g(t))}+\frac{ q E_{\Omega}}{ M}\left(c_{\gamma-\Omega t-\phi_0-\omega_0 t}+c_{\gamma+\Omega t-\phi_0-\omega_0 t}\right)\,,\label{g_of_t_equation}
\end{equation}
we get, at zero order in $E_\Omega$, the relation between the Newtonian orbital frequency and the characteristics of the binary,
\begin{equation}
\omega_0^2=\frac{2 q^2}{M r_{\circ}^3}\,.\label{omega_zero}
\end{equation}
Substituting $M$ obtained by the equation above in the first order expansion of Eq.~\eqref{g_of_t_equation}, we find a differential equation for $g(t)$.
\begin{align}
\ddot{g}(t)+\omega_0^2 g(t)+\frac{2r_{\circ}^3 \omega_0^4  c_{\gamma-\phi_0}}{q(\omega_0^2-\Omega^2)}=\frac{r_{\circ}^3 \omega_0^2 (\Omega-3\omega_0)c_{\gamma-\phi_0+t \Omega - t\omega_0}}{2q(\Omega-\omega_0)}+\frac{r_{\circ}^3 \omega_0^2 (\Omega+3\omega_0)c_{\gamma-\phi_0-t \Omega - t\omega_0}}{2q(\Omega+\omega_0)}
\,,\label{perturbationLaprroxint}
\end{align}
The equation above represents a driven harmonic oscillator with multiple resonant frequencies, whose solution is given by
\begin{align}
g(t)&=k_1 \cos(t\omega_0)+k_2 \sin(t\omega_0)\nn\\
&+\frac{r_\circ^3}{q}\left[\frac{2\omega_0^2  c_{\gamma-\phi_0}}{\Omega^2-\omega_0^2}-\frac{\omega_0^2 (\Omega-3\omega_0)c_{\gamma-\phi_0+t \Omega -t\omega_0}}{2\Omega(\Omega^2-3\Omega\omega_0+2\omega_0^2)}-\frac{\omega_0^2 (\Omega+3\omega_0)c_{\gamma-\phi_0-t \Omega -t\omega_0}}{2\Omega(\Omega^2+3\Omega\omega_0+2\omega_0^2)}\right].\label{gOftcircular}
\end{align}
in which $k_1$ and $k_2$ are integration constants. We set the two constants of integration to zero in the following. Finally, we can evaluate $Z_p$ considering the explicit solution for $g(t)$ given by Eq.~\eqref{gOftcircular} with $k_1=k_2=0$,
\begin{align}
Z_p&=\frac{r_\circ^2}{q}\bigg[\frac{3\omega_0^3 c_{\gamma-\phi_0}}{\omega_0^2-\Omega^2}-\frac{\omega_0^2 (\Omega^2-4\Omega\omega_0+6\omega_0^2)c_{\gamma-\phi_0+t\Omega-t\omega_0}}{2\Omega(\Omega-2\omega_0)(\Omega-\omega_0)}\nn\\
&+\frac{\omega_0^2 (\Omega^2+4\Omega\omega_0+6\omega_0^2)c_{\gamma-\phi_0-t\Omega-t\omega_0}}{2\Omega(\Omega+2\omega_0)(\Omega+\omega_0)}
\bigg]\,.\label{Zp}
\end{align}
The roots of the denominators in the solution for $g(t)$ are
\begin{align}
\bigg\{-2\omega_0,-\omega_0,0,\omega_0,2\omega_0\bigg\}\,.\nonumber
\end{align}
The negative values are solution because of the symmetry of the problem, but they are not adding any physics to the positive ones, so we will consider only $0,\omega_0,2\omega_0$. Let's evaluate the limit of $r(t)$ for these roots,
\begin{align}
\lim_{\Omega\rightarrow 0}r(t)&=r_{\circ}-\frac{2E_\Omega r_\circ^3 c_{\gamma-\phi_0}}{q}+\frac{E_\Omega r_\circ^3}{q}\bigg(\frac{7}{4}c_{\gamma-\phi_0-t \omega_0}\bigg)-\frac{E_\Omega r_\circ^3}{q}\bigg(\frac{3}{2}t \, \omega_0 s_{\gamma-\phi_0-t \omega_0}\bigg)\,,\nn\\
\lim_{\Omega\rightarrow \omega_0}r(t)&=r_{\circ}-\frac{E_\Omega r_\circ^3}{q}\bigg(\frac{1}{3}c_{\gamma-\phi_0- 2 t \omega_0}\bigg)+\frac{E_\Omega r_\circ^3}{q}\bigg(t\,\omega_0 s_{\gamma-\phi_0}\bigg)\,,\nn\\
\label{Omega_0limit}
\lim_{\Omega\rightarrow 2\omega_0}r(t)&=\infty\,.
\end{align}
In the high-frequency limit, the reasoning described before does not hold because the effect of the external field lives on a timescale much shorter than the one associated with the proper rotation of the binary, such that we can neglect the free motion of the system during one (or few) period of oscillation of the external electric field. So, we can just consider that 
\begin{align}
\lim_{\Omega\rightarrow \infty}r(t)=r_{\circ}\,.
\end{align}
From these results we see that resonant phenomena appear depending on the ratio between the incoming and the orbital frequency. Especially, in the $\Omega\rightarrow 0$ limit the radial motion of the reduced mass has a secular instability given by the last term of Eq.~\eqref{Omega_0limit}. This term can be understood thinking that the low frequency limit of our scattering corresponds to a perfect dipole inside a capacitor: in the large time limit, the two particles are dragged away from each other. In the $\Omega\rightarrow \omega_0$ case there is also such a secular term, but it can be set to zero with an appropriate choice of the initial condition. Finally, the $\Omega\rightarrow  2\omega_0$ case corresponds to a proper resonance, meaning that the amplitude of the motion for that value is infinite.

\subsection*{Scattered Fields}
Having solved the perturbed equations of motion, we can find the scattered electrical field, energy and the total cross section. 
When the system interacts with the external perturbation, the total field will contain a perturbed dipole term. In addition, the CM may contribute to the scattered field; we denote this contribution $\vec{E}_{\rm LW}$, where LW stands for Li\'enard–Wiechert,
\begin{equation}
\vec{E}_{\rm scattered}=\vec{E}_{2p}+\vec{E}_{\rm LW}\,.
\end{equation}
We express our results in the fixed observer frame, using
\begin{align}
\hat{\vec{R}}_0&=\cos\delta\cos\xi\,{\mathbf{P}}+\sin\delta\cos\xi\,{\mathbf{Q}}+\sin\xi\,{\mathbf{N}}\,,\\
\vec{n}(t)&=\cos{\phi(t)}\,{\mathbf{P}}+\sin\phi(t)\,{\mathbf{Q}}\,,\label{n_of_t}\\
\vec{\mathcal{E}}_{\Omega}(t)&=\cos\gamma\,{\mathbf{P}}+\sin\gamma\,{\mathbf{Q}}\,,
\end{align}
where $\delta$ and $\xi$ are the angles that characterize the position of the unitary vector $\hat{\vec{R}}_0$ with respect to the CM, in the fixed observer frame; $\phi(t)$ is given by Eqs.~\eqref{power in phi} and \eqref{Zp}, $\gamma$ is the direction of the linear polarization of the electric field in the orbital plane and $\vec{\mathcal{E}}_\Omega$ is the unitary vector in the direction of the external electric field ($\vec{E}_\Omega=E_\Omega \vec{\mathcal{E}}_\Omega$). Since we are considering the motion of a dipole in which the total charge is zero, the contribution from the CM acceleration is zero.

\subsection*{Radiation from the CM}
\label{Radiation from the CM}

The scalar and vector potentials produced by one accelerated particle are given by the Li\'enard–Wiechert potentials,
\begin{equation}
\Phi=\frac{q}{\left( {R}_0 -\frac{(\vec{v}\cdot \hat{\vec{R}}_0){R}_0}{c}\right)},\,\,\, \vec{A}=\frac{q \vec{v}}{c\left( {R}_0 -\frac{(\vec{v}\cdot \hat{\vec{R}}_0){R}_0}{c}\right)}\,.
\end{equation}
We can then find the electric and magnetic field for the accelerated charge in a relativistic context. For small velocities we get,
\begin{align}
\vec{E}&=\frac{q}{{R_0}^2}\hat{\vec{R}}_0+\frac{q}{c^2  {R_0}}\hat{\vec{R}}_0\times(\hat{\vec{R}}_0 \times \dot{\vec{v}})\,,\\
\vec{H}&=\frac{1}{ {R}_0}\hat{\vec{R}}_0\times \vec{E}\,.
\end{align}
Using Eq.~\eqref{CMacc}, we find
\begin{equation}
\vec{E}_{\rm LW}
=\frac{(q_1+q_2)^2 E_\Omega}{(m_1+m_2) \,c^2  {R}_0 }\hat{\vec{R}}_0\times (\hat{\vec{R}}_0 \times \vec{\mathcal{E}}_{\Omega})\label{ECM1p}\,.
\end{equation}
Finally, in the observer frame, one finds
\begin{align}
\vec{E}_{\rm LW}&=\frac{(q_1+q_2)^2 E_{\Omega} c_{\Omega t}}{(m_1+m_2) \,c^2 {R}_0 }\bigg[(-c_{\gamma} s^2_{\delta} c^2_{\xi }+s_{\gamma} s_{\delta} c_{\delta} c^2_{\xi}-c_{\gamma} s^2_{\xi}){\mathbf{P}}\nn\\
&+(s_{\gamma} \left(- c^2_{\delta}\right) c^2_{\xi}+c_{\gamma} s_{\delta} c_{\delta} c^2_{\xi}-s_{\gamma} s^2_{\xi}){\mathbf{Q}}+(c_{\gamma} c_{\delta} s_{\xi} c_{\xi}+s_{\gamma} s_{\delta} s_{\xi} c_{\xi}){\mathbf{N}}\bigg]\,.
\end{align}
As we can see from Eq.~\eqref{ECM1p}, this term is due to a non zero acceleration of the CM, that, naturally, depends on the wave perturbation. For two particles with opposite charge, the CM radiation is zero because of $q_1+q_2=0$, and the electric field will be only the one produced by the perturbed dipole.

The vector potential has contributions from the unperturbed dipole and a contribution from the perturbed part, induced by the incoming EM wave. Particularly, getting the electric field $\vec{E}$ from the vector potential $\vec{A}$, we find the same functional expression~\eqref{EandMfields}, but containing the acceleration of the dipole given by Eq.~\eqref{CMacc}. Therefore, using the definitions of dipole fields in Eq.~\eqref{EandMfields} and the expression for the radial separation \eqref{gOftcircular} we find,
\begin{align}
\vec{E}_{2p}(t)&=-\frac{ q r_{\circ}\omega_0^2}{c^2 R_0}\left[\left(\vec{n}(t)\times\hat{\vec{R}}_0\right)\times\hat{\vec{R}}_0\right]+\frac{E_{\Omega} r_{\circ}^3 \omega_0^2 c_{\Omega t}}{c^2 R_0}\left[ \left( \vec{\mathcal{E}}_{\Omega}\times\hat{\vec{R}}_0\right) \times \hat{\vec{R}}_0\right]\nn\\
&+\frac{4 E_{\Omega}  \omega_0^4 r_{\circ}^3 c_{\gamma-\phi_0}}{c^2 R_0 (\Omega^2-\omega_0^2)}\left[\left(\vec{n}(t)\times\hat{\vec{R}}_0\right)\times\hat{\vec{R}}_0\right]\nonumber\\
&-\frac{2 E_{\Omega} r_\circ^3 \omega_0^4}{c^2 R_0}\bigg[\frac{ (\Omega+3\omega_0)c_{\gamma-\phi_0-t\Omega-t\omega_0}}{2 \Omega (\Omega^2+3 \Omega \omega_0+2 \omega_0^2)}\nonumber\\
&+\frac{(\Omega-3\omega_0)c_{\gamma-\phi_0+t\Omega-t\omega_0}}{2 \Omega (\Omega^2-3 \Omega \omega_0 +2\omega_0^2)} \bigg]\left[\left(\vec{n}(t)\times\hat{\vec{R}}_0\right)\times\hat{\vec{R}}_0\right] \,.\label{E2pGeneral}\\
\vec{H}_{2p}(t)&=-\frac{ q r_{\circ}\omega_0^2}{c^2 R_0}\left[\vec{n}(t)\times\hat{\vec{R}}_0\right]+\frac{E_{\Omega} r_{\circ}^3 \omega_0^2 c_{\Omega t}}{c^2 R_0}\left[ \vec{\mathcal{E}}_{\Omega}\times\hat{\vec{R}}_0\right]+\frac{4 E_{\Omega}  \omega_0^4 r_{\circ}^3 c_{\gamma-\phi_0}}{c^2 R_0 (\Omega^2-\omega_0^2)}\left[\vec{n}(t)\times\hat{\vec{R}}_0\right]\nonumber\\
&-\frac{2 E_{\Omega} r_\circ^3 \omega_0^4}{c^2 R_0}\bigg[\frac{ (\Omega+3\omega_0)c_{\gamma-\phi_0-t\Omega-t\omega_0}}{2 \Omega (\Omega^2+3 \Omega \omega_0+2 \omega_0^2)}+\frac{(\Omega-3\omega_0)c_{\gamma-\phi_0+t\Omega-t\omega_0}}{2 \Omega (\Omega^2-3 \Omega \omega_0 +2\omega_0^2)} \bigg]\left[\vec{n}(t)\times\hat{\vec{R}}_0\right] \,,\label{H2pGeneral}
\end{align}
The first term describes the unperturbed dipole radiation, as we can see from a quick comparison with Eq.\til\eqref{EandMfields}. Once this term is expressed in the observer frame, $\vec{n}(t)$ also includes a term linear in the external perturbation, due to the first order Taylor expansion of the trigonometric functions in Eq.~\eqref{n_of_t}. The second term, that does not depend on $\vec{n}(t)$, is the only one that matters in the high frequency limit. The third term represents the modification to the dipole emission due to the external wave.

\subsection*{Cross section}

\begin{figure}
\centering
\includegraphics[width=0.6\textwidth]{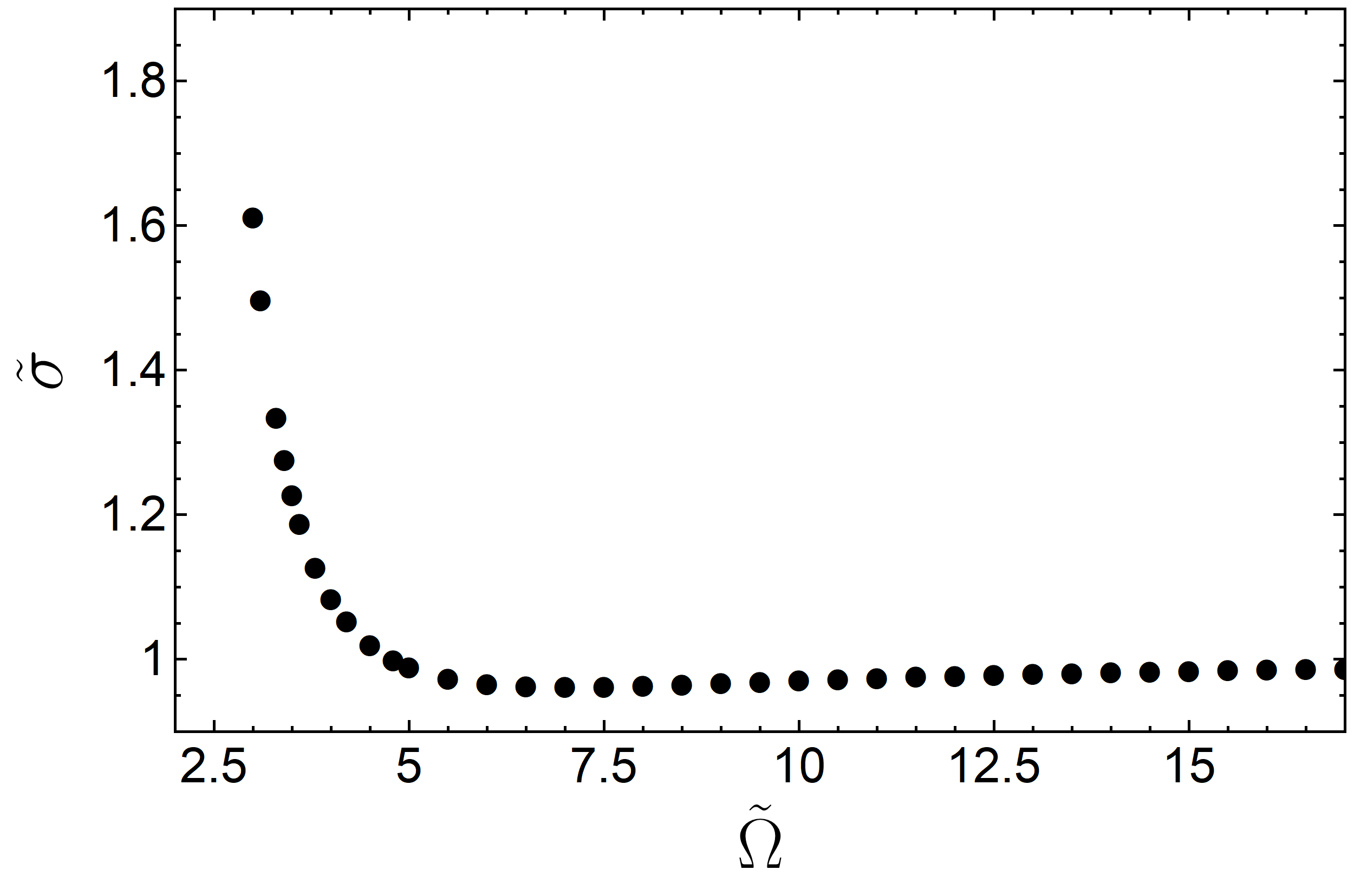}
\caption[Cross section GW scattering.]{Total scattering cross section for a dipolar wave as a function of the external frequency. Here, $\tilde{\sigma}\equiv \sigma/(A r_\circ^2)$ is shown as a function of $\tilde{\Omega}\equiv \Omega/\omega_0$. We set $c=1$ and the two angles $\gamma=\phi_0=0$. As expected, the cross section grows unboundedly for $\Omega=2\omega_0$. For large values of $\Omega$ we recover the standard high-frequency classical result~\eqref{EMOmegainfCrossSection}.}
\label{fig:cross_section}
\end{figure}

The scattering cross section is defined as the ratio between the energy emitted by the system in any given direction per unit of time, to the energy flux density of the incident radiation per unit of time. Considering that $\mathrm{d}I$ is the energy radiated per second by the binary into the solid angle $\mathrm{d}o$, we can define the differential cross section as
\begin{equation}
\mathrm{d}\sigma=\frac{\mathrm{d}I_{\rm scat}}{S_{\Omega}},
\end{equation}
where $S_\Omega$ is the modulus of the Poynting vector of the incoming wave.
Using the relation between intensity and Poynting vector and considering that the Poynting vector module is a time-varying quantity, we get
\begin{equation}
\frac{\mathrm{d}\sigma}{\mathrm{d}o} =\frac{\langle S_{\rm scat}\rangle R_0^2}{\langle S_{\Omega}\rangle},
\end{equation}
where the triangle brackets indicate a time average over one (or more) period and $\mathrm{d}o$ is the solid angle element given, with our choice of $\hat{\vec{R}}_0$, by
\begin{equation}
\mathrm{d}o=\cos\xi \mathrm{d}\xi \mathrm{d}\delta, \ \text{ with}\ \ \xi=[-\pi/2,\pi/2];\,\delta=[0,2\pi].
\end{equation}
In the high frequency limit, since the incoming wave is a monochromatic plane wave, its Poynting vector is
\begin{equation}
\vec{S}_{\Omega}=\left(\frac{c}{4\pi}E_{\Omega}^2 c^2_{\Omega t} \right){\mathbf{N}}\,.
\end{equation}
Its absolute value, averaged over one period of the EM wave ($2\pi/\Omega=T_{\Omega}$), is
\begin{equation}\label{Poynting_2p}
\langle S_{\Omega}\rangle=\frac{\Omega}{2\pi} \int_{T_{\Omega}}{S}_{\Omega} \mathrm{d}t= \frac{c E_{\Omega}^2}{8\pi}\,.
\end{equation}
To evaluate the Poynting vector of the scattered radiation we need to use the fields obtained in~\eqref{E2pGeneral} and~\eqref{H2pGeneral},
\begin{equation}
\langle S_{\rm scat}\rangle=\langle\frac{c}{4\pi}(\vec{E}_{2p}\times \vec{H}_{2p})\rangle=\frac{\Omega}{2\pi} \int_{T_{\Omega}}\frac{c}{4\pi}\lvert\vec{E}_{2p}\times \vec{H}_{2p}\rvert \mathrm{d}t\,.
\end{equation}
In the high frequency limit we can evaluate the differential scattering cross section using only the second term in Eq.~\eqref{E2pGeneral} and Eq.~\eqref{H2pGeneral},
\begin{equation}
\frac{\mathrm{d}\sigma}{\mathrm{d}o}=\left(\frac{q^2}{c^2 M}\right)^2 \left(2  c_{2 \xi} c^2_{\gamma -\delta}+c_{2 (\gamma -\delta )}-3\right)\,.
\end{equation}
The total high frequency scattering cross section is found by integrating the above, and yields
\begin{equation}\label{eq:thomson}
\sigma=\frac{32 \pi}{3}\left(\frac{q^2}{c^2 M}\right)^2\,,
\end{equation}
the standard Thomson result~\cite{Landau:1982dva}. Notice that $q^2/Mc^2$ is the classical charge radius. In the case of a circular orbit, the cross section in~\eqref{eq:thomson} can be given as a function of the unperturbed orbital frequency through~\eqref{omega_zero},
\begin{align}
\sigma=\frac{32 \pi}{3}\left(\frac{r_{\circ}^3 \omega_0^2}{2 c^2}\right)^2= A r_\circ^2\,\label{EMOmegainfCrossSection},
\end{align}
where $A=\frac{8 \pi}{3}\left(\frac{r_{\circ}^2 \omega_0^2}{c^2}\right)^2$. The total cross section for a wave with a generic frequency $\Omega$ is shown in Fig.~\ref{fig:cross_section} at $\Omega\gtrsim \omega_0$. 
From this plot, we can see that the cross section goes to infinity when the incoming frequency approaches twice of the orbital frequency and that in the high frequency limit it reaches the value given in Eq.~\eqref{EMOmegainfCrossSection}.

\section{Scattering of scalar waves}\label{Scattering of scalar waves}

For completeness, we now show that the previous results are straightforward to extend in the presence of a scalar interaction. Let's suppose that a binary system, made of two point-like scalar charges, interacts with a scalar wave and that is on a circular orbit of frequency $\omega_0$.

We consider the following action,
\begin{equation}
S=\int \mathrm{d}^4x \left[-\frac{1}{2}\partial^{\mu}\phi\partial_{\mu}\phi-\rho_q \phi\right]+S_m\,,
\end{equation}
where $\phi$ is the scalar field and $\rho_q$ is a general scalar charge density. The action for a free particle $S_m$ is as in \eqref{action1}. The scalar field is then governed by the Klein-Gordon equation
\begin{equation}\label{scalarfieldeq}
\partial^{\mu}\partial_{\mu} \phi=\rho_q\,.
\end{equation}
For a point-like charge $\rho_q=q\delta^3(\vec{x}-\vec{x}')  \frac{\mathrm{d}\tau}{\mathrm{d}t}$, in the non-relativist limit the static solution is given by the Green's function of the Laplace operator,
\begin{equation}
\phi(\vec{x})=-\frac{q}{4\pi}\frac{1}{\lvert\vec{x}-\vec{x}'\rvert}\,,\label{scalarpotential}
\end{equation} 
therefore, a Coulomb-like potential. From the Euler-Lagrange equation in a non-relativistic regime, one finds,
\begin{equation}
m_1 a_1^j=-q_1\left(\frac{\partial \phi_2}{\partial x_j}\right)_1,\,
m_2 a_2^j=-q_2\left(\frac{\partial \phi_1}{\partial x_j}\right)_2\,,
\end{equation}
in which the scalar field $\phi_i$ is the potential produced by the particle $i$.

Let's now consider the case in which a scalar wave impinges on the system. We call $\delta\phi=\phi_{\Omega}$ the perturbation in the scalar potential. The equations of motion are then altered to include the interaction of the binary with the wave,
\begin{align}
m_1 a_1^j&=-q_1\left(\frac{\partial \phi_2}{\partial x_j}\right)_1-q_1\left(\frac{\partial \phi_{\Omega}}{\partial x_j}\right)_1\,,\\
m_2 a_2^j&=-q_2\left(\frac{\partial \phi_1}{\partial x_j^2}\right)_2-q_2\left(\frac{\partial \phi_{\Omega}}{\partial x_j}\right)_2\,.
\end{align}
Using the explicit form of the potential \eqref{scalarpotential}, transforming the equations to the CM frame and assuming that the scalar field is homogeneous enough to be evaluated directly in the CM position, we find
\begin{align}
\ddot{\vec{R}}_{CM}&= -\left(\frac{q_1+q_2}{m}\right)\left(\frac{\partial \phi_{\Omega}}{\partial x_j}\right)_{CM}\,,\\
\ddot{\vec{r}}&= -\frac{1}{\mu}\frac{q_1 q_2}{ \lvert r\lvert^2}\vec{n} -\left(\frac{q_1}{m_1}-\frac{q_2}{m_2}\right)\left(\frac{\partial \phi_{\Omega}}{\partial x_j}\right)_{CM}\,.
\end{align}
These equations are formally equivalent to the EM counterpart, and no further calculation is necessary.

The existence of a background of light bosons may be motivated by the problem of dark matter, for instance (see e.g. \cite{Battaglieri:2017aum} and Chapter\til\ref{chapter:external_perturbers_NBS}). Their influence in binary systems was recently described in \cite{Blas:2016ddr,Boskovic:2018rub,LopezNacir:2018epg}. The scattering that we described here may be of interest to refine these studies.

  \cleardoublepage

\chapter{Scattering of gravitational waves by binary systems}\label{Scattering of gravitational wave}

\minitoc

Here we evaluate the effects of the scattering of a GW on a binary system, in the same fashion as the EM scattering depicted in Chapter\til\ref{chapter:scattering_processes}.

Broadly speaking, the scattering of GWs is suppressed by the smallness of the gravitational coupling constant. However, resonances between the impinging GW and a binary system may enhance the effects to measurable levels. This motivated a few studies in the past~\cite{Turner:1979yn,1978ApJ...223..285M}, both focusing on resonant interactions between a passing GW and a binary. The outcomes of this Chapter extend previous results in an important direction, by including dissipative terms and evaluating the scattered GW and the scattering cross section for two physical configurations: (i) for GWs propagating along the direction of the angular momentum of the system (i.e. oscillating in the orbital plane), and (ii) for GWs propagating perpendicularly to the angular momentum vector (i.e. GW travelling parallel to the orbital plane). 

The work developed here is also motivated by a study suggesting that the modes of oscillation of stars could be excited by passing GWs~\cite{McKernan:2014hha}. In fact, as it will be clear later, the master differential equation that rules these excitations is akin to our radial displacement in the binary, due to the incoming GW. Additionally, in both cases one may get resonances induced by the scattering process. In the single star case there are also reasons to expect that the GW signal can be the source of measurable deviations in the acoustic oscillations of the stars~\cite{Lopes:2014dba}. A still open issue concerns the study of scattering on binaries made of two stars. In this peculiar case, if the frequency of the excited mode is comparable with the proper orbital frequency, the scattering process can leave a signature both on the binary as a whole and on each single compact body in the couple.

\subsection*{The incoming gravitational wave}

Let us start defining the incoming GW properties. To coherently perform the PN expansion, in the following we restore the units of $G$ and $c$. We consider an incoming monochromatic GW propagating in the $z$ direction of a fixed basis $(\vec{e}_{x},\vec{e}_{y},\vec{e}_{z})$, at a frequency $\Omega$ (the employed geometrical conventions are described in detail in Section\til\ref{Geometrical conventions}). Additionally, the binary system is made of two compact stars (or BHs) modelled by two point particles of mass $m_1$ and $m_2$ orbiting at an orbital frequency $\omega_0$. In the transverse traceless (TT) gauge, the waveform is
\begin{equation}\label{Hij}
H_{ij} = \mathcal{P}_{ij}^{kl}\left\{H_{+}\,\cos\left(\Omega\,t -k\,z\right)\,e_{kl}^{+}+H_{\times}\,\sin\left(\Omega\,t -k\,z\right)\,e_{kl}^{\times}\right\} \,,
\end{equation}
where $e_{ij}^{+}=\vec{e}_{x}\otimes\vec{e}_{x}-\vec{e}_{y}\otimes\vec{e}_{y}$ and $e_{ij}^{\times}=\vec{e}_{x}\otimes\vec{e}_{y}+\vec{e}_{y}\otimes\vec{e}_{x}$ are the two polarization states of the GW~\cite{Maggiore:1900zz}, $\mathcal{P}_{ijkl}=P_{ik}P_{jl}-\frac{1}{2}P_{ij}P_{kl}$ is the TT projection operator, with $P_{ij}=\delta_{ij}-N_{i}N_{j}$ the projection onto the plane orthogonal to $\mathbf{N}$. As the wave is a solution of
\begin{equation}
\square H_{ij}=0,\,
\end{equation}
we have,
\begin{equation}
k=\pm\frac{\Omega}{c}\,,
\end{equation}
and $H_{+}=const$ and  $H_{\times}=const$. As a consequence, the spatial derivatives of $H_{ij}$ will be suppressed, $\partial_{k}H_{ij}=\mathcal{O}\left(\frac{1}{c}\right)$ (homogeneity condition), in particular $\partial_{j}H_{ij}=0$ (transverse), and we can assume that in the near zone of the compact binary one has $H_{ij}\sim (H_{ij})_{A}\sim (H_{jk})_{\mathrm{CM}}$. These conditions will be used in the following when we will perform a perturbative expansion of the solution.

\section{Post-Newtonian formalism}

To deal with the equations of motion for two point-like masses on a bound orbit, we shall treat the GW perturbation within a PN framework. This procedure highlights the non-linear character of the Einstein equations, but, given the relatively small spacetime curvatures in the system, allow us for a perturbative treatment. For GWs which are homogeneous on length-scales larger than the characteristic orbital distance between the masses, we find the same equations of motion as those described by Turner~\cite{Turner:1979yn} and Mashhoon~\cite{1978ApJ...223..285M}. Later, using an angle-action formalism to treat the variation of the orbital parameters, we find that the changes in the orbital parameters are linear in the incoming GW. 
Likewise, resonances between the binary and the incoming GW happen at certain discrete GW frequencies (integer multiples of the proper orbital frequency), in agreement with previous literature~\cite{Turner:1979yn,1978ApJ...223..285M}. 

\subsection{Einstein's equations}
We want to solve Einstein's equations
\begin{equation}
G^{\mu\nu} = \frac{8\pi G}{c^4} T^{\mu\nu}\,,
\end{equation}
where $G^{\mu\nu}$ is the Einstein tensor and $T^{\mu\nu}$ is the stress-energy tensor for point particles 
\begin{equation}\label{stress-energy tensor}
T^{\mu\nu}=\sum_{A=1,2}\frac{m_A}{\sqrt{-g}}\frac{v_A^{\mu}v_A^{\nu}}{\sqrt{-\left(g_{\rho\sigma}\right)_{A}\frac{v_{A}^{\rho}v_A^{\sigma}}{c^2}}}\delta^{(3)}(\mathbf{x}-\mathbf{x}_A)\,,
\end{equation}
where $g$ is the determinant of the metric $g_{\mu\nu}$. Defining the gothic metric $\mathfrak{g^{\mu\nu}}=\sqrt{-g}g^{\mu\nu}$ and the tensor $H^{\mu\alpha\nu\beta}=\mathfrak{g}^{\alpha\beta}\mathfrak{g}^{\mu\nu}-\mathfrak{g}^{\alpha\nu}\mathfrak{g}^{\beta\mu}$, we have the well-known identity~\cite{PoissonWill2014}
\begin{equation}\label{HtoG}
\partial_{\alpha\beta}H^{\mu\alpha\nu\beta} = (-g)\left(2G^{\mu\nu}+\frac{16 \pi G}{c^4} t^{\mu\nu}_{\mathrm{LL}}\right)\,,
\end{equation}
where $t^{\mu\nu}_{\mathrm{LL}}$ is the Landau-Lifshitz tensor~\cite{PoissonWill2014}. Next we define the gravitational field
\begin{equation}\label{def gothic}
l^{\mu\nu}=\mathfrak{g}^{\mu\nu}-\eta^{\mu\nu},
\end{equation}
where $\eta^{\mu\nu}=\mathrm{diag}(-1,1,1,1)$ is the Minkowski metric and we impose the harmonicity condition on the metric perturbation $l^{\mu\nu}$,
\begin{equation}
\partial_{\nu}l^{\mu\nu} = 0\,.
\end{equation}
Using~\eqref{HtoG}, we can rewrite the field equations as
\begin{equation}\label{relaxedEEqinH}
\square l^{\mu\nu}=\frac{16 \pi G}{c^4}T^{\mu\nu}+\Lambda^{\mu\nu}\,,
\end{equation}
where
\begin{equation}
\Lambda^{\mu\nu} = \frac{16 \pi G}{c^4} t^{\mu\nu}_{\mathrm{LL}} + \partial_{\rho}l^{\mu\sigma}\partial_{\sigma}l^{\nu\rho} -l_{\rho\sigma}\partial_{\rho\sigma}l^{\mu\nu}\,,
\end{equation}
is at least quadratic in the gravitational field. In our case, $l^{\mu\nu}$ is formed by two different terms, the perturbation $H^{\mu\nu}$ due to the incoming GW that we are superimposing on the original unperturbed gravitational field $h^{\mu\nu}$. Thus, at linear order we can write
\begin{equation}\label{h and htt}
l^{\mu\nu}=h^{\mu\nu}+H^{\mu\nu}\,.
\end{equation}
Since $\square H^{\mu\nu} =0$, the field equations~\eqref{relaxedEEqinH} can be rewritten as a d'Alembertian equation for $h^{\mu\nu}$,
\begin{equation}\label{eqnonso}
\square h^{\mu\nu}=\frac{16 \pi G}{c^4}T^{\mu\nu}[m,h^{\mu\nu},H^{\mu\nu}]+\Lambda^{\mu\nu}[h^{\mu\nu},h^{\mu\nu}]+\Lambda^{\mu\nu}[h^{\mu\nu},H^{\mu\nu}]+\Lambda^{\mu\nu}[H^{\mu\nu},H^{\mu\nu}]\,.
\end{equation}
The last term in \eqref{eqnonso} can be neglected when considering only the dominant, linear order in $H^{\mu\nu}$ terms. 

\subsection{Post-Newtonian iteration}
We perform the post-Newtonian iteration of the field equations in harmonic coordinates in the near-zone of the isolated source. As we are only interested in the effect of the external perturbation on the binary dynamics, we only need the lowest order PN expansion. We parametrize the metric by the usual PN potentials, using the variable $h^{00ii}\equiv h^{00}+h^{ii}$,
\begin{align}\label{hmunuPot}
& h^{00ii}=-\frac{4V}{c^2} + \mathcal{O}(c^{-4})\,,\\
& h^{0i}=-\frac{4V^{i}}{c^3} + \mathcal{O}(c^{-5})\,,\\
& h^{ij}= \mathcal{O}(c^{-4})\,.
\end{align}
Each potential obeys a flat space-time d'Alembertian equation sourced by the lowest order potentials and by some matter energy density components. We get
\begin{align}\label{eqnsVVi}
& \square V = -4\pi G \Upsilon -H^{ab}\partial_{ab}h^{00ii} \,,\\
& \square V^{i} = -4\pi G \Upsilon^{i} -H^{ab}\partial_{ab}h^{00ii} +\frac{1}{c}\partial_{t}H^{ia}\partial_{a}h^{00ii} \,,
\end{align}
where we have defined
\begin{equation}
\Upsilon = \frac{T^{00}+T^{ii}}{c^{2}} \,, \qquad \text{ and } \quad \Upsilon^{i}=\frac{T^{0i}}{c} \,.
\end{equation}
The first terms in the r.h.s. of Eqs.~\eqref{eqnsVVi} are of compact support, while the other terms are of non-compact support. We solve these equations perturbatively and up to linear order in $H$. The zeroth order corresponds to the Newtonian term and the equation becomes $\Delta V = -4\pi G \Upsilon$, with $\Upsilon = m_{1}\delta_{1}+m_{2}\delta_{2}$, and thus
\begin{equation}
V_{N}=\frac{Gm_{1}}{r_{1}}+\frac{Gm_{2}}{r_{2}}\,.
\end{equation}
We now decompose $V$ in a Newtonian part and a contribution linear in $H$, $V=V_{N}+V_{h}$. Inserting it into Eq.~\eqref{eqnsVVi}, we find
\begin{equation}
V_{h} = \Delta^{-1}\left[-H^{ab}\partial_{ab}\left(\frac{Gm_{1}}{r_{1}}+\frac{Gm_{2}}{r_{2}}\right)\right]\,.
\end{equation}
Using the fact that $H^{ij}\sim(H^{ij})_{CM}$, we see that the inverse Laplacian will not act on $H$. Further commuting it with the spatial derivatives, we get
\begin{equation}\label{Vh}
V_{h} = \frac{Gm_{1}}{r_{1}}H_{ij}n_{1}^{i}n_{1}^{j}+\frac{Gm_{2}}{r_{2}}H_{ij}n_{2}^{i}n_{2}^{j}\,.
\end{equation}
The potential $V^{i}$ can be obtained in a similar way.

\subsection{Geodesic equation}
The geodesic equations for point-particles is equivalent to the conservation of the matter stress-energy tensor, $\nabla_{\nu}T^{\mu\nu}=0$. We express the resulting equations for particle 1 as~\cite{Blanchet:2013haa}
\begin{equation}
\frac{\mathrm{d}\left(P^{i}\right)_{1}}{\mathrm{d}t} = \left(F^{i}\right)_{1} \,,
\end{equation}
with
\begin{align}
& P^{i} = \frac{g^i_{\mu}v^{\mu}}{\sqrt{-g_{\rho\sigma}v^{\rho}v^{\sigma}}}\,, \\
& F^{i} = \frac{1}{2}\frac{\partial^{i}g_{\mu\nu}v^{\mu}v^{\nu}}{\sqrt{-g_{\rho\sigma}v^{\rho}v^{\sigma}}}\,.
\end{align}
Using the expression of the metric as a function of the potentials, see Eq.~\eqref{hmunuPot}, we obtain at linear order in $H$
\begin{align}
P_{1}^{i} &= v_{1}^{i} -v_{1}^{j}(H^i_{j})_{1},\\
F_{1}^{i} &= \frac{1}{2}v_{1}^{j}v_{1}^{k}(\partial_{i}H_{jk})_{1} +(\partial^{i}V)_{1} \,.
\end{align}
Using the relation $V=V_{N}+V_{h}$, with $V_{h}$ given by Eq.~\eqref{Vh}, we finally obtain the acceleration of particle $1$,
\begin{equation}\label{eqacceleration}
a^i_1=-\frac{G m_2 }{r_{12}^{2}}\left(1+\frac{3}{2}H_{jk}n_{12}^{j}n_{12}^{k}\right)n_{12}^{i} +\frac{\mathrm{d}H_{ij}}{\mathrm{d}t}v_1^{j}\,.
\end{equation}
Here we ignored all higher order post-Newtonian corrections, since they will be sub-leading in the computation of the cross-section.

\subsection{Lagrangian formulation}
The equations of motion~\eqref{eqacceleration} can be derived from the Lagrangian
\begin{equation}
L =\frac{G m_1 m_2}{r_{12}}\left(1+\frac{1}{2}H_{ij}n_{12}^{i}n_{12}^{j}\right) +\frac{1}{2}m_1 v_1^2-\frac{1}{2}m_1H_{ij}v_1^i v_1^j +\frac{1}{2}m_2 v_2^2-\frac{1}{2}m_2H_{ij}v_2^i v_2^j\,.\label{lagrangian}
\end{equation}
Varying the Lagrangian with respect to the velocities, we obtain the linear momentum
\begin{equation}
P^i\equiv \sum_{A=1,2}\frac{\delta L}{\delta v_{A}^{i}}= m_1	v_1^i -m_1 H_{ij} v^j_1+m_2 v_2^i -m_2 H_{ij} v^j_2.\nonumber
\end{equation}
It is possible to see that the time-derivative of the momentum is zero. Then we get the energy associated with the binary motion,
\begin{align}
E &\equiv \sum_{A=1,2}\frac{\delta L}{\delta v_{A}^{i}}v_{A}^{i}-L\\
&=-\frac{G m_1 m_2}{r_{12}}\left(1+\frac{1}{2}H_{jk}n_{12}^{j}n_{12}^{k}\right)+\frac{1}{2}m_1 v_1^2-\frac{1}{2}m_1 H_{ij} v^i_1 v^j_1+\frac{1}{2}m_2 v_2^2-\frac{1}{2}m_2 H_{ij} v^i_2 v^j_2\,.
\end{align}
Similarly the angular momentum $J^i$ is given by
\begin{align}
J^i&\equiv \epsilon^{i}_{jk}\sum_{A=1,2}x_{A}^{j}\frac{\delta L}{\delta v_{A}^{k}}\nonumber\\
&=\epsilon_{ijk}\left[m_1\left(x_1^j v_1^k - H_{kl}x_1^j v_1^l\right)+m_2\left(x_2^j v_2^k - H_{kl}x_2^j v_2^l\right)\right]\,,
\end{align}
where $\epsilon^{i}_{jk}$ is the Levi-Civita tensor. Finally, we define the CM integral
\be
G^{i} = m_{1}x_{1}^{i} - m_{1}H^i_{j}x_{1}^{j} + [1\leftrightarrow 2]\,.
\ee
The conservation laws associated with all these quantities are
\begin{align}
\frac{\mathrm{d}P^{i}}{\mathrm{d}t}  &= 0 \,,\\
\frac{\mathrm{d}G^{i}}{\mathrm{d}t}  &= P^{i} - m_{1}\left(\dot{H}^i_{j}\right)_{1}x_{1}^{j}- m_{2}\left(\dot{H}^i_{j}\right)_{2}x_{2}^{j}\,,\\
\frac{\mathrm{d}E}{\mathrm{d}t} &= \frac{1}{2} m_1\left(\dot{H}_{ij}\right)_{\mathrm{CM}}v_{1}^{i}v_{1}^{j} +\frac{1}{2} m_2\left(\dot{H}_{ij}\right)_{\mathrm{CM}}v_{2}^{i}v_{2}^{j}-\frac{Gm_{1}m_{2}}{2r_{12}}\left(\dot{H}_{ij}\right)_{\mathrm{CM}}n_{12}^{i}n_{12}^{j}\,,\\
\frac{\mathrm{d}J^{i}}{\mathrm{d}t} &= \epsilon_{ijk}\bigg[-m_1v_{1}^{j}\left(H_{km}\right)_{\mathrm{CM}}v_{1}^{m} -m_2v_{2}^{j}\left(H_{km}\right)_{\mathrm{CM}}v_{2}^{m}+\frac{Gm_{1}m_{2}}{r_{12}}n_{12}^{j}\left(H_{km}\right)_{\mathrm{CM}}n_{12}^{m} \bigg]\,.
\end{align}
Unlike the Newtonian result, these quantities are not conserved, due to the incoming GW; the only conserved quantity here is the momentum $P^i$. 

We now wish to work in the CM coordinates. We define the total mass $m$, the symmetric mass ratio $\nu$ and the relative position $x^i$ and velocity $v^i$ as
\begin{align}
m&=m_1+m_2\,,\\
\nu&=\frac{\mu}{m}=\frac{m_{1}m_{2}}{m^{2}}\,,\\
x^{i}&=x_{1}^{i}-x_{2}^{i},\,\quad r=\vert\mathbf{x}\vert\,,\\
v^{i}&=v_{1}^{i}-v_{2}^{i}\,,\ a^{i}\equiv a_{1}^{i}-a_{2}^{i}\,.  
\end{align}
The CM coordinates are then obtained by solving the equation
\be
G^{i}=0 \,.
\ee
It implies the well-known Newtonian results, that are still valid at linear order in $H_{ij}$,
\begin{align}
& x_{1\,,\mathrm{CM}}^{i}=\frac{m_{2}}{m}x^{i},\,   x_{2\,,\mathrm{CM}}^{i}=-\frac{m_{1}}{m}x^{i},\\
& v_{1\,,\mathrm{CM}}^{i}=\frac{m_{2}}{m}v^{i},\,   v_{2\,,\mathrm{CM}}^{i}=-\frac{m_{1}}{m}v^{i}\,.
\end{align}
In the CM coordinates, the relative acceleration is given by
\begin{equation}
a^{i} = -\frac{Gm}{r^{2}}\left( 1+\frac{3}{2}H_{jk}n^{j}n^{k} \right)n^{i} +\dot{H}_{ij}v^{j}\,,
\end{equation}
and the conservation laws are now
\begin{align}
\frac{\mathrm{d}P^{i}}{\mathrm{d}t}  &= 0 \,,\\
\frac{\mathrm{d}E}{\mathrm{d}t} &=  \frac{1}{2} m\nu\dot{H}_{ij}v^{i}v^{j} -\frac{Gm^{2}\nu}{2r}\dot{H}_{ij}n^{i}n^{j} \,,\\
\frac{\mathrm{d}J^{i}}{\mathrm{d}t}& = \epsilon_{ijk}m\nu\bigg[ \frac{Gm}{r}n^{j}H_{km}n^{m}
-v^{j}H_{km}v^{m} \bigg] \,.
\label{ConsLaw}
\end{align}
%
\section{Hamiltonian formulation and angle-action variables}

To understand the gravitational problem, we will follow a different route from that we used in the EM example.
We will use an approach based on angle-action variables. The dynamics of a Keplerian orbit in Delaunay variables is well known, and we will use the powerful tool of perturbation theory in angle-action variables to describe the evolution of the perturbed system~\cite{BinneyTremaine}. The advantages of such an approach is that the calculations are simpler, notably because they capture the symmetries of the system. By promoting the integrals of motion to coordinate variables in the phase space, the dynamics of the system becomes very simple, as we will see. In particular, it allows a simple treatment of generic orbits and of the resonances that occur in such systems. However, in this work we will focus mostly on circular orbits and resonances will be absent from the final result. In Appendix~\ref{angleaction}, we review the Hamiltonian in the Delaunay variables, and explain the basics of perturbation theory in angle-action variables.

\subsection{Hamiltonian in the modified Delaunay variables}
The first step is to determine the Hamiltonian from the perturbed Lagrangian, and then to express it as a function of the modified Delaunay variables $\left(\theta_{1,2,3},\,J_{1,2,3}\right)$ (see Appendix~\ref{angleaction1}). We start from the reduced perturbed Lagrangian in the CM coordinates, in spherical coordinates,
\begin{align}
\tilde{L} &=  \frac{G m}{r}\left[1+\frac{1}{2}\,H_{ij}n^{i}n^{j}\right] + \frac{1}{2}\,\dot{r}^{2}\left[1-H_{ij}n^{i}n^{j}\right]+ \frac{1}{2}\,r^{2}\,\dot{\theta}^{2}\left[1-H_{ij}\theta^{i}\theta^{j}\right] \nonumber \\
&+ \frac{1}{2}\,r^{2}\,\sin^{2}\theta\,\dot{\varphi}^{2}\left[1-H_{ij}\varphi^{i}\varphi^{j}\right] - r\,\dot{r}\,\dot{\theta}\,H_{ij}n^{i}\theta^{j} - r\,\sin\theta\,\dot{r}\,\dot{\varphi}\,H_{ij}n^{i}\varphi^{j}- r^{2}\,\sin\theta\,\dot{\theta}\,\dot{\varphi}\,H_{ij}\theta^{i}\varphi^{j} \,,
\end{align}
we derive the conjugate momenta $p_{x}=\partial\tilde{L}/\partial\dot{x}$,
\begin{align}
p_{r}&=\dot{r}\left[1-H_{ij}n^{i}n^{j}\right] - r\,\dot{\theta}\,H_{ij}n^{i}\theta^{j} - r\,\sin\theta\,\dot{\varphi}\,H_{ij}n^{i}\varphi^{j} \,,\nonumber\\
p_{\theta}&=r^{2}\dot{\theta}\left[1-H_{ij}\theta^{i}\theta^{j}\right] - r\,\dot{r}\,H_{ij}n^{i}\theta^{j} - r^{2}\,\sin\theta\,\dot{\varphi}\,H_{ij}\theta^{i}\varphi^{j} \,,\nonumber\\
p_{\varphi}&=r^{2}\sin^2\theta\,\dot{\varphi}\left[1-H_{ij}\varphi^{i}\varphi^{j}\right] - r\,\sin\theta\,\dot{r}\,H_{ij}n^{i}\varphi^{j}-r^{2}\,\sin\theta\,\dot{\theta}\,H_{ij}\theta^{i}\varphi^{j} \,,
\end{align}
and then the reduced perturbed Hamiltonian
\begin{align}
\mathcal{\tilde{H}} & \equiv p_{r}\dot{r}+p_{\theta}\dot{\theta}+p_{\varphi}\dot{\varphi}-\tilde{L} \nonumber\\
& =-\frac{G m}{r}\left[1+\frac{1}{2}\,H_{ij}n^{i}n^{j}\right] + \frac{1}{2}p_{r}^{2}\left[1+H_{ij}n^{i}n^{j}\right]+ \frac{1}{2r^{2}}p_{\theta}^{2}\left[1+H_{ij}\theta^{i}\theta^{j}\right] \nonumber\\
& + \frac{1}{2r^{2}\sin^2\theta}p_{\varphi}^{2}\left[1+H_{ij}\varphi^{i}\varphi^{j}\right]+\frac{p_{r}\,p_{\theta}}{r}\,H_{ij}n^{i}\theta^{j} + \frac{p_{r}\,p_{\varphi}}{r\sin\theta}\,H_{ij}n^{i}\varphi^{j} +\frac{p_{\theta}\,p_{\varphi}}{\rho^{2}\sin\theta}\,H_{ij}\theta^{i}\varphi^{j}\,.
\end{align}
The perturbed Hamiltonian depends explicitly on time through the perturbation $H_{ij}$, given by Eq.~\eqref{Hij}.

The next step is to write the Hamiltonian as a function of the modified Delaunay variables. This can only be achieved with an expansion in the eccentricity $e$. In the following we will only consider the perturbation of a circular orbit. At this order we have that $J_{2}=0$ and $\theta_{2}$ is not defined. Our new set of variables is thus $\left\{\theta_{1,3},J_{1,3}\right\}$, and the relations between the old canonical variables and the angle-action ones are
\begin{equation}
r=\frac{J_{3}^{2}}{G m}\,, \theta=\frac{\pi}{2}-\arccos\left[1-\frac{J_{1}}{J_{3}}\right],\,
\varphi=-\theta_{1}+\theta_{3},\,
p_{r}=0,\, p_{\theta}=0\,, p_{\varphi}=J_{3}-J_{1}\,.
\end{equation}
The Hamiltonian that arises out of this procedure is
\begin{equation}
\tilde{\tilde{\mathcal{H}}}=-\frac{G^{2}m^{2}}{2J_{3}^{2}}\left[1+\left(H_{ij}n^{i}n^{j}\right)-\left(H_{ij}\lambda^{i}\lambda^{j}\right)\right] +\Omega\mathcal{T} \,,
\end{equation}
where we have introduced a new variable $\tau$ and its conjugate $\mathcal{T}$ to absorb the explicit dependence in time (cf. App.~\ref{angleaction}). In particular it depends not only on the actions but also on the angle variables. The dependence on the variable $\tau$ is only through the incoming GW and the only modes that contribute to the Fourier expansion are $k_{\tau}=\pm 1$. Then, as $\boldsymbol{\Omega}_{0}\left(\mathbf{J}\right) = \frac{\mathrm{d}\tilde{\tilde{\mathcal{H}}}_{0}}{\mathrm{d}\mathbf{J}} = \left(0,0,\frac{G^{2}m^{2}}{J_{3}^{3}},\Omega\right)$, we can see that the resonance occurs when
\begin{equation}
\Omega = \pm k_{3} n\,,
\end{equation}
where $n=\frac{G^{2}m^{2}}{J_{3}^{3}}$ is the orbital frequency of the binary and $k_{3}\in\mathbb{N}$.

We can also use the new set of angle-action variables $\left(\boldsymbol{\theta}',\,\boldsymbol{J}'\right)$, as constructed in Appendix~\ref{angleaction2}. The Hamiltonian for this set of variables is
\begin{equation}
\tilde{\mathcal{H}}'\left(\mathbf{J}'\right) = -\frac{G^{2}m^{2}}{{J'_{3}}^{2}} +\Omega\,\mathcal{T}' \,.
\end{equation}
The Hamilton equations are then
\begin{equation}
 \dot{\mathbf{J}'} = 0,\, \dot{\theta'_{1}} = 0,\, \dot{\theta'_{3}}=\frac{G^{2}m^{2}}{{J'_{3}}^{3}}  \text{ and }\quad \dot{\tau'} = \Omega \,.
\end{equation}

\subsection{Variation of the orbit elements}
We now relate the new set of variables to the orbit elements and obtain their evolution. From $J_{3}=\sqrt{Gma}$ (with $a$ the semi-major axis), we get
\begin{equation}
\frac{\mathrm{d}a}{\mathrm{d}t} = 2\sqrt{\frac{a}{Gm}}\frac{\mathrm{d}J_{3}}{\mathrm{d}t} = -2\sqrt{\frac{a}{Gm}}\frac{\partial\tilde{\mathcal{H}}_{1}}{\partial\theta_{3}}\,.
\end{equation}
From $J_{1}=J_{3}\left(1-\cos\iota\right)$ we get, using the previous relation,
\begin{equation}
\frac{\mathrm{d}\iota}{\mathrm{d}t} = \frac{\cos\iota-1}{\sin\iota}\frac{1}{a}\frac{\mathrm{d}a}{\mathrm{d}t}\,.
\end{equation}
We also have
\begin{equation}
\frac{\mathrm{d}\psi}{\mathrm{d}t} = -\frac{\mathrm{d}\theta_{1}}{\mathrm{d}t}\,, \quad \frac{\mathrm{d}\omega}{\mathrm{d}t} =-\frac{\mathrm{d}\psi}{\mathrm{d}t} \,, \text{ and }\quad \frac{\mathrm{d}l}{\mathrm{d}t} = \frac{\mathrm{d}\theta_{3}}{\mathrm{d}t}\,.\nonumber
\end{equation}
To obtain explicit results we specify to some specific configurations.

\subsection*{Parallel to the orbital plane: $\alpha=0$, $\beta=\pi/2$, $\kappa=\pi/2+\iota$}

We compute the variation of the energy, defined as $E\equiv H$, and get
\begin{equation}\label{dEdt1}
\frac{J_{3}^2}{G^2 m^2 \Omega}\frac{\mathrm{d}E}{\mathrm{d}t} = -\frac{H_{\times}}{2} \sin(2\iota) \cos(\Omega t) \sin(2\zeta)+\frac{H_{+}}{32} \sin(\Omega t) \left(\cos(4\iota)-17\right) \cos(2\zeta)\,.
\end{equation}
The variation of the semi-major axis is given by
\begin{equation}
\sqrt{\frac{a}{Gm}}\frac{\mathrm{d}a}{\mathrm{d}t} = -\frac{H_{+}}{8} \cos(\Omega t) \left(\cos(4\iota)-17\right)\sin(2\zeta)+ 2 H_{\times} \sin(2\iota) \sin(\Omega t)\cos(2\zeta)\,, 
\end{equation}
while the variation of the inclination angle is
\begin{equation}
\frac{a^{3/2}}{\sqrt{G m}}\frac{\mathrm{d}\iota}{\mathrm{d}t} = 2H_{\times} (1-\cos\iota)\cos\iota \sin(\Omega t) \cos(2\zeta)-\frac{H_{+}}{16} \cos(\Omega t) (\cos (4\iota)-17) \tan \left(\frac{\iota}{2}\right) \sin(2\zeta)\,.
\end{equation}
%

\subsection*{Perpendicular to the orbital plane: $\alpha=0$, $\beta=\pi/2$, $\kappa=\iota$}

The variation of the energy is
\begin{equation}\label{dEdt2}
\frac{J_{3}^2}{G^2 m^2 \Omega}\frac{\mathrm{d}E}{\mathrm{d}t} = -\frac{H_{+}}{16} \sin (\Omega t) (\cos (2\iota)+3)^2 \cos(2\zeta) - H_{\times} \cos(\Omega t) \cos^2\iota \sin(2\zeta)\,.
\end{equation}
The variation of the semi-major axis is given by
\begin{equation}
-\sqrt{\frac{a}{Gm}}\frac{\mathrm{d}a}{\mathrm{d}t} = 4H_{\times} \sin(\Omega t) \cos^2\iota\cos(2\zeta)+\frac{H_{+}}{4} \cos(\Omega t) (\cos (2\iota)+3)^2 \sin(2\zeta) ,
\end{equation}
and the variation of the inclination angle is
\begin{equation}
\frac{a^{3/2}}{\sqrt{G m}}\frac{\mathrm{d}\iota}{\mathrm{d}t} = 2H_{\times} \cos^{2}\iota \tan\left(\frac{\iota}{2}\right) \sin(\Omega t) \cos(2\zeta) +\frac{H_{+}}{8}\cos (\Omega t) (\cos (2\iota)+3)^2 \tan \left(\frac{\iota}{2}\right) \sin(2\zeta)\,.
\end{equation}

\section{Scattered gravitational wave}
The asymptotic waveform is given by~\cite{Blanchet:2013haa},
\begin{align}
l_{km}^{TT} &= \frac{2G}{c^2R}\mathcal{P}_{ijkm}\sum_{l=2}^{\infty}\frac{1}{c^l l!}\left\{N_{L-2}U_{ijL-2}\left(T-R/C\right)-\frac{2l}{(l+1)c}N_{aL-2}\varepsilon_{ab(i}V_{j)bL-2}\left(T-R/C\right)\right\}\nn\\
& + \mathcal{O}\left(\frac{1}{R^2}\right) \,,
\end{align}
where $\left(T,\,R\right)$ are the radiative coordinates. We recall that $\mathcal{P}_{ijkm}$ is the TT projection and the radiative moments $U_{L}$ and $V_{L}$ are related to the mass-type and current-type moments of the source. Here we suppose that the same relation still holds at linear order in $H_{ij}$, that is, $U_{L}(T)=M_{L}^{(l)}(T)$. The waveform is then given at our order by
\be\label{GWtotal}
l_{km}^{TT} = H^{\mathrm{TT}}_{km} + \frac{2G}{c^4R}\mathcal{P}_{ijkm}M_{ij}^{(2)} + \mathcal{O}\left(\frac{1}{R^2}\right)\,,
\ee
The contribution from the incoming GW, $H^{\mathrm{TT}}_{ij}$, has to be expanded at future null infinity, i.e. when $R\rightarrow+\infty$ keeping $T-\frac{R}{c}$ constant. We get\footnote{We did not prove this expression and only consider the general structure of a gravitational wave at infinity.}
\begin{equation}
\frac{2\Omega R}{c}H^{\mathrm{TT}}_{ij}=\mathcal{P}_{ijkl}\biggl[H_{+}e_{ij}^{+}\sin\left[\Omega \left(T-\frac{R}{c}\right)\right]-H_{\times}e_{ij}^{\times}\cos\left[\Omega \left(T-\frac{R}{c}\right)\right]\biggr]\,.
\end{equation}
Then, we have to link the canonical moments $M_{L}$ and $S_{L}$ to the real source moments $I_{L}$ and $J_{L}$, and then to figure out the expression of these source moments. We have the relation
\be
M_{L}(t)=I_{L}(t)+\delta I_{L}(t)\,,
\ee
In our case we are only interested in $M_{ij}$,  for which $\delta I_{ij}=0$. The source dipole moment is given by the usual formula,
\begin{equation}
I_{ij}(t) =\mathcal{FP}_{B=0}\int\mathrm{d}^{3}\mathbf{y}|\mathbf{y}|^{B}\left[\int_{-1}^{1}\mathrm{d}z \delta_{2}(z)\hat{y}_{ij}\bar{\Sigma}\left(\mathbf{y},t \right)\right] + \mathcal{O}({c^{-2}})\,,
\end{equation}
where $\mathcal FP$ is the finite part \cite{Blanchet:2013haa}. 
After some calculation we obtain,
\begin{align}
M_{ij} &= I_{ij} +\mathcal{O}\left(\frac{1}{c^2}\right)\,,\\
I_{ij} &= m_{1}y_{1}^{\langle ij\rangle} +m_{2}y_{2}^{\langle ij\rangle} -\frac{m_{1}}{7}\left(H^{\langle ij\rangle}y_{1}^{2} +4H^{\langle i}_{a}y_{1}^{j\rangle}y_{1}^{a}\right) -\frac{m_{2}}{7}\left(H^{\langle ij\rangle}y_{2}^{2} +4H^{\langle i}_{a}y_{2}^{j\rangle}y_{2}^{a}\right) \,.
\end{align}
The second term in the gravitational waveform~\eqref{GWtotal} is given by
\be
h_{km}^{\mathrm{TT}} = \frac{2G}{c^4 R} \mathcal{P}_{ijkm} I_{ij}^{(2)} \,.
\ee
The projection onto the plus and cross polarizations gives,
\begin{align}
h_{+} &= \frac{G}{c^4 R}\left(P_{i}P_{j}-Q_{i}Q_{j}\right)I_{ij}^{(2)} \,, \\
h_{\times} &= \frac{G}{c^4 R}\left(P_{i}Q_{j}+Q_{i}P_{j}\right)I_{ij}^{(2)}\,.
\end{align}
The explicit expression of the polarizations is given in the Section~\ref{app:polarization}. We can see that the amplitude of the scattered wave scales as $\omega_0^{-4/3}$.

\subsection{The polarizations of the scattered gravitational waves} \label{app:polarization}

\subsection*{Parallel to the orbital plane}

Similarly to the usual orbital frequency parameter $x= \left(\frac{Gm\omega_{0}}{c^{3}}\right)^{2/3}$, we define the parameter $X\equiv \left(\frac{Gm\Omega}{c^{3}}\right)^{2/3}$ related to the incoming frequency. When the wave is incident parallel to the orbital plane ($\alpha=0$, $\beta=\pi/2$, $\kappa=\pi/2+\iota$) we find

\begin{align}\label{hplusform}
h_{+} &= \frac{G \nu m x}{c^2 R} \Biggl\{ -(\cos (2\iota)+3) \cos(2\zeta) + H_{\times} \Biggl[\frac{X^{3/2}}{x^{3/2}} \cos(\Omega t) \Biggl(5 \sin^3\iota \cos\iota \cos(2\zeta)\nn\\
&+\frac{1}{48} \biggl(6 \sin(2\iota) (7 \cos(4\zeta)-25)+\sin (4\iota) (7 \cos(4\zeta)-17)\biggr)\Biggr)\nn\\
&  +\sin(\Omega t) \Biggl(\frac{X^3}{x^3} \biggl(\frac{1}{6} \sin\iota \cos\iota (\cos(2\iota)+3) \sin(4\zeta)-2 \sin^3\iota \cos\iota \sin(2\zeta)\biggr) \nn\\
&  +\frac{1}{4}\sin\iota \cos\iota \biggl(12 \sin^2\iota \sin(2\zeta)+\bigl(8 \cos\iota-3\cos(2\iota)-1\bigr) \sin(4\zeta)\biggr)\Biggr)\Biggr] \nn\\
& + H_{+} \Biggl[\cos (\Omega t) \Biggl(\frac{X^3}{2688 x^3} \biggl(2 \Bigl(661 \cos(2\iota)+18 \cos(4\iota)-21 (\cos(6\iota)+74)\Bigr) \cos(2\zeta) \nn\\
& \quad -7 \Bigl(-65 \cos(2\iota)+6\cos(4\iota)+\cos (6\iota)-198\Bigr) \cos (4\zeta)-448(\cos(2\iota)+3) \sin\zeta \nn\\
&+448 (\cos(2\iota)+3) \sin(3\zeta) +647 \cos(2\iota)+6 \cos(4\iota)-7 \cos(6\iota)-6\biggr) \nn\\
&+\frac{1}{2688} \biggl(7 \Bigl(67 \cos(2\iota)-3 \bigl(6 \cos(4\iota) +\cos(6\iota)-70\bigr)\Bigr) \cos^2(2\zeta) \nn\\
& +672 \sin^2\left(\frac{\iota}{2}\right) (\cos(2\iota)-3) \Bigl(\cos\iota \bigl((\cos (2\iota)+3) \cos(4\zeta)-4\bigr)-2 \sin ^2\iota\Bigr) \nn\\
&  -\cos(2\zeta) \Bigl(448 (\cos(2\iota)+3) \sin\zeta+3931 \cos(2\iota)+38 \cos(4\iota)-91 \cos(6\iota)-4774\Bigr)\biggr)\Biggr) \nn\\
&  +\frac{X^{3/2}}{1344 x^{3/2}} \sin (\Omega t) \Biggl(-6 \sin ^2\iota \biggl(8\cos(2\iota) +35 \cos(4\iota)-619\biggr) \sin(2\zeta) \nn\\
& 
-49 (\cos (2\iota)-3) (\cos(2\iota)+3)^2 \sin(4\zeta)\Biggr)\Biggr] \Biggr\}\,,
\end{align}
and
\begin{align}\label{hcrossform}
h_{\times} &= \frac{G \nu m x}{c^2 R} \Biggl\{ -4 \cos\iota \sin(2\zeta) + H_{\times} \Biggl[\frac{X^{3/2}}{x^{3/2}} \cos(\Omega t) \Biggl(\frac{7}{3} \sin\iota \cos^2\iota \sin(4\zeta)-\frac{1}{7} \sin^3\iota \sin(2\zeta)\Biggr) \nn\\
&  +\sin(\Omega t) \Biggl(\frac{X^3}{21 x^3} \biggl(2 \sin(3\iota)-2 \sin\iota \Bigl(7 \cos^2\iota \cos(4\zeta)+3 \sin^2\iota \cos(2\zeta)+5\Bigr)\biggr)  \nn\\
& + 8 \sin\iota \cos^2\iota \sin^4\zeta-\frac{8}{7} \cos(2\zeta) \biggl(\sin^3\iota-7 \sin\iota \cos\iota \sin^2\zeta\biggr)\Biggr)\Biggr] \nn\\
& + H_{+} \Biggl[\frac{X^{3/2}}{192 x^{3/2}} \Biggl(2 \cos\iota \biggl(7 (\cos(4\iota)-17) \cos(4\zeta)-409\biggr)+65 \cos(3\iota) \nn\\
& + 17 \cos(5\iota)\Biggr) \sin(\Omega t)+\cos(\Omega t) \Biggl(\frac{1}{48} \sin(2\zeta) \biggl(-24 \sin^2\left(\frac{\iota}{2}\right) (\cos(4\iota)-17) \sin^2\zeta+ \nn\\
& + \cos\iota \Bigl((35-3 \cos(4\iota)) \cos(2\zeta)  -23 \cos(2\iota)+\cos(4\iota)-32 \sin\zeta+38\Bigr)\nn\\
& -3 \sin\iota \Bigl(\cos(2\iota)+2 \cos (4\iota)-17\Bigr) \tan\iota\biggr) \nn\\
&  -\frac{X^3}{24 x^3} \cos\iota \sin (2\zeta) \biggl((\cos (4\iota)-33) \cos(2\zeta)-32 \sin\zeta+16\biggr)\Biggr)\Biggr] \Biggr\} \,.
\end{align}

\subsection*{Perpendicular to the orbital plane}

For perpendicular incidence ($\alpha=0$, $\beta=\pi/2$, $\kappa=\iota$), we have
\begin{align}\label{hplusform2}
h_{+} &= \frac{G \nu m x}{c^2 R} \Biggl\{ -(\cos (2\iota)+3) \cos(2\zeta) + H_{\times} \Biggl[\frac{X^{3/2}}{12 x^{3/2}} \cos^2\iota \cos(\Omega t) \Biggl(-7 (\cos(2\iota)+3) \cos(4\zeta)\nn \\
& -60 \sin^2\iota \cos(2\zeta) +17 \cos(2\iota)+75\Biggr) +\sin(\Omega t) \Biggl(\frac{X^3}{6 x^3} \cos^2\iota \biggl(24 \sin^2\iota \sin\zeta \cos\zeta\nn\\
& -(\cos (2\iota)+3) \sin(4\zeta)\biggr)  + \cos^2\iota \biggl(-3 \sin^2\iota \sin(2\zeta)\nn\\
&-\frac{1}{4} \Bigl(-8 \cos\iota+5 \cos(2\iota)+7\Bigr) \sin(4\zeta)\biggr)\Biggr)\Biggr] \nn\\
& + H_{+} \Biggl[\frac{X^{3/2}}{x^{3/2}} \sin(\Omega t) \Biggl(\frac{1}{112} \sin^2\iota (35 \cos(2\iota)+109) (\cos(2\iota)+3) \sin(2\zeta) \nn\\
& +\frac{7}{192} (\cos (2\iota)+3)^3 \sin(4\zeta)\Biggr)+\cos(\Omega t) \Biggl(\frac{X^3}{2688 x^3} \biggl(-448 (\cos(2\iota)+3) \cos\zeta\nn \\
& -448 (\cos(2\iota)+3) \cos(3\zeta)+14 (\cos(2\iota)+3) \Bigl(12 \cos(2\iota) +\cos (4\iota)+51\Bigr) \cos(4\zeta) \nn\\
& -24 \sin^2\iota \Bigl(92 \cos(2\iota)+7 \cos(4\iota)+157\Bigr) \cos(2\zeta)-448 (\cos(2\iota)+3) \sin\zeta \nn\\
& +448 (\cos(2\iota)+3) \sin(3\zeta)+457 \cos(2\iota)+78 \cos(4\iota)+7 \cos(6\iota)+738\biggr)  \nn\\
&  +\frac{1}{5376} \biggl(448 (\cos(2\iota)+3) \cos\zeta +448 (\cos(2\iota)+3) \cos(3\zeta)+14 (\cos(2\iota)+3) \Bigl(-168 \cos\iota \nn\\
&+132 \cos(2\iota) -24 \cos(3\iota)+15 \cos(4\iota)+109\Bigr) \cos(4\zeta) \nn\\
&+16 \sin^2\iota (\cos(2\iota)+3) (91\cos(2\iota)+345) \cos(2\zeta)\nn\\
&  +448 (\cos(2\iota)+3) \sin\zeta-448 (\cos(2\iota)+3) \sin(3\zeta) +8400 \cos\iota-4837 \cos(2\iota) \nn\\
&+21 \Bigl(104 \cos(3\iota)-30 \cos(4\iota)+8 \cos(5\iota)+\cos(6\iota)-82\Bigr)\biggr)\Biggr)\Biggr] \Biggr\}\,,
\end{align}
and
\begin{align}\label{hcrossform2}
h_{\times} &= \frac{G \nu m x}{c^2 R} \Biggl\{-4 \cos\iota \sin(2\zeta) + H_{\times} \Biggl[\frac{X^{3/2}}{x^{3/2}} \cos(\Omega t) \Biggl(\frac{1}{7} \sin^2\iota \cos\iota \sin(2\zeta)-\frac{7}{3} \cos^3\iota \sin(4\zeta)\Biggr) \nn\\
&  +\sin(\Omega t) \Biggl(\frac{2 X^3}{21 x^3} \cos\iota \biggl(7 \cos^2\iota \cos(4\zeta)+3 \sin^2\iota \cos(2\zeta)-2 \cos(2\iota)+4\biggr)\nn \\
&  + \biggl(\frac{8}{7} \cos\iota \cos(2\zeta) \Bigl(14 \sin^2\left(\frac{\iota}{2}\right) \cos\iota \sin^2\zeta+\sin^2\iota\Bigr)-2 \cos^3\iota \sin^2(2\zeta)\biggr)\Biggr)\Biggr] \nn\\
& + H_{+} \Biggl[\frac{X^{3/2}}{192 x^{3/2}} \Biggl(-2 \cos\iota \biggl(14 (\cos(2\iota)+3)^2 \cos(4\zeta)+593\biggr)-269 \cos(3\iota)\nn\\
&  -17 \cos(5\iota)\Biggr) \sin(\Omega t) +\cos(\Omega t) \Biggl(\frac{X^3}{192 x^3} \csc\iota \biggl(\Bigl(101 \sin(2\iota)+12 \sin(4\iota)+\sin(6\iota)\Bigr) \sin(4\zeta)\nn \\
&  -64 \sin(2\iota) \Bigl(\sin\zeta+\sin(3\zeta)-\cos\zeta+\cos(3\zeta)\Bigr)\biggr) +\frac{1}{192} \biggl(3 \Bigl(-48 \cos(2\iota)\nn \\
&  +39 \cos(3\iota)-4 \cos(4\iota)+3 \cos(5\iota)-76\Bigr) \sin(4\zeta)+\cos\iota \Bigl(386 \sin(4\zeta)-64 \cos\zeta\nn\\
&  +64 \bigl(\sin\zeta+\sin(3\zeta)+\cos(3\zeta)\bigr)\Bigr)+4 \sin^2\left(\frac{\iota}{2}\right) \Bigl(538 \cos\iota +136 \cos(2\iota)\nn\\
&+153 \cos(3\iota)+14 \cos(4\iota)+13 \cos(5\iota)+170\Bigr) \sec\iota \sin(2\zeta)\biggr)\Biggr)\Biggr] \Biggr\} \,.
\end{align}

\subsection{The energy balance equation}
Before going further, we want to check that the energy balance equation is verified,
\be
\langle\frac{\mathrm{d}E}{\mathrm{d}t}\rangle = -\langle\mathcal{F}\rangle\,,
\ee
where the brackets stand for the angular average over one orbital period and the left-hand side has been computed in Eqs.~\eqref{dEdt1}-\eqref{dEdt2}. The gravitational flux is defined as,
\begin{align}\nn
\mathcal{F} &\equiv  c\int_{\mathcal{S}}(-g)t_{\mathrm{LL}}^{0i}\mathrm{d}S_{i} \\
&= \frac{c^{3}}{16\pi G}\int_{\mathcal{S}} \partial_{\tau}H_{ij}\partial_{\tau}h^{ij}\,R^{2}\mathrm{d}\Sigma\,,
\end{align}
where $\mathrm{d}S_{i}=N^{i}R^{2}\mathrm{d}\Sigma$ is the surface element of the two-dimensional surface $\mathcal{S}$. Inserting the expressions for $H^{\mathrm{TT}}_{ij}$ and $h^{\mathrm{TT}}_{ij}$ we get for the first configuration,
\begin{equation}
\mathcal{F} = \frac{128 a^2 \nu m \omega_{0}^6 H_{+} \sin ^3\left(\frac{\pi\Omega }{\omega_{0}}\right) \cos \left(\frac{\pi  \Omega }{\omega_{0}}\right)}{15\pi \Omega \left(\Omega^2-4\omega_{0}^2\right)} ,
\end{equation}
and for the second configuration,
\begin{equation}
\mathcal{F} = \frac{8 a^2 \nu  m \omega_{0}^5 \sin ^3\left(\frac{\pi  \Omega }{\omega_{0}}\right) \cos \left(\frac{\pi  \Omega }{\omega_{0}}\right) (14 H_{+} \omega_{0}-5 H_{\times} \Omega )}{15 \pi \Omega \left(\Omega ^2-4 \omega_{0}^2 \right)} .\ \qquad
\end{equation}
After averaging over one orbital period the variation of the energy~\eqref{dEdt1}-\eqref{dEdt2}, we can see that the quadrupole formula is respected,
\be
\langle\frac{\mathrm{d}E'}{\mathrm{d}t}\rangle = -\langle\mathcal{F}\rangle\,,
\ee
where we have defined the modified energy
\begin{align}\nn
E' &= E +\frac{Gm^2\nu}{2r}\left(H_{ij}n^{i}n^{j}\right)+\frac{1}{2}m\nu\left(H_{ij}v^{i}v^{j}\right)\\
&= -\frac{Gm^2\nu}{2r}\left[1-\frac{1}{2}\left(H_{ij}n^{i}n^{j}\right)\right]+\frac{1}{2}m v^{2}\,.
\end{align}

\subsection{The cross section}

Using the previous results, the scattering cross section can be computed generically for any binary orientation and any incoming gravitational wave. We find that, for a $+$-polarized wave, for instance, the scattering cross section depends on the angle with which the wave hits the binary. However, in the limit when $\Omega\rightarrow\infty$, the cross section for either edge- or head-on configurations is the same,
\be
\sigma = \frac{11408\,\pi\nu^2}{2205}\left(\frac{v}{c}\right)^4\frac{a^6}{\lambda_{GW}^4} \,.\label{crosssection1}
\ee
where we have introduced the orbital radius of the binary system as $a=\frac{v}{\omega_0}$, as well as the wavelength of the incoming GW, $\lambda_{GW}\equiv\frac{c}{\Omega}$.
From this formula we can see that at equal total mass, the effect is larger the slower the system. Also, it highlights a structure similar to that of Rayleigh scattering of light. 

Note that the cross section is not allowed to grow unboundedly, since that would take us away from the perturbative regime we work in. In particular, we must require that the scattered wave is always of much smaller amplitude than the incoming GW. Evaluating the scattered wave a wavelength away from the scatterer, we find that the cross-section can be expressed as
\be\label{crosssection2}
\sigma=C_1^2\,\frac{11408\,\pi\,a^2}{2205}\,,
\ee
with $C_1\ll 1$ the ratio between scattered and incident GWs. In addition, we must require that $C_2\equiv H\Omega/\omega_0\ll 1$, to ensure that the back-reaction on the binary is small. This condition prevents the cross section from getting arbitrarily large for weakly-bound binaries.

\section{Conclusions}

After a period of consolidation of the detection of 
gravitational waves, entering the period of {\it precision} GW physics, we will require a better control on the possible effects that affect the production and propagation of GWs. This will also be essential to use GWs as an ultimate proof of the constituents of the Universe. As an example of possible effects that might affect the propagation of GWs, in this Chapter we studied the GW scattering over a binary of compact objects.

Our results show that gravitationally-bound binaries are able to scatter incoming gravitational radiation. We have computed the scattered field for general configurations, two special cases are shown in Section~\ref{app:polarization}. 
Consider now the binary neutron-star systems: PSR J1411+2551~\cite{Martinez:2017jbp} (orbital period $2.6$ days, total mass $2.5\mathrm{M}_{\odot}$) and the ultra-relativistic pulsar PSR J1946+2052~\cite{Stovall:2018ouw} (orbital period $0.078$ days, total mass $2.5\mathrm{M}_{\odot}$). For a GW incoming at a frequency $f=\Omega/(2\pi)=200\mathrm{Hz}$, we find from the expressions in Sec.~\ref{app:polarization} that the order of magnitude of the correction in the amplitude -- due to scattering -- is, for the first binary pulsar $h_{+,\times}\sim 10^{-5}\,\mathrm{H}_{+,\times}$, while for the ultra-relativistic pulsar we have $h_{+,\times}\sim 10^{-7}\,\mathrm{H}_{+,\times}$. As expected from Eq.~\eqref{crosssection2} the effect is larger for slowly rotating pulsars. We also find that the effect is slightly stronger for the configuration of an incoming gravitational wave parallel to the orbital plane.

The numbers above are small, but not desperately small as to be discouraging. However, how likely is such an event? The magnitude of the scattered wave is insensitive (to this order) to the structure of the compact objects forming the binary. Thus, stellar mass black-hole binaries of similar periods will give rise to similar scattered amplitudes. The number of stellar mass black holes in the central parsec of our own Galactic nucleus was recently reported to be significant, of order $N={\cal O} (10^4)$~\cite{2018Natur.556...70H}. If this finding generalizes to other supermassive black holes in galactic nuclei, this implies that there exists a substantial screen of potential scatterers around supermassive black holes. Such screen may give rise to detectable levels of scattered radiation, either from binaries (the object of the current study) or from isolated objects (e.g. Ref.~\cite{McKernan:2014hha}). For example, consider GWs generated by stellar-mass black hole binaries in the last stages of the inspiral and merger, and (barely) detectable by LIGO, $f=\Omega/(2\pi) \sim$ 20 {\rm Hz}. The emitting-binary is close to the galactic center, and the emitted GWs will now cross a screen of binaries, which we assume have parameters close to the binary pulsar above (orbital period $2.6$ days, total mass $2.5\mathrm{M}_{\odot}$).
Using the scattering cross-section~\eqref{crosssection1}, and the number density found in the galactic center $n$, the mean free path of the GW is $\sim 1/(n\sigma)\sim 100\,{\rm Mpc}$, a number which is clearly too big to be of relevance. These estimates assume quasi-circular motion, we do not expect any qualitatively important change to occur when eccentricity is included.

Finally, it is worth to mention that an analysis for secular effects of a stochastic background of GWs was already performed in \cite{Hui:2012yp}. Even in this case, with the accuracy achieved by current GW observatories, the smallness of the gravitational constant makes a detection far from feasible. However, on a positive note, pulsar timing holds the promise to overcome these limitations and detect minor variations in the time of arrival of the radio-wave from a background of GWs~\cite{Khmelnitsky:2013lxt}.

	\chapter{The close limit approximation of compact objects}\label{chapter:CLAP}

\minitoc

The ability to model the evolution of compact binaries from the inspiral to the coalescence is central to GW astronomy. Current waveform catalogues are built from vacuum binary BH (BBH) models, by evolving Einstein equations numerically and complementing them with knowledge from slow-motion and small curvatures expansions. Much less is known about the coalescence process in the presence of matter, or in theories other than GR. In this final Chapter of Part {\bf II}, we explore the Close Limit Approximation as a powerful tool to understand the coalescence process in general setups.  In particular, we study the head-on collision of two equal-mass, compact but horizonless objects. Afterwards, we will apply the CLAP to investigate the effect of colliding black holes on surrounding scalar fields.

\subsection*{A non-standard way to generate gravitational waves}

The GWs generation due to the coalescence of compact objects is now well understood. In order to describe the system during its evolution, one can separate the entire motion in different stages, each of which corresponds to different physical configurations. The inspiral part, coincides with a large separation between the sources and it is modeled using slowly moving and weakly stressed
sources\footnote{The sources that are suitable for a Post Newtonian expansion are the ones for which $\varepsilon={\rm max}\{\lvert U/c^2\rvert^{1/2},\lvert T^{0i}/T^{00}\rvert,\lvert T^{ij}/T^{00}\rvert^{1/2}\}\ll 1$, where $T^{\mu\nu}$ are the components of the matter stress-energy tensor and $U$ is its Newtonian potential.}~\cite{Blanchet:2013haa,Poisson_will_2014}. During this inspiral motion, the orbit of the binary shrinks because of GWs emission, up to a point in which the above approximations are not enough to coherently describe the coalescence of compact objects. In fact, given the highly non-linear character of Einstein's equations, when the bodies are sufficiently close to each other, one needs full numerical methods to account for such non-linearities. This stage is called {\it the merger} and requires full numerical simulations, commonly referred to ``numerical relativity'' (NR)~\cite{Pretorius:2005gq,Campanelli:2005dd,Baker:2005vv,Buchman:2012dw}. Once the final object is formed, it will continue to release energy and momenta during, what is called, the {\it ringdown}. Similarly to the case of a bell hit by an hammer, one can use spacetime perturbation
theory to model this final relaxation mechanism~\cite{Regge:1957td,Zerilli:1971wd,Kokkotas:1999bd,Ferrari:2007dd,Berti:2009kk}. Perturbation theory, extended to
include self-force effects~\cite{Barack:2009ux,Poisson:2011nh}, can also be used to describe binaries characterized by an extreme mass ratio between the masses of its componentes. 

These models are built in the framework of GR, and rely on the hypothesis that all compact objects in the Universe (with masses $\gtrsim(2.5-3)\,M_\odot$) are BHs. Hence, to test GR and the BH hypothesis beyond null or biased tests~\cite{Yunes:2009ke}, one needs to construct the inspiral, merger and ringdown not assuming GR, nor BHs as the sources. In particular, extended models of the merger phase might be the key to perform the aforementioned tests in the strong-field regime. However, the extension of NR to modified gravity theories and to ECOs is not an easy task, since they require the time-evolution problem to be {\it well-formulated} and {\it well-posed}~\cite{Sotiriou:2008rp,Delsate:2014hba,Papallo:2017qvl,Ripley:2019irj,Kovacs:2020ywu,Witek:2020uzz,East:2020hgw}. With the former it is commonly intended that the equation of motion can be rewritten as a system which contains only first order derivatives both in time and space; the latter instead refers to specific conditions on the partial derivative equations, that ensure for the accuracy of the results (e.g. being symmetric or strongly hyperbolic). Notably, the conditions above are not easy to achieve in theories with extra fields and couplings. We will not go further in this vast topic, since in this thesis we are only interested in physical systems for which perturbation theory is applicable.

A possible way to circumvent the above problem is to use an alternative approach, in particular the CLAP~\cite{Price:1994pm,Abrahams:1995wd,Nicasio:1998aj,Khanna:1999mh,Gleiser:1996yc,Gleiser:1998rw,Sopuerta:2006wj,LeTiec:2009yf}. In this approach, the slice of spacetime describing the last stages of the merger is described using standard initial data of NR simulations. This metric turns out to be a small deformation of the stationary spacetime of the final BH, and can then be studied using perturbation theory; the perturbation parameter, in this case, is the separation between the two BHs. Thus, we do not need NR to evolve these initial data: it is sufficient to solve the perturbation equations (e.g. the Zerilli equation) to obtain the GWs emission from the merger and ringdown stages. Since the CLAP showed a remarkable agreement with NR simulations~\cite{Anninos:1993zj,Anninos:1998wt}, its extension beyond GR and beyond the BH hypothesis could be a valuable tool to model the merger stages of compact binary coalescence in an extended framework.

In the following, Greek indices refer to quantities defined on a spacetime- four-dimensional ($4D$) manifold, Latin indices to their three-dimensional spatial part ($3D$).

\section{The formalism}\label{sec:CLAP}
We shall here review the standard CLAP approach for BBH coalescences in GR.  This approach is based on the assumption
that in the final stages of the coalescence, when the two BHs are sufficiently close to each other, the spacetime is a
small deformation of a single BH spacetime, and thus can be studied using the techniques of spacetime perturbation
theory. To this aim, BBH initial data, originally developed in the context of NR, are recast as a perturbation of Kerr spacetime. In this Chapter we shall consider, as a first step, the case of a head-on collision of non-rotating BHs, which
leads to a non-rotating BH. Thus, the BBH initial data are recast as a perturbation of a Schwarzschild BH.

\subsection{The $3+1$ decomposition}\label{sec:3+1}
We briefly recall the basic concepts of the $3+1$ decomposition in NR. We refer the interested reader to one of
the several excellent books and reviews on the
subject\,\cite{Gourgoulhon:2007ue,alcubierre2008introduction,baumgarte2010numerical,shibata2015numerical} for further
information and details.

In order to formulate GR (or any other gravitational theory) as a time evolution problem, we first decompose the $4D$
spacetime in a set of $3D$ spatial hypersurfaces $\Sigma_t$, labeled by a time parameter $t$,
each of them having a $3$-metric $\gamma_{ij}$ given by the space components of
\begin{equation}
\gamma\mndd=g\mndd-n_\mu n_\nu\,.
\end{equation}
Here, $g\mndd$ is the $4D$ spacetime metric, and $n^\mu$ is the unit timelike ($n^\mu n_\mu=-1$) vector normal to the
hypersurfaces $\Sigma_t$.  Thus, we can write the spacetime metric as
\begin{align}
ds^2&=g\mndd dx^\mu dx^\nu \nn\\
&=-(\alpha^2-\beta_i\beta^i)dt^2+2\gamma\ijdd\beta^i dt dx^j+\gamma\ijdd dx^idx^j\,,
\end{align}
where $\alpha$, $\beta^i$ are called lapse function and shift vector, respectively, and contain the information about
how the coordinate system changes from a slice to another. The choice of the lapse and shift, thus, corresponds to the
choice of the foliation of the spacetime. The embedding of the hypersurfaces in the $4D$ spacetime is described by the
extrinsic curvature
\begin{equation}
\label{eq:extrinsic_curv}
K\ijdd=-\frac{1}{2}\mathcal{L}_{\bm n}\gamma\ijdd\,,
\end{equation}
where $\mathcal{L}_{\bm n}=\alpha^{-1}(\partial_t-\mathcal{L}_\beta)$ is the Lie derivative along the unit vector
$n^\mu$. If a scalar field $\Phi$ is also present, we define its momentum as,
\begin{equation}
  \label{eq:scalar_momentum}
  \Pi=-\mathcal{L}_{\rm n}\Phi\,.
\end{equation}
Then, all $4D$ quantities such as the Ricci scalar $R$, the Ricci tensor $R\mndd$, etc. can be decomposed in terms
of the $3D$ metric $\gamma_{ij}$. The $3D$ Ricci tensor and Ricci Scalar, $\prescript{3}{}{R}_{ij}$,
$\prescript{3}{}{R}$, the extrinsic curvature and $K\ijdd$, its trace $K=\gamma^{ij}K\ijdd$, and the scalar
field momentum $\Pi$ follow from this $3D$ metric. We denote with $\nabla_\mu$ the covariant derivative with respect to the $4D$ metric $g\mndd$,
and with $D_i$ the covariant derivative with respect to the $3D$ metric $\gamma\ijdd$.

With this decomposition, Einstein equations (possibly supplemented with the scalar field dynamical equation) lead to
two sets of equations: (i) the evolution equations, which are (with a careful choice of variables) hyperbolic equations
giving the time evolution of the $3D$ quantities; (ii) the constraint equations, which are ellyptic equations which have
to be satisfied at any $3D$ slice $\Sigma_t$.

The evolution equations are given by\til\cite{Okawa:2014nda}
\begin{align}
\left(\partial_t-\mathcal{L}_\beta\right)K_{ij}&=-D_i D_j \alpha+\alpha\left(R_{ij}-2K^k_i K_{jk}+K K_{ij}\right)+4\pi\alpha\left(\gamma_{ij}\left(S-\rho\right)-2S_{ij}\right)\,,\\
\left(\partial_t-\mathcal{L}_\beta\right)\Pi&=\alpha\left(-D^i D_i \Phi+K\Pi-D^i \alpha D_i \Phi\right)\,,
\end{align}
where $\mathcal{L}_\beta$ is the Lie derivative along the shift vector, $\rho$ is the matter energy density and $S_{ij}$ is the $3+1$ decomposition of the matter energy-momentum tensor, for which explicit expressions are given in the case of scalar fields in Eqs.\til\eqref{eq:rhostt}-\eqref{eq:Sstt}.

The constraint equations are formed by an {\it Hamiltonian constraint equation}
\begin{equation}
\label{eq:HGR}
\mathcal{H}^{\rm GR}\equiv\prescript{3}{}{R}+K^2-K\ijdd K\ijuu=0\,,
\end{equation}
and by three {\it momentum constraint equations}
\begin{equation}
\label{eq:MGR}
\mathcal{M}^{\rm GR}_i\equiv D_jK^j_i-D_i K=0\,.
\end{equation}
In presence of matter, the constraint equations are
\begin{align}
  \label{eq:HGRphi}
\mathcal{H}^{\rm GR}&=16\pi\rho\,,\\
  \label{eq:MGRphi}
\mathcal{M}^{\rm GR}_i&=8\pi j_i\,,
\end{align}
where $j_i$ is its energy-momentum flux. For a massless scalar field (with vanishing self-potential) the scalar-related quantities take the form
\begin{align}
\label{eq:rhostt}
\rho&=\frac{1}{2}\Pi^*\Pi+\frac{1}{2}D^i\Phi^*D_i\Phi\,,\\
\label{eq:jstt}
j_i&=\frac{1}{2}\left( \Pi^*D_i\Phi+\Pi D_i\Phi^* \right)\,,\\
\label{eq:Sstt}
S_{ij}&=\frac{1}{2}\left(D_i \Phi^* D_j \Phi+D_i\Phi D_j\Phi^*\right)+\frac{1}{2}\gamma_{ij}\left( \Pi^*\Pi- D^k \Phi^* D_k \Phi \right)\,,
\end{align}
where $\Pi$ is the momentum of the scalar as defined in Eq.\til\eqref{eq:scalar_momentum}. The above quantities are computed from the scalar stress-energy tensor given in Eq.\til\eqref{StressEnergy}, in the massless scalar limit.

The general procedure to find the evolution of physical systems strictly depends on the specific numerical implementation. However, for the purpose of this work, we may assume that in NR one first solves the constraint equations at the initial time $t=t_0$, finding the {\it initial data} on $\Sigma_{t_0}$ of the system. Then, one solves the evolution equations, finding the spacetime at all times
$t\ge t_0$. Within the CLAP, instead, the evolution of physical initial data is determined through the use of perturbation theory, therefore solving a linearized version of the evolution equations (see below).
\subsection{Initial data for binary black holes in general relativity}\label{subsec:BL}

Solving the constraint equations~\eqref{eq:HGR}-\eqref{eq:MGR} and finding initial data appropriate to study a given
system is a subject of study on its
own\,\cite{Cook:2000vr,alcubierre2008introduction,baumgarte2010numerical,shibata2015numerical}. The original formulation
of the CLAP~\cite{Price:1994pm} used the Misner initial data\,\cite{Misner:1960zz} (a common choice for NR simulations
decades ago) to describe BBH head-on collisions. Subsequently, it was found that the Brill-Lindquist (BL) initial
data\,\cite{Brill:1963yv} are more appropriate for NR simulations, because they have a simpler form and are easier to be
extended to the more realistic non-head-on case (the Bowen-York initial data~\cite{Bowen:1980yu}). Thus, later
applications of the CLAP employ the BL (or Bowen-York) initial
data~\cite{Abrahams:1995wd,Andrade:1996pc,Khanna:1999mh,Sopuerta:2006wj}.  Here we shall use BL initial data and their extensions.

In the case of a head-on collision of two non-rotating compact objects starting from rest, the extrinsic curvature
identically vanishes. Additionaly, considering that in vacuum $\Phi=0$, the constraint
equations\,\eqref{eq:HGR}-\eqref{eq:MGR} reduce to $\prescript{3}{}{R}=0$. The BL three-metric $\gamma_{ij}$ is a
conformally flat solution of this equation describing two BHs. It has the form
\begin{equation} \label{eq:eq:BL_id}
\prescript{3}{}{ds}^2_{\rm BL}=\varphi_{\rm BL}^4 \prescript{3}{}{d\eta}^2=\varphi_{\rm BL}^4 \left( dR^2+R^2d\Omega^2 \right)\,,
\end{equation}
where $d\Omega^2=d\theta^2+d\phi^2\sin^2\theta$ , $\prescript{3}{}{d\eta}^2$ is the flat three-metric, and
$\varphi_{\rm BL}(R,\theta,\phi)$ is the conformal factor. With this choice, the Hamiltonian constraint
$\prescript{3}{}{R}=0$ reduces to $\nabla^2\varphi_{\rm BL}=0$. The BL solution of this equation is
\begin{equation} \label{eq:BL_id_newtonian}
\varphi_{\rm BL}=1+\frac{m_1}{2|{\bm R}-{\bm R_1}|}+\frac{m_2}{2|{\bm R}-{\bm R_2}|}\,,
\end{equation}
where the vector ${\bm R}$ is the position vector in the flat three-space, and ${\bm R_i}$ is the position of the $i$-th
BH; $M=m_1+m_2$ is the ADM mass of the entire spacetime. Note that $m_1$ and $m_2$ are not the ADM masses of the two
BHs, which are~\cite{Brill:1963yv}
\begin{equation}\label{eq:bare_masses}
M_1=m_1\left(1+\frac{m_2}{2d}\right)\,,~~M_2=m_2\left(1+\frac{m_1}{2d}\right)\,,
\end{equation}
and $d=|{\bm R_2}-{\bm R_1}|$. We choose the origin of the coordinate system in the center-of-mass: $m_1{\bm
  R_1}+m_2{\bm R_2}={\bm 0}$. In the region of spacetime where $R>R_1,R_2$, the conformal factor can be expanded in
terms of Legendre polynomials:
\begin{equation}
\varphi_{\rm BL}=1+\frac{M}{2R}+\sum_{\ell=1}^{\infty}\xi_\ell\left( \frac{M}{R} \right)^{\ell+1} P_\ell\left(\cos\theta\right)\,.
\end{equation}
The coefficients $\xi_\ell$ are given by~\cite{Andrade:1996pc}
\begin{equation}
\xi_\ell=\left(\frac{R_1}{M}\right)^{\ell}\frac{m_1}{2M}+\left(\frac{R_2}{M}\right)^{\ell}\frac{m_2}{2M}\,.\label{eq:defxi0}
\end{equation}
In the following we shall consider equal-mass binary systems; thus $m_1=m_2=M/2$. By choosing the $Z$-axis aligned with
the motion, ${\bm R_{1/2}}=(0,0,\pm Z_0)$ with $Z_0=d/2$. Thus $\xi_1=0$, and Eq.~\eqref{eq:defxi0} reduces to
\begin{equation} \label{eq:xi_M1eqM2}
\xi_\ell=\frac{1}{2}\left(\frac{Z_0}{M}\right)^{\ell}\,,\;\;\; \text{for }\ell=2,4,6,\dots \, .
\end{equation}
When the coefficients $\xi_\ell$ vanish, $\varphi_{\rm BL}=1+M/(2R)$ and the metric~\eqref{eq:eq:BL_id} describes the
$t=const.$ slices of the Schwarzschild's background. Indeed, by defining the Schwarzschild radial coordinate $r$ in terms
of the isotropic radial coordinate $R$ by
\begin{equation}
\label{eq:R_isotropic_to_Schw}
R=\frac{1}{4}\left(\sqrt{r}+\sqrt{r-2M}\right)^2\,,
\end{equation}
Eq.~\eqref{eq:eq:BL_id} gives
\begin{equation}
\prescript{3}{}{ds}^2_{\rm BL}=\mathcal{F}_{\rm BL}^4\left(f^{-1}dr^2+r^2d\Omega^2\right)\,,
\end{equation}
where
\begin{align}
\label{eq:f_of_Schwa}
f(r)&=1-\frac{2M}{r}\,,\\
\mathcal{F}_{\rm BL}&\equiv\varphi_{\rm BL}\left( R,\theta\right) \left( 1+M/2R\right)^{-1}\,.
\end{align}
If $\varphi_{\rm BL}=1+M/(2R)$,
$\mathcal{F}_{\rm BL}=1$ and by defining the $4D$ spacetime metric as 
\begin{equation}
\label{eq:pert_Schw_CLAP_GR_BL}
ds^2=-f dt^2+\prescript{3}{}{ds}^2_{\rm BL}\,,
\end{equation}
(corresponding to an appropriate choice of the shift vector and lapse function, i.e. to an appropriate gauge choice), it
coincides with the Schwarzschild geometry. If, instead, the coefficients $\xi_\ell$ are small but non-vanishing, the
metric~\eqref{eq:eq:BL_id} is a perturbation of Schwarzschild's three-metric, and Eq.~\eqref{eq:pert_Schw_CLAP_GR_BL} is
a perturbation of Schwarzschild's spacetime.
Therefore the BL three-metric, in the coordinates $(r,\theta,\phi)$, has the form
\begin{equation} \label{eq:dsBL_M1M2z0}
\prescript{3}{}{ds}^2_{\rm BL}= \left( 1+\frac{1}{1+\frac{M}{2R}} \sum_{\ell=2,4,\dots}^{\infty}
\xi_\ell \left(\frac{M}{R}\right)^{\ell+1}P_\ell\left(\cos\theta\right)\right)^4 \left(f^{-1} dr^2+r^2d\Omega^2\right)\,.
\end{equation}
If $\xi_\ell\ll1$ we can linearize in the parameters $\xi_\ell$, and Eq.~\eqref{eq:dsBL_M1M2z0} gives:
\begin{equation}
\label{eq:dsBL_pert}
\prescript{3}{}{ds}^2_{\rm BL}= \left( 1+\frac{4}{1+\frac{M}{2R}}\sum_{\ell=2,4,\dots}^{\infty}
\xi_\ell\left(\frac{M}{R}\right)^{\ell+1} P_\ell\left(\cos\theta\right)\right)\left(f^{-1} dr^2+r^2d\Omega^2\right)\,.
\end{equation}
The parameter $Z_0$ in Eq.~\eqref{eq:xi_M1eqM2} describes the (initial) distance
between the BHs in the isotropic frame. This quantity does not have a direct physical interpretation; thus, several
authors characterize the initial BH separation with the proper distance $L$ between the apparent horizons of the two
BHs, given by~\cite{Andrade:1996pc}
\begin{equation}
  L=\int_{Z_1}^{Z_2}\left[1+\frac{M}{4}\left(\frac{1}{Z_0+Z}+\frac{1}{Z_0-Z}\right)\right]^2dZ\,.\label{eq:defL}
\end{equation}
The extrema $Z_1$, $Z_2$ are the intersections of the apparent horizons of the two BHs with the $Z$-axis, and can be
found by numerical integration of the equations describing the apparent
horizons~\cite{bishop1982closed,bishop1984horizons}; the procedure to compute these quantities is described in detail,
for instance, in~\cite{Sopuerta:2006wj}. This computation shows that, for example, $L=3M$ for $Z_0\simeq0.5M$, $L=3.5M$
for $Z_0\simeq0.7M$, $L=4M$ for $Z_0\simeq0.85M$.

A comparison with NR computations in the head-on case shows that the CLAP is accurate, in the equal-mass case, for
$L\lesssim4M$ (see~\cite{Gleiser:1998rw} and the discussion in~\cite{Sopuerta:2006wj}), corresponding to
$Z_0\lesssim0.85M$ and thus $\xi_2\lesssim0.36$; as noted in~\cite{Price:1994pm}, it is remarkable that the agreement
extends far beyond the region $\xi_\ell\ll1$ in which the CLAP is expected to be applicable. Therefore, we shall apply
Eq.~\eqref{eq:dsBL_pert} and, more generally, perturbation theory, also to initial separations for which the condition
$\xi_\ell\ll1$ is only marginally satisfied.
%
\subsection{Perturbations and their time evolution}\label{subsec:recast_BH}
The metric in Eq.~\eqref{eq:pert_Schw_CLAP_GR_BL} describes the Schwarzshild spacetime, which we consider as the
background, with a perturbation with even parity, at a fixed ``initial'' time $t=t_0$. Thus, it can be recast in the
form
\begin{equation}
g_{\mu\nu}=g^{(0)}_{\mu\nu}+h_{\mu\nu}\,,
\end{equation}
where
\begin{equation}\label{eq:Schwarzschild_metric}
g^{(0)}_{\mu\nu}={\rm  diag}(-f,f^{-1},r^2,r^2\sin^2\theta)
\end{equation}
is the Schwarzschild spacetime (with $f$ defined in Eq.\til\eqref{eq:f_of_Schwa}), and $h_{\mu\nu}(t,r,\theta,\phi)$ is
the perturbation, whose evolution in the wave zone $r\gg M$ can be described in terms of the Zerilli function
$Q$~\cite{Zerilli:1971wd}. The original definition of the Zerilli function is appropriate for the study of oscillating
solutions; in this case, we want to study the evolution of a given set of initial data at the time $t=t_0$; thus, it is
more appropriate the definition of the Zerilli function in terms of gauge-invariant quantities given
in~\cite{Moncrief:1974am,Cunningham:1979px}.

We first note that the only non-vanishing perturbations in the
metric~\eqref{eq:pert_Schw_CLAP_GR_BL}-\eqref{eq:dsBL_pert} are 
\begin{align}
  h_{rr}(r,\theta)&=f^{-1}\sum_{\ell=2,4,\dots}g_\ell(r) \xi_\ell P_\ell(\cos\theta)\,,\nonumber\\
  h_{\theta\theta}(r,\theta)&=\frac{h_{\phi\phi}}{\sin^2\theta}=r^2\sum_{\ell=2,4,\dots}g_\ell(r)\xi_\ell P_\ell(\cos\theta)\,,
\label{eq:BLrecast0}
\end{align}
where
\begin{equation}
g_\ell=4\left(1+\frac{M}{2R}\right)^{-1}\frac{M^{\ell+1}}{R^{\ell+1}}\,.
\end{equation}
Using the notation of~\cite{Cunningham:1979px} for polar parity, axially symmetric perturbations,
Eqs.~\eqref{eq:BLrecast0} correspond to the perturbation functions $H_2^\ell(t,r)$, $K^\ell(t,r)$ given by 
\begin{align}
h_{rr}(t,r,\theta)&=f^{-1}\sum_\ell H_2^\ell(t,r)P_\ell(\cos\theta)\,,\nonumber\\
h_{\theta\theta}(t,r,\theta)&=r^2\sum_\ell K^\ell(t,r)P_\ell(\cos\theta)\,,\label{eq:BLrecast}
\end{align}
which implies that at the initial time $t=t_0$,
\begin{equation}\label{eq:H2_Kl_equality}
H_2^\ell(t_0,r)=K^\ell(t_0,r)=g_\ell\xi_\ell,~~{\text{with }}\ell=2,4,6,\dots\,,
\end{equation}
while the other perturbation functions ($H_0^\ell,H_1^\ell$) and ($G^\ell,h_0^\ell,h_1^\ell$) identically vanish at $t=t_0$.
Since the leading contribution comes from the quadrupole perturbations, we shall consider the $\ell=2$ contribution only. 
Then, following~\cite{Cunningham:1979px}, we define the $\ell=2$ Zerilli function at $t=t_0$,
\begin{equation}
  \psi(t_0,r)=\sqrt{\frac{4\pi}{5}}{\lambda}^{-1}Q(r)\xi_2\,,\label{eq:zerillit0}
\end{equation}
where $\lambda=1+3M/(2r)$ and (leaving implicit the index $\ell=2$ and the arguments of $H_2(t_0,r)$ and $K(t_0,r)$)
\begin{align}
  Q(r)&=\frac{2rf^2}{\xi_2}
  \left[\frac{H_2}{f}-\frac{1}{\sqrt{f}}\frac{d}{dr}\frac{rK}{\sqrt{f}}\right]+\frac{6r}{\xi_2}K\nonumber\\
 &= 2r f^2\left[\frac{g}{f}    -\frac{1}{\sqrt{f}}\frac{d}{dr}\frac{r g}{\sqrt{f}}\right]+6rg\label{eq:defQ}\,,
\end{align}
with 
\begin{equation}
g=4\left(1+M/(2R)\right)^{-1}M^3/R^3\,,
\end{equation}
and $R$ defined in Eq.\til\eqref{eq:R_isotropic_to_Schw}. Thanks to the considerations above, it is possible to show that the perturbation equations, that comes from a variational principle in a gauge-invariant fashion\til\cite{Moncrief:1974am,Cunningham:1979px}, can be written as a wave equation for the Zerilli function,
\begin{equation}
-\frac{\partial^2\psi}{\partial t^2}+\frac{\partial^2\psi}{\partial r_*^2}-V_Z\psi=0\label{eq:zer}\,,
\end{equation}
where $r_*=r+2M\log\left|\frac{r}{2M}-1\right|$ is the tortoise coordinate, and
\begin{equation}\label{eq:zer_potential}
  V_Z=f\left\{\frac{1}{{\lambda}^2}\left[\frac{9M^3}{2r^5}-\frac{3M}{r^3}\left(1-\frac{3M}{r}\right)\right]
  +\frac{6}{r^2\lambda}\right\}\,,
\end{equation}
is the Zerilli potential. Eq.~\eqref{eq:zer} is called the Zerilli equation~\cite{Zerilli:1971wd}.

In most implementations of the CLAP, the Zerilli equation~\eqref{eq:zer} with initial condition~\eqref{eq:zerillit0} is solved in the time domain. Alternatively, it can be solved in the frequency domain. Indeed, it can be shown (see
e.g.~\cite{Lousto:1996sx}) that if the function $\psi(t,r_*)$ is the solution of the Zerilli equation~\eqref{eq:zer} with
initial conditions~\eqref{eq:zerillit0}  and $\dot\psi(t_0,r_*)=0$ (i.e., assuming stationarity of the initial data), then,
choosing the time coordinate such that $t_0=0$, 
the Laplace transform
\begin{equation}\label{eq:laplace_psi}
  \tilde\psi(\omega,r_*)=\int_{0}^\infty dt\psi(t,r_*)e^{i\omega t}\,,
\end{equation}
is the solution of the Zerilli equation with source
\begin{equation}
\frac{\partial^2\tilde\psi}{\partial r_*^2}+(\omega^2-V_Z)\tilde\psi=S\,,\label{eq:zerS}
\end{equation}
where
\begin{equation}\label{eq:zer_source}
S(\omega,r_*)=i\omega\psi(t=0,r_*)\,,
\end{equation}
and boundary conditions of ingoing wave at the horizon, outgoing wave at infinity (see Eq.~\eqref{eq:bcsomm1}).
Finally, the total energy emitted in GWs in the collision is given by~\cite{Price:1994pm}
\begin{equation}
E=\frac{1}{384\pi}\int_0^\infty \left\lvert \frac{\partial \psi}{\partial t}\right\rvert^2 dt\,.\label{total_E}
\end{equation}
We have solved Eq.~\eqref{eq:zerS} using two different approaches, reproducing in both cases the results in the literature for head-on BBH collisions with the CLAP
approach~\cite{Price:1994pm}. The total energy obtained with our computation agrees with that of
Ref.~\cite{Price:1994pm} within one percent.

Finally, in this Chapter we only consider head-on collisions of non-rotating, equal mass compact objects starting from rest as
in~\cite{Price:1994pm}; however, it is worth mentioning that the CLAP approach has been extended to more realistic
setups, considering BHs with initial velocity~\cite{Baker:1996bt}, unequal masses~\cite{Andrade:1996pc} and non-head-on
binary inspirals~\cite{Khanna:1999mh}.

\subsection{Numerical integration of the Zerilli equation with source} \label{app:Zerilli_solver}
We shall discuss here the numerical integration of the Zerilli equation with source in the frequency domain, Eq.~\eqref{eq:zerS}
(see also~\cite{Lousto:1996sx,Campanelli:1997un}):
\begin{equation} \label{eq:Zerilli_model}
\frac{\partial^2 \tilde{\psi}}{\partial r_*^2}+ \left(\omega^2-V_Z\right)\tilde{\psi}=S\,,
\end{equation}
where $\tilde\psi(\omega,r)$ is the Laplace transform~\eqref{eq:laplace_psi} of the Zerilli function
$\psi(t,r)$,
\begin{equation}
  \tilde\psi(\omega,r_*)=\int_{0}^\infty dt\psi(t,r_*)e^{i\omega t}\,,\label{eq:laplace_psi2}
\end{equation}
and
\begin{equation}
S(\omega,r)=i\omega\psi(t=0,r)=i\omega\sqrt{\frac{4\pi}{5}}\frac{1}{1+\frac{3M}{2r}}Q(r)\xi_2\,,\label{eq:sourceterm}
\end{equation}
where $Q(r)$ is given in Eq.~\eqref{eq:defQ} and $\xi_2=Z_0^2/(2M^2)\lesssim1$.

If the Zerilli equation~\eqref{eq:Zerilli_model} describes perturbations of a Schwarzschild BH, it is defined in
$-\infty<r_*<+\infty$. In this case, the source term~\eqref{eq:sourceterm} does not vanish at the horizon,
$S(r_*\to-\infty)=i\omega\psi(t=0,r_*\to\infty)=\bar S\neq0$. The ingoing wave boundary conditions at the horizon
$r_*\to-\infty$, $\partial\psi/\partial r_*=\partial\psi/\partial t$\,\footnote{Note that these conditions are consistent with $\dot\psi(t=0)=0$ because in the BL initial data $\partial\psi/\partial r_*=0$ at the horizon.} translate in the Laplace transform space into
$\partial\tilde{\psi}/\partial r_*(\omega)=-i\omega\tilde{\psi}(\omega)-\psi(t=0)$ . Therefore, the boundary conditions of
$\tilde\psi(\omega,r_*)$ are:
\begin{align}
\tilde\psi(\omega,r_*)&=A^He^{-i\omega r_*}+\frac{\bar S}{\omega^2}~~~~~(r_*\to-\infty)\,,\nonumber\\
\tilde\psi(\omega,r_*)&=A^\infty e^{i\omega r_*}~~~~~~~(r_*\to+\infty)\,,
  \label{eq:bcsomm1}
\end{align}
with $A^H$, $A^\infty$ constants to be determined. The constant term is related to the fact that the BL initial data do
not vanish at the horizon, and do not affect the GW emission at infinity.

We find the solution of Eq.~\eqref{eq:Zerilli_model} satisfying ingoing boundary conditions at the horizon,
outgoing boundary conditions at infinity, by employing two different approaches: the Green function approach and a
shooting method, finding the same results.

\subsection*{Green's function method}

The Green function approach consists in finding two independent solutions of the homogeneous Zerilli equations:
$\tilde{\psi}^H$, satisfying ingoing wave conditions at the horizon, and $\tilde{\psi}^\infty$, satisfying outgoing wave
conditions at infinity, i.e.
\begin{align}
& \hspace{-0,5cm}\tilde{\psi}^H= \begin{cases}
       & e^{-i \omega r_*}, \qquad \qquad \qquad \qquad r_*\rightarrow -\infty\,,\\
       & D_{\rm in}e^{-i \omega r_*}+D_{\rm out} e^{+i \omega r_*},\,\,\,\, r_*\rightarrow +\infty\,,
    \end{cases} \nn\\
  & \label{eq:homogeneous_sols_BBH}
  \hspace{-0,5cm}\tilde{\psi}^\infty= \begin{cases}
       & B_{\rm in}e^{-i \omega r_*}+B_{\rm out} e^{+i \omega r_*},\,\,\, r_*\rightarrow -\infty\,,\\
       & e^{i \omega r_*},  \qquad \qquad \qquad \,\,\, \qquad r_*\rightarrow +\infty\,.
    \end{cases} 
\end{align}
The solution of the Zerilli equation with source, Eq.~\eqref{eq:Zerilli_model}, satifying the boundary
conditions~\eqref{eq:bcsomm1}, is then:
\begin{equation} \label{eq:Zerilli_gensol}
  \tilde{\psi}(\omega,r_*)=\frac{\tilde{\psi}^\infty}{W} \int_{-\infty}^{r_*}  S
  \tilde{\psi}^H dr'_*+\frac{\tilde{\psi}^H}{W} \int_{r_*}^\infty  S \tilde{\psi}^\infty dr'_*\,,
\end{equation}
where $W=\tilde{\psi}^H\left( \partial\tilde{\psi}^{\infty}/\partial r_*\right)-\tilde{\psi}^\infty\left( \partial
\tilde{\psi}^{H}/\partial r_*\right)$
is the constant Wronskian of the homogeneous equation. By imposing the boundary conditions at infinity~\eqref{eq:bcsomm1}
we find
\begin{equation}
  A^\infty(\omega)=\frac{1}{W}\int_{-\infty}^{+\infty}S\tilde{\psi}^Hdr_*\label{eq:zerinf}\,.
\end{equation}

Apparently, the integral~\eqref{eq:zerinf} is not well defined at the lower bound, where the integrand reduces to the
oscillating term $\bar Se^{-i\omega r_*}$. This is due to the fact that, strictly speaking, the Laplace
transform~\eqref{eq:laplace_psi2} is well defined in the upper complex plane; the inverse Laplace transform can be
computed along a path $\omega=\omega_R+i\epsilon$ with $\epsilon\ll1$ and $-\infty<\omega_R<+\infty$. Thus, along this
path $e^{-i\omega r_*}\to0$ as $r_*\to-\infty$ and the oscillating term disappears. As suggested
in~\cite{Lousto:1996sx}, we can compute the integrals for real values of $\omega$, as long as we subtract the ill-valued
contribution at the horizon:
\begin{equation}
  A^\infty(\omega)=\frac{1}{W}\int_{\bar{r}_*}^{+\infty}S\tilde{\psi}^Hdr_*+\frac{i}{\omega}\frac{\bar{S}}{W}e^{-i\omega \bar{r}_*}\,,
  \label{eq:zerinf_ren}
\end{equation} 
where $\bar{r}_*$ is negative and very large. We have computed $A^\infty(\omega)$, by evaluating the integrals from
$\bar{r}_*=-44\,M$ to the extraction radius $r_*^{\rm extr}=400\,M$.

The time-domain Zerilli function at infinity can then be computed, as a function of the retarded time $u=t-r_*$, as
\begin{equation}
\psi(u)=\frac{1}{2\pi}\int_{-\infty}^{+\infty}A^\infty(\omega) e^{-i\omega u}d\omega\,.\label{eq:invlapl}
\end{equation}

\subsection*{Shooting method}

The shooting approach, instead, consists in the numerical integration of Eq.~\eqref{eq:Zerilli_model}, from
$\bar{r}_*=-44\,M$ to the extraction radius $r_*^{\rm extr}=400\,M$, by imposing the boundary conditions~\eqref{eq:bcsomm1}
and matching the solution at $r_*^{\rm extr}$ with an analytic expression obtained by an asymptotic expansion of
Eq.~\eqref{eq:Zerilli_model}.
For each value of the frequency $\omega$, we performed the numerical integration of Eq.~\eqref{eq:Zerilli_model} for
different values of $A^H$ until we obtained an outgoing wave at infinity, as in Eq.~\eqref{eq:bcsomm1}. In this
way we computed the function $A^\infty(\omega)$ and then, by Eq.~\eqref{eq:invlapl}, the Zerilli function at infinity.

The results of the two approaches perfectly agree with each other, and they also agree with the results of~\cite{Price:1994pm} (see Fig.\til\ref{fig:BBHvsBECO}).

\section{Close limit of extreme compact objects}\label{sec:BH_mimickers}
We shall now apply the CLAP to describe the head-on collision of two non-rotating, equal mass ECOs. We model an ECO (see
e.g.~\cite{Cardoso:2019rvt} and references therein) as a spherically symmetric compact body with mass $M$ and a
surface at $r=r_0$,
\begin{equation}
  r_0=2M(1+\epsilon)\,,
\end{equation}
with $\epsilon\ll1$.
One of the motivation to consider these class of objects is to mimic the effects of a theory where BHs are absent (via, for example, high-energy phenomena). We assume for example, that a theory of quantum gravity forbids the existence of horizons via Planck-scale physics\til\cite{Giddings:1992hh,Mazur:2004fk,Mathur:2005zp,Mathur:2008nj,Giddings:2009ae,Unruh:2017uaw}. We do not wish to build such a theory but merely to investigate some of its consequences. Furthermore, we do assume that departures from GR occur only close to the Schwarzschild radii.

In addition to the above motivation, ECO models of the above kind are also interesting from the perspective of testing some of the unique features related with BH spacetimes. In fact, a rather (a priori) simple question, what happens in the $\epsilon \rightarrow 0$ limit of ECOs, might give non-trivial outcomes, as the appearance of echoes will show later in this Chapter.

A fundamental difference between the surface of an ECO and the horizon of a BH is that an incoming wave is partially
reflected by the ECO surface, while it is totally absorbed by a BH horizon. Thus, given a (scalar, gravitational, etc.)
test field $\phi(t,r_*)\sim\tilde\phi(r_*)e^{i\omega t}$, its boundary condition near the surface $r=r_0$ is
\begin{equation}
  \tilde\phi(r_*)\sim e^{-i\omega r_*}+\mathfrak{R}e^{i\omega r_*}\,.\label{bceco0}
\end{equation}
The parameter $\mathfrak{R}$, which in general depends on the frequency $\omega$ and on the spin of the field, is called
{\it reflectivity coefficient} of the ECO.
\subsection{Initial data for extreme compact objects}\label{subsec:BL-ECO}
We shall first introduce the spacetime metric of a single, isolated, spherically symmetric ECO; then, using the
procedure of Brill and Lindquist discussed in Sec.~\ref{subsec:BL}, we shall define initial data corresponding to two
ECOs starting from rest (this approach can naturally be extended to the case of ECOs with initial momentum, using the
procedure of Bowen and York~\cite{Bowen:1980yu}).

The GW signal produced by the collision of ECOs will consist of waves caused by excitations of the exterior
spacetime~\cite{Vishveshwara:1970zz,Berti:2009kk}, as well as contributions which probe the interior of the
object~\cite{Cardoso:2016rao,Cardoso:2016oxy,Cardoso:2017cqb,Abedi:2016hgu,Mark:2017dnq,Testa:2018bzd,Maggio:2018ivz}. For
a wide class of ECOs, the lapse in its interior is very small, leading to large delays in signals which have probed the
interior geometry~\cite{Ferrari:2000sr,Cardoso:2017cqb,Cardoso:2019rvt}. Thus, in the following we ``freeze'' the inner
region, which means that in practice we cut it off from our domain, and all the information about the interior is
replaced by boundary conditions at the surface.

Let us first consider a single, isolated (spherically symmetric) ECO. Throughout this Chapter we always assume that the exterior of ECO spacetimes can be described through vacuum GR equations. Hence, we consider that small-scale departures from GR, that should be needed to form ECOs in the first place, are irrelevant in what follows. Working in vacuum GR, due to Birkhoff theorem, the exterior of the ECO
$r>r_0$ is described by Schwarzschild's geometry (Eq.\til\eqref{eq:Schwarzschild_metric}). Since $\epsilon\ll 1$, $r_0<3M$ and thus the light ring $r=3M$, where null circular geodesics are defined, lies in the exterior of the ECO. This plays a crucial role in the GW emission, as we discuss below. Then, we define the isotropic radial coordinate as in Schwarzschild metric~\eqref{eq:R_isotropic_to_Schw}:
\begin{equation}
R=\frac{1}{4}\left( \sqrt{r}+\sqrt{r-2M}\right)^2\,.\label{eq:R_isotropic_to_ECO}
\end{equation}
In terms of the coordinate $R$, the location of the surface is $R_0=R(r_0)=M/2(1-2\sqrt{\epsilon})$.
The interior of the ECO depends on the profile of its energy density $\rho(R)$. As discussed above, under
our assumptions the emitted GW signal does not depend on the structure of the interior; it only depends on the location
of the surface. Yet, for consistency we shall model the interior with a specific example of energy density profile $\rho(R)$,
vanishing for $R>R_0$ (see Appendix~\ref{app:ECO_initialdata}).

Let us now consider two spherically symmetric ECOs, with equal masses and total ADM mass $M$,
initially at rest.  We perform a $3+1$ decomposition of the spacetime, as in Sec.~\ref{sec:3+1}. The three-metric of
each slice is solution of the Hamiltonian constraint equation~\eqref{eq:HGRphi}:
\begin{equation} 
\mathcal{H}^{\rm GR}= \prescript{3}{}{R}+K^2-K\ijdd K\ijuu=16 \pi \rho\,.\label{eq:H_ECO}
\end{equation}
Since the ECOs are initially at rest, the initial extrinsic curvature vanishes and Eq.~\eqref{eq:H_ECO} reduces to
$\prescript{3}{}{R}= 16 \pi \rho$. In the exterior of both ECOs, $\rho=0$ and the metric can be written in the BL 
form, Eq.~\eqref{eq:dsBL_pert}:
\begin{equation}
\label{eq:dsBL_pert_ECO}
\prescript{3}{}{ds}^2_{\rm BL-ECO}= \left( 1+\frac{4}{1+\frac{M}{2R}} \sum_{\ell=2,4,\dots}^{\infty}
\xi_\ell P_\ell\left(\cos\theta\right)\right)\left(f^{-1} dr^2+r^2d\Omega^2\right)\,,
\end{equation}
with $\xi_\ell=\left(Z_0/M\right)^\ell/2$ as in Eq.~\eqref{eq:xi_M1eqM2}. In the isotropic frame, the two ECOs move
along the $z$-axis and are located at $R=\pm Z_0$, with $0<Z_0\lesssim 0.85M$ (see Sec.~\ref{subsec:BL} for a detailed discussion on the meaning of $Z_0$ and how it is related to physical distances). We call Eq.~\eqref{eq:dsBL_pert_ECO} the BL-ECO initial data.

We remark that although the BL initial data for BBHs in Eq.~\eqref{eq:dsBL_pert} and the BL-ECO initial data for binary
ECOs in Eq.~\eqref{eq:dsBL_pert_ECO} are formally identical, the former are defined outside the BH horizon,
while the latter are defined outside the ECO surface. This difference, as we shall show below, leads to a difference in
the GW emission. Note also that, as discussed above, the region inside the surfaces of the ECOs (explicitly computed for
a specific case of energy density profile in Appendix~\ref{app:ECO_initialdata}) does not contribute to the GW signal, by assumption.

Finally, let us stress that, as it happens for the case of BBHs in GR, the initial data to describe two ECOs are not unique. Our approach has been chosen by mere mathematical convenience, under what we believe a reasonable physical assumption (e.g. small lapse). Yet, this simplified choice allow us to unveil some of the properties of ECOs collision. However, only a comparison with full non-linear simulations involving ECO binaries will assess the degree of validity of our approach.

\subsection{Head-on collision}\label{subsec:ECO-collision}
%
We shall now recast the BL-ECO initial data in Eq.~\eqref{eq:dsBL_pert_ECO} as a perturbation of a single object,
\begin{equation}
\label{eq:eco_exp}
g_{\mu\nu}=g^{(0)}_{\mu\nu}+h_{\mu\nu}\,.
\end{equation}
Then, we shall evolve them using the tools of perturbation theory.

Since we are not dealing with vacuum spacetimes, the outcome of the ECO collision depends, in principle, on the
features of the ECOs, i.e. on their internal structure. Two outcomes are possible: either the collision leads to a
single BH (which, in the head-on collision of non-rotating ECOs, is described by the Schwarzschild metric), or it leads to an
ECO. In the former case the initial data should be recast as a perturbation of a Schwarzschild solution; in the latter,
as a perturbation of an ECO. We shall treat these two cases separately.

\subsubsection{Formation of a black hole}
If the final object is a Schwarzschild BH, the background metric $g^{(0)}_{\mu\nu}$ in Eq.~\eqref{eq:eco_exp} is
Schwarzschild's metric. The procedure of recasting the BL-ECO initial data~\eqref{eq:dsBL_pert_ECO} as a perturbation of
Schwarzschild metric is formally equivalent to the derivation in Sec.~\ref{subsec:recast_BH}, leading to an $\ell=2$
Zerilli function at the initial time $t=t_0$ given by Eqs.~\eqref{eq:zerillit0}-\eqref{eq:defQ}.

Since the background is Schwarzschild's metric, the Zerilli equation~\eqref{eq:zer} has to be solved with the same
boundary condition as in the BBH case. Therefore, the gravitational waveform emitted in the
collision of two ECOs into a Schwarzschild BH, computed with the CLAP (and neglecting the effect of perturbations inside
the ECO's surface) coincides with the waveform emitted in a BBH collision.

\subsubsection{Formation of an extreme compact object}
If the final object is an ECO, the background metric $g^{(0)}_{\mu\nu}$ is Schwarzchild metric only for
$r>r_0$. Again, the procedure of recasting the BL-ECO initial data~\eqref{eq:dsBL_pert_ECO} as a perturbation of the ECO
metric follows the derivation in Sec.~\ref{subsec:recast_BH}, leading to the Zerilli function at the initial time
$t=t_0$ given in Eqs.~\eqref{eq:zerillit0} and \eqref{eq:defQ}. However, the Zerilli function $\psi(t,r_*)$ is only defined
for $r_*>r_{0*}=r_0+2M\log|r_0/(2M)-1|$, and it satisfies boundary conditions different from those of a BH.

Using the Laplace transform approach discussed in Sec.~\ref{subsec:recast_BH}, one finds the Zerilli equation with source
given in Eqs.~\eqref{eq:zerS} and \eqref{eq:zer_source}. Although the equation is the same as for BBH collisions, the
boundary conditions are different: since the surface is partially reflecting, they are (see Eq.~\eqref{bceco0})
\begin{align} \label{eq:BC_ECO}
  \tilde{\psi}&\sim e^{-i \omega  \left( r_*- r_{0*}\right)}+ \mathfrak{R}\,
  e^{i \omega \left( r_*- r_{0*}\right)}~~~~~~(r_*\to r_{0*})\nn\\
\tilde{\psi}&\sim  e^{i \omega r_*} \hspace{4.2cm} (r_*\to\infty)\,,
\end{align}
where $\mathfrak{R}$ is the reflectivity coefficient of the ECO. In general $\mathfrak{R}$ is a function of $\omega$; however, for simplicity, we shall assume it to be a constant.

We solved the Zerilli equation~\eqref{eq:zerS} with boundary conditions\,\eqref{eq:BC_ECO} using a shooting method, for different values of $\epsilon$ and of $\mathfrak{R}$, finding the gravitational waveform emitted in the merger and ringdown phases of the collision. Since we only consider the leading quadrupolar contribution, the Zerilli function is proportional to $\xi_2$, and thus to $Z_0^2$ (cf. Eq.~\eqref{eq:zerillit0}). Moreover, it has (in geometric units) the dimensions of length, hence
since the only dimensionful quantity characterizing the ECO is its ADM mass $M$, $\psi(t,r_*)\propto Z_0^2/M$.

\subsection{Results of the numerical integration}\label{app:solveZerECO}
If the outcome of the collision is an ECO, the Zerilli equation~\eqref{eq:Zerilli_model} describes perturbations of the
ECO spacetime which, as discussed in Sec.~\ref{sec:BH_mimickers}, coincides with Schwarzschild's spacetime with the
domain restricted to $r_{0*}<r_*<+\infty$. Moreover, we impose at $r_*\to r_{0*}$ the partially reflecting boundary
conditions ~\eqref{eq:BC_ECO}:
\begin{align}
  \tilde{\psi}&=A^H(e^{-i \omega(r_*-r_{0*})}+\mathfrak{R}e^{i \omega(r_*-r_{0*})})+\frac{\hat S}{\omega^2} ~~~(r_*\to r_{0*})\label{eq:BC_ECO2}\\
  \tilde{\psi}&=A^\infty e^{i \omega r_*}~~\hspace{4.7cm}(r_*\to\infty)  \nonumber
\end{align}
with $\mathfrak{R}$ reflectivity coefficient which we assume, for simplicity, to be constant, and the constant term in
the ingoing boundary conditions is due to the fact that $S(r_{0*})=i\omega\psi(t=0,r_{0*})=\hat S\neq0$.  Note that
since the tortoise coordinate does not extend to $-\infty$, the Laplace transform $\tilde\psi(\omega,r_*)$ is
well-defined for real frequencies, and we do not need to worry about ill-defined contributions to the integrals.

In this case we only perform the integration using the shooting approach, i.e. we integrate
Eq.~\eqref{eq:Zerilli_model}, from $\bar{r}_*=-44\,M$ to the extraction radius $r_*^{\rm extr}=400\,M$, by imposing the
boundary conditions~\eqref{eq:BC_ECO2}, for different values of $A^H$, until we obtained an outgoing wave at
infinity. Then, using Eq.~\eqref{eq:invlapl} we compute the Zerilli function at infinity.


\begin{figure}
\centering
\includegraphics[width=0.5\textwidth,keepaspectratio]{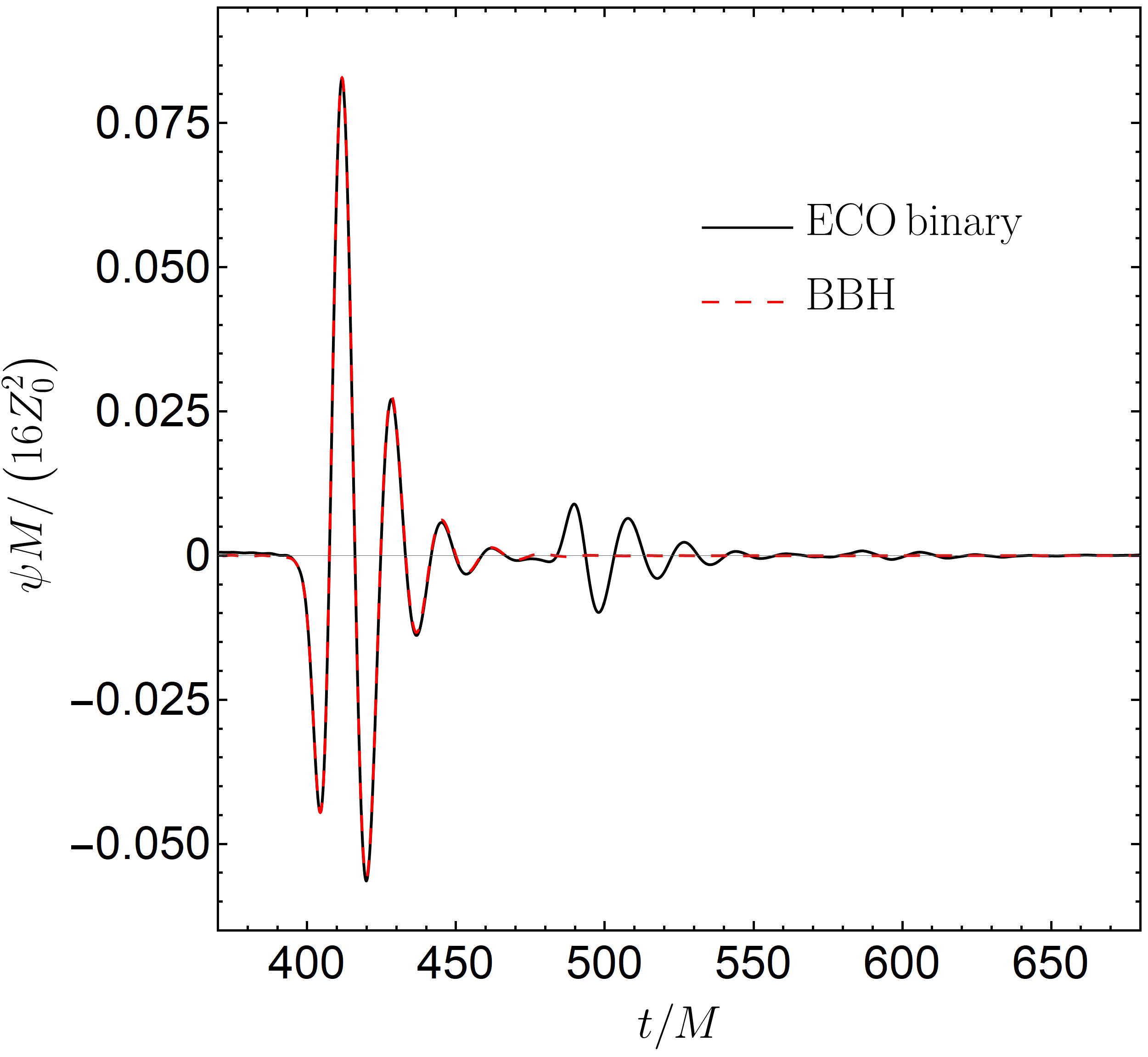}
\caption[GW head-on collision of equal-mass BHs/ECOs.]{Gravitational waveform for the head-on collision of two equal-mass, spherically symmetric BHs/ECOs starting
  from rest, evaluated at the extraction radius $r_*^{extr}=400\,M$, for two possible outcomes. If
  the final object is a BH, the waveform is identical to that resulting from the collision of two BHs: a sharp burst
  followed by ringdown (caused by the relaxation of the light ring). When the final product is itself an ECO, we observe
  a similar initial stage, followed at late times by echoes of the initial
  burst~\cite{Cardoso:2016oxy,Cardoso:2019rvt}. Here, the final object is taken to have a reflectivity of
  $\mathfrak{R}=0.1$ and a surface $r_0=2M(1+\epsilon)$ with $\epsilon=10^{-10}$.}
\label{fig:BBHvsBECO}
\end{figure}

\begin{figure}[ht]
\centering
\includegraphics[width=0.6\textwidth,keepaspectratio]{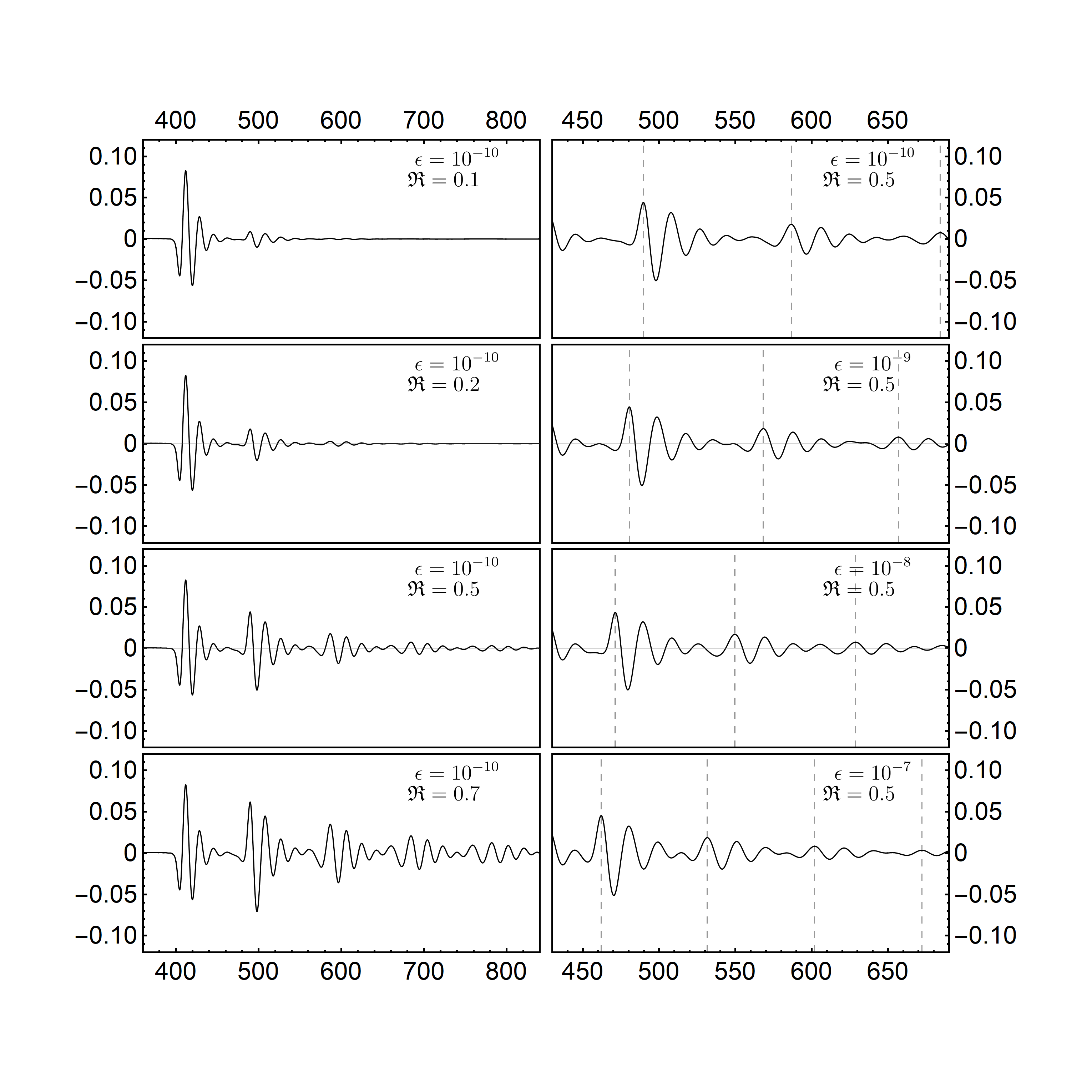}
\caption[GW head-on collision of equal-mass ECOs varying $\mathfrak{R}$ and $\epsilon$.]{Gravitational waveforms for head-on collisions of two equal-mass, spherically symmetric ECOs starting from rest,
  leading to an ECO, for different values of the surface parameter $\epsilon$ and of the reflectivity $\mathfrak{R}$,
  evaluated at the extraction radius $r_*^{extr}=400\,M$ and rescaled by $16Z_0^2/M$. The vertical dashed lines correspond to the
  peak value of each echo.
%
  The amplitude of the echoes increases with $\mathfrak{R}$, while the time delay between echoes (i.e. between the dashed
  lines) decreases when $\epsilon$ increases.}
%
\label{fig:Echoes_plot}
\end{figure}

Figure~\ref{fig:BBHvsBECO} shows the Zerilli quadrupolar waveform generated by the head-on collision of two spherically
symmetric compact objects starting from rest, evaluated at the extraction radius $r_*^{\rm extr}=400\,M$. The waveforms are rescaled by $16Z_0^2/M$, and thus they do not depend neither on $Z_0$ nor on $M$. We compare processes leading to a Schwarzschild BH (dashed curve) and to an ECO with $\epsilon=10^{-10}$ and $\mathfrak{R}=0.1$ (solid line).
The early-time component of the signals parts are identical, while they differ at late times, since the ECO collision signal is characterized by a series of ``echoes'', which are a characteristic
feature of GW signals from ECOs (see
e.g.~\cite{Cardoso:2016rao,Cardoso:2016oxy,Mark:2017dnq,Cardoso:2017cqb,Correia:2018apm,Cardoso:2019rvt}). We stress
again that, in this approximation, the signal depends on the nature of the final object, but it does not depend on
whether the colliding objects are BHs or ECOs.

A simple interpretation of the echo structure is the following. The light ring excitation by the initial data is followed by its
``relaxation''. Since both BHs and ECOs have equivalent geometries close to the light ring, they both relax in the same
way~\cite{Cardoso:2016rao,Cardoso:2016oxy,Cardoso:2017cqb,Cardoso:2019rvt}.  This relaxation produces an outgoing wave,
which corresponds to the early part and ringdown of the signal.  But there's also an ``ingoing'' wave which interacts
with the ECO via boundary conditions. Due to a non-zero reflectivity, the ingoing pulse is partly reflected back, and
interacts with the light ring again. The process continues and produces a sequence of distorted copies of the original
burst, i.e. the echoes. 

Figure~\ref{fig:Echoes_plot} displays the dependence of the Zerilli waveform on the ECO parameters $\epsilon$ and $\mathfrak{R}$. We can
see that the echo delay increases with $\epsilon$, while the amplitude of the echoes is larger for larger values of the
reflectivity $\mathfrak{R}$.

We find that the time separation between echoes is well described (with a relative error smaller than $2\%$) by the
following analytical fit,
\begin{equation}
\Delta t_{\rm ECO}\sim 4.3 \left\lvert \log\epsilon \right\rvert M\,,\label{eq:echo_delay}
\end{equation}
for $10^{-10}<\epsilon <10^{-6}$. This logarithmic behaviour is consistent with our knowledge of echoes in the GW signals~\cite{Cardoso:2019rvt}. We only solved the Zerilli equation for
$\epsilon\ge10^{-10}$; for smaller values of $\epsilon$, our numerical approach loses accuracy; however, since this
limitation is of computational nature only, we expect the fit~\eqref{eq:echo_delay} to hold for smaller values of
$\epsilon$ as well.

\begin{figure}
\centering
\includegraphics[width=0.4\textwidth,keepaspectratio]{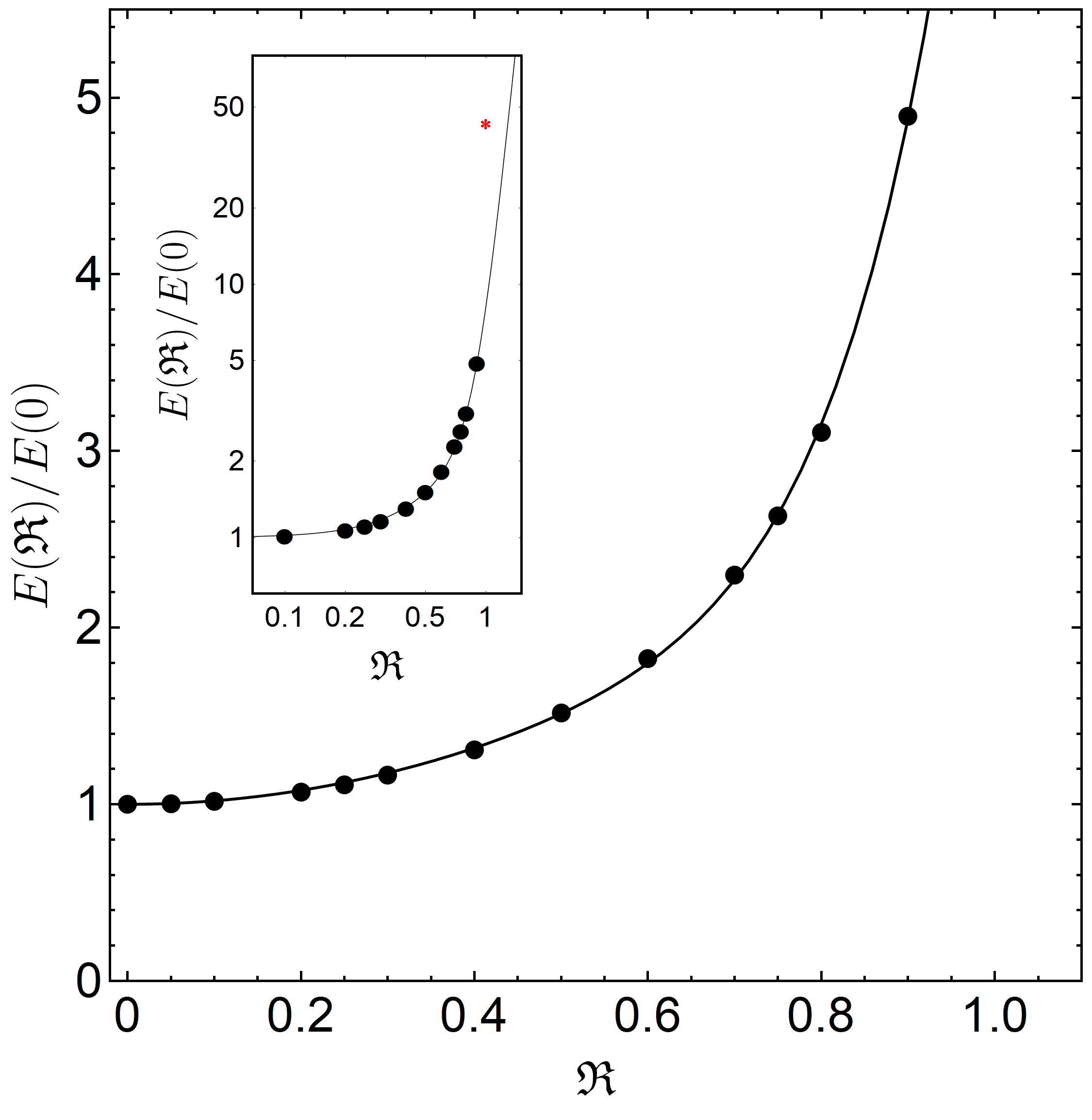}
\caption[Energy in the collision of ECOs vs $\mathfrak{R}$.]{Total energy emitted in the head-on collision from rest of two equal-mass, spherically symmetric compact objects. Fixing the ECOs/BHs initial separation $Z_0$, the ratio of energy emitted is independent of such quantity. The final object is assumed to be an ECO with $\epsilon=10^{-10}$ and the results are shown as a function of the reflectivity $\mathfrak{R}$. {\bf Inset:} log-log scale of the same results, including $\mathfrak{R}=1$. This also corresponds to the formation of a wormhole~\cite{Cardoso:2016rao} (see main text).}
\label{fig:E_vs_reflectivity}
\end{figure}

Finally, we compared the total energy $E$ emitted in GW during a compact object collision leading to an ECO, with that emitted in
a BBH collision. It can be computed in terms of the Zerilli function from Eq.~\eqref{total_E}. We find that, for
$\epsilon<10^{-3}$, the energy radiated is very weakly dependent on $\epsilon$, while it increases with the reflectivity
$\mathfrak{R}$. This is due to the fact that the presence of echoes contributes to the energy loss of the system in GWs.

In Fig.~\ref{fig:E_vs_reflectivity} we show the total energy $E$ emitted in GWs as a function of $\mathfrak{R}$,
normalized by the energy emitted for $\mathfrak{R}=0$ (which coincides with the result for a BBH collision~\cite{Price:1994pm}). 
This shows which part of the emitted energy is due to echoes. The function $E(\mathfrak{R})$ can be
described by the following fit (accurate within $1.5\%$ in the range $0\le \mathfrak{R}\le 0.8$):
\begin{equation}
\frac{E}{M}\approx 10^{-6}\left(6.14+  \mathfrak{R}^{2} (1.29+3.26 \mathfrak{R}^{6})\right)\frac{256Z_0^4}{M^4}\,.\label{eq:rad_fit}
\end{equation}

One can use the CLAP formalism to understand, in particular, the radiation given away during the formation of
wormholes~\cite{Morris:1988tu,Visser:1995cc}. A class of these objects can be considered as ECOs with
$\mathfrak{R}=1$, when the ``throat'' is made of a rigid shell~\cite{Cardoso:2016rao}. This corresponds to taking
$\mathfrak{R}=1$ in the above framework.  The inset of Fig.~\ref{fig:E_vs_reflectivity} shows the total energy computed
with the CLAP for head-on collisions forming one of these wormholes.  The total radiated energy is slightly off the
predictions of the fit \eqref{eq:rad_fit} because points $\mathfrak{R}\gtrsim 0.8$ were not used in the fit.  Notice
that the total radiated energy is over one order of magnitude larger than the corresponding process forming a BH: 
substantial amount of energy is released in late-time echoes.

Furthermore, a spectral analysis shows that the ECO formation excites certain characteristic frequencies, which
correspond to the quasinormal frequencies of the final object.  In the particular case of a thin-shell wormhole
described above, most of the radiation is in fact contained in such modes: the energy spectrum shows clear peaks at such
quasinormal frequencies, which are in excellent agreement with the spectral findings of Ref.~\cite{Cardoso:2016rao}.

\section{Binary black holes and scalar fields}\label{sec:CLAP_STT}
From the first observation of the Higgs boson by the ATLAS Collaboration\til\cite{Aad:2012tfa}, a growing recognition
has been given in studying the effects of scalar particles, both at a cosmological and an astrophysical
level\til\cite{Ikeda:2018nhb,Boskovic:2018lkj,Isi:2018pzk,Bernard:2019nkv,Sun:2019mqb,Berti:2019wnn,Cardoso:2020hca,Ikeda:2020xvt}.
Moreover, most of the modifications of GR which have been proposed so far can be reformulated in terms of couplings between gravity and extra fields, the simplest of which are scalar fields\,\cite{Berti:2015itd}. In this Section we study how the presence of scalar fields may affect a BBH collision, and whether they can leave
observable imprints during the GWs generation.

We shall consider gravity minimally coupled with a complex scalar field. Since we are interested in BH solutions, we do not include matter fields in the model. Thus, the action is:
\begin{equation}
S=\int
d^4x\sqrt{-g}\left(\frac{R}{16\pi}-\frac{1}{2}g^{\mu\nu}\partial_\mu\Phi^*\partial_\nu\Phi\right)\,.\label{eq:action_minimal}
\end{equation}
The field equations obtained from this action are Einstein's equations coupled with the Klein-Gordon equations:
\begin{align}
\label{eq:Einstein_eq}
R_{\mu\nu}-\frac{1}{2}g_{\mu\nu}R&=8\pi T_{\mu\nu}\,,\\
\label{eq:KG_eq}
\square\,\Phi&=0\,,
\end{align}
where 
\begin{equation}
  T_{\mu\nu}=-\frac{1}{2}g_{\mu\nu}\left(\partial_\lambda\Phi^*\partial^\lambda \Phi\right)+
  \frac{1}{2}\left(\partial_\mu\Phi^*\partial_\nu\Phi+\partial_\mu\Phi\partial_\nu\Phi^*\right)\,,\nonumber
\end{equation}
is the scalar field stress-energy tensor. A wide class of modified gravity theories in which gravity is non-minimally coupled with a scalar field -- the so-called Bergmann-Wagoner scalar-tensor theories (see e.g.~\cite{fujii2003scalar,Berti:2015itd} and references therein) -- is formally equivalent to the theory in Eq.~\eqref{eq:action_minimal}. If restricted to vacuum spacetimes in fact, a conformal rescaling of the metric maps one theory into the other~\cite{fujii2003scalar}. Thus, the scalar field $\Phi$ can be interpreted either as a fundamental ``matter'' field in GR, or as a gravitational degree-of-freedom in a
modified gravity theory.

The $3+1$ decomposition of the action~\eqref{eq:action_minimal} has been discussed in Sec.~\ref{sec:3+1}.
In particular, the constraint equations have the form
\begin{align}
\label{eq:Hstt1}
&\mathcal{H}^{\rm GR}-16\pi \rho=0\,,\\
\label{eq:Mstt1}
&\mathcal{M}^{\rm GR}_i-8\pi j_i=0\,,
\end{align}
where the energy density $\rho$ and the energy-momentum flux $j_i$ of the scalar field are
given in Eqs.~\eqref{eq:rhostt}-\eqref{eq:jstt}.

As extensively discussed in the Introduction, there are various no-scalar hair theorem in theories including an extra scalar degree-of-freedom. Some of those focus on time-independent real fields\til\cite{Bekenstein:1972ky,Hawking:1972qk}, others on static BH spacetimes\til\cite{Sotiriou:2011dz}. Theories described by the action~\eqref{eq:action_minimal} also satisfy a no-scalar hair theorem: stationary BH
solutions are described by the Kerr metric, and thus they have vanishing scalar
field (see~\cite{Cardoso:2016ryw} and references
therein). Therefore, we know that the remnant of a BBH collision becomes --~in the timescale of the QNM oscillations,
i.e. of $\sim1-10\,M$~-- a stationary BH solution, with vanishing scalar field. The no-hair theorem does not tell us what happens before reaching the final stationary configuration. However, a similar
result applies to the inspiral part of a BBH coalescence. Indeed, an analysis in the PN approximation~\cite{Blanchet:2013haa,Poisson_will_2014}, which accurately describes the BBH inspiral, shows that the binary dynamics in the theory~\eqref{eq:action_minimal}, up to $2.5$ order in the PN expansion, is the same as in GR~\cite{Will:1989sk}. As argued before, similar results hold also in the a PN treatment of Bergmann-Wagoner scalar-tensor gravity.  However, we still do not know if the scalar field significantly affect the dynamics during the merger and ringdown stages.

In order to address this problem, we shall study the QNMs of the scalar field during the merger and ringdown of a BBH (head-on) collision. We do not expect the scalar field to grow large before the collision, thus we shall treat it as a
perturbation of the BBH spacetime; we define a perturbation parameter $\epsilon\ll1$, such that $\Phi=O(\epsilon)$.
Therefore, $T_{\mu\nu}=O(\epsilon^2)$ and, to linear order in the perturbation, we can neglect the scalar field from
Einstein's equations~\eqref{eq:Einstein_eq} (and in particular from the constraint equations~\eqref{eq:Hstt1}-\eqref{eq:Mstt1}). We shall then study the linearized ($O(\epsilon)$) field equation of the scalar field in the BBH spacetime, which is modeled using the CLAP. For such configurations we shall compute the scalar field QNMs. Then, by
comparing the scalar field QNMs in a BBH spacetime with the scalar field QNMs in a stationary BH spacetime, we will
assess whether the binary dynamics significantly affects the scalar field dynamics.

The linearization process just described will provide a wave equation in a non-trivial spacetime, that needs to be solved with suitable boundary conditions. Despite not being impossible in principle, this assumptions might already be an indication of the absence of scalar instabilities in BBH spacetimes. However, only the numerical solution of the above problem will quantify {\it how much} QNMs change due to the interaction energy present in the binary (even foreseeing that their imaginary part will not change sign).
\subsection{The background}
We describe the BBH spacetime (neglecting the scalar field, as discussed above) using BL initial data. Therefore, we recast the BBH spacetime as a perturbation of the Schwarzschild metric. Including the leading-order quadrupolar contribution, as
discussed in Sec.~\ref{subsec:recast_BH}, the total spacetime can be written as $g_{\mu\nu}=g^{(0)}_{\mu\nu}+h_{\mu\nu}$ where $g^{(0)}_{\mu\nu}={\rm
  diag}(-f,f^{-1},r^2,r^2\sin^2\theta)$ and $h_{\mu\nu}$ is given by Eqs.~\eqref{eq:BLrecast}-\eqref{eq:H2_Kl_equality}:
\begin{align}
h_{rr}&=f^{-1}gP_2(\cos\theta)\xi_2\,,\nonumber\\
h_{\theta\theta}&=r^2g P_\ell(\cos\theta)\xi_2\,,\label{eq:BLrecast1}
\end{align}
where $g=4\left(1+M/(2R)\right)^{-1}M^3/R^3$, the isotropic coordinate $R$ is defined in
Eq.\til\eqref{eq:R_isotropic_to_Schw}, and $\xi_2=Z_0^2/(2M^2)\lesssim1$.
We stress that we are assuming the background to be stationary, thus neglecting the motion of the BHs, in the timescale of the
oscillation ($t\sim(1-10)M$). This is a crude approximation, since the BH separation changes, and their
velocities become non-negligible in this timescale. Thus, the result of this computation should be considered as an
order-of-magnitude estimate of the effect of the BH dynamics on the scalar QNMs.

\subsection{Scalar perturbations' master equation}

Since $h_{\mu\nu}\propto Z_0^2$, we can expand the D'Alembertian operator $\square=\frac{1}{\sqrt{-g}}\partial_\mu\left(
g^{\mu\nu}\sqrt{-g}\partial_\nu\right)$ in Eq.~\eqref{eq:KG_eq}, for small separations, as
\begin{equation} \label{eq:box_expansion}
\square = \square^{(0)}+ Z_0^2 \, \square^{(1)} + \mathcal{O}(Z_0^3)\,,
\end{equation}
where the explicit form of the operator $\square^{(1)}$ is given in Appendix~\ref{app:separateKG}.
Expanding the scalar field as
\begin{equation}\label{eq:spher_harm_decomp}
\Phi\left( t,r,\theta,\phi\right)=\frac{1}{r}\sum_{\ell,m}\psi_{\ell m}\left( t,r\right) Y^{\ell m}\left(\theta,\phi\right)\,,
\end{equation}
we get a non-separable equation, where the harmonic component $\psi_{\ell m}$ couples with the components
$\psi_{\ell\pm2\,m}$. As discussed in Appendix~\ref{app:separateKG}, we can follow the same approach used to study
scalar field perturbations around rotating BHs (see e.g.~\cite{Cano:2020cao,Pierini:2021jxd}); indeed, the leading-order
rotational corrections are quadrupolar as well. Remarkably, the $\ell\leftrightarrow\ell\pm2$ couplings do not affect
the QNM frequencies at leading order in the perturbations (see also~\cite{kojima1993normal,Pani:2012bp}), and thus they
can be neglected, leading to a decoupled, Schr\"{o}dinger-like equation:
\begin{equation} \label{eq:KG_notortoise}
  \frac{\partial^2 \psi_{lm}}{\partial t^2}+\frac{\partial^2 \psi_{lm}}{\partial r^2}
  \left( U_0+Z_0^2\tilde{U}_0\right)+\frac{\partial \psi_{lm}}{\partial r}  \left( U_1+Z_0^2\tilde{U}_1\right)+\psi_{lm} \left(  W_0 +Z_0^2  W_1\right)=0\,,
\end{equation}
where $\ell\ge1$. The monopolar $\ell=0$ perturbations are not affected by the $Z_0^2$ corrections (see  Eq.\til\eqref{eq:q12lm}), hence the $\ell=0$ QNMs are the same as in the single BH case. The derivation of Eq.~\eqref{eq:KG_notortoise} and the explicit form of the functions $U_A(r)$, $\tilde U_A(r)$, $W_A(r)$ ($A=0,1$) are given in
Appendix~\ref{app:separateKG}.

In order to find the QNMs, we have solved Eq.~\eqref{eq:KG_notortoise} through direct integration with outgoing boundary
conditions at infinity and with ingoing boundary conditions at the horizon. The boundary conditions have been determined as a perturbative, polynomial expansion at each boundary, whose coefficients have been found solving
Eq.~\eqref{eq:KG_notortoise} order by order, as explained, e.g. in Ref.\til\cite{Pani:2013pma}.

\begin{figure}
\centering
\includegraphics[width=0.6\textwidth,keepaspectratio]{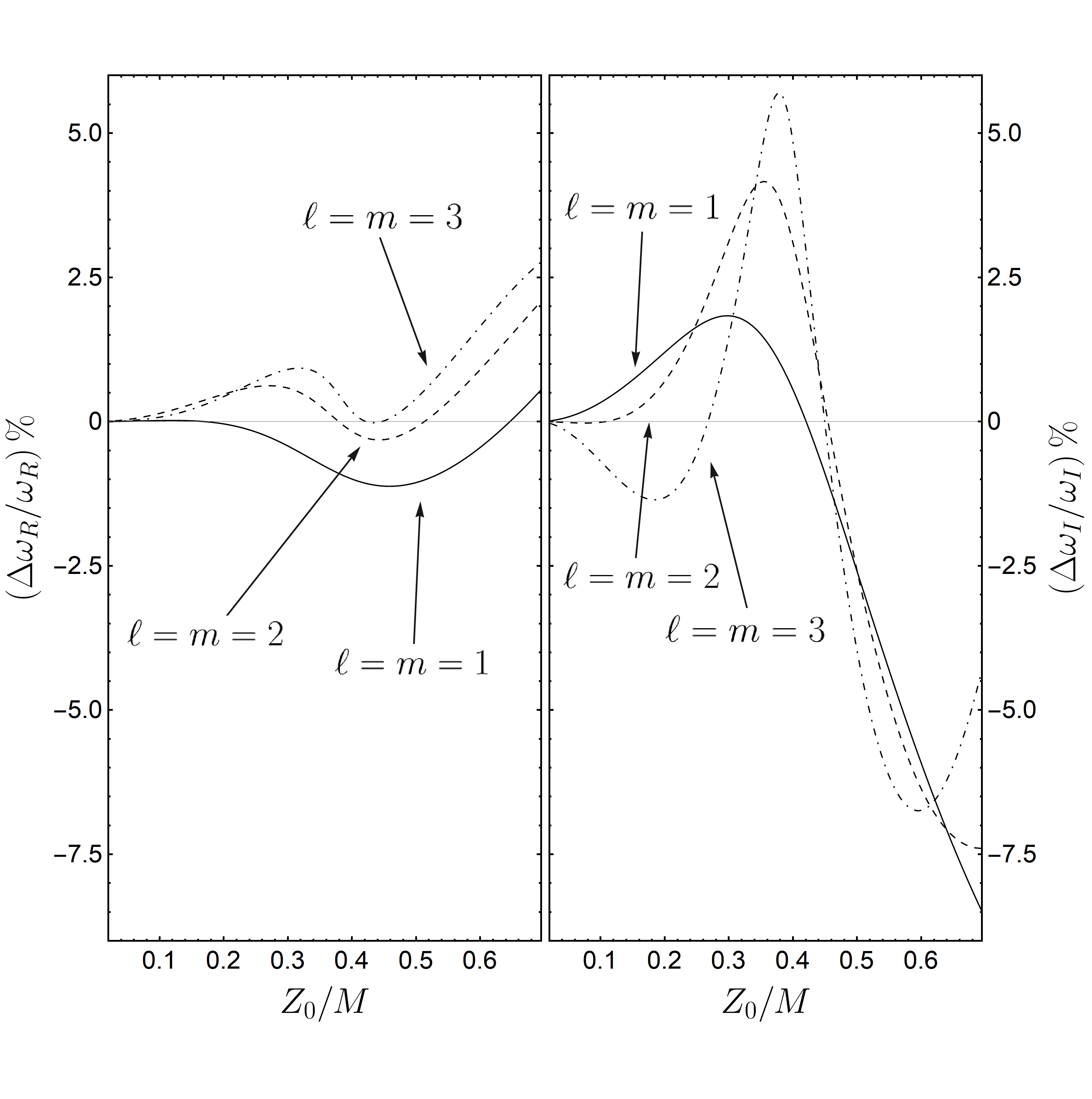}
\caption[Binary BHs scalar QNMs deviations.]{Fractional percentage shifts $\Delta\omega_R/\omega_R$ (left panel) and $\Delta\omega_I/\omega_I$ (right
  panel), as defined in Eq.~\eqref{eq:shifts}, for $\ell=1,2,3$ and $m=\ell$, as functions of the BHs initial separation
  $Z_0/M$.}
\label{fig:BBH_QNM}
\end{figure}
\subsection{Quasi normal modes}

To validate our integration method, as a first step we computed the fundamental scalar QNMs of a Scwharzschild BH,
setting $Z_0=0$. Our results agree with those in the literature (e.g.~\cite{GRITJHU}) within $0.2\%$. Then, we
computed the scalar QNMs in BBH collisions for different values of $\ell\ge1$ and of $Z_0$. Figure~\ref{fig:BBH_QNM} shows the fractional percentage shift of real and imaginary parts of the QNMs with respect to those in Schwarzschild BHs:
\begin{equation}
\frac{\Delta\omega_{R/I}}{\omega_{R/I}}=\frac{\omega_{R/I}-\omega_{R/I}^{\rm (Schw)}}{\omega_{R/I}^{\rm (Schw)}}\,,\label{eq:shifts}
\end{equation}
for $\ell=1,2,3$ and $m=\ell$, as functions of $Z_0/M\le0.7$.

As we can see from Fig.~\ref{fig:BBH_QNM}, the QNMs shifts have a non-trivial dependence on the BH separation $Z_0$, but
each mode is shifted by just a few percent from the corresponding mode of an isolated BH. Despite the approximations
discussed above, this result provides a strong indication that the BH dynamics does not significantly affect the
behaviour of the scalar perturbations, at linear order. Thus, we expect that, like in the inspiral and in the late
ringdown, the scalar field does not play a relevant role in the BH dynamics. If, instead, a significant
growth of the scalar field takes place in a BBH coalescence, there would be an instability at the linearized level as
well; we provide evidence against this scenario, because the mode frequencies do not change enough in this stage to
become unstable.

Thus, our results indicate that scalar fields in GR, or in modified gravity theories in which the no-hair theorem
applies, do not significantly affect the coalescence of BBHs.

These results are also interesting in light of a completely different question: at what time the GW signal from a BBH
coalescence can be described as a superposition of QNMs? This problem, i.e. the determination of the starting time of
the ringdown, is widely
debated~\cite{Berti:2007fi,Baibhav:2017jhs,Bhagwat:2017tkm,Ota:2019bzl,Giesler:2019uxc,Forteza:2020hbw}. Indeed, the
procedure of constructing a GW template is based on joining different approximations, from the regime where the BHs
separation is large (PN approximation) to the ringdown oscillations (perturbation theory), passing through the highly
non-linear merger process that requires numerical approaches. Our results indicate that an observer measuring the scalar
oscillations during a BBH collision may extend the validity of the ringdown treatment closer to the merger, where, in
principle, only full numerical studies are accurate and reliable. This is an indication that the same may hold true for
the gravitational waveform.

\section{Conclusions}\label{sec:Conclusions_clap}
GW astronomy has the potential to answer crucial questions regarding the correct description of
gravity. The full exploitation of such potential requires knowledge about the dynamics of compact objects in a generic
theory of gravity.  While NR is the tool of excellence for this, the evolution of a single binary within the context of a modified theory can take months to perform on supercomputers,
and may require years of careful study of the relevant partial differential equations and associated well-posedness.

In this Chapter, we explored the close limit approximation as a ``quick-and-dirty'' tool to understand non-linear coalescence processes. Its remarkable agreement with full non-linear simulations is an important benchmark. In fact,
albeit it is a perturbative scheme, it uses constraint-satisfying initial data, and their evolution works accurately
even when the premises of the model are only partially satisfied. Consequently, this provides some confidence that this
technique works well also when extending those studies beyond GR or BH spacetimes. The main requirement to use the CLAP
consists in having solutions of the constraint equations. These can be solved, as we showed, also in the presence of
fundamental fields (see also Ref.~\cite{Okawa:2014nda,Zilhao:2015tya}).

Further effort is required here, the payoff is significant: with much less computational time and effort one is able to
investigate setups that, in principle, should be only described through non-linear simulations. We showed how the CLAP can work for the coalescence of
equal-mass, compact, horizonless objects, and how it too predicts the existence of echoes in GWs. This is a significant result in that it extends and complements other past perturbative
calculations~\cite{Cardoso:2016rao,Cardoso:2016oxy,Cardoso:2019rvt}.
Moreover, we studied scalar fields minimally coupled with gravity in BH spacetimes (which are
equivalent to non-minimally coupled scalar fields in a Bergmann-Wagoner scalar-tensor theory), estimating the scalar
modes in the merger of a BH binary, and showing that they are very similar to those in a stationary BH
spacetime. 
In future works, we will further use this tool to investigate compact objects collisions in other - and perhaps more complicated, alternative theories, as EsGB for instance, and we will investigate its usage also for non head-on binaries.

  \cleardoublepage
\epigraphhead[450]{{\it This part is based on Refs.\til\cite{Annulli:2021lmn,Annulli:2019fzq}}}
\part{Unstable processes in alternative theories}

  \cleardoublepage

	\chapter{Spontaneous scalarization in binary black holes spacetimes}\label{chapter:EsGB}

\minitoc

Studying the stability of a physical configuration plays a fundamental role in Physics. In the case of alternative theories of gravitation, for instance, it is crucial to understand the viability of the theory itself and of the solutions of its field equations. Finding the onset of an instability may lead to a great opportunity. In fact, thanks to such mechanisms one could constrain a theory  or a coupling constant if the observations do not meet the predictions. In addition, they furnish a natural framework to develop new BHs or stars solution, different from the typical ones of GR, that might arise as the result of the instability.

An important example of such processes appears when extra degrees of freedom are coupled with curvature or matter fields. In non-minimally coupled scalar-tensor theories, one may find that, for certain coupling strengths, a star solution of GR is unstable and provokes an instability. This process may lead to compact objects with non-trivial charges and it is dubbed {\it spontaneous scalarization}~\cite{Damour:1992we,Ruffini:1971bza,Hawking:1972qk,Sotiriou:2011dz,Sotiriou:2013qea,Sotiriou:2014pfa,Silva:2017uqg,Doneva:2017bvd,Witek:2018dmd,Silva:2018qhn,Minamitsuji:2018xde,Doneva:2019vuh,Fernandes:2019rez,Minamitsuji:2019iwp,Cunha:2019dwb,Andreou:2019ikc,Ikeda:2019okp}, as it is akin to the more famous spontaneous magnetization of ferromagnetic materials. As this process may be an important smoking gun for some alternative theories, in this Chapter we use results from the CLAP\til\cite{Price:1994pm,Anninos:1995vf,Abrahams:1995wd,Andrade:1996pc} (see Chapter\til\ref{chapter:CLAP}) to model unstable processes in binary BH spacetimes. This Chapter is also motivated by recent works on the scalarization of multi-body systems, that showed how signatures of this non-perturbative mechanism can emerge dynamically\til\cite{Barausse:2012da,Palenzuela:2013hsa,Shibata:2013pra,Silva:2020omi,Cardoso:2020cwo}. 

Starting from theories allowing for spontaneous scalarization of isolated BHs, we investigate dynamical scalarization processes in binary BH spacetimes, specializing our computations in the context of EsGB. Working with non-trivial couplings between the scalar and the spacetime curvature, we wish to highlight the effects of the scalar field dynamics in a two-body spacetime.

\section{Einstein-scalar-Gauss-Bonnet gravity}
In the following we consider Einstein-scalar-Gauss-Bonnet gravity as a specific example of scalar-tensor theory of the above class. EsGB emerges naturally in the low-energy limit of string theories~\cite{METSAEV1987385,Kanti:1995vq,Charmousis:2014mia}, and is the only alternative theory that includes an extra scalar degree-of-freedom, coupled to a quadratic curvature term constructed from the spacetime metric, which equations of motion are second (differential) order. For EsGB, a 3+1 decomposition of the field equations has been recently performed\til\cite{Witek:2020uzz,Julie:2020vov}.

The action of EsGB is given by
\be \label{eq:EsGB_action}
S=\frac{1}{16\pi}\int d^4x\sqrt{-g}\left[R-\frac{1}{2}\left(\nabla \Phi\right)^2+\frac{\eta}{4} \mathfrak{f}(\Phi)\rgb\right]\,,
\ee
where $\eta$ is the dimensionful coupling constant of the theory and $\mathfrak{f}\oo\Phi\cc$ is a generic coupling function between the scalar field and the Gauss-Bonnet invariant $\rgb$, that is
\begin{equation}
\rgb=R^2-4R\ijdd R\ijuu +R_{ijkl}R^{ijkl}\,,
\end{equation}
with $R\oo R_{ij}\cc$ being the Ricci scalar (tensor) and $R_{ijkl}$ the Riemann tensor. The equations of motion obtained from the action\til\eqref{eq:EsGB_action} are given by
\begin{align}
\label{eq:EsGB_einsteineq}
G_{\mu\nu}&=\frac{1}{2} T\mndd-\frac{1}{8}\eta\,\mathcal{G}\mndd\,,\\
\label{eq:EsGB_scalarKG}
\square\Phi&=-\frac{\eta}{4} \frac{\partial \mathfrak{f}\oo\Phi\cc}{\partial\Phi}\rgb\,,
\end{align}
where $G_{\mu\nu}$ is the usual Einstein tensor and
\begin{equation}
\mathcal{G}\mndd= 16 R^\alpha_{(\mu}\mathcal{C}_{\nu)\beta} +8\mathcal{C}^{\alpha\beta}\oo R_{\mu\alpha\nu\beta}-g\mndd R_{\alpha\beta} \cc -8 \mathcal{C} G\mndd -4 R \mathcal{C}\mndd\,,
\end{equation}
with 
\begin{equation}
\mathcal{C}\mndd= \nabla_\mu \nabla_\nu \mathfrak{f}\oo\Phi\cc=\mathfrak{f}'\nabla_\mu\nabla_\nu\Phi+\mathfrak{f}''\nabla_\mu\Phi\nabla_\nu\Phi\,,
\end{equation}
and the scalar field stress-energy tensor is defined as,
\begin{equation}
T_{\mu\nu}=\partial_\mu\Phi\partial_\nu\Phi-\frac{1}{2}g\mndd \partial^\alpha\Phi\partial_\alpha\Phi\,.
\end{equation}
%

In order to study the evolution of any physical configuration in EsGB, one needs to find consistent initial data. This consists in solving the EsGB constraint equations coming directly from Eqs.\til\eqref{eq:EsGB_einsteineq}-\eqref{eq:EsGB_scalarKG}.
A solution to these equations, in general, includes complicated functions of the scalar $\Phi$ and the scalar momentum density $K_{\Phi}$\footnote{$K_{\Phi}$ is defined as the Lie derivative of the scalar field with respect to the normal vector to the initial hypersurface of foliation.}. However, in this work we are only interested in understanding if, and how, BBHs in vacuum might be unstable in EsGB. In order to assume a trivial scalar field profile $\Phi=0$ and momentum density $K_{\Phi}=0$ on the initial hypersurface, we restrict to theories obeying $d \mathfrak{f}/d \Phi\rvert_{\Phi=0}=0$. With this assumption, we rule out theories allowing only for BH solutions with scalar hair, as the ones due to an exponential coupling function (see Ref.\til\cite{Kanti:1995vq}). Further considering BHs initially at rest, the momentum constraint equations are identically satisfied (and therefore not shown here), while the Hamiltonian reads as in vacuum GR
\be \label{eq:GR_reduced_ID_eq}
\prescript{3}{}{R}=0\,,
\ee
where $\prescript{3}{}{R}$ is the Ricci scalar evaluated on the initial three-spacelike hypersurface of foliation.
\section{Binary black hole background}
\label{section:EsGB_BBH_background}
In order to model a BBH spacetime, one needs to account for their interaction energy. The CLAP formalism of BBHs in GR succeeded to consistently describe such configurations\til\cite{Price:1994pm,Anninos:1995vf,Abrahams:1995wd,Andrade:1996pc,Annulli:2021dkw}, and we will use this approximation in what follows. Remarkably, this perturbative method was used to find the ringdown waveforms produced by the head-on collision of BHs binaries in GR (see Chapter\til\ref{chapter:CLAP}). This approach is based on having initial data describing BBHs that are solutions of the Hamiltonian constraint equation\til\eqref{eq:GR_reduced_ID_eq}. Such solution is not unique: different initial data\til\cite{Misner:1960zz,Brill:1963yv,Bowen:1980yu} may be used within the CLAP. The ones that we employ in this work are given by the Brill-Lindquist initial data\til\cite{Brill:1963yv}. These are conformally flat, time symmetric initial data representing two BHs initially at rest. 

Let us focus on equal-mass binaries, of total ADM mass $M$. In isotropic cartesian coordinates, we place the BHs on the $Z$-axis (${\bm R_{1/2}}=(0,0,\pm Z_0)$, where ${\bm R_i}$ is the position of each BH in this reference), therefore the origin of the reference frame is in the CM of the system. As shown in detail in Refs.\til\cite{Abrahams:1995wd,Andrade:1996pc,Sopuerta:2006wj,Annulli:2021dkw}, using the CLAP of BBHs, we can recast the $4D$ initial spacetime as a perturbation of the Schwarzschild metric. Thus, including for the sake of simplicity only the leading-order quadrupolar contribution\til\cite{Price:1994pm}, the spacetime can be written as 
\begin{equation}\label{eq:pert_Schw_CLAP_GR_BL_esgb}
g_{\mu\nu}=g^{(0)}_{\mu\nu}+h_{\mu\nu}\,,
\end{equation}
where 
\begin{align}
g^{(0)}_{\mu\nu}&={\rm
  diag}(-f,f^{-1},r^2,r^2\sin^2\theta)\,,\nn\\
f&=1-2M/r\,,
\end{align}
and, using the Legendre polynomial $P_2 \left(\cos\theta\right)$, $h_{\mu\nu}$ is given by 
\begin{align}
h_{rr}&=f^{-1}gP_2(\cos\theta)\frac{Z_0^2}{2M^2}\,,\nonumber\\
h_{\theta\theta}&=r^2g P_2(\cos\theta)\frac{Z_0^2}{2M^2}\,,\label{eq:BLrecast1_esgb}
\end{align}
with 
\begin{equation}
g=4\left(1+M/(2R)\right)^{-1}M^3/R^3\,,
\end{equation}
and the isotropic coordinate $R$ is defined in terms of the Schwarzschild radial coordinate $r$ as
\begin{equation}
\label{eq:R_isotropic_to_Schw_esgb}
R=\frac{1}{4}\left(\sqrt{r}+\sqrt{r-2M}\right)^2\,.
\end{equation}
%

The parameter $Z_0$ in Eq.~\eqref{eq:BLrecast1_esgb} represents the initial separation between the BHs in the isotropic frame. For $Z_0=0$ there is just a single BH of mass $M$ in the initial slice. When $0<Z_0\lesssim 0.4$ one single common horizon appears\til\cite{Anninos:1995vf,Abrahams:1995wd,Andrade:1996pc}. In this regime, the spacetime can be thought to represent two BHs close to one another, enveloped by a common distorted horizon~\cite{Price:1994pm}. Moreover, it is worth to note that $Z_0$ itself is only a parameter and not a physical quantity. However, it is possible to establish a realation between $Z_0$ and the physical distance between the apparent horizons of the initial colliding BHs ($L$)~\cite{Andrade:1996pc,Annulli:2021dkw,bishop1982closed,bishop1984horizons,Gleiser:1998rw,Sopuerta:2006wj}: an explicit computation gives $L=3M$ for $Z_0\simeq0.5M$, $L=3.5M$ for $Z_0\simeq0.7M$, $L=4M$ for $Z_0\simeq0.85M$.

The metric in Eq.\til\eqref{eq:pert_Schw_CLAP_GR_BL_esgb} shows how the colliding BHs spacetime can be seen as a time-dependent perturbation of a Schwarzchild background. Hence, in the CLAP, the time evolution of this small (even) gravitational perturbations can be achieved by gauge-invariant perturbations techniques~\cite{Moncrief:1974am,Cunningham:1978zfa,Cunningham:1979px}. Notably, as shown in Ref.\til\cite{Price:1994pm}, the gravitational perturbation equations can be cast in a single Zerilli equation for one unknown function (the Zerilli function)\til\cite{Zerilli:1970se}. Solutions of such equation provide GW signals remarkably similar to the results obtained using full numerical simulations\til\cite{Anninos:1993zj}.

Instead, in the following, we use the metric in Eq.\til\eqref{eq:pert_Schw_CLAP_GR_BL_esgb} only as the background spacetime in which evolving the scalar field, thus neglecting the motion of the BHs in the timescale of the oscillation. This is a severe approximation. First because astrophysical BHs in binaries move at large velocities when close to one another. Furthermore, on a timescale of order $M$, the BHs collide, hence the extrinsic spacetime curvature will take non-zero values, changing the background spacetime in which scalar perturbations propagate. However, albeit an approximation, restricting to a {\it frozen} background still shows the main feature of the onset of instabilities in binary spacetimes, as we shall see later.
 
A CLAP treatment allowing for spontaneous scalarization {\it during} the collision (or the inspiral) of BHs in EsGB is left for future work.

\section{Scalar instabilities}
To test the onset of scalar instabilities in BBHs geometries, we study the behaviour of small linear scalar fluctuations in backgrounds described by Eq.\til\eqref{eq:pert_Schw_CLAP_GR_BL_esgb}. These vacuum configurations have been chosen since EsGB allows also for BH solutions identical to GR.


Small scalar perturbations can be mathematically expressed replacing $\Phi \rightarrow \epsilon\Phi$ in Eqs.\til\eqref{eq:EsGB_einsteineq}-\eqref{eq:EsGB_scalarKG}, with $\epsilon$ a small bookkeeping parameter. Thus, one can linearize the Einstein-KG system up to $\mathcal{O}(\epsilon)$. In this limit, the KG equation decouples from Einstein's equations. Hence, the background spacetime is not affected by the scalar perturbations. Our perturbation scheme will eventually breakdown at sufficiently late times: the exponentially growing scalar gives rise to an exponentially growing stress-tensor, the backreaction of which on the geometry can no longer be neglected. Here, we focus solely on the early-time development of the instability. 

What we are left to solve is the KG equation
\begin{equation} \label{eq:KG_equation}
\square\Phi=-\frac{\eta}{4} \frac{\partial \mathfrak{f}\oo\Phi\cc}{\partial\Phi}\rgb\,,
\end{equation}
where the box operator ($\square=\frac{1}{\sqrt{-g}}\partial_\mu\oo g^{\mu\nu}\sqrt{-g}\partial_\nu\cc$) is defined on the BBH background in Eq.\til\eqref{eq:pert_Schw_CLAP_GR_BL_esgb}. In the following we further assume that the Gauss-Bonnet coupling is such that can be well approximated by a quadratic function (linearizing it around an extremum for instance), hence,
\begin{equation} \label{eq:coupling_function}
\mathfrak{f}\oo\Phi\cc=\frac{\Phi^2}{2} \,.
\end{equation}
As shown in Refs.\til\cite{Silva:2017uqg,Doneva:2017bvd}, in this class of theories the KG equation admits solutions composed by a constant scalar around spacetimes satisfying GR equations. Furthermore, a linear stability analysis showed that, for certain values of the coupling constant $\eta$, GR solutions may be unstable. To find the endpoint of this instability, one needs to solve the equation of motion including the backreaction of the scalar on Einstein's equations. This eventually leads to scalarized (or hairy) BHs or stars. For new detailed studies of the properties of rotating and non-rotating scalarized BHs in EsGB we also refer the reader to Refs.\til\cite{Blazquez-Salcedo:2018jnn,Collodel:2019kkx,Berti:2020kgk}.

Conversely, here we are interested in the effect on scalar fluctuations due to the presence of a binary. Hence, both the box operator and $\mathcal{R}_{\rm GB}$ in Eq.\til\eqref{eq:EsGB_scalarKG} depend on the perturbed BBH spacetime, and in the CLAP, we may expand them in powers of the small BHs separation $Z_0$,
\begin{align} \label{eq:box_expansion_esgb}
\square &= \square^{(0)}+ Z_0^2 \, \square^{(1)} + \mathcal{O}(Z_0^3)\,,\nn\\
\rgb &= \rgb^{(0)}+ Z_0^2 \, \rgb^{(1)} + \mathcal{O}(Z_0^3)\,.
\end{align}
Decomposing the scalar in spherical harmonics as,
\begin{equation} \label{eq:Phi_ansatz}
\Phi\oo t,r,\theta,\varphi\cc=\frac{1}{r}\sum_{\ell ,m}\psi_{\ell m}\oo t,r\cc Y^{\ell m}\oo\theta,\varphi\cc\,,
\end{equation}
the KG equation is non-separable because it couples different components of the index $\ell$. The mathematical details of the procedure to separate (perturbatively) Eq.\til\eqref{eq:EsGB_scalarKG} are given in  Appendix\til\ref{app:separateKG}. Let us summarize here the most important passages to arrive to the master equation that describes scalar perturbations in EsGB, in the axisymmetric stationary BBH spacetime. 

The procedure is similar to the one described in Ref.\til\cite{Annulli:2021dkw,Cano:2020cao}. Let us expand the scalar field using the spherical harmonics base. The key point is that, for each spherical harmonics index $\ell$, the KG equation becomes separable in the limit $Z_0 \rightarrow 0$. Hence, using the specific ansatz in Eq.\til\eqref{eq:Phi_ansatz_2}, for each $\ell\geq 1$ one gets a Schr\"{o}dinger-like equation that includes corrections in $Z_0^2$,
\begin{align}
\label{eq:KG_notortoise_esgb}
&\frac{\partial^2 \psi_{\ell m}}{\partial t^2}+\frac{\partial^2 \psi_{\ell m}}{\partial r^2} \oo U_0+Z_0^2\tilde{U}_0\cc+\frac{\partial \psi_{\ell m}}{\partial r}  \oo U_1+Z_0^2\tilde{U}_1\cc\nn\\
&+\psi_{\ell m} \oo \oo W_0+\frac{\eta}{4}\tilde{W}_0\cc +Z_0^2 \oo W_1+ \frac{\eta}{4} \tilde{W}_1\cc\cc=0\,,
\end{align}
where all the potentials are listed in Eq.\til\eqref{eq:potentials_notortoise}. The monopolar $\ell =0$ perturbations are not affected by the $Z_0^2$ corrections (see Eq.\til\eqref{eq:q12lm}), hence the $\ell =0$ modes are the same as in the single BH case. Setting $\eta/M^2=Z_0/M=0$ in Eq.\til\eqref{eq:KG_notortoise_esgb}, one gets the perturbations describing scalar perturbations in a static Schwarzschild spacetime\til\cite{Berti:2009kk}. Additionally, one may notice that the scalar field fluctuations are independently affected both by $\eta$ and $Z_0$. This means that there might be non-trivial effects on the scalar QNMs of oscillation of a BBH even in pure GR (setting $\eta=0$ and $Z_0\neq 0$). For such scenario, we refer the interested reader to Chapter\til\ref{chapter:CLAP} Section\til\ref{sec:CLAP_STT}. 

In the following we strictly focus on EsGB (hence $\eta \neq0$).

\subsection{Boundary conditions}
%
To find unstable modes, we start with an harmonic time dependent scalar field, 
\begin{equation} \label{eq:phi_harmonic_timedep}
\psi_{\ell m}\oo t,r\cc=\Psi\oo \omega,r\cc e^{-i \omega t}\,,
\end{equation}
where we dropped the subscript $_{\ell m}$ in the r.h.s.. Substituting the ansatz\til\eqref{eq:phi_harmonic_timedep} in Eq.\til\eqref{eq:KG_notortoise_esgb}, an unstable mode is found when a bounded regular solution of the KG equation
\begin{equation} \label{eq:KG_BBH}
\frac{\partial^2 \Psi}{\partial r^2}  \oo U_0+Z_0^2\tilde{U}_0\cc+\frac{\partial \Psi}{\partial r}\oo U_1+Z_0^2\tilde{U}_1\cc+\Psi \oo \oo W_0+ \frac{\eta}{4}\tilde{W}_0\cc+Z_0^2 \oo W_1+ \frac{\eta}{4} W_2\cc -\omega^2\cc=0\,,
\end{equation}
with potentials in Eq.\til\eqref{eq:potentials_notortoise}, possesses a frequency that satisfies 
\begin{equation}
\omega=\omega_R+i\omega_I,\;\;\text{with }\omega_I>0	.
\end{equation}

%
%
Being interested in the onset of the instability, without loss of generality, we might look for solutions with purely imaginary frequencies ($\omega_R=0$). The asymptotic behaviours of Eq.\til\eqref{eq:KG_BBH} provide us  the proper boundary conditions to be imposed. Especially, asking for regularity both at the horizon and at spatial infinity, we get %
\begin{align} \label{eq:BC_boundstates}
&\Psi \oo r \sim 2M\cc = 
\oo r-2M\cc^{\frac{2  M  \omega_I}{\sqrt{1-2 \oo Z_0/M\cc^2 q^{(1)}_{\ell m}} }} \sum_{n=0}^N a_n \oo r-2M\cc^n\,,\nn\\
&\Psi \oo r \sim\infty\cc = \frac{e^{- r \omega_I}}{r^l}\sum_{n=0}^N b_n r^{-n}\,,
\end{align}
where the coefficients $a_n,b_n$ have to be found substituting Eq.\til\eqref{eq:BC_boundstates} in  Eq.\til\eqref{eq:KG_BBH} and solving it order by order. For each configuration, the value of $N$ has to be increased until the boundary conditions\til\eqref{eq:BC_boundstates} do not converge to fixed values\til\cite{Pani:2013pma}.

\subsection{Isolated black hole scalar bound states}
\begin{figure}[ht]
\centering
\includegraphics[width=0.5\textwidth,keepaspectratio]{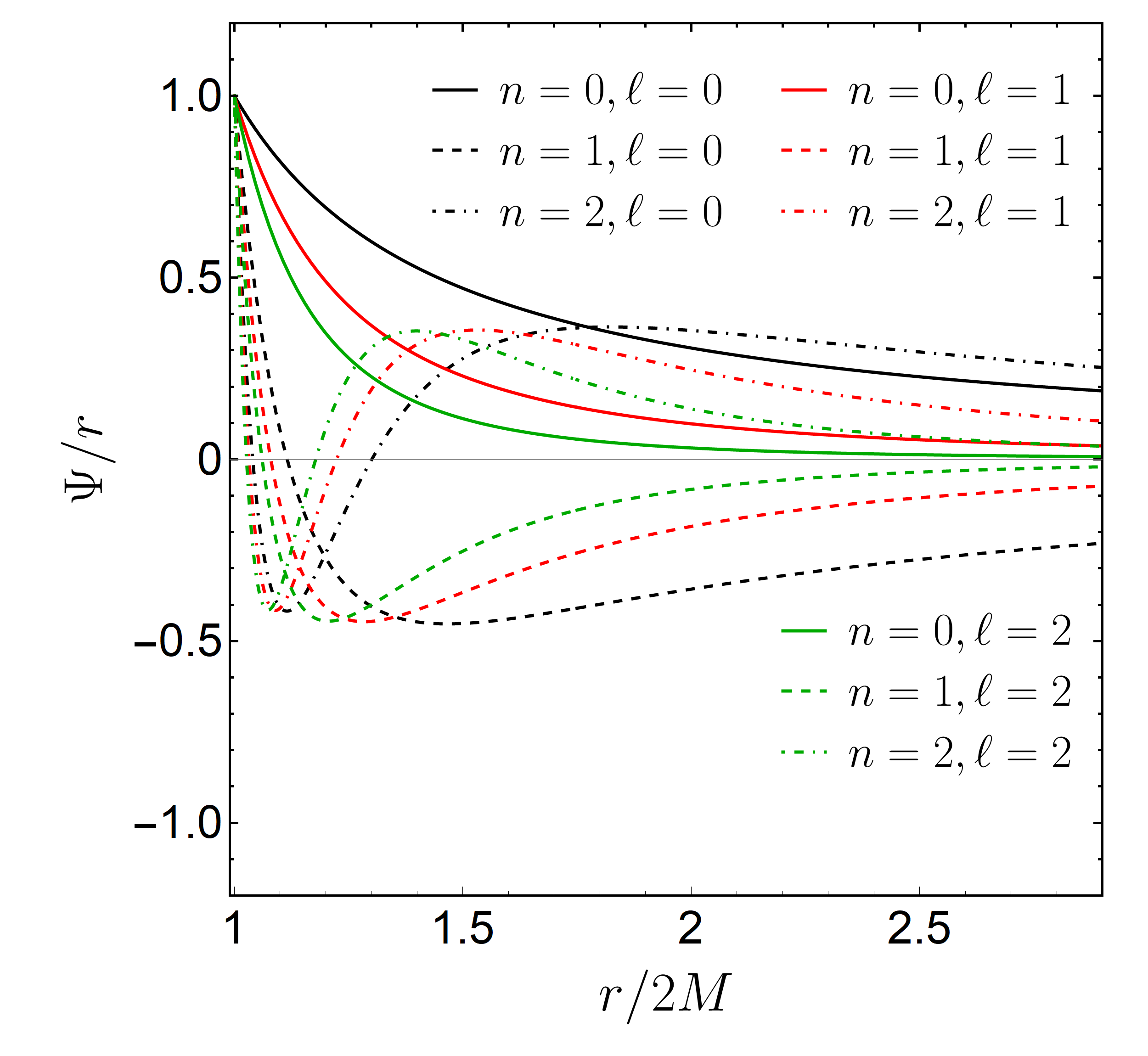} 
\caption[Single BH scalar unstable bound state.]{Scalar profiles for different values of $\ell$, for the first three scalarized solutions around an isolated static BH. Solid, dashes and dotdashed lines correspond to zero, one or two nodes solution respectively. The black lines ($\ell=0$) match with previous literature results\til\cite{Silva:2017uqg}. Because of spherical symmetry, scalar perturbations of an isolated BH in EsGB are insensitive to the specific values of $m$. Thus, each curve correspond to a specific, single value of $\ell$, regardless of the value of $m$.}	
\label{fig:BH_spscal_n012}
\end{figure}
As a consistency check, we first integrate Eq.\til\eqref{eq:KG_BBH} for a single static BH ($Z_0=0$), searching for static bound states, as the ones found in\til\cite{Silva:2017uqg,Doneva:2017bvd}. This means that in the following we seek only for solutions with 
\begin{equation}
\omega=0\,.
\end{equation}
Considering the quadratic coupling function in Eq.\til\eqref{eq:coupling_function}, a comparison with the results in Ref.\til\cite{Silva:2017uqg} is straightforward. In Fig.\til\ref{fig:BH_spscal_n012} we show different scalar bound states that correspond to unstable solutions around Schwarzschild BHs for the first three scalarized solutions, for $\ell=0,1,2$. Not all the values of $\eta/M^2$ provide static scalar non-trivial solutions. In fact, these bound states correspond only to a specific set of $\eta/M^2$. The corresponding values of the coupling parameter are summarized in table\til\ref{table:eta_values}.
\begin{table}[th] 
\centering
	\begin{tabular}{c||c}
		\hline
		\hline &
	\multicolumn{1}{c}{$\left(\eta/M^2\right)^{n\ell m}_{Z_0=0}$}\\
		\hline
		$\ell$ &  n=0\,\;\;\;\;\; n=1 \;\;\,\, n=2 \\ 
		\hline
		\hline
		0 & $2.902\;\;\;\,\,\,   19.50\;\;\;\,\,     50.93$\\
		1 & $8.282\,\,\;\;\;\,\,     29.82\;\;\;\,\,   65.84$\\
		2 & $16.30\,\,\;\;\;\,\,  42.97\;\,\,   \;\;  83.82$\\
		\hline
		\hline
	\end{tabular} 
	\caption[Set of $\eta$ for  single BH scalar bound states.]{Values of the coupling constant $\eta$ corresponding to the static scalar bound states solutions around isolated BHs. Each value of $\eta/M^2$ refers to a different curve in Fig.\til\ref{fig:BH_spscal_n012}. The values for $\ell=0$ agree with the literature\til\cite{Silva:2017uqg}.}
	\label{table:eta_values}
\end{table}
Compare to previous literature\til\cite{Silva:2017uqg}, we evaluate the static unstable bound states also for $\ell>0$. These solutions will serve as benchmarks for the bound states solution in the BBH case, as we shall see in the next section. 

Each entry in table\til\ref{table:eta_values} corresponds to a parabola in a $\left(\eta,M\right)$ plane. Non-linear studies including the scalar field backreaction on the spacetime geometry showed how hairy BHs solutions, end points of the tachyonic scalar instability, belong only to an infinite set of narrow bands in the $\left(\eta,M\right)$ plane\til\cite{Silva:2017uqg}. The values in table\til\ref{table:eta_values}, computed through a linear analysis, coincide only with one of the two ends of each band.

\subsection{Scalarization in binaries spacetimes}
Let us turn now to the case of two BHs in a binary. Hence, we solve Eq.\til\eqref{eq:KG_BBH} for $Z_0\neq0$. As clear from the coefficients in Eq.\til\eqref{eq:q12lm}, scalar monopolar perturbations vanishes when $Z_0\neq 0$. Hence, the results obtained for isolated BHs hold when $\ell=0$. 

\begin{figure}[ht]
\centering
\includegraphics[width=0.6\textwidth,keepaspectratio]{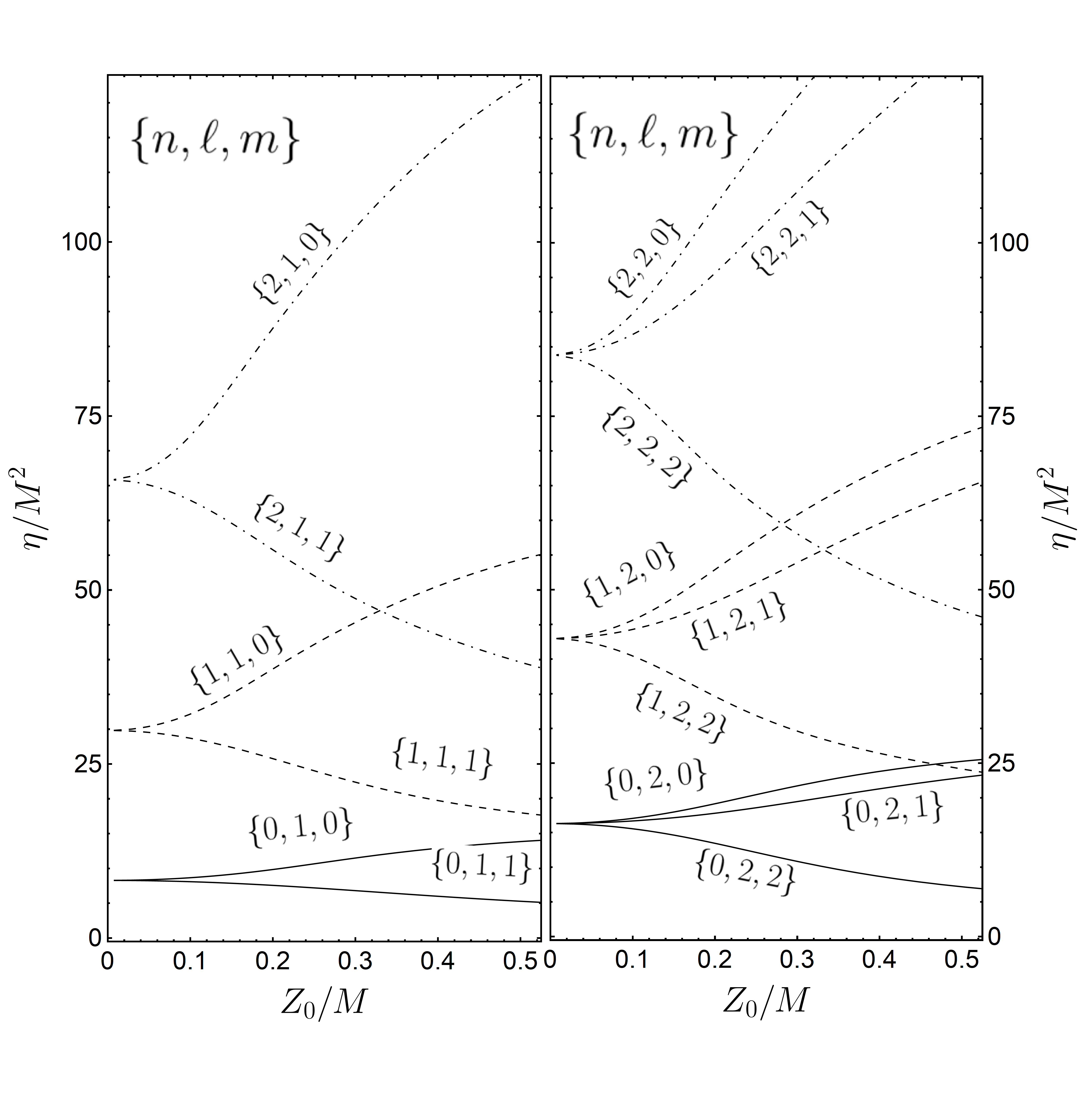}
\caption[$\eta$ vs BH separation for binary BHs.]{Existence lines for the coupling constant of EsGB gravity, corresponding to static bound states solutions of Eq.\til\eqref{eq:KG_BBH}, as a function of the normalized geometrical BHs separation ($Z_0/M$). Each curve is labelled for different values of $\{n,\ell,m\}$. In both panels, the solid lines correspond to zero node solutions ($n=0$), the dashed to one node ($n=1$) and dotdashed to two nodes ($n=2$). Results for negative values of $m$ coincide with their positive $m$ counterpart, and therefore are not explicitly shown in the legend. All the different branches depart, respectively, from each value shown in Tab.\til\ref{table:eta_values}, previously evaluated for $Z_0=0$. {\bf Left panel}:   bound states associated with $\ell=1$. {\bf Right panel}: bound states associated with $\ell=2$.}	
\label{fig:BH_spscal_binary}
\end{figure}

For $\ell\geq 1$ instead, we compute how the specific values of $\eta/M^2$ shown in Tab.\til\ref{table:eta_values} vary as a function of the BHs separation. Results are summarized in Fig.\til\ref{fig:BH_spscal_binary}.
Different branches for the same $\ell$ refer to different values of the spherical 
harmonic index $m$. From Eq.\til\eqref{eq:q12lm} we may notice that each branch in Fig.\til\ref{fig:BH_spscal_binary} departs from the single BH value ($Z_0=0$) to larger values of $\eta/M^2$ if $q_1^{(\ell m)}>0, q_2^{(\ell m)}<0$, and to smaller ones if $q_1^{(\ell m)}<0, q_2^{(\ell m)}>0$. 
All the branches in Fig.\til\ref{fig:BH_spscal_binary} for which the value of $\eta/M^2$ decreases when $Z_0$ increases can be approximated by the following fit
\begin{equation} \label{eq:fit_eta_Z0}
\frac{\eta}{M^2} \approx  \oo\frac{\eta}{M^2}\cc_{Z_0=0}^{n\ell m}-a^{n\ell m}\oo\frac{Z_0}{M}\cc^{3/2}\,,
\end{equation}
accurate within $1\%$ for $0\leq Z_0/M\leq 0.4$. In the above fit, the first term on the r.h.s corresponds to each specific entry in Tab.\til\ref{table:eta_values} and $a^{n\ell m}$ is a constant that depends on the the number of nodes and on the angular indices. As an example, some of its values are $a^{011}=8.74,a^{022}=31.64,$ etc.. 

Finally, given the assumptions made to build the binary spacetime in Section\til\ref{section:EsGB_BBH_background}, we stress that the results summarized in Fig.\til\ref{fig:BH_spscal_binary}, obtained for stationary backgrounds, have to be intended only as an indication of what happens to scalar fields in BBHs geometries, even when the BHs are left free to collide.
\section{Conclusions}
As depicted in Fig.\til\ref{fig:BH_spscal_binary}, BBH spacetimes in EsGB might suffer field instabilities. These results indicate that this process can happen {\it before} the final object is formed. Specifically, we showed that asymmetric configurations describing a BBH can scalarize due to different perturbation modes. Notably, fixing a value of $\ell>0$, this unstable mechanism can be enhanced by BBH spacetimes, for smaller values of the coupling constant compare to the corresponding scalarization threshold value of the (final) isolated BH. 

Nonetheless, in a realistic scenario the effect of the velocity of the colliding BHs might change the picture just described. Furthermore, the assumption that the BHs in the initial slice are simple Schwarzschild BHs might fail. In fact, each component of the binary might have already individually scalarized because of the $\ell=0$ modes shown in Tab.\til\ref{table:eta_values}, that appear for lower values of $\eta/M^2$ with respect to modes with $\ell\geq 1$. However, our findings are not completely ruled out, since spherically symmetric scalarized BHs exist only for specific bands\footnote{As an example, let us assume that a binary is composed by two Schwarzschild BH of mass $M_{\rm BH}$ each and that $\eta/M_{\rm BH}^2>19.50$. In this case none of the two initial BHs can be scalarized due of monopolar instabilities, because being out of a scalarization band\til\cite{Silva:2017uqg}. However, for an initial BH distance of $Z_0/\left(2M_{\rm BH}\right)\sim0.5$, the $\{n,\ell,m\}=\{0,1,\pm 1\}$ mode of the binary grows unboundedly if, for instance, $\eta/M_{\rm BH}^2\sim 20$.}, that depend on the value $\eta/M^2$. However, we recognize that the observation of the binary instability, described by the results in Fig.\til\ref{fig:BH_spscal_binary}, would require some undesirable {\it ad hoc} fine-tuning of the parameters. To conclude, the above results remark once more the fundamental role that the strong field regime possesses during BHs collisions and coalescences: in order to perform consistent tests of alternative theories, we need waveforms that properly accounts for backreacting effects when high spacetime curvatures are involved. Again, this work is a first step towards the study of the GWs produced by merging BHs in EsGB through the CLAP formalism.
%

	\chapter{Spontaneous vectorization of compact stars}
\label{chapter:spontaneous_vectorization}

\minitoc

Let us now turn our attention to unstable processes in the context of theories with an extra vectorial degree-of-freedom. The vector field discussed here can be interpreted in different ways, either as (i) the well-known
electromagnetic field, or as (ii) a still unknown vector field, which is ``hidden'' since it is weakly coupled with the standard model.

Within the interpretation (i), we are studying strong-gravity modifications of the coupling between the gravitational and the
electromagnetic field. We remark that the effects we are seeking only show up in the presence of a very large spacetime curvature, such as those in the core of NSs, or near the horizon of BHs. Therefore, despite the enormous accuracy of existing experimental data on the electromagnetic field, the effects studied in this Chapter are
not ruled out by current observations.  In particular, we mainly study the effects of the inclusion of a coupling
$\sim RX_\mu X^\mu$ (where $R$ is the spacetime curvature) in the action of the theory. This coupling resembles a mass term (but with a non-constant and non-uniform ``mass''). We note that even the existence of a photon mass has not been definitely ruled out \cite{Eidelman:2004wy,Pani:2012vp}; a
photon-curvature coupling is more elusive, since it shows up only in strong curvature regions.

Within the interpretation (ii), one tries to enlarge the standard model with as many fields as possible, and question which of those fields can be constrained with experiments. In this context, the theory treated in this Chapter arises naturally, in the sense that (hidden, with small couplings to the standard model) vectors are a generic prediction of string theory~\cite{Polchinski:1998rq}, and are promising dark matter candidates~\cite{Essig:2013lka}. Generalized theories with vector fields, avoiding ghosts and other pathologies, have recently been studied in Ref.~\cite{Heisenberg:2014rta}. 

As previously discussed in Chapter\til\ref{chapter:EsGB}, even in this framework we look for smoking-gun effects of such new fields and couplings. Notably, we will find and describe an example of {\it vectorized} compact stars.

\section{Hellings-Nordtvedt gravity\label{Hellings}}

In the Hellings-Nordtvedt (HN) gravity theory~\cite{Hellings:1973zz},
a single massless vector field is non-minimally coupled to the gravitational field.
The action for the HN theory is,
\begin{equation}
S=  \int d^4x\frac{\sqrt{-g}}{16\pi}(R - F_{\mu\nu}F^{\mu\nu}- {\Omega} X_{\mu} X^\mu{} R- \eta X^{\mu} X^{\nu} R_{\mu\nu})+ S_{\rm M}\,,\label{eq:action}
\end{equation}
where $X_{\mu}$ is a massless vector field, $F_{\mu\nu}=X_{\nu;\mu}-X_{\mu;\nu}$, $R_{\mu\nu}$ (the subscript $_;$ represents a covariant derivative) and $R$ are the Ricci
tensor and scalar, respectively, $\Omega$ and $\eta$\,\footnote{We choose the signs convention for the coupling constants
different from those used in Ref.~\cite{Hellings:1973zz}. Our conventions are consistent with those used in studies of scalar-tensor
theories.} are dimensionless coupling constants, and $S_{\rm M}$ is the matter fields action. The
action~\eqref{eq:action} yields the field equations \cite{Will:1993ns}
\begin{align}
\label{eq:einstein_equation}
R_{\mu\nu}-\frac{1}{2}g_{\mu\nu}R-\Omega \Theta^{(\Omega)}_{\mu\nu}-
\eta\Theta^{(\eta)}_{\mu\nu}+\Theta^{(F)}_{\mu\nu}&=8\pi G T_{\mu\nu}\,,\\
\label{eq:vector_field}
F^{\mu\nu}_{\,;\nu}+\frac{1}{2}\Omega X^{\mu}R+\frac{1}{2}\eta X^{\nu}R^{\mu}_{\nu}&=0\,,
\end{align}
where
\begin{align}
\Theta_{\mu\nu}^{(\Omega)}&=X_\mu X_\nu R +X_\alpha X^\alpha R_{\mu\nu}-\frac{1}{2}g_{\mu\nu}X_\alpha X^\alpha R-(X_\alpha X^\alpha)_{;\mu\nu}+g_{\mu\nu}\square(X_\alpha X^\alpha)\,,\\
\Theta_{\mu\nu}^{(\eta)}&=2X^\alpha X_{(\mu} R_{\nu)\alpha}-\frac{1}{2}g_{\mu\nu}
X^\alpha X^\beta R_{\alpha\beta}-(X^\alpha X_{(\mu})_{;\nu)\alpha}
+\frac{1}{2}\square(X_\mu X_\nu)+\frac{1}{2}g_{\mu\nu}(X^\alpha X^\beta)_{;\alpha\beta}\,\\
\Theta^{(F)}_{\mu\nu}&=-2(F^{\alpha}_{\,\,\mu}F_{\nu\alpha}-\frac{1}{4}g_{\mu\nu}F_{\alpha\beta}F^{\alpha\beta})\,,
\end{align}
and $T_{\mu\nu}$ is the matter stress-energy tensor. 

As already noticed in Ref.\til\cite{Will:1993ns}, the divergence of Eq.\til\eqref{eq:vector_field} leads to the constraint equation,
\begin{equation}\label{eq:constr}
\left(\Omega X^\mu R+\frac{1}{2}\eta X^\nu R^\mu_\nu\right)_{;\mu}=0\,,
\end{equation}
that comes directly from the fact that the action is not fully gauge invariant (not being invariant under $X_\mu \rightarrow X_\mu + \lambda_{,\mu}$, with $\lambda$ being a scalar function). However, this extra constraint equation is trivially satisfied in our case, since we will restrict to vector fields of the form $X_\mu=\{X_0,0,0,0 \}$, when dealing with non-rotating stars (as explained in Sec.\til\ref{sec:formalism_star_HN} in detail). 

Before diving into the calculations of star configurations in HN gravity, let us highlight at least two fundamental issues of this theory. The first is related with the fact that the vector-curvatures couplings in the action break the gauge invariance of the theory. Being the strength of this violation spacetime dependent provides an uncertainty about how many degrees-of-freedom the theory possesses\til\cite{Eardley:1973zuo}. Furthermore, HN gravity is also known to be plagued by ghost instabilities\til\cite{Esposito-Farese:2009wbc}. These problematic fields spoil the validity of the theory, unless new terms are included to cure such fundamental issue. Said so, let us stress that the choice of working with HN gravity has been pursued mostly for its mathematical simplicity. Thanks only to two extra coupling terms, we can highlight some peculiar features, that might appear also in more complicated -- and better justified -- vector-tensor theories\til\cite{Heisenberg:2018vsk,Heisenberg:2014rta,BeltranJimenez:2013btb,Kase:2020yhw}. Hence, HN gravity is mostly used as a proxy to get insight on what happens when an extra vector field is embedded in high curvature spacetimes.
\section{Linearized fluctuations of stars}\label{Lin_pert}

We consider perturbations of static, spherically symmetric stars in HN gravity, composed by a perfect fluid. The
background is thus described by a spacetime metric with the form
\begin{equation}\label{eq:line_element}
ds^2=-Fdt^2+\frac{1}{G}dr^2+r^2d\theta^2 +r^2\sin^2\theta d\phi^2\,,
\end{equation}
where $F(r)$ and $G(r)$ are general functions of the radial coordinate $r$, and
by a stress-energy tensor with the form
\begin{equation}\label{stress_energy_tensor}
T_{\mu\nu}=(p+\rho)u_{\mu}u_{\nu}+g_{\mu\nu}p\,,
\end{equation}
where
\begin{equation}
u^{\mu}=\left(F^{-1/2},0,0,0\right)
\end{equation}
is the four-velocity of the fluid, $p(r)$ is its pressure, and $\rho(r)$ is its energy density.

We study two different equations of state (EOS) for the fluid composing the star. The first is a constant
density (CD) EOS (see, e.g.~\cite{Shapiro:1983du}) with radius $R$ and mass $M$, where $\rho=3M/(4\pi R^3)=const$, and
\begin{align}\label{eq:axial_flat_star}
G&=1-\frac{8}{3} \pi  r^2 \rho\,,\nonumber\\
p&= \left(\frac{\rho \left(G^{1/2}-\sqrt{1-\frac{2 M}{R}}\right)}{3 \sqrt{1-\frac{2 M}{R}}-G^{1/2}}\right)\,,\nonumber\\
F&= \left(\frac{3}{2} \sqrt{1-\frac{2 M}{R}}-\frac{1}{2} G^{1/2}\right)^2 \,.
\end{align}
The second is the polytropic (Poly) EOS which has been used in~\cite{Damour:1993hw} to study
spontaneous scalarization in scalar-tensor theories,
\begin{equation}\label{eq:EOS}
\rho(p)=\left(\frac{p}{K n_0 m_b}\right)^{\frac{1}{\Gamma}} n_0 m_b+ \frac{p}{\Gamma-1}\,.
\end{equation}
where $n_0=0.1 \rm fm^{-3}=10^{53} \rm km^{-3}$, $m_b=1.66\times 10^{-24} \rm g=1.23 \times 10^{-57} \rm km$ is the average baryon mass, $\Gamma=2.34$ and $K=0.0195$ are dimensionless parameters.

When $X_\mu=0$ the field equations~\eqref{eq:einstein_equation}- \eqref{eq:vector_field} reduce to those of GR.
Therefore, all vacuum or matter solutions of GR are also solutions of the HN theory, including those describing
spherically symmetric compact stars in GR. We shall now study the stability of these solutions, considering a small
vector field perturbation:
\begin{equation}
X_\mu= \varepsilon \xi_\mu\,,
\label{eq:Xeps}
\end{equation}
where $\varepsilon\ll 1$ is a dimensionless bookkeeping parameter. At first order in $\varepsilon$,
Eq.~\eqref{eq:einstein_equation} reduces to GR Einstein's equations, and the vector field equation~\eqref{eq:vector_field}
can be written as
\begin{equation}
F^{\mu\nu}_{\,;\nu}-4\pi G\Omega X^{\mu}T+4\pi G\eta X^{\nu}\left(T^\mu_\nu-\frac{1}{2}\delta^\mu_\nu T\right) =0\,. \label{eq:vector_field_linear}
\end{equation}
The vector perturbation $\xi_\mu$ can be expanded in vector spherical harmonics,
\begin{equation}
\xi_{\mu}=\sum_{l} 
\begin{pmatrix}
 \begin{bmatrix}0 \\ 0 \\ a_{l}(r)(\sin\theta)^{-1}\partial_\phi Y_{l} \\
  -a_{l}(r)\sin\theta\partial_\theta Y_{l}  \end{bmatrix}& +\begin{bmatrix}
  f_{l}(r) Y_{l} \\  h_{l}(r) Y_{l} \\  k_{l}(r) \partial_\theta Y_{l} \\ k_{l}(r) \partial_\phi Y_{l} \end{bmatrix}
\end{pmatrix}e^{-i \omega t}\,,\label{eq:vector_perturbation_def}
\end{equation}
where $Y_{lm}(\theta,\phi)$ are scalar spherical harmonics, and, since the perturbation equations do not depend on the
azimuthal index $m$, we leave that index implicit. The perturbations $a_l(r)$ (with
$l\ge1$) have axial parity, i.e. they transform as $(-1)^{l+1}$ for a parity transformation
$\theta\rightarrow\pi-\theta$, $\phi\rightarrow\phi+2\pi$, while the perturbations $f_l(r)$ and $h_l(r)$ with $l\ge0$,
and $k_l(r)$ with $l\ge1$, have polar parity, since they transform as $(-1)^l$ for a parity transformation. These two
classes of perturbations can be studied separately, because they are decoupled in the perturbations equations.

We note that when $T_{\mu\nu}=0$,
Eq.~\eqref{eq:vector_field_linear} reduces to Maxwell's equations, while the equations for the gravitational
field~\eqref{eq:einstein_equation} coincide with Einstein's equations plus terms quadratic in the vector field.
Therefore, at first order in the perturbations HN gravity coincides with Einstein-Maxwell theory for BH spacetimes. Thus, since BHs are linearly stable in Einstein-Maxwell theory\til\cite{Regge:1957td,Moncrief:1974ng,Moncrief:1974gw,Chandrasekhar:1975zza,Chandrasekhar:1985kt,Whiting:1988vc,Dias:2015wqa}, they are also stable against linear perturbations in
HN gravity. 

\subsection{Instabilities and spontaneous vectorization in the axial sector}
The harmonic decomposition of the linearized vector field (Eq.~\eqref{eq:vector_field_linear}) yields a system of
ordinary differential equations for the perturbation functions.
For the axial part we get (for $l\ge1$),
\begin{equation}
F G a_{l}''+\frac{1}{2}  \left(G F'+F G'\right)a_{l}'+\left[\omega^2-F\left(\frac{l (l+1)}{r^2}\right)\right]a_{l}-2 \pi  F \left[\eta  \left( \rho - p \right)+2 \Omega  \left(\rho -3 p\right)\right]a_{l}=0\,,\label{eq:axial} 
\end{equation}
where a prime denotes a derivative with respect to the coordinate $r$. The term 
\begin{equation}
\eta  \left( \rho - p \right)+2 \Omega  \left(\rho -3 p\right)\,,\label{effective_mass}
\end{equation}
in Eq.~\eqref{eq:axial} behaves as an effective mass (squared) for the vector field. When it is negative, one expects GR
configurations to be unstable against radial perturbations. For instance, in theories with $\eta=0$, this is the case
when the coupling constant $\Omega$ is negative and $\rho>3p$, or when $\Omega$ is positive and $\rho<3p$. A similar
approach has been used for a qualitative study of the stability properties of scalar-tensor theories
in~\cite{Lima:2010na,Pani:2010vc}.
\subsection*{Numerical solutions}\label{app:Axial perturbation integration}
We have solved numerically Eq.~\eqref{eq:axial}, for CD and Poly stars, as an eigenvalue problem
for the frequencies $\omega$.  In both cases, we have used direct integration to search for
instabilities~\cite{Macedo:2016wgh,GRITJHU}, looking for unstable solutions with purely imaginary frequency $\omega_{I}>0$,
and imposing regularity at the center of the star and at infinity.  
In order to enforce a regular behavior near the center of the star, we perform an asymptotic expansion of the
axial perturbation equation \eqref{eq:axial}, by expanding the perturbation function as
%
\begin{equation}
a_{l}=\sum_{i=0}^{N}a_l^i r^i\,.
\end{equation}
We truncate the expansion at $N=4$ because we found that further coefficient does not affect significantly the results.
Thus, we find the values of the coefficients $a_l^{i>0}$ in terms of $a_l^0$ (which can be set to an arbitrary
value). In terms of these coefficients we can compute $a_l(r_0)$ and $a_{l,r}(r_0)$ at $r_0\ll R$, with $R$ being the surface of the star. We then numerically
integrate Eq.~\eqref{eq:axial} from $r_0$ to $R$ and, imposing regularity of the
perturbations, from the surface to $r\gg R$.

Unstable modes have frequency $\omega=\omega_R+ i \,\omega_I$, with $\omega_{I}>0$. Since we look for the onset of
the instability, we look for solutions with purely imaginary frequency, i.e., $\omega_R=0$, by matching the solution far
away from the star with
%
%
%
%
\begin{equation}
a_{l}(r)\approx e^{\omega_{\text{\rm{I}}}  r}c_1+e^{-\omega_{\text{\rm{I}}} r}c_2\,,
\end{equation}
where $c_1$ and $c_2$ are two constants of integration. Finally, since we require an asymptotically flat spacetime, we
impose $c_1=0$. We thus find a perturbation which grows in time and regular at spatial infinity, behaving asymptotically as
%
\begin{equation}
a_{l}(t,r)=c_2e^{-\omega_{\text{I}} r} e^{\omega_{\text{I}} t}\,.
\end{equation}
Later, in order to solve the perturbation equations with polar parity~\eqref{eq:dipolar_one} and \eqref{eq:dipolar_two}, we will follow
the same approach. The only difference is that, for each value of the harmonic index $l$, we have two perturbation
functions, $f_l(r)$ and $k_l(r)$.
\begin{figure}
\centering
\includegraphics[width=0.6\textwidth,keepaspectratio]{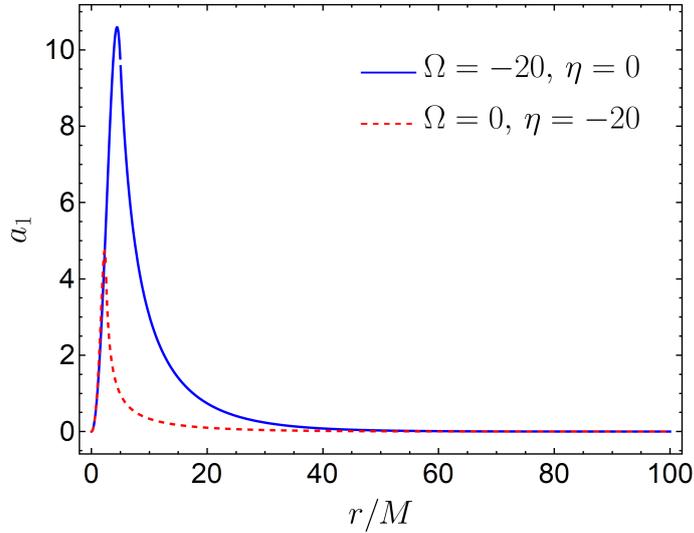}
\caption[Axial unstable dipolar profile for a CD star.]{Unstable dipolar vector perturbation profile for a CD star. The solid blue line
  ($\eta=0,\Omega=-20$) corresponds to an instability rate $M\omega=0.094 i$ for a CD star compactness $M/R=0.2$. The dashed red line
  ($\Omega=0,\eta=-20$) corresponds to a rate $M\,\omega=0.071\,{i}$ for a CD star compactness $M/R=0.4$.}\label{gr:plot_alm_example}
\end{figure}
\begin{figure*}
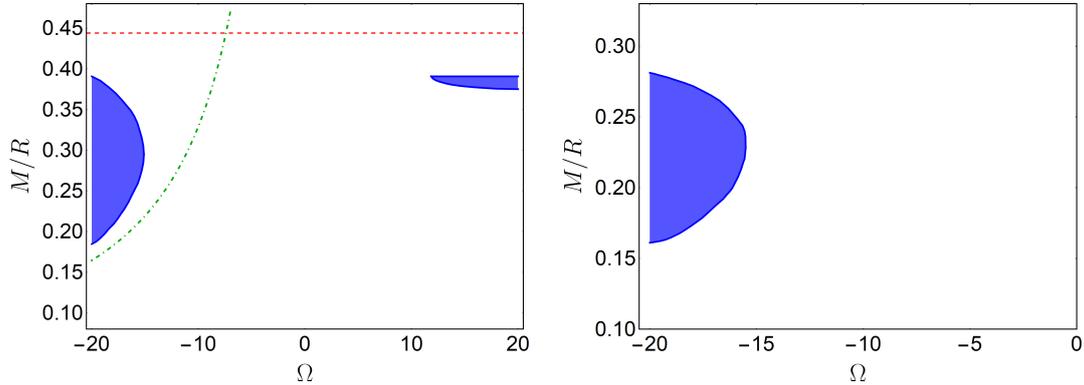

\begin{tabular}{cc}
\centering
\includegraphics[width=0.45\textwidth]{plot_phase_space}&
\includegraphics[width=0.45\textwidth]{plot_phase_space_EOS}
\end{tabular}
\caption[Phase space for axial instabilities in compact stars.]{The shaded regions represent CD ({\bf left panel}) and Poly ({\bf right panel}) star configurations with different values of $\Omega$ and $M/R$, which are unstable under axial perturbations in HN gravity with $\eta=0$. The
green dot-dashed curve is the Newtonian solution for a CD star, Eq.~\eqref{eq:newtonian_axial}. The dashed red line
corresponds to the Buchdal limit on the compactness of a CD star ($M/R<4/9\approx0.444$).
}
\label{gr:plot_phase_space}
\end{figure*}

We found that unstable modes are present for some configurations, the properties of which are summarized in
Figs.~\ref{gr:plot_alm_example} and \ref{gr:plot_phase_space}. Figure~\ref{gr:plot_alm_example} shows the radial profile of
dipolar ($l=1$) unstable modes for a CD star with compactness $M/R=0.2$ and couplings constants $(\Omega,\eta)=(-20,0)$
and for a CD star with compactness $M/R=0.4$ and coupling constants $(\Omega,\eta)=(0,-20)$.

When $\Omega, \eta >0$, we find unstable solutions as well. For each choice of $\eta$ and $\Omega$ we find a sequence of
characteristic frequencies corresponding to unstable solutions with nodes.

In Fig.~\ref{gr:plot_phase_space} we show the stability diagram for different values of the coupling constant $\Omega$,
assuming $\eta=0$, obtained by considering dipolar axial perturbations of stellar configurations with different values
of the compactness $M/R$.  The left panel refers to CD stars, while the right panel refers to Poly stars. The shaded region
corresponds to configurations which are unstable under axial perturbations. Strictly speaking, these regions correspond
to instability to dipolar ($l=1$) perturbations, but we find strong evidence that the configurations unstable to $l>1$
axial perturbations are also unstable to dipolar ones.  The dotted horizontal line represents the Buchdal limit
$\frac{M}{R}<\frac{4}{9}$, which we verified to be satisfied in HN theory, while the dot-dashed curve corresponds to the
Newtonian configurations for CD stars (see discussion below).  Note that, as discussed above, for negative couplings
$\Omega, \,\eta$, even Newtonian stars can become unstable. We find that unstable configurations also exists in the case
of $\eta\neq0$.

The separation between the stable and unstable regions, i.e. the boundaries of the shaded regions in
Fig.~\ref{gr:plot_phase_space}, correspond to zero-mode solutions, i.e., static regular solutions with non-vanishing
vector field. In order to improve our understanding of this boundary, we shall now consider zero-mode solutions in the
Newtonian limit (i.e., $\frac{M}{R}\ll 1$) for a CD star. In this limit Eq.~\eqref{eq:axial} reduces to
\begin{equation}
a_{l}''-\left(\frac{l(l+1)}{r^2}+\mu^2\right) a_{l}=0\,,\label{eq:axnewt}
\end{equation}
where $\mu^2=2 \pi \rho(2\Omega+\eta)$ is the effective mass of the vector inside the star. Imposing regularity at the
origin, the general solution of Eq.~\eqref{eq:axnewt} inside the star is (modulo an arbitrary multiplicative constant)
$a_l=\sqrt{r}J_{l+1/2}(-i\mu r)$, with $J_\nu$ Bessel function.  Outside the star $\rho=0$ and imposing regularity at
infinity Eq.~\eqref{eq:axnewt} gives $a_l\propto r^{-l}$. Matching the interior and the exterior solution at the radius
of the star $r=R$ we find that a regular solution exists only for $I_{l-1/2}(-i\mu R)=0$, where $I_\nu$ is the modified
Bessel function, i.e., for $\mu R=i\pi$, which corresponds to
\begin{equation}\label{eq:newtonian_axial}
-3 M \left(\eta +2 \Omega \right)= 2 \pi ^2 R\,.
\end{equation}
When $\eta=0$, this equation admits a non-trivial solution for negative values of $\Omega$. Thus, for $\eta=0$,
$\Omega<0$ we expect the presence of unstable solutions.

The Newtonian prediction~\eqref{eq:newtonian_axial} is shown in Fig.~\ref{gr:plot_phase_space} (green dot-dashed
curve). We note that this line is close to the boundary of the unstable region, and it is closer for smaller values of
the compactness, as expected.  At fixed negative coupling constant $\Omega$, as the compactness increases to large
values, the quantity $\rho-3P$ decreases. For Poly stars, at $M/R\sim 0.27$ it becomes negative. The effective
squared mass of the vector \eqref{effective_mass} is then positive, and the star is stable. Thus, all the main features
of Fig.~\ref{gr:plot_phase_space} can be understood in simple terms. For the same reasons, unstable solutions lying on
the right side of the plot (positive coupling constants) exist for very large values of the compactness, when the
effective mass squared is again negative.

It is worth noting that our results resemble those of scalar-tensor theories (compare our Fig.~\ref{gr:plot_phase_space}
with Fig. 1 of Ref.~\cite{Pani:2010vc}). The root of the mechanism is the same: a tachyonic instability that is either
triggered by a ``wrong'' sign of the coupling constants or by the wrong sign of the trace of the stress-energy tensor.
Despite these similarities, we are here discussing the dipolar axial sector excitations of the vector field, which have
a very different behavior from those of the scalar field. The end state of this instability is unknown to us.

Including backreaction on Einstein's equations, the axial perturbations give a contribution to
the $(\theta,\theta)$ component of Einstein's equations, which can be considered as an effective stress-energy tensor.
This suggests that  the star will be made to rotate as a result of such instability. Another outcome is possible: that
the star exits the instability window through mass shedding. The fate of stars on the unstable branch remains an open
issue.

\subsection{Spontaneous and induced vectorization in the polar sector}

Since we are interested in static and spherically symmetric solutions of the full non-linear field equations in HN
gravity, we shall now study linear vector field perturbations with polar parity in this theory.

\subsection*{Dynamical case}

To begin with, let us consider monopolar ($l=0$) perturbations. 
In the exterior of the star, we find that the $l=0$ polar perturbation equations reduce to those of GR, i.e.
\begin{align}
(i\omega h_0+f_0')\left(rFG'+4FG-rGF'\right)+2rGF\left(i\omega h_0+f_0'\right)'&=0\,,\nonumber\\
\omega\left(i\omega h_0+f_0'\right)&=0\,.\label{eq:l0gr}
\end{align}
Since the radial electric field $E_r$ is proportional to $i \omega h_0+f_0'$, when $\omega\neq0$ the second of
Eqs.~\eqref{eq:l0gr} implies that $E_r=0$, and thus the wave is pure gauge: there are no spherically symmetric electromagnetic waves
with radial electric fields in the exterior of the star, in HN gravity as in GR. Then, since the solution inside the
star has to match the exterior solution, $E_r$ is pure gauge in the entire spacetime. In other words, there is no
dynamical {\it linear} instability for spherically symmetric modes.

Let us now consider the polar perturbations with the $l\ge1$ case. By replacing the
expansion~\eqref{eq:vector_perturbation_def} in the $r$ component of Eq.~\eqref{eq:vector_field_linear} we find:
\begin{equation}
h_l=\frac{-l(l+1) F k_l'+i r^2 \omega f_l'}{F \left(2 \pi  r^2
   ((6 \Omega +\eta ) p-(\eta +2 \Omega ) \rho )-l(l+1)\right)+r^2 \omega^2}.\label{eq:l1r}
\end{equation}
Replacing Eq.~\eqref{eq:l1r} in the $t$ and $\theta$ components of Eq.~\eqref{eq:vector_field_linear} we obtain a system
of coupled ordinary differential equations (ODEs) in $f_l$ and $k_l$,
\begin{align}\label{eq:dipolar_one}
&\frac{G f_l''  (1-\frac{r^2 \omega^2}{F (2 \pi  r^2
   ((6 \Omega +\eta ) p -(\eta +2 \Omega ) \epsilon  )-l (l+1))+r^2
   \omega^2})}{F}\nonumber\\
&+\frac{i l (l+1) \omega G
   k_l'' }{F (l (l+1)+2 \pi  r^2 ((-\eta -6 \Omega ) p -(-\eta
   -2 \Omega ) \epsilon  ))-r^2 \omega^2}\nonumber\\
&-\frac{f_l  (l (l+1)+2 \pi  r^2 (3 (-\eta -2 \Omega ) p +(-\eta +2 \Omega )
   \epsilon  ))}{r^2 F}-\frac{i l (l+1) \omega k_l }{r^2
   F}\nonumber\\
&+\frac{i l (l+1) \omega k_l' }{2 r F (F (2 \pi 
   r^2 ((6 \Omega +\eta ) p +(-\eta -2 \Omega ) \epsilon  )-l (l+1))+r^2   \omega^2)^2} (r F (-G F' 
   (l (l+1)\nonumber\\
   &+2 \pi  r^2 ((-\eta -6 \Omega ) p -(-\eta -2 \Omega ) \epsilon
    ))-r^2 \omega^2 G' )+r^3 (-\omega^2) G
   F' +F^2 (r G'  (l (l+1)\nonumber\\
&+2 \pi  r^2 ((-\eta -6
   \Omega ) p -(-\eta -2 \Omega ) \epsilon  ))+4 G (l (l+1)+\pi
    r^3 ((6 \Omega +\eta ) p' +(-\eta -2 \Omega ) \epsilon
   ' ))))\nonumber\\   
&-\frac{f_l' }{2 r F (F (2 \pi  r^2 ((6 \Omega
   +\eta ) p +(-\eta -2 \Omega ) \epsilon  )-l (l+1))+r^2 \omega^2)^2} (r F (G (l^2 (l+1)^2
   F' \nonumber\\
   &+4 \pi  r^2 ((-\eta -6 \Omega ) p  (F'  (l
   (l+1)-2 \pi  r^2 (-\eta -2 \Omega ) \epsilon  )+2 r \omega^2)\nonumber\\
   &-(-\eta -2 \Omega
   ) \epsilon   (l (l+1) F' +2 r \omega^2)+\pi  r^2 (-\eta -6 \Omega
   )^2 p ^2 F' +\pi  r^2 (-\eta -2 \Omega )^2 \epsilon  ^2
   F' \nonumber\\
   &+r^2 \omega^2 ((-\eta -6 \Omega ) p' -(-\eta -2 \Omega ) \epsilon
   ' )))+r^2 \omega^2 G'  (l (l+1)+2 \pi  r^2 ((-\eta -6
   \Omega ) p \nonumber\\
   &-(-\eta -2 \Omega ) \epsilon  )))+r^3 \omega^2 G
   F'  (l (l+1)+2 \pi  r^2 ((-\eta -6 \Omega ) p -(-\eta -2 \Omega )
   \epsilon  ))\nn\\
   &-F^2 (r G' +4 G)
   (l (l+1)+2 \pi  r^2 ((-\eta -6 \Omega ) p -(-\eta -2 \Omega ) \epsilon
    ))^2)=0\,,
    \end{align}
    \begin{align}
    \label{eq:dipolar_two}
&\frac{i \omega
   G  f_l'' }{F  (2 \pi  r^2 ((6 \Omega +\eta )
   p -(\eta +2 \Omega ) \epsilon  )-l (l+1))+r^2 \omega^2}+\frac{i \omega f_l }{r^2 F }\nonumber\\
   & -\frac{G  k_l''  (2 \pi  F  ((6 \Omega +\eta )
   p +(-\eta -2 \Omega ) \epsilon  )+\omega^2)}{F  (2 \pi  r^2 ((6
    \Omega +\eta ) p +(-\eta -2 \Omega ) \epsilon  )-l (l+1))+r^2 \omega^2}\nn\\
    &+\frac{k_l
    (2 \pi  F  ((-\eta -6 \Omega ) p -(-\eta -2 \Omega )
   \epsilon  )-\omega^2)}{r^2 F }\nonumber\\
   &+\frac{i \omega f_l'  }{2 r F 
   (F  (2 \pi  r^2 ((6 \Omega +\eta ) p +(-\eta -2 \Omega )
   \epsilon  )-l (l+1))+r^2 \omega^2)^2}(r F  (G  F'  (l
   (l+1)\nonumber\\
   &+2 \pi  r^2 ((-\eta -6 \Omega ) p -(-\eta -2 \Omega ) \epsilon  ))+r^2
   \omega^2 G' )+r^3 \omega^2 G  F' +F ^2 (4
   G  (\pi  r^3 ((-\eta -6 \Omega ) p' \nonumber\\
   &-(-\eta -2 \Omega ) \epsilon
   ' )-l (l+1))-r G'  (l (l+1)+2 \pi  r^2 ((-\eta -6 \Omega
   ) p -(-\eta -2 \Omega ) \epsilon  ))))\nonumber\\
    & -\frac{k_l'  }{2 r F  (F  (2 \pi  r^2 ((6
   \Omega +\eta ) p +(-\eta -2 \Omega ) \epsilon  )-l (l+1))+r^2 \omega^2)^2} (r \omega^2 F  (G  F'  (l
   (l+1)\nonumber\\
   &+4 \pi  r^2 ((6 \Omega +\eta ) p +(-\eta -2 \Omega ) \epsilon  ))+r^2
   \omega^2 G' )-F ^2 (-2 \pi  r G  F' 
   ((-\eta -6 \Omega ) p \nonumber\\
   &-(-\eta -2 \Omega ) \epsilon  ) (l (l+1)+2 \pi  r^2
   ((-\eta -6 \Omega ) p -(-\eta -2 \Omega ) \epsilon  ))+r \omega^2 G' 
   (l (l+1)\nonumber\\
   &+4 \pi  r^2 ((-\eta -6 \Omega ) p -(-\eta -2 \Omega ) \epsilon
    ))+4 l (l+1) \omega^2 G )+r^3 \omega^4 G  F' +2
   \pi  F ^3 (r G'  ((-\eta -6 \Omega ) p \nonumber\\
   &-(-\eta -2 \Omega )
   \epsilon  ) (l (l+1)+2 \pi  r^2 ((-\eta -6 \Omega ) p -(-\eta -2 \Omega )
   \epsilon  ))+2 l (l+1) G  (r (-\eta -6 \Omega ) p'  \nonumber\\
   &+2 (-\eta   -6 \Omega ) p-r (-\eta -2 \Omega ) \epsilon ' -2 (-\eta -2 \Omega ) \epsilon   )))=0\,.
\end{align}

The perturbation equations have simpler expressions in the exterior of the star. Indeed, as $\rho=p=0$ 
Eqs.~\eqref{eq:dipolar_one} and \eqref{eq:dipolar_two} can be cast as a single ``master equation'' in terms of the quantity
\begin{equation}
\psi = f_l + i \omega k_l\,,
\end{equation}
which is:
\begin{equation}\label{eq:schwarz_polar}
 \frac{\left(l^2+l\right) \psi }{2 M r-r^2}+\frac{l (l+1) (2 M-r) \psi '' }{l (l+1)   (2 M-r)+r^3 w^2}+\frac{2 l (l+1)  \left(l (l+1) (r-2 M)^2-M r^3
   w^2\right) \psi '}{r \left(l (l+1) (2 M-r)+r^3 w^2\right)^2}=0\,.
\end{equation}
\begin{figure}
\centering
\includegraphics[width=0.6\textwidth,keepaspectratio]{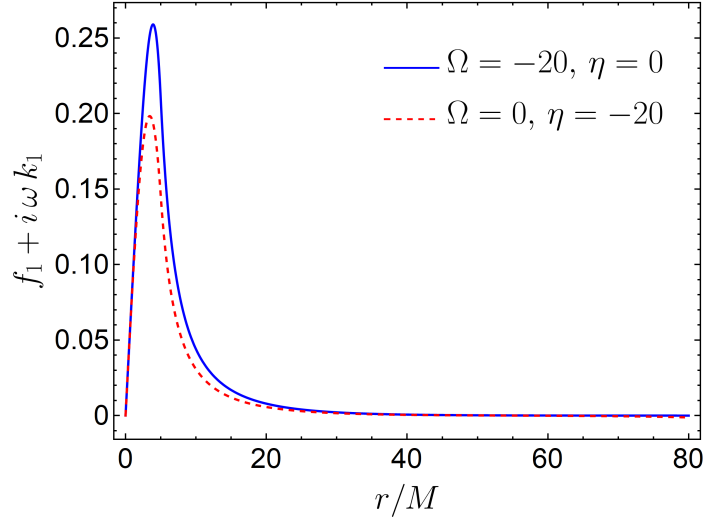}
\caption[Polar unstable dipolar profile for a CD star.]{Unstable dipolar vector perturbation profile for a constant density star with compactness $M/R=0.2$. The solid blue line ($\eta=0,\Omega=-20$) corresponds to an instability rate $M\omega=0.103 i$. The dashed red line ($\Omega=0,\eta=20$) corresponds to a rate $M\,\omega=0.0989 i$.}\label{gr:plot_psi_example}
\end{figure}
Solving the Cauchy problem given by Eqs.~\eqref{eq:dipolar_one} and \eqref{eq:dipolar_two}, with appropriate initial
conditions (as discussed in Section \ref{app:Axial perturbation integration}) and matching at the boundary of the star
with Eq.~\eqref{eq:schwarz_polar}, we find unstable configurations for CD stars. In Fig.~\ref{gr:plot_psi_example},
for instance, we show the radial profile of an unstable mode with $l=1$ for a constant density star of compactness
$M/R=0.2$, for $\Omega=-20, \eta=0$ or $\eta=20, \Omega=0$.
As in the case of axial perturbations, we can construct the instability diagram in the space ($\Omega, M/R$). The static
zero-mode solutions in the Newtonian limit $M/R\ll1$ yields
\begin{equation}\label{eq:newtonian_polar_lone}
3 M \left(\eta -2 \Omega \right)= 2 \pi ^2 R\,.
\end{equation}
In Fig.~\ref{gr:plot_phase_space_polar_CDS} we show the instability region (shaded region) in the ($\Omega, M/R$) plan,
for $\eta=0$ and negative values of the coupling constant $\Omega$. The horizontal dotted line represents the Buchdal
limit, and the dot-dashed curve represents the Newtonian zero-mode solutions corresponding to
Eq.~\eqref{eq:newtonian_polar_lone}.
\begin{figure}
\centering
\includegraphics[width=0.6\textwidth,keepaspectratio]{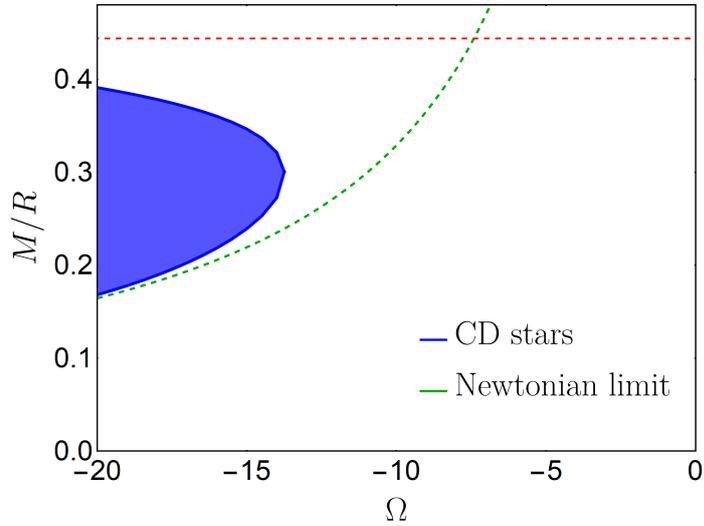}
\caption[Phase space for polar instabilities in compact stars.]{Instability diagram for CD stars, for polar perturbations with $l=1$. The shaded region represents solutions
  which are unstable under polar perturbations in HN gravity with $\eta=0$. The green dot-dashed curve describes
  zero-frequency modes in the Newtonian regime, Eq.~\eqref{eq:newtonian_polar_lone}. The dashed red line corresponds to
  the Buchdal limit ($M/R<4/9\approx0.444$).}\label{gr:plot_phase_space_polar_CDS}
\end{figure}

An interesting feature of these results is that the contribution of the coupling $\eta$ to the effective mass squared
has opposite sign from that of axial perturbations: large negative $\eta$ make Newtonian stars unstable against axial
perturbations, and large positive $\eta$ turn Newtonian stars unstable against polar perturbations.

\subsection*{Static case}

\begin{figure}
\centering
\includegraphics[width=0.6\textwidth,keepaspectratio]{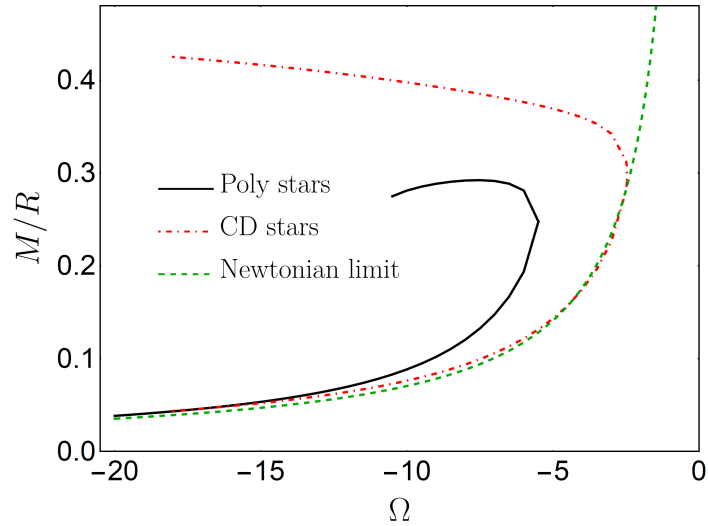}
\caption[(Polar) Phase space in the Newtonian limit]{Linear static vector field solutions for a NS background (solid black line) and for a CD star (dot-dashed red line)
 for $\eta=0$. The dashed green line corresponds to the Newtonian analytic solution for the polar sector in the CD star
 background.}
\label{gr:plot_linearsol_CDS_EOS}
\end{figure}

We showed that spherically symmetric polar modes have no interesting dynamics. 
However, there is still room for the existence of non-trivial static ($\omega=0$) solutions. Replacing the
expansion~\eqref{eq:vector_perturbation_def} in the field equations~\eqref{eq:vector_field_linear} we find: 
\begin{equation}
h_0=0\,,
\end{equation}
and\footnote{It is possible to show that $k_0$ can be canceled out by the use of an appropriate combination of the independent  components of the modified Maxwell equations.},
\begin{align}
&\big[-f_0 '  \left(2 r \left(m' -2\right)+r (r-2 m ) \nu ' +6 m \right)+2 r (r-2 m ) f_0 '' +4 \pi  r^2 (2 \Omega  (3 p-\rho )\nn\\
&+\eta  (3 p+\rho ))f_0\big] \frac{e^{-\nu  }}{2 r^2}=0\,.\label{eq:static_linear_polar_equation}
\end{align}
Solving Eq.\eqref{eq:static_linear_polar_equation} we find a class of linear static vector field solutions, for both the
CD and the Poly star configurations. Equation~\eqref{eq:static_linear_polar_equation} also implies an analytical relation
between the compactness and the coupling constant, in the Newtonian regime for a CD star. Indeed, since $h_0=0$, we can
assume the following form for the vector field:
\begin{equation}
X_\mu=(f_0(r)/r,0,0,0)\,.\label{eq:one_component_assumption}
\end{equation}
In the limit $M/R\ll1$, for a CD star Eq.~\eqref{eq:static_linear_polar_equation} reduces to
\begin{equation}
f_{0}''-\mu^2 f_{0}=0\,,
\end{equation}
where $\mu^2=2 \pi \rho(2\Omega-\eta)$ is the effective mass of the vector inside the star. Imposing regularity at the
origin we find $f_0=e^{\mu r}-e^{-\mu r}$ inside the star, and imposing regularity at infinity we find $f_0=const$ in
the exterior. By matching the interior and exterior solutions at the radius of the star we find
\begin{equation}
6M(\eta -2 \Omega)=\pi ^2 R\,.
\end{equation}
The static solutions are summarized in Fig.\til\ref{gr:plot_linearsol_CDS_EOS}. These solutions might be said to be {\it induced},
rather than arising spontaneously as the end product of an instability: they arise as the end product of (perhaps special) initial conditions.
Such vectorized solutions have no parallel in scalar-tensor theory and do not exist in the axial sector of HN gravity
itself. Note again that $\eta$ contributes with the opposite sign, relative to axial perturbations in Eq.~\eqref{eq:axnewt}.
From these solutions we can conclude that in GR, a vector field coupled with the curvature of the spacetime can have a
non-trivial profile around compact stars. Moreover, this result suggests that vectorized stars can appear even in
full non-linear HN gravity. 

We should mention that, generically, the vector $X_\mu$ will tend to grow {\it all} its components. Thus, with the exception
of a measure-zero set of initial conditions, finding only a non-zero time component is impossible. In other words, the spherically symmetric state that occurs at linear (and non-linear, as we show below) level is not generic and should always be accompanied by the linear instability of the non-symmetric modes. However, from a purely mathematical level the distinction between induced and spontaneous processes can be made and we have adopted such nomenclature here.

\section{Static, vectorized neutron stars}
We shall now determine full non-linear, stationary and spherically symmetric NS configurations in HN theory, solutions
of Eqs.~\eqref{eq:einstein_equation} and \eqref{eq:vector_field}. In other words, we show that the induced vectorized
solutions, found above at a linear level, do indeed exist at full non-linear level.  Hereafter, we assume $\eta=0$. 

\subsection{Formalism and structure equations}\label{sec:formalism_star_HN}
For
convenience, we rewrite the line element of Eq.~\eqref{eq:line_element} defining $F=e^{\nu(r)}$ and
$G=1-\frac{2m(r)}{r}$. A spherically symmetric vector field can only have non-vanishing $t$- and $r$-
components. Moreover, the $r$ component of the vector field equation reduces to
\begin{equation}
\label{eq:X1}
\frac{X_r}{4 r^3}\Big[\Omega (r-2 m)  (-2 m' (r \nu '+4)+(4 r-6 m) \nu ')+2 r (r-2 m) \nu ''+r (r-2 m) \nu '^2\Big]=0\,,
\end{equation}
which implies $X_r\equiv0$. 
Therefore, all the space components of the vector field identically vanish:
\begin{equation}
X_{\mu}=\left\{X(r),0,0,0\right\}\,.
\end{equation}
The structure equations for the star are given by the $(t,t)$, $(r,r)$ components of the Einstein equations
\eqref{eq:einstein_equation}, the vector field equation \eqref{eq:vector_field} and the conservation of the
stress-energy tensor:
\begin{equation}\label{eq:stress_energy}
\nabla_{\nu}T^{\mu\nu}=0\,.
\end{equation}
We note that Eq.~\eqref{eq:stress_energy} holds in HN gravity because the GR modifications do not affect the matter
section of the action~\eqref{eq:action}; therefore, as explicitly shown in Ref.~\cite{Will:1993ns}, the four-divergence
of the stress-energy tensor vanishes in this theory, as in GR. We thus obtain a system of four ODEs in the variables
\begin{equation}
\left\{m(r),\nu(r),p(r),X(r)\right\}\,.
\end{equation}
These modified Tolman-Oppenheimer-Volkoff (TOV) equations are explicitly given by,
\begin{align}\label{eq:TOV_system}
  &\frac{e^{-\nu}}{2 r^2} \big[-4 e^{2 \nu} (4 \pi  r^2 \rho-m')-4 r^2 X'^2+-e^{\nu }
    \Omega  X^2 (4 m' (r \nu '+1)+4 (3 m-2 r) \nu '\nonumber\\
    &+r (r-2 m) (\nu '^2-4 \nu ''))-e^{\nu } (4 \Omega  X (r (r-2 m) X''-X' (r m'+2 r (r-2 m) \nu '+3 m-2r))\,,\nonumber\\
    &+2 r (2
    \Omega -1) (r-2 m) X'^2)\big]=0\,,\nonumber\\
  &\frac{e^{-\nu }}{2 r^2 (r-2 m)} \big[2 m (-2 r^2 (4 e^{\nu }-1) X'^2-2 e^{\nu }
    (r \nu '+1)-2 r \Omega  X (r \nu '+4) X'\nonumber\\
& +\Omega  X^2(r \nu ' (r \nu '+2)-2))   +16 r m^2 e^{\nu }X'^2+r^2 (-16 \pi  r p e^{\nu }+2 \Omega  X (r \nu '+4) X'\nn\\
&+2 e^{\nu }    (\nu '+2 r X'^2)-2 r X'^2-\Omega  X^2 \nu ' (r \nu '+2))\big]=0\,,\nonumber\\
&\frac{e^{-\nu }}{4 r^2} \big[-2 X' (2 r (m'-2)+r (r-2 m) \nu '+6 m)\nonumber\\
&+\Omega  X (-2 m' (r \nu '+4)+2 (2 r-3 m) \nu '+r (r-2 m) (2 \nu
   ''+\nu '^2))+4 r (r-2 m) X''\big]=0\,,\nonumber\\
&\frac{(r-2 m) (2 p'+(p+\rho) \nu ')}{2 r}=0\,.
\end{align}
%
This system is invariant under the transformation
\begin{equation}\label{eq:invariance_TOV}
\{\nu(r)\rightarrow\nu_0+\tilde{\nu}(r),\, X\rightarrow e^{\frac{\nu_0}{2}}\tilde{X}\}\,,
\end{equation}
where $\nu_0$ is an arbitrary constant.
%
%
%
%
\subsection{Expansions at the center}\label{app:exp_center}
The asymptotic expansion near the center of the star of the modified TOV equations~\eqref{eq:TOV_system} is:
%
\begin{align}
\rho&= \rho_c+\rho_1 r +\frac{\rho_2}{2}r^2\,,\nonumber\\
p (r)&= p_c+p_1 r +\frac{p_2}{2}r^2\,,\nonumber\\
\nu (r)&= \nu_c+\nu_1 r +\frac{\nu_2}{2}r^2\,,\label{eq:nu_expansion}\\
X (r)&= {X}_{c}+{X}_{1} r +{X}_{2}\frac{r^2}{2}\,,\nonumber\\
m (r)&= m_3 r^3\nonumber\,.
\end{align}
Performing the transformation~\eqref{eq:invariance_TOV}, we solve the modified TOV equations for $\tilde{\nu}=\nu-\nu_c$
and $\tilde{X}$. Comparing order by order, the non-vanishing coefficients of the expansion~\eqref{eq:nu_expansion} are
\begin{align}\label{eq:central_values}
  m_3 &= \frac{4 \pi  \left(3 p_c \tilde{X}_{c}^2 \Omega  (\Omega +2)+\rho_c
    \left(\tilde{X}_{c}^2 \Omega  (2 \Omega +1)-1\right)\right)}{9 \tilde{X}_{c}^2
    \Omega ^2 \left(\tilde{X}_{c}^2 (\Omega -1)+1\right)-3}\,,\nonumber\\
  \tilde{\nu}_2 &= \frac{8 \pi  \Big(3 p_c \left(\tilde{X}_{c}^2 \Omega  (2 \Omega +1)
    -1\right)}{9 \tilde{X}_{c}^2 \Omega ^2 \left(\tilde{X}_{c}^2 (\Omega -1)+1\right)-3}+\frac{8 \pi  \rho_c \left(\tilde{X}_{c}^2 \Omega  (4 \Omega -1)-1\right)\Big)}{9
    \tilde{X}_{c}^2 \Omega ^2 \left(\tilde{X}_{c}^2 (\Omega -1)+1\right)-3}\,\nonumber,\\
  \tilde{X}_{0,2}&=\frac{4 \pi  \tilde{X}_{c} \Omega  \left(3 \tilde{X}_{c}^2 \Omega
    (3 p_c+\rho_c)+3 p_c-\rho_c\right)}{9 \tilde{X}_{c}^2 \Omega ^2 \left(\tilde{X}_{c}^2 (\Omega -1)+1\right)-3}\,,\nonumber\,\\
  p_2&=-\frac{4 \pi  (p_c+\rho_c) \Big(3 p_c \left(\tilde{X}_{c}^2 \Omega  (2 \Omega +1)
    -1\right)}{9 \tilde{X}_{c}^2 \Omega ^2 \left(\tilde{X}_{c}^2 (\Omega -1)+1\right)-3}-\frac{4 \pi  (p_c+\rho_c) \rho_c \left(\tilde{X}_{c}^2 \Omega  (4 \Omega -1)-1\right)
    \Big)}{9 \tilde{X}_{c}^2 \Omega ^2 \left(\tilde{X}_{c}^2 (\Omega -1)+1\right)-3}\,.
\end{align}
In the limit $\Omega=0$ these coefficients reduce to those of the TOV equations in GR. Moreover, we note that as
\begin{equation}
9 \tilde{X}_{c}^2    \Omega ^2 \left(\tilde{X}_{c}^2 (\Omega -1)+1\right)-3=0\label{eq:den}
\end{equation}
all the coefficients 
defined in \eqref{eq:central_values}
diverge. Thus, when Eq.~\eqref{eq:den} admits solution $r=\bar r$ inside the star, i.e. $0\le\bar r\le R$, then the
modified TOV equations do not allow for a regular vectorized solution. Since, as the compactness of the star decreases, the root $\bar r$ becomes smaller, this is the reason for the existence of a threshold compactness under which the vectorized solution disappears (see e.g. Fig.~\ref{gr:plot_vectorizedsol_chargevscompactness}).
\subsection{Vectorized stars}\label{Vectorized stars}

We numerically solve the modified TOV equations, assuming the polytropic EOS introduced in Eq.~\eqref{eq:EOS}. At the
surface of the star (where the pressure vanishes) we evaluate the components of the spacetime metric, of the vector
field and of its first derivative. Then, we numerically integrate the equations in the exterior, which correspond to
the modified TOV equations with $\rho=p=0$, from the stellar surface to infinity.
With this procedure, we have a unique solution of the modified TOV equations for any choice of the quantities
\begin{equation}
\left\{p_{c},X_{c},\Omega\right\}\,,
\end{equation}
i.e. for any choice of the pressure and of the (time component of the) vector field at the center of the star, and for
any value of the coupling constant $\Omega$. At infinity, the vector field has the form
\begin{equation}
X_0(r\gg R)=X_{\rm \infty}+\frac{\alpha}{r}\,,\label{eq:field_at_inf}
\end{equation} 
where $\alpha$ is a constant which can be considered as a sort of vector charge (although it is not a Noether
charge, as in the case of the scalar charge in scalar-tensor theories~\cite{EspositoFarese:2004cc}), and $X_{\rm\infty}$
is the asymptotic value of (the time component of) the vector field.

We search for solutions of the modified TOV equations with a non-trivial vector field configuration, and with the same
asymptotic behavior as GR solutions, i.e., $X_{\rm\infty}=0$.
Is it worth noting that the mass function does not remain constant in the exterior of the star, due to the energy
contribution of the non-trivial vector field.  The gravitational mass that a far away observer can measure, i.e., the Arnowitt-Deser-Misner
mass of the spacetime, is the asymptotic value of $M(r)$. This is the definition of gravitational mass that we are going
to use in the rest of the Chapter.

The baryonic mass of the NS is defined as~\cite{Misner:1974qy}
\begin{equation}
\bar{m}=m_b \int d^3x \sqrt{-g}u^0 n(r)\,,
\end{equation}
where $u^0$ is the time component of the four-velocity and $n(r)$ is the number density of baryons, which is related to
the pressure by
\begin{equation}
p(r)=K n_0 m_b \left(\frac{n(r)}{n_0}\right)^{\Gamma}\,.
\end{equation}
For each vectorized solution, we can evaluate the normalized binding energy of the stars defined as,
\begin{equation}
\frac{E_b}{M}=\frac{\bar{m}}{M}-1\,.\label{eq:Ebm}
\end{equation}
In order to have a bound object, we need $E_b$ to be positive. Moreover, the dependence of the gravitational mass on the
central density often conveys information on the stability of the configuration. Indeed, in GR a necessary condition for
radial stability of a stellar configuration is $dM/d\rho_c>0$, or equivalently
$dM/dR<0$~\cite{Friedman:1988er,Shapiro:1983du}. The condition for stability in generalized theories depends on the
number of extra fields, and becomes more complicated~\cite{Brito:2015yfh}.
\begin{figure}
\centering
\includegraphics[width=0.6\textwidth,keepaspectratio]{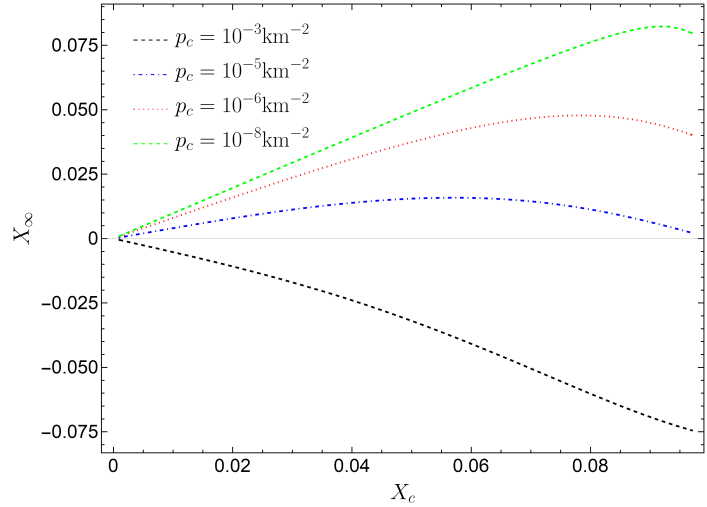}
\caption[$X_\infty$ vs field in the center.]{Time component of the vector field at infinity, as a function of its value at the center of the star, for
  $\Omega=-5$ and for different values of the central pressure. From top to bottom,
  $p_c=10^{-8},10^{-6},10^{-5},10^{-3}\,{\rm Km}^{-2}$.}\label{gr:plot_field_inf_diff_pressure}
\end{figure}
In Fig.~\ref{gr:plot_field_inf_diff_pressure} we show the results of the numerical integration of
the modified TOV equations in the case of $\Omega=-5$. 

In the figure, the asymptotic value of the vector field,
$X_{\rm\infty}$, is shown as a function of the vector field at the center, $X_c$. Each curve corresponds to a different
value of the central pressure $p_c$. The ``physical'' solutions, i.e., those with the same asymptotic behavior as the
GR solutions, correspond to the intersections of the curves with the $X_{\rm\infty}=0$ axis. We see that all curves
intersect the $X_{\rm\infty}=0$ axis at the origin (corresponding to the GR solution), but some of them also have a
second intersection, which corresponds to the vectorized solutions.

%
\begin{figure}
\centering
\includegraphics[width=0.55\textwidth,keepaspectratio]{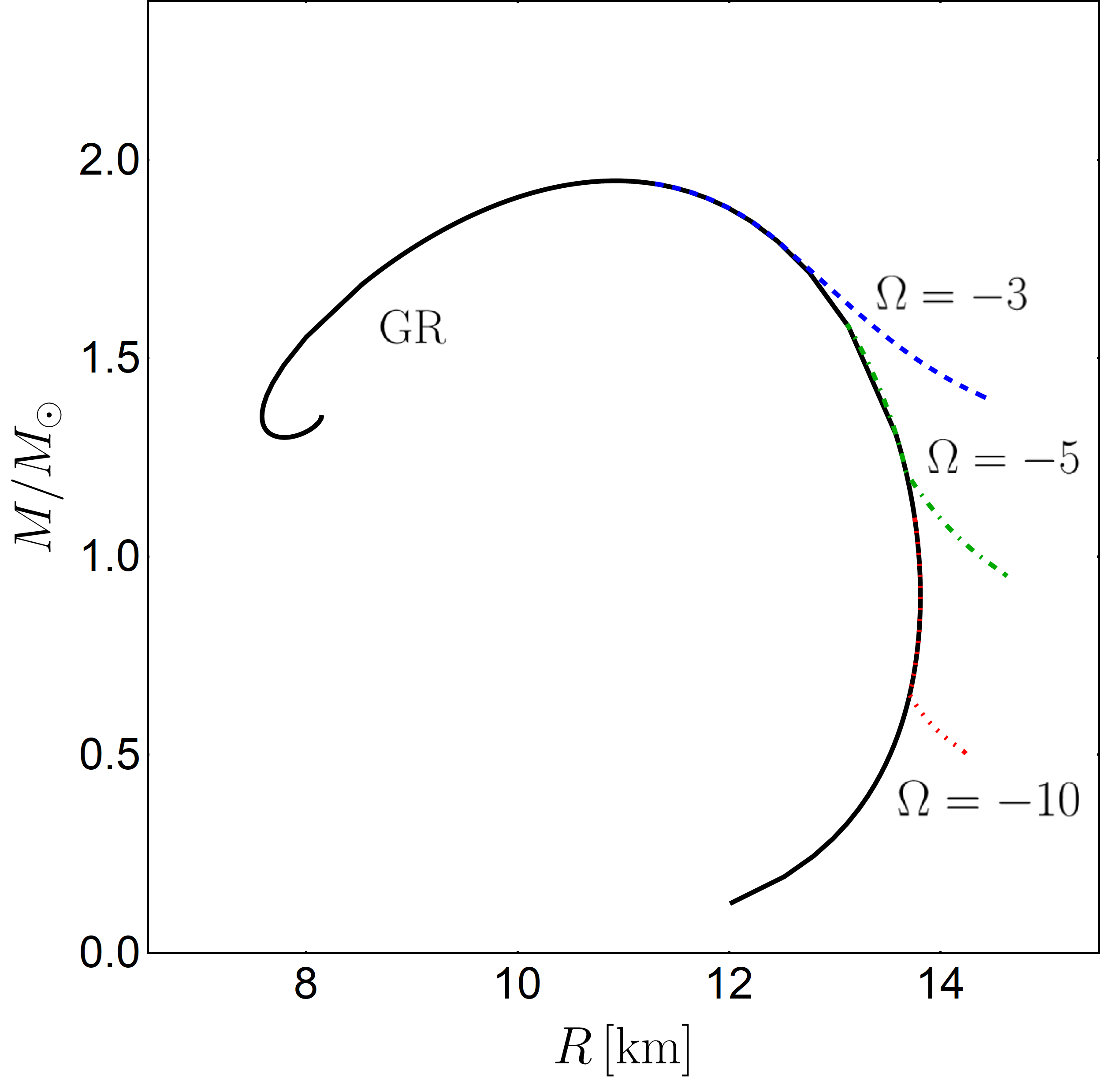}
\caption[Mass-radius diagram for vectorized stars.]{Mass-radius configurations for different coupling constants. The longest line (black) represents the solution
  for NSs in GR given the EOS in Eq.~\eqref{eq:EOS}, while the other branches correspond to vectorized
  solutions.}\label{gr:mass_radius}
\end{figure}
\begin{figure}
\centering
\includegraphics[width=0.6\textwidth,keepaspectratio]{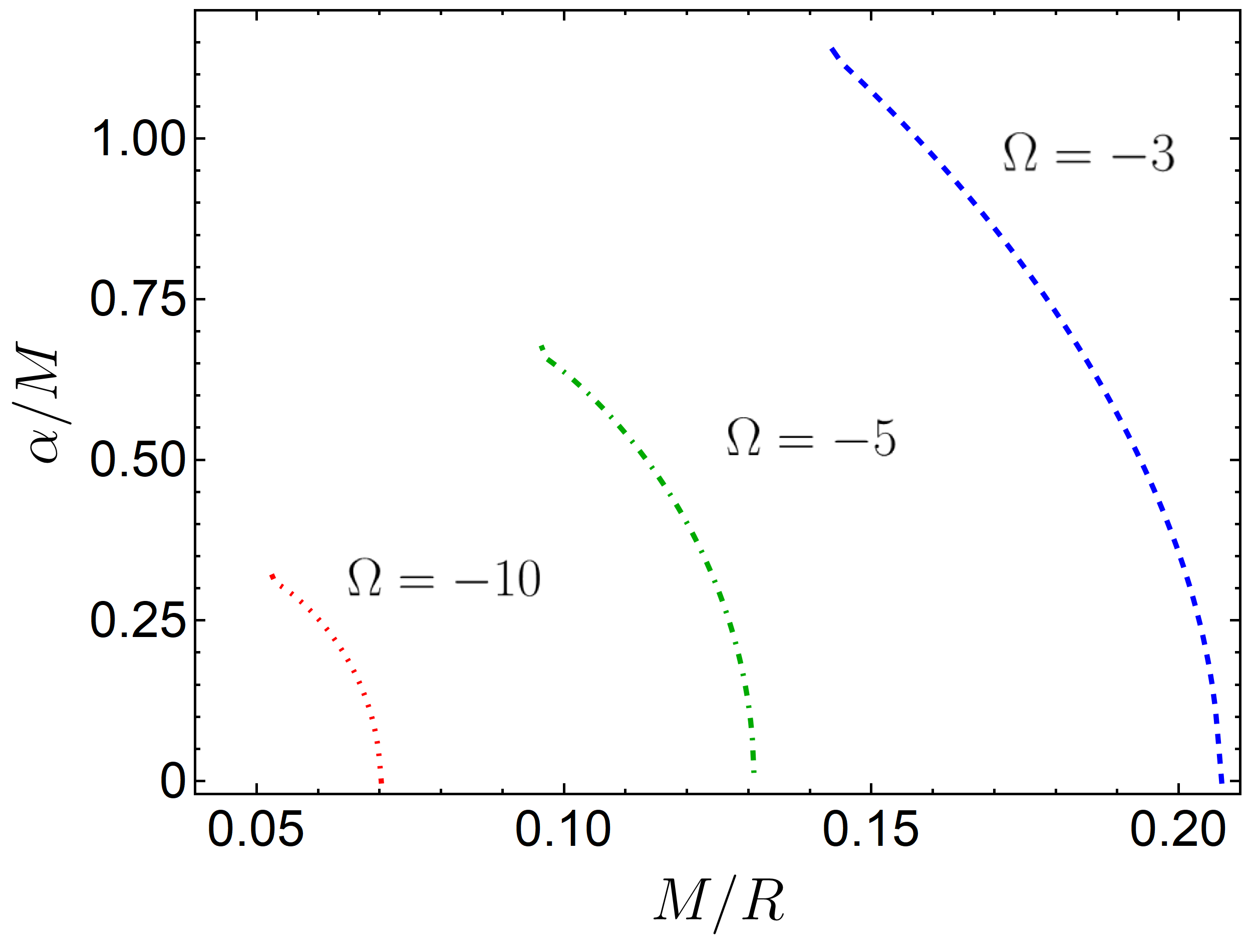}
\caption[Vector charge vs compactness.]{Vector field charge as a function of the compactness, for different stellar configurations. All the stars carrying a zero charge ($\alpha=0$) are always a solution of the theory, although they are not explicitly shown in the figure. In fact, in the range in which there are stars with a non-zero charge, the solutions are always two, as it was clear from Fig.\ref{gr:plot_field_inf_diff_pressure}.}\label{gr:plot_vectorizedsol_chargevscompactness}
\end{figure}
We computed the vectorized solutions for a wide range of values of the central pressure and of the
coupling constant $\Omega$.  The masses and the radii of these configurations are shown in Fig.~\ref{gr:mass_radius};
the corresponding values of the vector charge $\alpha$ (see Eq.~\eqref{eq:field_at_inf}) is shown in
Fig.~\ref{gr:plot_vectorizedsol_chargevscompactness} as a function of the compactness. 
We can see that comparing different compact star configurations (for a given value of the coupling $\Omega$), as the
compactness {\it decreases} there is a smooth transition between GR stars and vectorized solutions. Then, below a
threshold value of the compactness, the vectorized solution does not exist anymore. As discussed in Section\til\ref{app:exp_center}, this behavior is due to the fact that, when the compactness reaches the threshold value, the
modified TOV equation are not well behaved near the center of the star, since they become degenerate. Specifically, below the
threshold value, the weak energy condition is violated, leading to unphysical objects.
%
%
For positive values of $\Omega$, our linearized results (see Sec.~\ref{Lin_pert}) suggest that there is more than one
solution corresponding to every value of the central pressure, with different numbers of nodes in the profile of $X_0(r)$.


%
\begin{figure}
\centering
\includegraphics[width=0.55\textwidth,keepaspectratio]{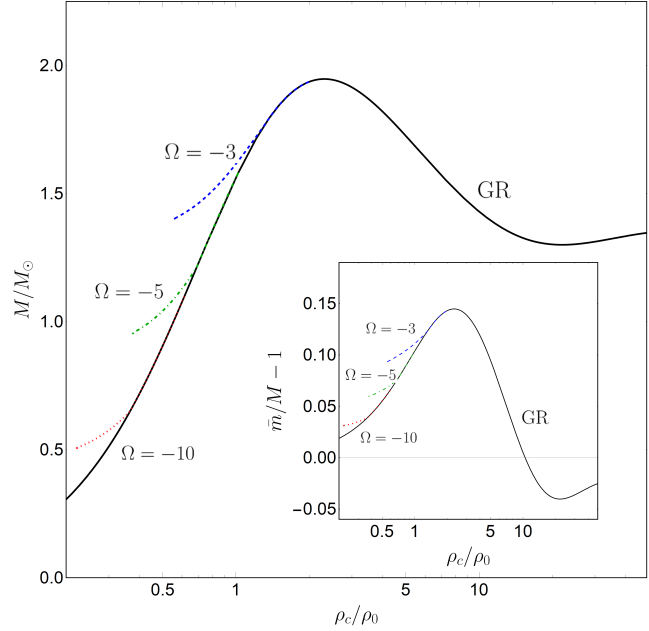}
\caption[Mass vs density for vectorized stars.]{Gravitational mass as a function of the (normalized) central baryonic density. This shows again that the solutions for $\Omega<0$ are stable solutions of the theory, since they are associated with $d M/d\rho_c>0$. {\bf Inset:} the normalized binding energy as a function of the central energy density.}\label{gr:plot_massenergy}
\end{figure}
We did not perform a dynamical stability analysis of such vectorized solutions. However, important information is conveyed by the dependence of the total (normalized) gravitational mass on the central energy density ($\rho_c=m_b n(0)$). This function is shown in  Fig.~\ref{gr:plot_massenergy}.
In GR, a criterium for stability is that $d M/d\rho_c>0$. Such criterium holds only approximately in modified
theories~\cite{Harada:1998ge,Brito:2015yfh}. We will use this as merely indicative, as more sophisticated analysis tools include a dynamical evolution or the analysis done in Ref.~\cite{Brito:2015yfh}. With such criterium, all the vectorized
solutions associated with a negative coupling constant are stable: they are in the stable branch of the $dM/d\rho_c$
curve. Moreover, in the inset of Fig.~\ref{gr:plot_massenergy} we can see how the solutions in vector-tensor theory are associated with larger binding energy than in GR, indicating that they are the preferred
configuration. For positive couplings, however, the behavior is the opposite.

Finally, we note that the instability of solutions in the positive $\Omega$ semiplane is consistent with previous results in scalar-tensor gravity~\cite{Mendes:2016fby}. Thus, none of the solutions associated with a positive coupling constant are stable and most likely do not play any astrophysical role.

Somewhat surprisingly, our results show that the changes in the NS structure with respect to the GR are smaller for {\it
  larger values of the coupling constants}. This finding is consistent with previous reported results in a related
theory~\cite{Ramazanoglu:2017xbl}. In fact, it is apparent from Fig.~\ref{gr:plot_vectorizedsol_chargevscompactness}
that the charge (and so the field) inside vectorized stars is larger for smaller (absolute) values of the
compactness. This is probably due to the role of the coupling constant $\Omega$ in the modified TOV system. In fact, it resembles
the role of a mass for the scalar field in scalar-tensor gravity: the larger the scalar mass, the smaller are the
modifications from GR.  Taking into account this, from the mass-radius plot one can quantify the range of values of
$\Omega$ that allow vectorized stars to exist,
\begin{equation}
\Omega\approx [-12,-2]\,.
\end{equation}

Finally, let us stress that there is no {\it linear} instability of spherically symmetric stars in GR.
Thus, there is no linear mechanism to drive a GR star to these new vectorized states that we just described.
Such vectorized spherical stars must therefore arise out of non-linear effects (such as selected initial conditions).

\section{Conclusions}

In this final Chapter, we have shown novel spherically symmetric star configurations in HN gravity. Particularly, when a vector field is non-minimally coupled to gravity, compact star solutions with a non-trivial, asymptotically vanishing vector field configuration may arise. Our results suggest that such vectorized solutions belong to two classes. One is spherically symmetric and ``induced'' by non-trivial initial conditions or environments. We build fully non-linear spacetimes describing such stars. These stars carry a non-zero electric charge and give rise to dipolar electromagnetic radiation when accelerated. The calculation of such fluxes and its use in astrophysical
observations to constrain the coupling constants is an interesting open problem. The second family may arise as the end-state of the instability of GR solutions, and are thus ``spontaneously vectorized'' stars. Actually, we are unable at this point to follow the evolution of such instability or even to understand its end state, since it drives a non-spherically symmetric mode. The end state could be a star with a non-trivial vector field, but it could also simply be a GR solution away from the instability region.

As a final remark, in the case in which the extra vector field is interpreted as a hidden vector field, decoupled from the other matter fields, the action of HN gravity is analogous to to the so-called ``Jordan frame'' action of scalar-tensor theory\til\cite{fujii2003scalar}. However, while in scalar-tensor theories the Jordan frame representation is equivalent to an ``Einstein frame'' representation, in
which the scalar field is minimally coupled to gravity and coupled to matter, there is no reason to believe that a similar correspondence exists in vector-tensor theories, such as HN gravity. Hence, the theory studied here might not admit an Einstein frame representation.  Recently, a theory with a massive scalar field minimally coupled to gravity and non-minimally coupled to matter (i.e., an Einstein frame vector-tensor theory) has been studied in~\cite{Ramazanoglu:2017xbl}. Again, there is no fundamental reason to believe that a Jordan frame representation of such theory exists. See, however, Ref.~\cite{Ramazanoglu:2019tyi} for a thorough discussion on this issue.

   \cleardoublepage
      \epigraphhead[450]{}
      
      \chapter{Final remarks}

In this thesis we exploited different theoretical models, trying either to challenge General Relativity or to have a better understanding of unknown matter content. The fundamental motivation towards the work developed in each Chapter comes from the prospect that gravitational wave detections will strengthen our ability to test compact objects and the environment in which they live. We believe that investigating configurations and events that might produce detectable gravitational waves will provide new valuable knowledge. As a common thread, in each Part of this manuscript we tried to address aspects of two of the most important unknown entities: the nature of dark matter and the properties of the gravitational interaction. Future detections will assess the validity of some of the models that we developed. However, even non-measurements of our predictions might be useful. Actually, this scenario happens  more often than a direct observation of new physics. Given a certain measurement, the possibility to constrain new astrophysical structure, or extra parameters of an alternative theory, is an important tool on its own. In fact, current and future constraints can guide theorists when modifying gravity or looking for non-trivial star or black hole configurations. 

Enlarging our predictive power will help assessing at least some of the major unresolved mysteries of the universe. Just think of one of the oldest open problems in Physics: the dark matter composition. Despite believed to not interact directly with the particle constituting the Standard Model, the gravitational interaction between dark matter and massive black holes might provide favourable scenarios to test the properties of this unknown form of matter. Parts of this thesis quantified some of the possible observables that might rise when black holes plunge or orbitate inside a dark matter halo, modelled through ultralight scalars. These fuzzy dark matter models are among the most popular current dark matter candidates. Remarkably, we have self-consistently evaluated, for the first time, the dynamical friction exerted by the scalar medium on massive particles moving in such environment. 

On a slightly different note, we also looked at systems that require a full general relativistic framework to be comprehensively described. In these cases, the underlying idea is to search for smoking guns for new physics, that might arise from slight modifications of General Relativity. For instance, we quantified the structure of compact stars in theories with non-trivial vector-tensor couplings. Notably, we found a new class of neutron stars that do not have a direct connection with known instabilities. Additionally, the behaviour of black hole binaries has been studied searching for possible unstable mechanisms. In particular, we focused on theories with direct couplings between a scalar field and the Gauss-Bonnet invariant. In this context, we quantified the onset of the instability, when a scalar field propagates in binary black hole spacetimes.

To date, there are formulations of quantum theories of gravity in which there are mechanisms that prevent collapsing stars to form a black hole. The outcome of this interrupted collapse is sometimes called an extreme (or exotic) compact object. Providing high curvature spacetimes, but avoiding all the pathologies coming from the presence of singularities, these compact objects acquired an increasing interest in the last decades. Despite not focusing on the different theories that can produce them, we looked at the process of gravitational wave generation in these horizonless spacetimes. To investigate the properties of gravitational waves emitted by extreme compact bodies we used a tool called the close limit approximation. Starting with two extreme compact objects with sufficiently small initial separation, the energy emitted by the collision of such objects can been quantified through perturbation theory. The outstanding agreement that this method had when compared with full numerical simulations of colliding black holes, makes us confident that it can be also employed to study the collision of more exotic configurations. The initial part of a gravitational wave emitted during the collision of such extreme compact bodies resembles the one produced by colliding black holes, hence the name black hole mimickers. However, the late part of the emitted signal is characterized by repeating echoes, due to the presence of a surface rather than an horizon. Remarkably, the energy deposited in these echoes alone might reach that produced by standard black holes collisions. 

Finally, the gravitational wave detections by current and future interferometers rely on the assumption of negligible interactions between the propagating waves and the medium in which they travel. In a different Chapter of this thesis, we provided the cross section that quantifies the interaction between gravitational waves and binaries. Based on the current values of the population of compact objects in the central part of a nearby galaxy, this interaction is in fact irrelevant for the present gravitational wave observatories. However, considering the increasing sensitivity of future gravitational wave detectors, as well as the possibility of having a greater number of binaries in other galaxies, our findings will acquire more relevance.

   \cleardoublepage
      \epigraphhead[450]{}

\part{Appendix}
\appendix

  \cleardoublepage

\chapter{Scalar Q-balls}\label{ScalarQ}
In this Appendix we show how to apply the theoretical framework developed in Chapter\til\ref{chapter:Theoretical_framework} to study the existence and the small perturbations of another class of scalar-made objects: Q-balls. In this case gravity is absent and Q-balls are kept together by scalar self-interactions.

\section*{Background configurations}
%
\begin{figure}	
\centering
\includegraphics[width=0.5\textwidth,keepaspectratio]{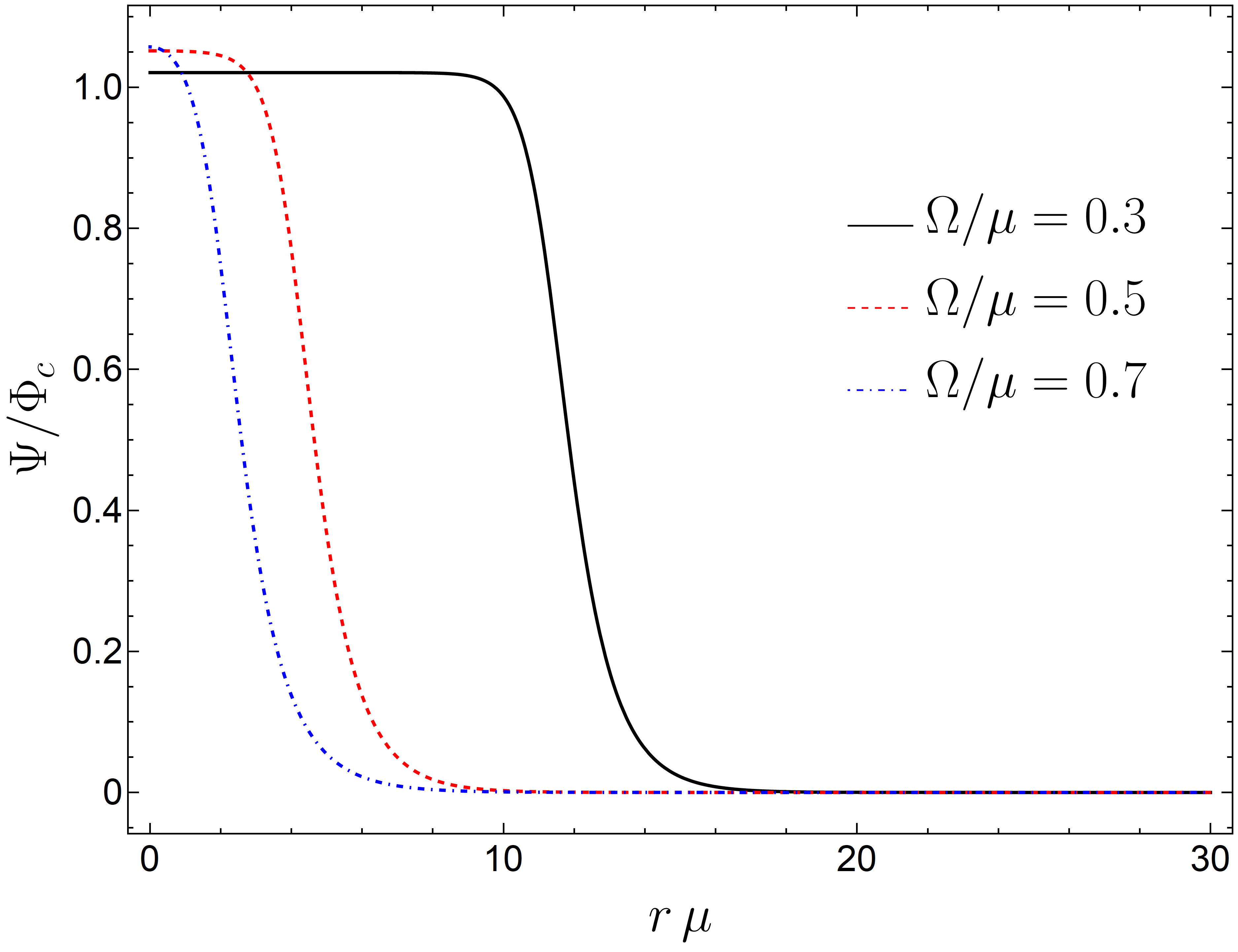} 
\caption[Q-balls radial profiles.]{Three radial profiles $\Psi(r)/\Phi_c$ obtained through numerical integration of Eq.~\eqref{EOM_Qball_radial1} with appropriate boundary conditions ($\Psi(\infty)\rightarrow 0$ and $\partial_r\Psi(0)=0$). Each curve corresponds to a different Q-ball. }\label{fig:QBalls}
\end{figure}
The field equation for $\Phi$ is obtained through the variation of action~\eqref{action} with respect to $\Phi^*$ and reads
\begin{align}
&\nabla^\mu\partial_\mu  \Phi-2 \frac{d \mathcal{U}_{\rm Q}}{d|\Phi|^2} \Phi=0\,,\label{EOM_Qball}
\end{align}
where we used $g_{\mu \nu}=\eta_{\mu \nu}$ and the potential $\mathcal{U}_{\rm Q}$ defined in Eq.\eqref{Potential_Qball}.
We now look for localized solutions of this model with the form~\eqref{BKG_ansatz} -- the so-called Q-balls. This ansatz yields the radial equation
\begin{equation} 
\partial^2_r \Psi+\frac{2}{r}\partial_r \Psi+\left[\Omega^2-2 \frac{d \mathcal{U}_{\rm Q}}{d|\Phi|^2}\right]\Psi=0\,.\label{EOM_Qball_radial}
\end{equation}
For the class of non-linear potentials~\eqref{Potential_Qball}, the last equation becomes  
\begin{equation} 
\partial^2_r \Psi+\frac{2}{r}\partial_r \Psi+\left[\Omega^2-\mu^2 \left(1-\frac{\Psi^2}{\Phi_c^2}\right)\left(1-3\frac{\Psi^2}{\Phi_c^2}\right)\right]\Psi=0\,.\label{EOM_Qball_radial1}
\end{equation}
According to the results of Ref.~\cite{Coleman:1985ki}, there exist stable Q-ball solutions for any $0<\Omega<\mu$, independently of the free parameter $\Phi_c$. 
Additionally, it is known that, in the limit $\Omega/\mu \ll1$, the radial function $\Psi$ mimics an Heaviside step function (the so-called \textit{thin-wall} Q-ball)~\cite{Coleman:1985ki,Ioannidou,Tsumagari2008}. On the other hand, in the regime $\Omega/\mu \sim 1$, the function $\Psi$ starts to fall earlier and drops very slowly (\textit{thick-wall} Q-ball)~\cite{Ioannidou,Tsumagari2008}. In particular, using the results of Ref.~\cite{Tsumagari2008} one can show that, in the thin-wall limit,
\begin{equation}\label{Psi_thin}
\Psi(r) \simeq\Phi_c\left[1+\left(\frac{\Omega}{2 \mu}\right)^2\right]\Theta\left(\frac{\mu}{\Omega^2}-r\right)\,.
\end{equation}
Notice that the Q-ball radius is approximately given by $R_Q\simeq\mu/\Omega^2$.

\begin{figure}	
\centering
\includegraphics[width=0.45\textwidth,keepaspectratio]{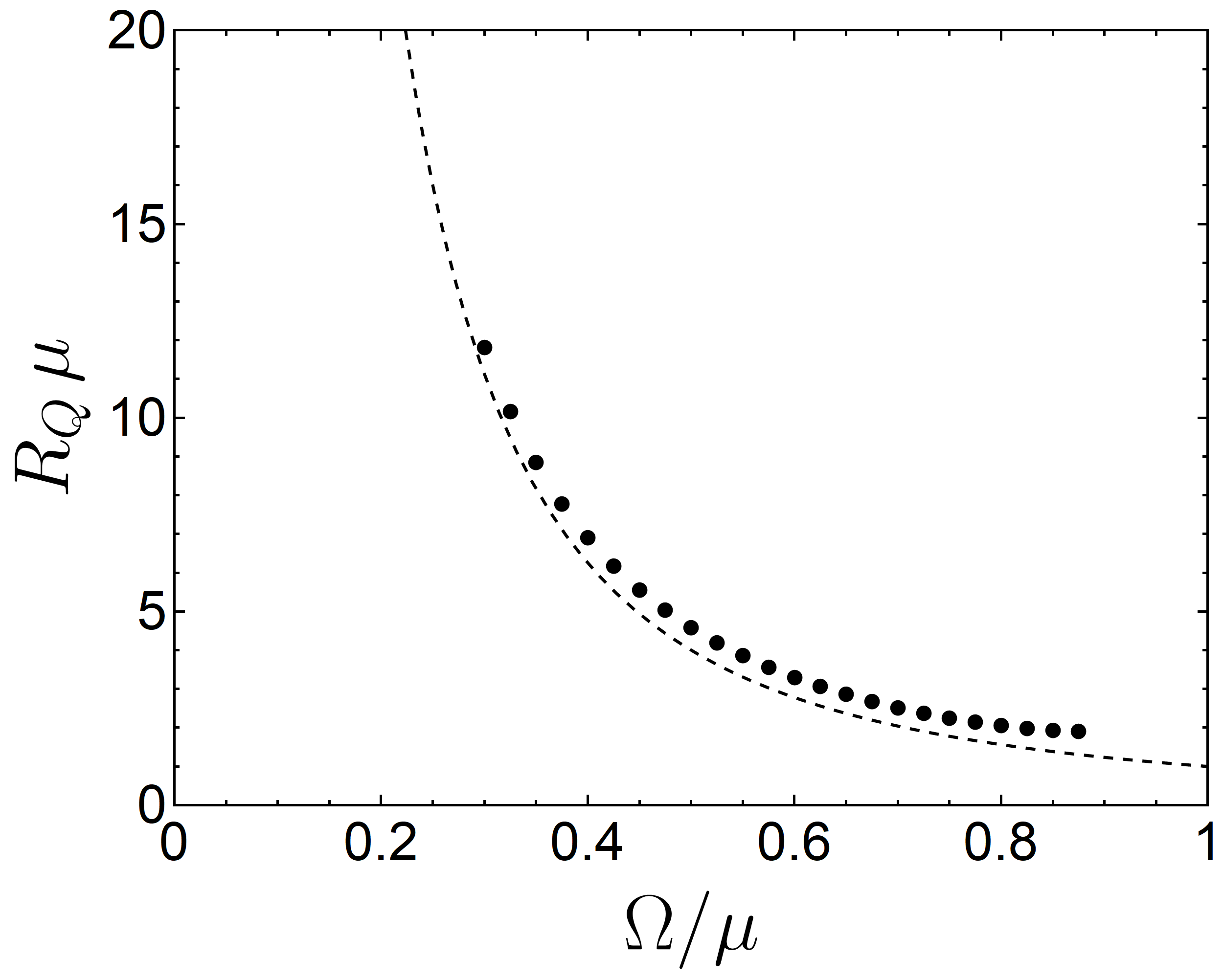} 
	\caption[Q-balls radius-frequency relation.]{Numerical results for the dependence of the Q-ball radius $R_Q \mu$ on the internal frequency $\Omega/\mu$, obtained through direct integration of Eq.~\eqref{EOM_Qball_radial1}. The dashed line is the thin-wall limit prediction, Eq.~\eqref{Psi_thin}. A fit on the numerical results gives $R_Q\sim 1.08 \mu \Omega^{-2}$, within $2\%$ of error, showing a good accordance with the predicted behaviour, Eq.~\eqref{Psi_thin}.
	} \label{fig:Radius_Omega}
\end{figure}

A few examples of radial profiles $\Psi(r)$ constructed numerically from Eq.~\eqref{EOM_Qball_radial1} are shown in Fig.~\ref{fig:QBalls}. 
From these results it is already evident that, when $\Omega/\mu \to 0$, the scalar does acquire a Heaviside-type profile. In such a limit the scalar drops to zero on the outside, on a lengthcale
$\sim 1/\mu$. These results also indicate that the radius of the Q-ball grows when $\Omega/\mu \to 0$. This is made more explicit in
Fig.~\ref{fig:Radius_Omega}, showing the numerical results for the dependence of the Q-ball radius $R_Q$  on the frequency $\Omega$. We defined the Q-ball radius $R_Q$ to be such that $\dfrac{\Psi(R_Q)}{\Psi(0)}=1/2$. The dashed line, corresponding to the thin-wall limit \eqref{Psi_thin}, agrees remarkably well with the numerics. 

The Q-ball charge $Q$ and mass $M_Q$ are obtained through~\eqref{NoetherCharge} and \eqref{Energy}, respectively, and read  
\begin{align}
& Q=4\pi \Omega \int dr\, r^2 \Psi^2(r)\,,\\
& M_Q=\frac{1}{2} Q \Omega+4\pi \int dr\,r^2\left(\frac{(\partial_r \Psi)^2}{2}+\mathcal{U}(\Psi^2)\right)\,.
\end{align}
For thin-wall Q-balls these become
\begin{align}
Q&=\frac{4 \pi}{3} \frac{\mu^3}{\Omega^5} \Phi_c^2\,,\\
M_Q&=\frac{2 \pi}{3} \frac{\mu^3}{\Omega^4} \Phi_c^2\,.
\end{align}
We are using a flat background spacetime, which requires that $M_Q/R_Q \ll 1$. In the thin-wall limit, this corresponds to
\begin{equation}
\Omega/\mu \gg \sqrt{\frac{2\pi}{3}} \Phi_c\,.
\end{equation}
\section*{Small perturbations}

We now wish to understand the effect of a small perturbation on such Q-ball configurations. They can be considered either as sourceless small deformations of the background, or sourced by an external particle. Such perturber could be another Q-ball or simply some charge, piercing the Q-ball or orbiting around it. In the following, the external probe is modelled as pointlike, which means that our results are valid only for objects whose spatial extent are $\ll R_{\rm Q}$.
We consider an interaction between the perturber and the Q-ball described by the action
\begin{equation}
\mathcal{S}_\text{int}\equiv -\int d^4x \sqrt{-g} \,{\rm Re}\left(\Phi\right) T_p\,,\label{coupling_Qball}
\end{equation}
with $T_p\equiv g_{\mu \nu} T_p^{\mu \nu}$ being the trace of the particle's stress-energy tensor defined in Eq.~\eqref{Stress_energy_particle}. This coupling
allows for equations of motion that are simultaneously simple enough to be handled via our perturbation scheme, described in Chapter~\ref{chapter:Theoretical_framework}, and it shows interesting dynamical features, as we shall see later. In the present analysis, we neglect the backreaction on the particle motion, therefore, the particle's world line $x_p^\mu(\tau)$ is considered to be known. 

An external particle sources a scalar field fluctuation of the form~\eqref{Perturbation} in the Q-ball background, which satisfies the linearized equation
\begin{align}
&-\partial_t^2\delta\Psi+\nabla^2 \delta\Psi +\left[\Omega^2-\mu^2\left(1- 8\frac{\Psi^2}{\Phi_c^2}+9 \frac{\Psi^4}{\Phi_c^4}\right)\right]\delta\Psi+2 i \Omega \partial_t \delta\Psi+2\mu^2 \frac{\Psi^2}{\Phi_c^2} \left(2-3 \frac{\Psi^2}{\Phi_c^2}\right)\delta\Psi^*\nonumber\\
&=T_p e^{i \Omega t}\,,
\label{eq:Qball_pert_eq}
\end{align}
and its complex conjugate. The sourceless case, corresponding to small Q-ball deformations, is simply recovered by setting $T_p=0$.
Decomposing the particle stress-energy trace as
\begin{equation}
T_p e^{i \Omega t}=\sum_{l,m} \int \frac{d \omega}{\sqrt{2 \pi} r}\Big[T_1^{\omega l m} Y_l^m e^{-i \omega t}+\left(T_2^{\omega l m }\right)^*\left(Y_l^m\right)^* e^{i \omega t}\Big]\,,\label{MatterDecomposition}
\end{equation}
where $T_1^{\omega l m}$ and $T_2^{\omega l m}$ are radial complex-functions defined by ,
\begin{align}
&T_1^{\omega l m}\equiv \frac{r}{2 \sqrt{2 \pi}}\int dt d\theta d\varphi \sin \theta\, T_p e^{i (\omega+ \Omega)t} \left(Y_l^m\right)^*\,, \hspace{0.5cm} \label{eq:source_decomp_T1}\\
&T_2^{\omega l m}\equiv \frac{r}{2 \sqrt{2 \pi}}\int dt d\theta d\varphi \sin \theta\, T_p e^{i (\omega- \Omega)t} \left(Y_l^m\right)^*\,,\hspace{0.5cm}\label{eq:source_decomp_T2}
\end{align}
and plugging the decompositions~\eqref{Decomposition} and~\eqref{MatterDecomposition} in Eq.~\eqref{eq:Qball_pert_eq}, one obtains the matrix equation
\begin{equation} 
\partial_r \boldsymbol{Z} -V_Q(r) \boldsymbol{Z}=\boldsymbol{T}\,,\label{Qball_Perturbation_Matrix_Sourced}
\end{equation}
where the vector $\boldsymbol{Z}\equiv (Z_1, Z_2, \partial_r Z_1, \partial_r Z_2)^T$, the matrix $V_Q$ is given by
\begin{equation*}
	V_Q\equiv
	\begin{pmatrix} 
	0 & 0 & 1 & 0 \\
	0 & 0 & 0 & 1 \\
	V_s-(\omega+\Omega)^2 & V_c & 0 & 0  \\
	V_c & V_s-(\omega-\Omega)^2 & 0 & 0  
	\end{pmatrix}\,,
\end{equation*}
and we defined the radial potentials
\begin{align}
V_s(r)&\equiv \frac{l(l+1)}{r^2}+\mu^2\left(1- 8\frac{\Psi_0^2}{\Phi_c^2}+9 \frac{\Psi_0^4}{\Phi_c^4}\right)\,, \\
V_c(r)&\equiv - 2\mu^2 \frac{\Psi_0^2}{\Phi_c^2} \left(2-3 \frac{\Psi_0^2}{\Phi_c^2}\right)\,,
\end{align}
and the source term\footnote{Again, to simplify the notation, we omit the labels $\omega$, $l$ and $m$ in the functions $T_1^{\omega l m}$ and $T_2^{\omega l m}$.}
\begin{equation}
\boldsymbol{T}(r)\equiv \big(0,0,T_1,T_2\big)^T\,.\label{stress_energy_decomposition_qball}
\end{equation}
The functions $Z_1$ and $Z_2$ are clearly independent of the azimuthal number $m$. The symmetry of Eq.\til\eqref{Qball_Perturbation_Matrix_Sourced} implies that the radial functions satisfy $Z_2(\omega,l;r)=Z_1(-\omega,l;r)^*$. 

To solve the small perturbations problem, either in the sourced or sourceless case, we need to establish suitable boundary conditions. We require regular solutions at the origin,
\begin{equation}
\boldsymbol{Z}(r \to 0)\sim \left(a r^{l+1},b r^{l+1},a (l+1)r^l,b (l+1)r^l\right)^T\,,\nonumber
\end{equation} 
with (complex) constants $a$ and $b$, and satisfying the Sommerfeld radiation condition at infinity
\begin{equation}
\label{Qball_FO_Inf}
\boldsymbol{Z}(r \to \infty)\sim \left(Z_1^\infty e^{i k_1 r}, Z_2^\infty e^{i k_2 r},
i k_1 Z_1^\infty e^{i k_1r},
i k_2 Z_2^\infty e^{i k_2r}\right)\,,
\end{equation} 
with 
\begin{align}
k_1&\equiv \epsilon_1 \sqrt{\left(\omega+\Omega\right)^2-\mu^2}\,, \\ 
k_2&\equiv \epsilon_2 \left(\sqrt{\left(\omega-\Omega\right)^2-\mu^2}\right)^*\,.
\end{align}
where we are using the principal complex square root.

Consider then the set of independent solutions $\{\boldsymbol{Z_{(1)}},\boldsymbol{Z_{(2)}},\boldsymbol{Z_{(3)}},\boldsymbol{Z_{(4)}}\}$ uniquely determined by
\begin{align}
\boldsymbol{Z_{(1)}}(r \to 0)&\sim \Big(r^{l+1},0,(l+1)r^l,0\Big)^T\,,\nonumber \\
\boldsymbol{Z_{(2)}}(r \to 0)&\sim \Big(0,r^{l+1},0,(l+1)r^l\Big)^T\,,\nonumber \\
\boldsymbol{Z_{(3)}}(r \to \infty)&\sim \Big(e^{i k_1 r},0,i k_1 e^{i k_1 r},0\Big)^T\,,\nonumber \\
\boldsymbol{Z_{(4)}}(r \to \infty)&\sim \Big(0,e^{i k_2 r},0,i k_2 e^{i k_2 r}\Big)^T\,.
\end{align}
The $4 \times 4$ matrix $F(r)\equiv\big(\boldsymbol{Z_{(1)}},\boldsymbol{Z_{(2)}},\boldsymbol{Z_{(3)}},\boldsymbol{Z_{(4)}}\big)$ is the fundamental matrix of the system~\eqref{Qball_Perturbation_Matrix_Sourced}. As it is shown in Section\til\ref{section:Small_perturbations}, for a system of the form~\eqref{Qball_Perturbation_Matrix_Sourced}, the determinant ${\rm det}(F)$ is independent of $r$.
\subsection*{Sourceless perturbations}
Free oscillations of Q-ball configurations are regular scalar fluctuations satisfying the Sommerfeld radiation condition at infinity. They correspond to scalar perturbations of the form 
\begin{equation}
\delta \Psi=\frac{1}{\sqrt{2 \pi} r} \left[Z_1 Y_l^m e^{-i\omega t}+Z_2^*\left(Y_l^m\right)^* e^{i\omega^* t}\right]\,,\label{field_decomposition_QNM_qball}
\end{equation}
where $Z_1$ and $Z_2$ are solutions of system~\eqref{Qball_Perturbation_Matrix_Sourced}, with $\boldsymbol{T}=0$. For complex-valued $\omega$, the free oscillations are QNMs. For a real $\omega$, these are termed normal modes. Notice that for the discrete set $\{\omega_{\rm QNM}\}$ of QNM frequencies, the solutions $\{\boldsymbol{Z_{(1)}},\boldsymbol{Z_{(2)}},\boldsymbol{Z_{(3)}},\boldsymbol{Z_{(4)}}\}$ are not linearly independent. In fact, it is easy to see that the condition ${\rm det}(F)=0$ holds if and only if $\omega$ is a QNM frequency (\textit{i.e.,} $\omega \in \{\omega_{\rm QNM}\}$).

\subsection*{External perturbers}
Let us turn now to the perturbations induced by an external particle, whose interacting with the background scalar field. How is such a body exciting the Q-ball, how much radiation does the interaction give rise to, what backreaction does the Q-ball
exert on the perturber? These are all questions that can be raised in this context, and that we wish to answer. 

To obtain physical observable quantities one needs to find the solutions of system \eqref{Qball_Perturbation_Matrix_Sourced} that are regular at the origin and satisfy the Sommerfeld condition at infinity. These can be obtained through the method of variation of parameters 
\begin{align}
Z_1(r)&=\sum_{k=3}^4 \Bigg[\sum_{n=1}^2 F_{1,n}(r) \int_\infty^r dr' F^{-1}_{n,k}(r') \boldsymbol{T}_k(r')+\sum_{n=3}^4 F_{1,n}(r) \int_0^r dr' F^{-1}_{n,k}(r') \boldsymbol{T}_k(r') \Bigg]\,,\\
Z_2(r)&=\sum_{k=3}^4 \Bigg[\sum_{n=1}^2 F_{2,n}(r) \int_\infty^r dr' F^{-1}_{n,k}(r') \boldsymbol{T}_k(r')+\sum_{n=3}^4 F_{2,n}(r) \int_0^r dr' F^{-1}_{n,k}(r') \boldsymbol{T}_k(r') \Bigg]\,.
\end{align}
The total energy, linear and angular momenta radiated during a given process can be found using solely the amplitudes $Z_1^\infty$ and $Z_2^\infty$. These are given by
\begin{align}
Z_1^\infty&=\sum_{k=3}^4 \int_{0}^{\infty} dr' F^{-1}_{3,k}(r')\boldsymbol{T}_k(r') \,,\label{Z1inf} \\
Z_2^\infty&=\sum_{k=3}^4 \int_{0}^{\infty} dr' F^{-1}_{4,k}(r')\boldsymbol{T}_k(r') \,. 
\end{align} 
Let us now apply our framework to two physically relevant setups: a particle plunging into a Q-ball configuration, and a particle in a circular orbit around the Q-ball.
\paragraph*{Plunging particle.}
Consider a particle moving at a constant velocity $\boldsymbol{v}=-v\boldsymbol{e}_z$ (with $v>0$), plunging into a Q-ball, and crossing its center at $t=0$. 
In this case, the trace of the particle's stress-energy tensor reads
\begin{equation}
T_p=-\left[\delta\left(r+v t\right)\delta\left(\theta\right) \Theta(-t)+ \delta\left(r-v t\right)\delta\left(\theta-\pi\right) \Theta(t)\right] m_p \,\delta(\varphi)\sqrt{1-v^2}/(r^2 \sin \theta)\,.
\label{T_p_plunging}
\end{equation}
Therefore, the source decompositions in Eqs.~\eqref{eq:source_decomp_T1}-\eqref{eq:source_decomp_T2} read as
\begin{align}
T_1=&-\left[\cos\left((\omega+\Omega)r/v\right) \delta_l^\text{even}-i \sin\left((\omega+\Omega)r/v\right) \delta_l^\text{odd}\right] m_p \, Y_l^0(0,0) \delta_m^0\sqrt{1-v^2}/(\sqrt{2\pi}r v)\,,	\\
T_2=&-\left[\cos\left((\omega-\Omega)r/v\right) \delta_l^\text{even}-i \sin\left((\omega-\Omega)r/v\right) \delta_l^\text{odd}\right] m_p \, Y_l^0(0,0) \delta_m^0\sqrt{1-v^2}/(\sqrt{2\pi}r v)\,.
\end{align}
These satisfy the property 
\begin{equation}
T_2(\omega,l,0;r)= T_1(-\omega,l,0;r)^*\,.
\end{equation}
Thus, due to the form of the system~\eqref{Qball_Perturbation_Matrix_Sourced}, one has
\begin{align}
Z_2(\omega,l,0;r)&=Z_1(-\omega,l,0;r)^*\,,\\
Z_2^\infty(\omega,l,0)&=Z_1^\infty(-\omega,l,0)^*\,.\label{property_Z1Z2_plunge_qball}
\end{align}
Finally, the spectral fluxes~\eqref{Energy_flux}, \eqref{Momentum_flux} and \eqref{AngularMomentum_flux} become, respectively,
\begin{align}
\frac{d E^{\rm rad}}{d \omega}&=4 \left|\omega+\Omega\right|{\rm Re}\left[\sqrt{(\omega+\Omega)^2-\mu^2}\right] \sum_l \left|Z_1^\infty(\omega,l,0)\right|^2\,, \label{Energy_flux_qball}\\
\frac{d P_z^{\rm rad}}{d \omega}&=\sum_{l}\frac{8(l+1) \Theta\left[\left(\omega+\Omega\right)^2-\mu^2\right]\left|(\omega+\Omega)^2-\mu^2\right|}{\sqrt{(2l+1)(2l+3)}} \text{Re}\left[Z_1^\infty(\omega,l,0) Z_1^\infty(\omega,l+1,0)^*\right]\,,\label{Momentum_flux_qball}\\
\frac{d L_z^{\rm rad}}{d \omega}&=0\,.
\end{align}
%
\paragraph*{Orbiting particle}
The next setup is composed by a particle describing a circular orbit of radius $r_{\rm orb}$ and angular frequency $\omega_{\rm orb}$ inside a Q-ball and in its equatorial plane. The trace of the particle's stress-energy tensor is
\begin{equation}
T_p=-\frac{m_p}{r_{\rm orb}^2}\sqrt{1-\left(\omega_{\rm orb} r_{\rm orb}\right)^2}\delta(r-r_{\rm orb})\delta\left(\theta-\frac{\pi}{2}\right)\delta(\varphi-\omega_{\rm orb} t)\,, \label{T_p_orbiting}
\end{equation}
which implies
\begin{equation}
T_{1,2}=-m_p \sqrt{\pi/2}\, Y_l^m\left(\pi/2,0\right) \sqrt{1-\left(\omega_{\rm orb} r_{\rm orb}\right)^2}/r_{\rm orb}\ \delta\left(r-r_{\rm orb}\right)\delta\left(\omega \pm\Omega-m \omega_{\rm orb}\right)\,. \label{T12_orbiting}
\end{equation}
%
%
Notice that $T_2(\omega,l,m)=(-1)^m T_1(-\omega,l,-m)$, hence due to the form of system~\eqref{Qball_Perturbation_Matrix_Sourced}, we have
\begin{align}
Z_2(\omega,l,m;r)&=(-1)^m Z_1(-\omega,l,-m;r)^*\,,\\
Z_2^\infty(\omega,l,m)&=(-1)^m Z_1^\infty(-\omega,l,-m)^*\,.
\end{align}
Then, the emission rate expressions~\eqref{Energy_flux_rate} and~\eqref{AngularMomentum_flux_rate} imply, omitting the arguments $(\omega,l,m)$, 
\begin{align}
&\dot{E}^{\rm rad}= \frac{2}{\pi}\int d\omega \left|\omega+\Omega\right| {\rm Re}\left[\sqrt{(\omega+\Omega)^2-\mu^2}\right]\sum_{l,m} \left|Z_1^\infty\right|^2  \,, \nn\\
&\dot{L}_z^{\rm rad}=\frac{2}{\pi}\int d\omega \,\epsilon_1(\omega){\rm Re}\left[\sqrt{(\omega+\Omega)^2-\mu^2}\right]\sum_{l,m} m \left|Z_1^\infty\right|^2. \nn
\end{align}
where we remind that $\epsilon_1\equiv \text{sign}(\omega+\Omega+ \mu)$. Re-writing expression~\eqref{T12_orbiting} in the form
\begin{equation}
T_{1,2}=\widetilde{T}(\omega_{\rm orb},r_{\rm orb}) \,\delta\left(r-r_{\rm orb}\right) \delta\left(\omega\pm\Omega-m \omega_{\rm orb}\right)\,,
\end{equation}
the previous expressions for the rate of emission read
\begin{align}
\dot{E}^{\rm rad}&=\frac{2}{\pi}  \sum_{l,m}\widetilde{T}^2\Big[a_1\left|F_{3,3}^{-1}\left(m\omega_{\rm orb}-\Omega;\,r_{\rm orb}\right)\right|^2a_2\left|F_{3,4}^{-1}\left(m\omega_{\rm orb}+\Omega;\,r_{\rm orb}\right)\right|^2\Big]\,,\label{Energy_flux_orbiting}\\
\dot{L}_z^{\rm rad}&=\frac{2}{\pi}  \sum_{l,m}m \widetilde{T}^2\Big[\epsilon_1 a_1 \left|F_{3,3}^{-1}\left(m\omega_{\rm orb}-\Omega;\,r_{\rm orb}\right)\right|^2+\epsilon_1 a_2\left|F_{3,4}^{-1}\left(m\omega_{\rm orb}+\Omega;\,r_{\rm orb}\right)\right|^2\Big]\,.\label{AngularMomentum_flux_orbiting}
\end{align}
where 
\begin{align}
&a_1=|m \omega_{\rm orb}|{\rm Re}\left[\sqrt{\left(m \omega_{\rm orb}\right)^2- \mu^2}\right]\,,\nn \\
& a_2=\left|m \omega_{\rm orb}+2\Omega\right|{\rm Re}\left[\sqrt{\left(m \omega_{\rm orb}+2\Omega\right)^2- \mu^2}\right].
\end{align}
\section*{Free oscillations}
%
%
\begin{table*}
\centering
	\begin{tabular}{ccccc}
		\hline
		\hline
		$l$ &  \multicolumn{4}{c}{$\omega_{\rm QNM}/\mu$} \\ 
		\hline
		\hline
        0 &     $0.439$ & $0.689$ & $0.931 - 1.2 \times 10^{-4} i$& $1.153 - 1.6\times 10^{-2} i$ \\
		1 & $0.300$  &  $0.555$  &   $0.806 - 9.8 \times 10^{-4} i$ &  $1.04 - 3.3\times 10^{-3}i$\\
		\hline
		\hline
	\end{tabular} 
	\caption[Q-ball QNMs.]{Some QNM frequencies of a Q-ball configuration with $\Omega/\mu=0.3$, for $l=\{0,1,2\}$. Note that the first column corresponds to normal modes, with $\omega<\mu$, hence screened from distant observers: they are confined to a spatial extent $\sim R_{\rm Q}$, the radius of the Q-ball (these modes are the counterpart of the NBS modes in Table \ref{table:QNM_BS_invariant}). There is an infinity of QNM frequencies, parametrized by an integer overtone index $n$. At large $n$, ${\rm Re}\left(\omega_{\rm QNM}\right)\sim 0.22n\sim \pi n/R_{\rm Q}$, as might be anticipated by a WKB analysis. Our results for the imaginary part of $\omega_{\rm QNM}$ carry a large uncertainty, and should be taken as order of magnitude estimate only.
	}
	\label{table:QNM_Qball}
\end{table*}
The numerical search for QNM frequencies for Q-balls is summarized in Table~\ref{table:QNM_Qball}, for the particular configuration with $\Omega=0.3\mu$. Whenever $\omega_{\rm QNM}$ are pure real numbers, they refer to normal modes of the object. For a mode to be normal, it must not be dissipated to infinity, hence the condition $\omega<\mu-\Omega$ is necessary, which also implies that such modes are screened from far-away observers, by the Q-ball background itself. This means that perturbations associated with the real-valued frequencies in Table~\ref{table:QNM_Qball} do not reach spatial infinity. Such modes are the analogs of the NBS modes, which were {\it all} normal (cf. Table~\ref{table:QNM_BS_invariant}). Q-balls, in addition to such modes also have quasinormal modes, which decay in time since they are sufficiently large energy to propagate at large distances.

\section*{Particles plunging into Q-balls\label{sec:Plunging_particle}}

%
\begin{figure}
\centering
\includegraphics[width=0.5\textwidth,keepaspectratio]{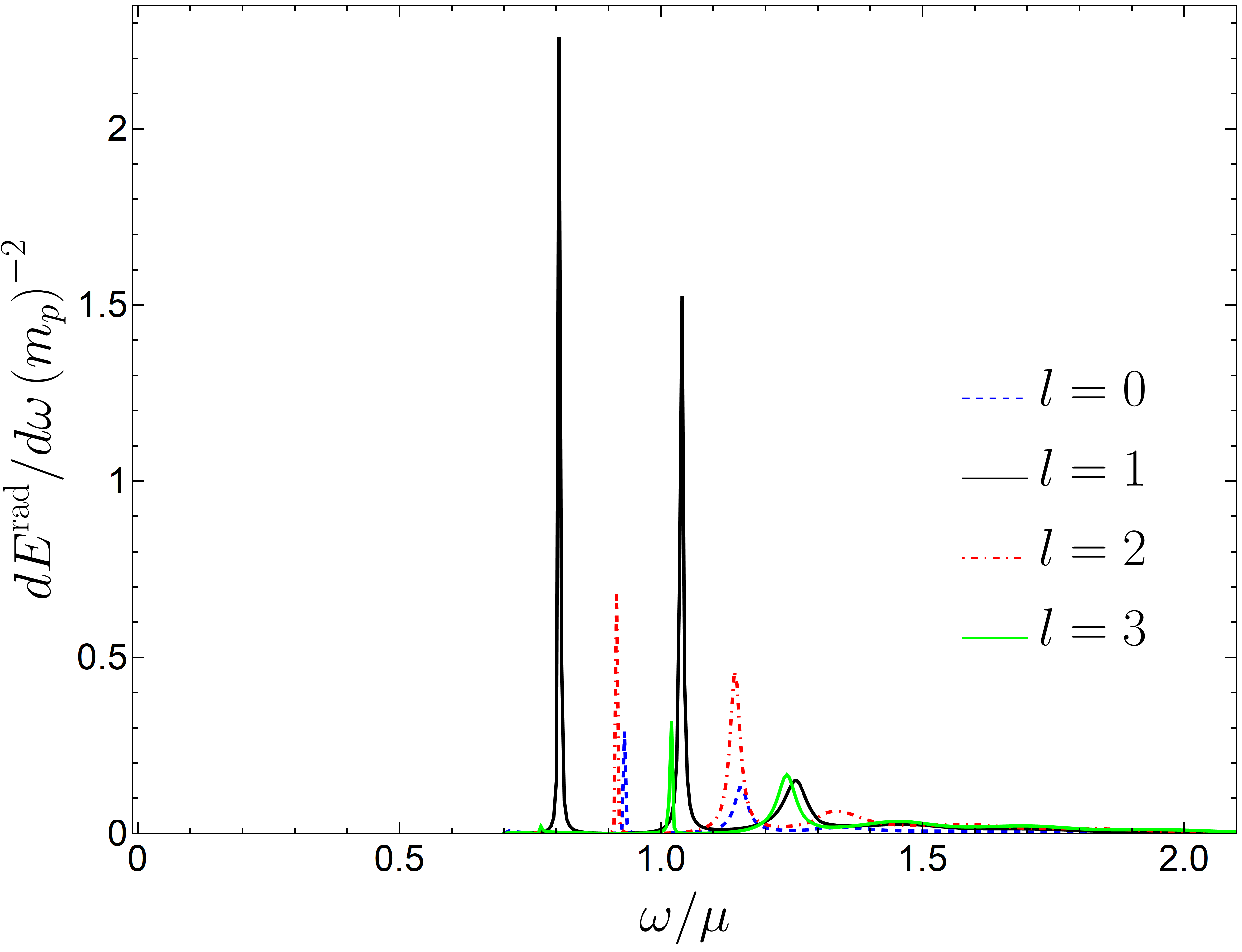} 
\caption[Spectra particle plunging in a Q-ball.]{Energy spectra of scalar radiation emitted when a particle of rest-mass $m_p$
plunges through a Q-ball with $\Omega=0.3\mu$ with a large velocity $v=0.8c$.
The spectrum was decomposed into multipoles (cf. Eq.~\eqref{Energy_flux_qball}). 
The sharp peaks correspond to the excitation of QNM frequencies $\omega_{\rm QNM}$ (see~Tab.~\ref{table:QNM_Qball}).
}
\label{fig:Plunging_Spectra_Qball}
\end{figure}
\begin{figure}
\centering
\includegraphics[width=0.5\textwidth,keepaspectratio]{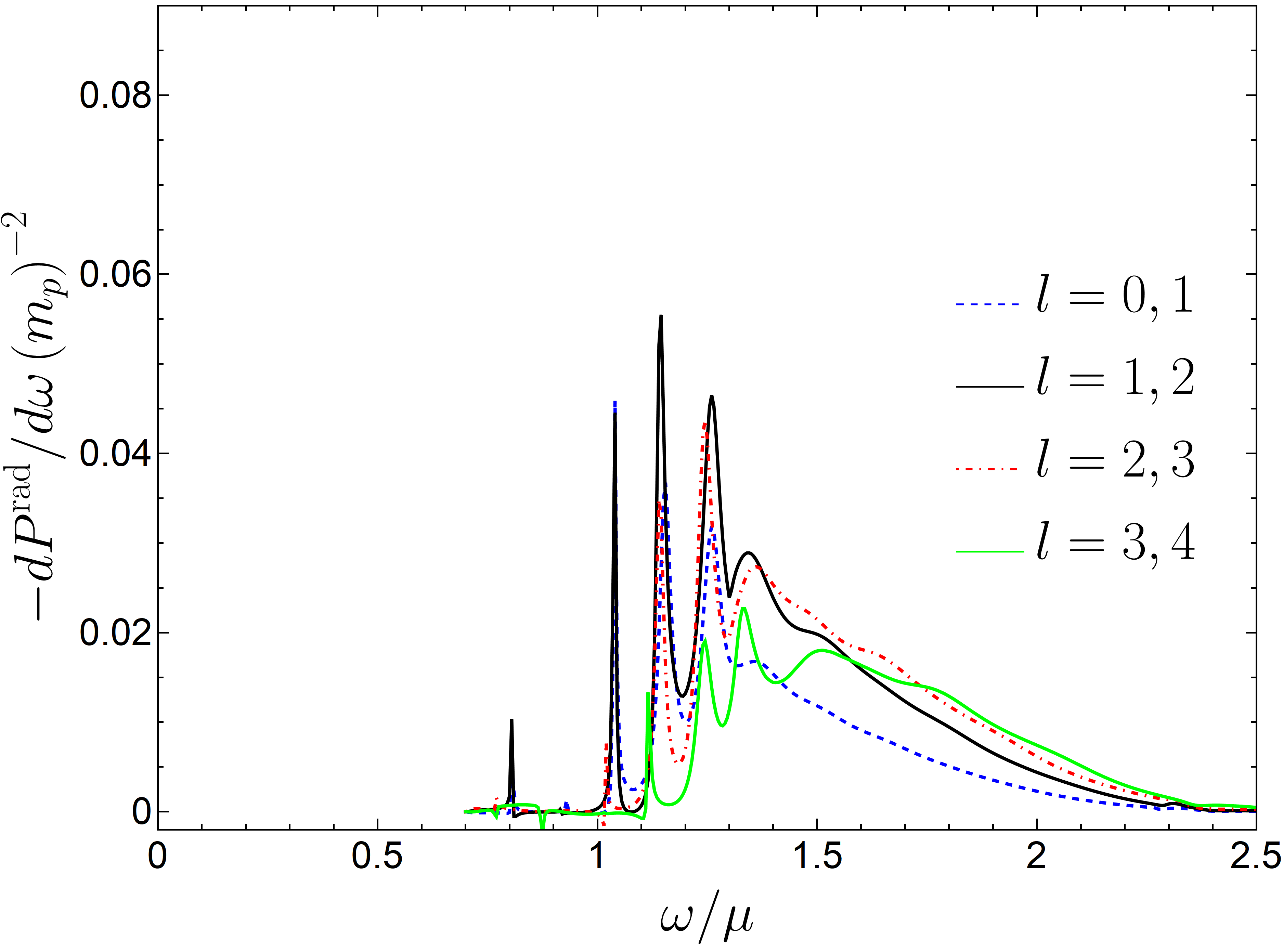}
\caption[Momenta particle plunging in a Q-ball.]{Linear momentum radiated when a particle plunges through a Q-ball (described by $\Omega=0.3\mu$) with a velocity $v=0.8$. 
Different lines correspond to the different multipolar cross terms in Eq.~\eqref{Momentum_flux_qball}.}
\label{fig:Plunging_momentum_Qball}
\end{figure}

For concreteness, here we restrict the discussion to a large-velocity plunge $v=0.8c$.
The multipolar energy spectrum $d E^{\rm rad}_l/d \omega$ of radiation released during such process is shown in Fig.~\ref{fig:Plunging_Spectra_Qball} for the first lowest multipoles, obtained through numerical evaluation of Eq.~\eqref{Energy_flux_qball}. Just like a hammer hitting a bell excites its characteristic vibration modes, the effect of a plunging particle is to excite the QNMs of a Q-ball.
Figure~\ref{fig:Plunging_Spectra_Qball} illustrates this feature very clearly, the peaks in the energy spectrum are all coincident with the QNMs, some of them identified in Table~\ref{table:QNM_Qball}. This feature was absent in the dynamics of NBS, simply because the modes of NBS (Table \ref{table:QNM_BS_invariant}) are all normal and confined to the NBS itself: they do not propagate to large distances. Most of the radiation is dipolar, also apparent in Fig.~\ref{fig:Plunging_Spectra_Qball}, but a substantial amount is emitted in other multipoles as well.
For example, the $l=4$ mode still carries roughly $10\%$ of the total radiated energy. Our results are compatible with an exponential suppression at large $l$, of the form
$E^{\rm rad}_l\sim 0.085 e^{-0.39 l}$. We can use this to sum over multipoles, and find the total energy radiated, 
\begin{equation}
E^{\rm rad}\sim 0.188 \, m_p^2 \,\mu\,.
\end{equation}

The emitted radiation carries momentum, which is caused by an interference term between multipoles (cf. Eq.~\eqref{Momentum_flux_qball}). For radiation entirely emitted in one single direction, the linear momentum $P^{\rm rad}=E^{\rm rad}/c$. However, this is in general only a (poor) upper bound on the radiated linear momentum, as a number of multipoles are involved in the process.
Figure~\ref{fig:Plunging_momentum_Qball} shows the contribution of the modes $l\leq 4$ to the spectral flux of linear momentum $d P_z^{\rm rad}/d \omega$, obtained through numerical evaluation of~\eqref{Momentum_flux_qball}. Again, most of the contribution comes from the excitation of the Q-ball's QNMs. Note the interesting aspect that in some frequency ranges and for some interference terms, the momentum is positive, i.e., along the direction of the motion. 
We observed numerically that the total flux of linear momentum $P_z^{\rm rad}$ converge exponentially in $l$, for sufficiently large $l$. The total radiated momentum is negative, and thus represents a slowing-down of the moving point particle. Using a similar fitting procedure to sum over multipoles, we find for this particular configuration,
\begin{equation}
P^{\rm rad}\sim-0.088 \, m_p^2\, \mu\,.
\end{equation}
%

\section*{Orbiting particles\label{sec:Orbiting_particle}}
%
\begin{figure*}
\centering
\begin{tabular}{c}
\includegraphics[width=0.4\textwidth,keepaspectratio]{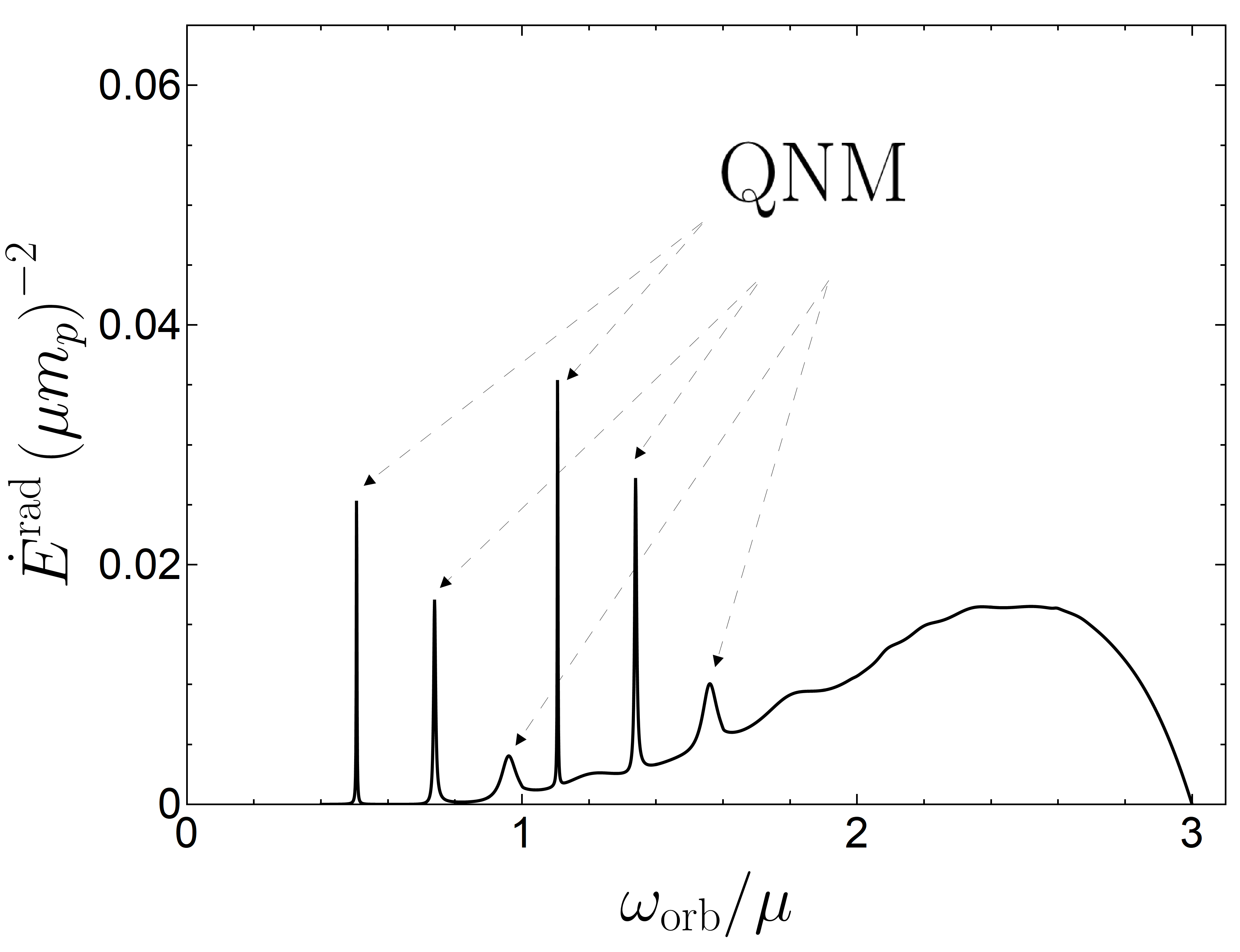}\,\,\, \includegraphics[width=0.4\textwidth,keepaspectratio]{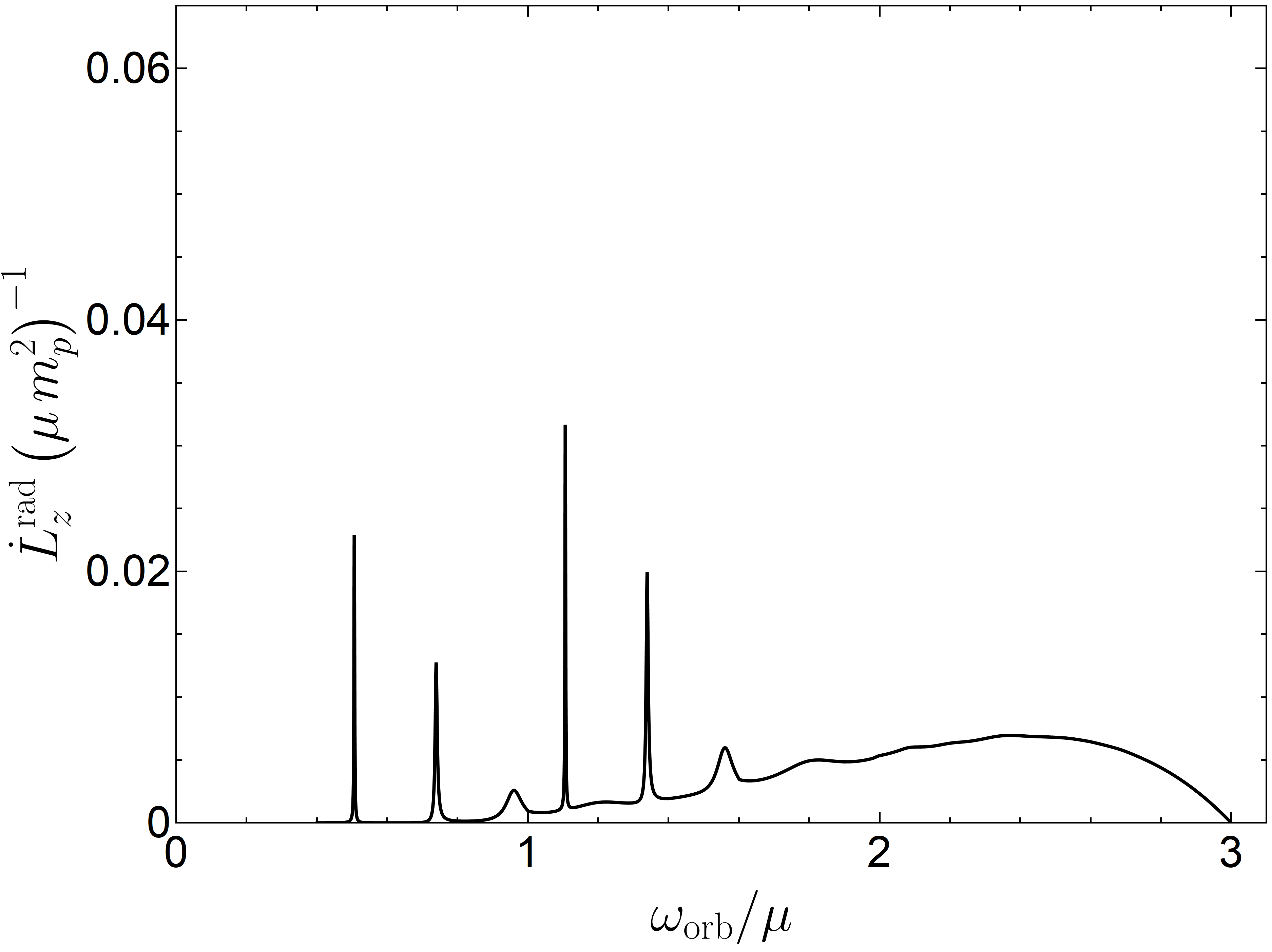}
\end{tabular}
\caption[Energy particle in circular orbit inside a Q-ball.]{Average dipolar ($l=1$, including $m=\pm1$) rate of energy ({\bf left}), and angular momentum ({\bf right}) radiated by a particle describing a circular orbit around a Q-ball with $\Omega=0.3\mu$, at radius $r_{\rm orb}\mu=1/3$ and with orbital frequency $\omega_{\rm orb}$. The peaks are associated with the excitation of QNM frequencies $\omega_{\rm QNM}$ for $\omega_{\rm orb}=\text{Re}\left(\omega_{\rm QNM}\right)\pm \Omega$ -- each QNM frequency is excited by two different $\omega_{\rm orb}$ spaced by $2\Omega$. The excitation of the QNM frequencies with $\text{Re}(\omega_{\rm QNM})=\{0.806, \,1.04$ (in Tab.~\ref{table:QNM_Qball})$,\,1.298\}\mu$ is clearly seen from these plots. 
}
	\label{fig:OrbitingFluxesQball}
\end{figure*}

\begin{figure}
\centering
\includegraphics[width=0.5\textwidth,keepaspectratio]{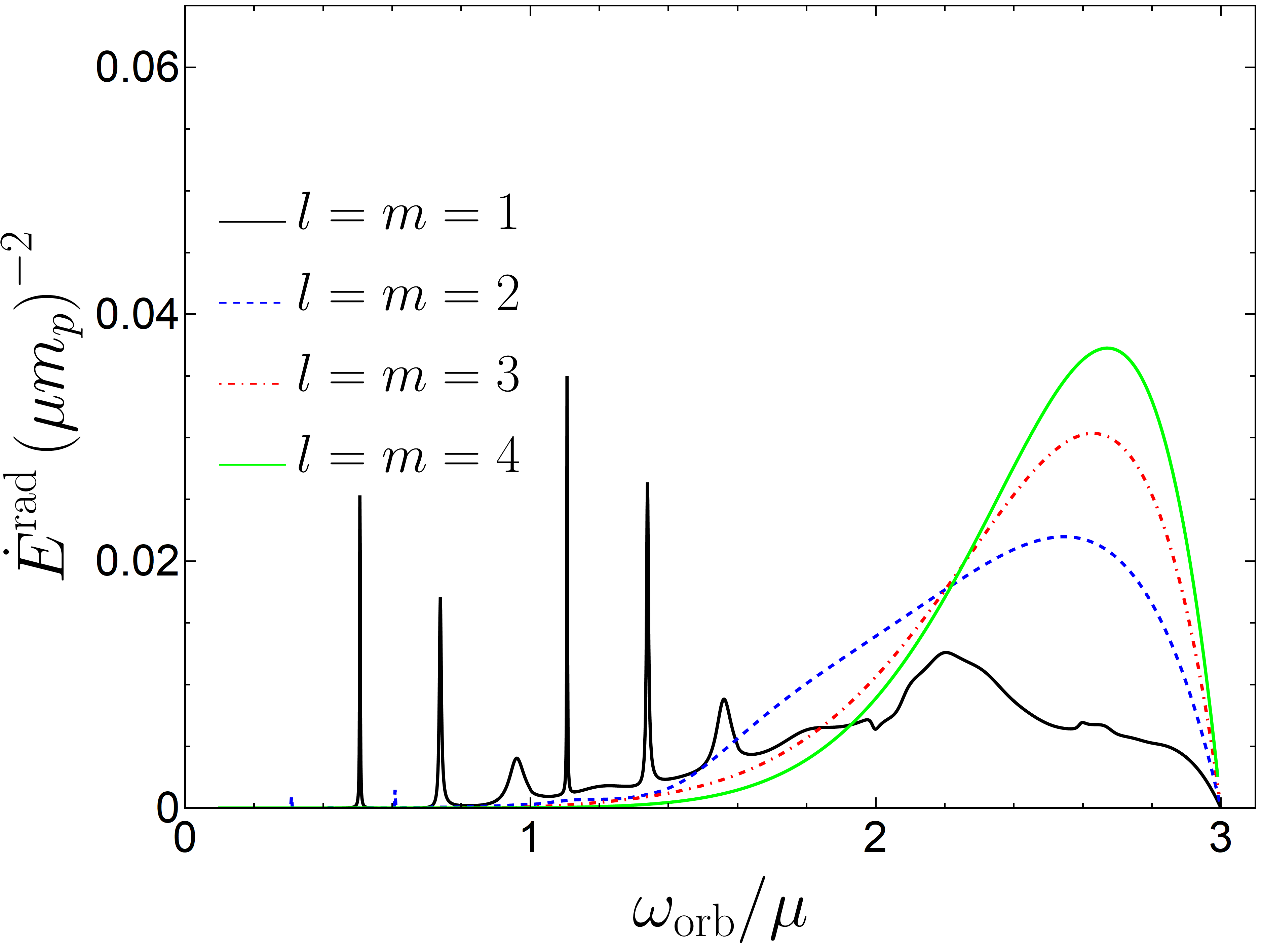} \caption[Energy particle in circular orbit inside a Q-ball, different $l$s.]{Average rate of energy radiated by a particle describing a circular orbit around a Q-ball with $\Omega=0.3\mu$, at a radius $r_{\rm orb}\mu=1/3$ and with orbital frequency $\omega_{\rm orb}$ for different values of $l=m$. At low frequencies the radiation is mostly dipolar. At large orbital frequencies the radiation is synchrotron-like and peaked at large $l=m$. In the high-frequency regime, there is a critical multipole $m$ beyond which the energy radiated decreases exponentially
(see main text for further details). 
There are QNM peaks for all multipoles, but they are visible only for the dipolar and quadrupolar.} 	
\label{fig:OrbitingFluxesQballMorels}
\end{figure}
The average dipolar flux of energy and angular momentum, emitted by a particle in circular orbit inside a Q-ball ($\Omega=0.3\mu$), at an orbital distance $r_{\rm orb}\mu=1/3$, 
are shown in Figs.~\ref{fig:OrbitingFluxesQball}. The pointlike source is assumed to be orbiting due to some external force, and its orbital frequency is varied, scanning possible resonant behavior of the Q-ball. As expected, and verified numerically, the quantity $\dot{E}^{\rm rad}$ is an even function of $\omega_{\rm orb}$, whereas $\dot{L}_z^{\rm rad}$ is an odd one. A few features are apparent in the results above (obtained evaluating Eqs.~\eqref{Energy_flux_orbiting}-\eqref{AngularMomentum_flux_orbiting}).
The fluxes have clear peaks, which correspond to the resonant excitation of the QNMs of the Q-ball. It's worth to note that for each QNM frequency listed in Tab.~\ref{table:QNM_Qball} there are two peaks associated with different orbital frequencies separated by a distance $2\Omega$: the resonances now occur at $\omega_{\rm orb}=\Omega\pm\omega_{\rm QNM}$. This is directly due to the decomposition in Eq.~\eqref{MatterDecomposition}.

In flat space, a scalar charge on a circular orbit also emits radiation~\cite{Cardoso:2007uy,Cardoso:2011xi}. For small orbital frequencies and massless fields, the flux is dipolar and of order $\dot{E}\sim q^2r_{\rm orb}^2\omega_{\rm orb}^4/(12\pi)$~\cite{Cardoso:2007uy,Cardoso:2011xi} (given the interaction \eqref{coupling_Qball}, the scalar charge $q=m_p$).
This explains the rise of the dipolar flux when the orbital frequency increases. However, at large frequencies, the radiation becomes of synchrotron type, and the radiation is emitted preferentially in higher multipoles~\cite{Misner:1972jf,Breuer}. This is apparent in Fig.~\ref{fig:OrbitingFluxesQballMorels} where we show the contribution of higher multipoles to the flux. 
Note that all other multipoles also have resonant peaks, but these are less pronounced than the dipolar. At large Lorentz factors $\gamma$, there is a critical $m$ mode after which the fluxes 
becomes exponentially suppressed. The critical multipole is of order $m_{\rm crit}\propto \gamma^2$~\cite{Misner:1972jf,Breuer}. Thus an evaluation of a large number of multipoles is necessary to have an accurate estimate of fluxes at large velocities. Our results are consistent with such a prediction. We find that as $\omega_{\rm orb}$ increases, the flux peaks at higher and higher $m$, but there's always a threshold $m$ beyond which the radiation output is exponentially suppressed. Finally, since this process is not axially symmetric, one cannot use expression~\eqref{Momentum_flux} to compute the flux of linear momentum along $z$. Nevertheless, it is straightforward to show that the average rate of linear momentum radiated $\dot{P}_z^{\rm rad}$ vanishes.

\begin{figure}[ht]
\centering
\includegraphics[width=0.5\textwidth,keepaspectratio]{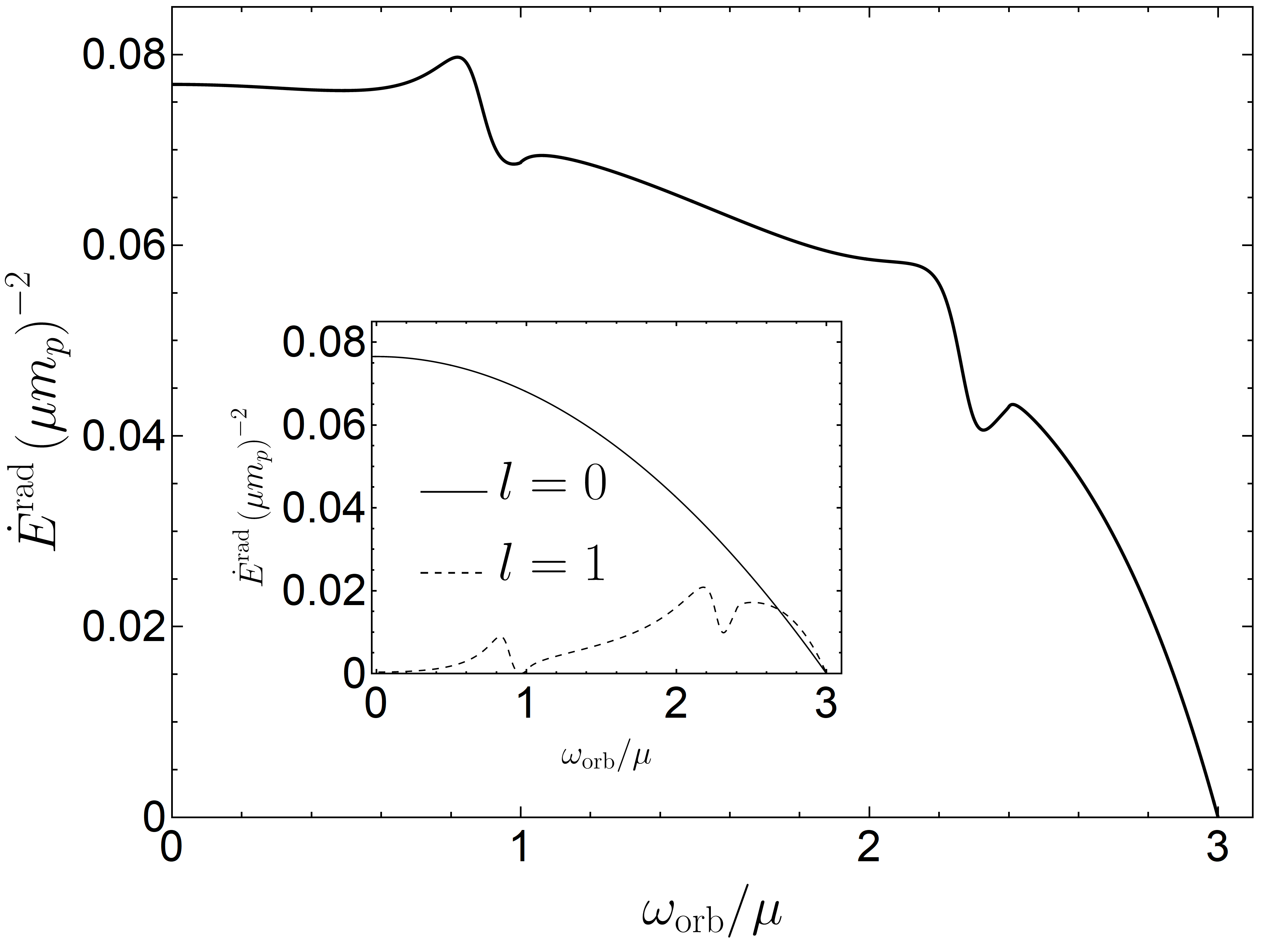} \caption[Energy particle in circular orbit inside a Q-ball, $\Omega=0.7$]{Average rate of energy radiated by a particle describing a circular orbit around a Q-ball with $\Omega=0.7\mu$, at radius $r_{\rm orb}\mu=1/3$ and with orbital frequency $\omega_{\rm orb}$. For such a scalar configuration there is radiation emitted also in the monopole mode, and it dominates the emission, as seen in the inset.} 	\label{fig:OrbitingFluxesQballOmega0p7}
\end{figure}
\begin{figure}
\centering
\includegraphics[width=0.5\textwidth,keepaspectratio]{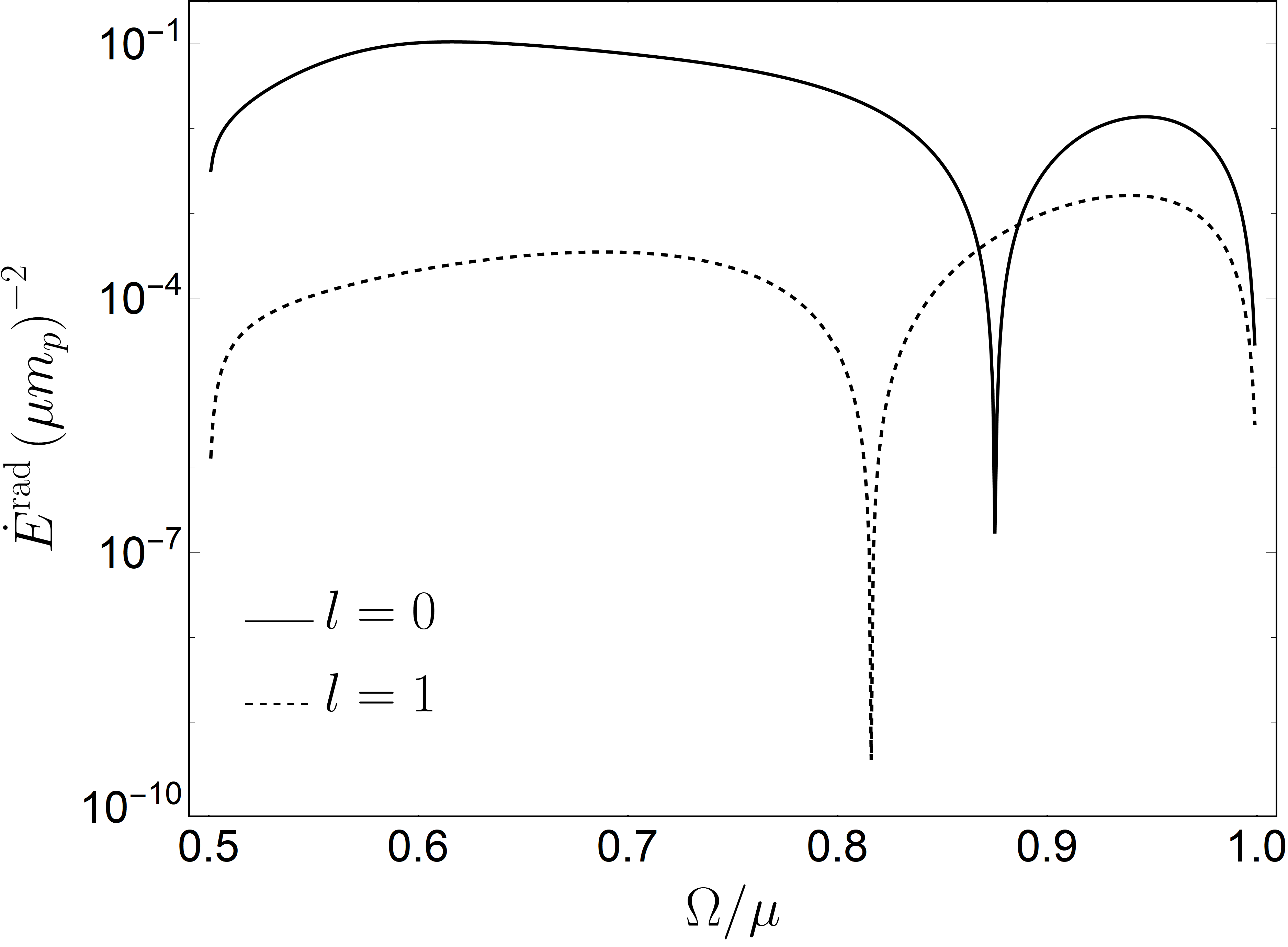} \caption[Energy particle standing inside a Q-ball.]{Average rate of energy radiated in the case of a particle standing at a fixed radius $r_{\rm orb}\mu=1/3$ as function of $\Omega/\mu$. It is shown the dominant contributions from the modes $l=0$ and $l=1$. The average rate of angular momentum radiated in this case vanishes.} 	\label{fig:NotOrbitingFluxesQball}
\end{figure}
One interesting aspect, not seen in the study of NBSs, concerns {\it monopolar} emission and emission from particles {\it at rest}. Both features are usually absent.
It follows from Eq.~\eqref{Energy_flux_orbiting}, that for Q-ball configurations with $\Omega\leq \mu/2$ there is no emission of $l=0$, and the first mode contributing to the radiation is $l=1$. For these objects there is no radiation emitted if the particle is at rest, with $\omega_{\rm orb}=0$. However, for Q-balls
with $\Omega/\mu>1/2$ there is indeed emission of $l=0$ modes, contributing more than (or, at least as much as) the $l=1$ modes to the radiation (see Fig.~\ref{fig:OrbitingFluxesQballOmega0p7}). Interestingly, for these Q-balls there is also radiation emitted even when the particle is at rest (see Fig.~\ref{fig:NotOrbitingFluxesQball}). This type of behavior is due to the coupling~\eqref{coupling_Qball} between two dynamical entities: the external perturber (through $T_p$) and the Q-ball configuration (through $\Phi$). The different coupling considered for NBSs, led to the absence of these features. 

  \cleardoublepage

\chapter{Post-Newtonian expansion of the Einstein Klein Gordon system} \label{app:Newtonian}
In this Appendix we show how the Einstein-Klein-Gordon system reduces to the Schr\"{o}dinger-Poisson system in the Newtonian limit. As follows, we obtain the equations describing a perturbation to the Newtonian fields up to first post-Newtonian corrections. Finally, we consider perturbations caused by a point-like particle. In this section we follow closely the treatment in Chapter 8.2 of Ref.~\cite{Poisson_will_2014}.  

The Einstein-Klein-Gordon system is the set of field equations for $\Phi$ and $g_{\mu \nu}$, which is obtained through the variation of the action~\eqref{action} with respect to $\Phi^*$ and $g_{\mu \nu}$, and reads
\begin{align} 
\frac{1}{\sqrt{-g}}\partial_{\mu}\left(\sqrt{-g}g^{\mu\nu}\partial_{\nu}\Phi\right)&= \mu^2\Phi\,, \nn \\
R_{\mu\nu}&= 8\pi \widetilde{T}_{\mu\nu}^S\,,\label{EOM_BosonSapp}
\end{align}
where the Einstein equations are written in an alternative form using the trace-reversed stress-energy tensor of the scalar field
\begin{equation}
\widetilde{T}^S_{\mu \nu }\equiv T^S_{\mu \nu}-\frac{1}{2}T^S g_{\mu \nu}=\partial_{(\mu}\Phi^* \partial_{\nu)}\Phi+\frac{1}{2}g_{\mu \nu}\mu^2|\Phi|^2  \,.\nonumber
\end{equation}
In the last equations we used $\mathcal{U}\sim \mu^2|\Phi|^2/2$, since we want to consider a (Newtonian) weak scalar field $|\Phi|\ll1$.
More precisely, in our perturbation scheme we consider that $\Phi \sim \mathcal{O}(\epsilon)$, with $\epsilon\ll1$. Moreover, in the Newtonian limit, we consider the spacetime metric ansatz
\begin{align}
	g_{tt}&=-1-2U +\mathcal{O}(\epsilon^4)\,,\nonumber \\
	g_{tj}&=\mathcal{O}(\epsilon^3),\,	g_{jk}=\mathcal{O}(\epsilon^2)\,,
\end{align}
with $j,k=\{x,y,z\}$ and where $U(t,x,y,z)\sim \mathcal{O}(\epsilon^2)$. This gives the Ricci tensor components
\begin{align}
	R_{tt}&= \nabla^2U+ \mathcal{O}(\epsilon^4)\,, \nonumber \\
	R_{tj}&=\mathcal{O}(\epsilon^3),\, R_{jk}= \mathcal{O}(\epsilon^2)\,,
\end{align}
where we are considering that
\begin{equation}
\partial_{t}U\sim \mathcal{O}(\epsilon^3),\, \partial_{t}^2U\sim \mathcal{O}(\epsilon^4)\,.
\end{equation}
%
%
The non-relativistic limit of the scalar field $\Phi$ is incorporated in our perturbation scheme by considering that~\footnote{This can be shown rigorously by doing an expansion in powers of ($1/c$). It corresponds to the assertion that, in the non-relativistic limit, the energy-momentum relation is $E\sim \mu+\frac{1}{2 \mu}p^2+\mu U$, with $p^2 \ll \mu^2$ and $|U|\ll 1$.}
\begin{equation}
	\partial_j \Phi \sim \mathcal{O}(\epsilon^2),\, \partial_t \widetilde{\Phi} \sim \mathcal{O}(\epsilon^3)\,,
\end{equation}
where we introduced an auxiliary scalar field $\widetilde{\Phi}$ such that
\begin{align}
\Phi= \frac{1}{\sqrt{\mu}} e^{-i\mu t}\widetilde{\Phi}\,.
\end{align}
Then, the components of the trace-reversed stress-energy tensor of the scalar field are
\begin{align}
	\widetilde{T}_{tt}^S&=\frac{1}{2} \mu |\widetilde{\Phi}|^2+\mathcal{O}(\epsilon^4)\,,\nonumber \\
	\widetilde{T}_{tj}^S&=\mathcal{O}(\epsilon^3),\,\widetilde{T}_{jk}^S= \mathcal{O}(\epsilon^2)\,.
\end{align}
Therefore, at Newtonian order, the Einstein equations reduce to the Poisson equation
\begin{equation}
	\nabla^2U=4\pi \mu |\widetilde{\Phi}|^2\,.
\end{equation}
On the other hand, it is possible to show that, at leading order $\mathcal{O}(\epsilon^3)$, the Klein-Gordon equation reduces to the Schr\"{o}dinger equation
\begin{equation}
	i \partial_t \widetilde{\Phi}=-\frac{1}{2 \mu} \nabla^2\widetilde{\Phi}+\mu U \widetilde{\Phi}\,.
\end{equation}
So, we have showed that, in the Newtonian limit, the Einstein-Klein-Gordon system for $\Phi$ and $g_{\mu \nu}$ reduces to the Schr\"{o}dinger-Poisson system for $\widetilde{\Phi}$ and $U$.

Let us now extend our perturbation scheme to first post-Newtonian order. We start by considering the spacetime metric ansatz
\begin{align}
g_{tt}&=-1-2U-2\delta U -2\left(\psi+ U^2\right)+\mathcal{O}(\epsilon^6)\,,\nonumber \\
g_{tj}&=-4 U_j+\mathcal{O}(\epsilon^5)\,,\nonumber \\
g_{jk}&=\left(1-2U\right) \delta_{jk}+\mathcal{O}(\epsilon^4)\,,
\end{align}
with the post-Newtonian terms $U_j(t,x,y,z)\sim \mathcal{O}(\epsilon^3)$, $\psi(t,x,y,z)\sim \mathcal{O}(\epsilon^4)$ and the perturbation $\delta U(t,x,y,z)\sim \mathcal{O}(\xi)$, where $\mathcal{O}(\epsilon^6)<\mathcal{O}(\xi)<\mathcal{O}(\epsilon^2)$. This results in the Ricci tensor components
\begin{align}
R_{tt}=& \nabla^2U+\nabla^2 \delta U+ 3 \partial_t^2 U+4 U \nabla^2 U+ \nabla^2 \psi+ \mathcal{O}(\epsilon^6)\,, \nonumber \\
R_{tj}=&2 \nabla^2U_j+ \mathcal{O}(\epsilon^5)\,, \nonumber \\
R_{jk}=&\nabla^2U\delta_{jk}+\mathcal{O}(\epsilon^4)\,,
\end{align}
where we imposed the harmonic coordinate condition, which results in
\begin{equation}
	\partial_tU+\partial_j U^j=0\,.
\end{equation}
Now, we introduce a perturbation $\delta \Phi$ to the Newtonian scalar field, such that
\begin{equation}
	\delta \Phi=\frac{1}{\sqrt{\mu}}e^{-i \mu t}\delta \widetilde{\Phi}\,,
\end{equation}
treated in our perturbation scheme with
\begin{equation}
\delta \Phi\sim \mathcal{O}(\xi/\epsilon),\,	\partial_j\delta \Phi \sim \mathcal{O}(\xi),\,\partial_t \delta \widetilde{\Phi} \sim \mathcal{O}(\xi \,\epsilon)\,.
\end{equation}
Then, the components of the trace-reversed stress-energy tensor of the scalar field are
\begin{align}
	\widetilde{T}_{tt}^S&=\frac{1}{2} \mu |\widetilde{\Phi}|^2+{\rm Im}\left(\widetilde{\Phi}\,\partial_t \widetilde{\Phi}^* \right)-\mu U|\widetilde{\Phi}|^2+\mu {\rm Re}\left(\widetilde{\Phi}^* \delta \widetilde{\Phi}\right)+\mathcal{O}(\epsilon^6)\,,\nonumber \\
	\widetilde{T}_{tj}^S&={\rm Im}\left(\widetilde{\Phi}\,\partial_j \widetilde{\Phi}^*\right)+\mathcal{O}(\epsilon^5)\,, \nonumber \\
	\widetilde{T}_{jk}^S&=\frac{1}{2} \mu |\widetilde{\Phi}|^2+\mathcal{O}(\epsilon^4)\,.
\end{align}
Thus, it is possible to show that, at first post-Newtonian order, the Einstein equations reduce to
\begin{align}
\nabla^2 \psi&=8\pi \Big[{\rm Im}\left(\widetilde{\Phi}\, \partial_t \widetilde{\Phi}^*\right)-3\mu U |\widetilde{\Phi}|^2\Big]\,,\nonumber \\
\nabla^2U_j&=4 \pi \,{\rm Im}\left(\widetilde{\Phi}\,\partial_j \widetilde{\Phi}^*\right)\,, \nonumber\\
\nabla^2 \delta U&=8 \pi \mu\, {\rm Re} \left(\widetilde{\Phi}^* \delta \widetilde{\Phi}\right)\,,\label{PN_Einstein}
\end{align}	
where we used the equations that are satisfied at Newtonian order and we assumed $\partial^2_t U=0$, since this happens to be always the case in this work.
On the other hand, until order $\mathcal{O}(\epsilon^5)$, the Klein-Gordon equation reduces to
\begin{equation}
i \partial_t \delta \widetilde{\Phi}=-\frac{1}{2 \mu} \nabla^2 \delta \widetilde{\Phi}+\mu U \delta \widetilde{\Phi}+\mu \widetilde{\Phi} \,\delta U +\frac{1}{2 \mu} \partial_t^2 \widetilde{\Phi}+ i U \partial_t \widetilde{\Phi}+\mu \psi \widetilde{\Phi} -\frac{1}{\mu} U \nabla^2 \widetilde{\Phi}- 4i\, U^j \partial_j \widetilde{\Phi}\,.\label{KG_all}
\end{equation}	
Finally, note that, in the case $\mathcal{O}(\epsilon^4)<\mathcal{O}(\xi) <\mathcal{O}(\epsilon^2)$, the last equation becomes simply 
\begin{equation}
	i \partial_t \delta \widetilde{\Phi}=-\frac{1}{2 \mu} \nabla^2 \delta \widetilde{\Phi}+\mu U \delta \widetilde{\Phi}+\mu \widetilde{\Phi} \,\delta U\,.
\end{equation}

In the case of a perturbation caused by a point-like particle, one just needs to include the trace-reversed stress energy tensor of the point-like particle, Eq.~\eqref{Stress_energy_particle}, in the Einstein equation~\eqref{EOM_BosonSapp}. This is given by
\begin{equation}
\widetilde{T}^p_{\mu \nu}\equiv T^p_{\mu \nu}-\frac{1}{2}T^p g_{\mu \nu}=\frac{m_p}{2 u^0} \left(2 u_\mu u_\nu+g_{\mu \nu}\right) \frac{\delta (r-r_p)}{r^2} \frac{\delta(\theta-\theta_p)}{\sin \theta}\delta(\varphi-\varphi_p)\,,\nonumber
\end{equation}
with the particle's 4-velocity $u^\mu\equiv d x^\mu/d \tau$.
We consider that $m_p \sim \mathcal{O}(\xi)$ and that the particle is non-relativistic, so that $u^i\sim \mathcal{O}(\epsilon)$ in our perturbation scheme.
Then, the components of the trace-reversed stress-energy tensor of the particle are
\begin{align}
	\widetilde{T}_{tt}^p&=\frac{m_p}{2} \frac{\delta (r-r_p)}{r^2} \frac{\delta(\theta-\theta_p)}{\sin \theta}\delta(\varphi-\varphi_p)+\mathcal{O}(\epsilon^4)\,,\nonumber \\
    \widetilde{T}_{tj}^p&=\mathcal{O}(\epsilon^3),\,\widetilde{T}_{jk}^p=\mathcal{O}(\epsilon^4)\,.
\end{align}
Thus, we conclude that we just need to add an extra term to the last equation in~\eqref{PN_Einstein}, which becomes
\begin{equation}
	\nabla^2 \delta U=4\pi \left[2\mu\, {\rm Re}\left(\widetilde{\Phi}^* \delta \widetilde{\Phi}\right)+P\right]\,,
\end{equation}
with
\begin{equation}
	P(t,r,\theta,\varphi)\equiv m_p \frac{\delta (r-r_p)}{r^2} \frac{\delta(\theta-\theta_p)}{\sin \theta}\delta(\varphi-\varphi_p)\,.\nonumber
\end{equation}

Let us now consider the case of a non-relativistic point-like particle sourcing ultra-relativistic scalar perturbations to the Newtonian background. In our perturbation scheme, we consider
\begin{equation}
\delta \Phi\sim \mathcal{O}(\xi \epsilon^3),\,\partial_j\delta \Phi \sim \mathcal{O}(\xi \epsilon^2),\, \partial_t \delta \Phi \sim \mathcal{O}(\xi \epsilon^2)\,,\nonumber
\end{equation}
where, in the ultra-relativistic limit, the energy-momentum relation becomes instead $E\sim p$, with $E\gg \mu $.
with $\mathcal{O}(\epsilon^4)<\mathcal{O}(\xi)<\mathcal{O}(\epsilon^2)$. 
So, at Newtonian order, the perturbation in the scalar field does not enter in the Einstein equations, since we have
\begin{equation}
\widetilde{T}_{tt}^S=\mathcal{O}\left(\epsilon^4\right),\,\widetilde{T}_{tj}^S=\mathcal{O}(\epsilon^3),\, \widetilde{T}_{jk}^S=\mathcal{O}(\epsilon^2)\,.\nonumber
\end{equation}
In the case of a non-relativistic point-like particle, at Newtonian order, the Einstein equations describing the perturbation reduce to the Poisson equation
\begin{equation}
	\nabla^2\delta U= 4\pi P\,. \label{eq_UR_1b}
\end{equation} 
Note that the assumption of a non-relativistic perturber sourcing an ultra-relativistic scalar perturbation is consistent as long as the scalar is sufficiently light. Finally, at leading order, the Klein-Gordon reduces to
\begin{align}
-\partial^2_t \delta \Phi +\nabla^2 \delta \Phi =2 \mu^2 \Phi\, \delta U\,.\label{eq_UR_2b}
\end{align}
%

\chapter{Scalar fluxes, newtonian boson stars and black holes} \label{app:incoming_flux}

In this Appendix we illustrate a number of toy-model that capture the important features associated with the fluxes of scalar fields inside DM structure.

\section*{Flux of energy inside a newtonian boson star} 

We start computing the incoming flux of energy over a tiny spherical surface at the center of a fundamental NBS. Consider a stationary NBS of the form
\begin{equation}
\Phi= \Psi(r)e^{-i\left(\mu- \gamma\right) t}\,,
\end{equation}
where $\Psi$ is a solution of system~\eqref{EOM_BS_radial}.
This stationary field can be written as a sum of incoming and outgoing parts $\Phi=\Phi_{\rm in}+\Phi_{\rm out}$ where
\begin{align}
	\Phi_{\rm in}&\equiv e^{-i\left(\mu- \gamma\right) t} \int_{-\infty}^{0}ds\,\overline{\Psi}(s) e^{i s r}\,, \nonumber \\
	\Phi_{\rm out}&\equiv e^{-i\left(\mu- \gamma\right) t} \int_{0}^{+\infty}ds\,\overline{\Psi}(s) e^{i s r}\,,	
\end{align}
with
\begin{equation}
	\overline{\Psi}(s)=\frac{1}{2 \pi} \int_{-\infty}^{+\infty}dr\, \Psi(r)e^{-i s r}\,,
\end{equation}
and where we are using an even extension of $\Psi$ to negative values of $r$. Note that $\overline{\Psi}$ is a real-valued function, since $\Psi$ is real-valued. 
Now, the incoming flux of energy over a tiny spherical surface of radius $r_+\ll R$ is given by
\begin{equation}\label{Ein}
	\dot{E}_{\rm in}\simeq 4 \pi r_+^2 T_{tr}^{\rm in}(r=0)\,.
\end{equation}
At leading order, one has
\begin{align}
T_{tr}^{\rm in}(r=0) &\simeq \mu\, {\rm Im}\left(\Phi_{\rm in}\partial_r \Phi_{\rm in}^*\right) \nonumber \\
&=-\frac{\mu}{2} \int_{-\infty}^{0} ds' \int_{-\infty}^{0} ds \left(s'+s\right)\overline{\Psi}(s') \overline{\Psi}(s)\,.\nonumber
\end{align}
Numerical evaluation of the last expression for a fundamental NBS gives
\begin{equation}
	T_{tr}^{\rm in}(r=0) \sim 2.69\times 10^{-4}\,\mu^7 M_{\rm NBS}^5\,.
\end{equation}
Finally, the incoming flux of energy is
\begin{equation}
	\dot{E}_{\rm in}\sim 3.38\times 10^{-3}\,r_+^2 \mu^7 M_{\rm NBS}^5\,.
\end{equation}

\section*{Introducing a dissipative boundary} 
This Section looks at two toy models, aimed at understanding the evolution
of an NBS with a small BH at its center. The main effect that the BH produces is, naturally,
dissipation at the horizon. 
\subsection*{A string absorptive at one end} \label{app:string_toy}
Here, we wish to study a one-dimensional model of absorption of a scalar structure
when the boundary conditions suddenly change. Consider then a string, initially fixed at $x=0,\,L$, described by the wave equation
\begin{equation}
\partial^2_x\Phi-\partial^2_t\Phi=0\,.
\end{equation}
A normal mode satisfying $\Phi(x=0)=\Phi(x=L)=0$ is
\begin{align}
\Phi&=e^{-i\omega_n t}\sin\omega_n x\,,\\
\omega_n&=\frac{(n+1)\pi}{L}\,,n=0,1,2...\,.
\end{align}
We take a configuration with $\omega_n=\omega_0$ and use this as initial data for a problem where the boundary condition
at the origin becomes absorptive. In particular, Laplace-transform the wave equation to find,
\begin{align}
\frac{d^2\Psi}{dx^2}+\omega^2\Psi&=-\dot{\Phi}(0,x)+i\omega\Phi(0,x)\,,\label{eq_inh}\\
\Psi(\omega,x)&=\int dt e^{i\omega t}\Phi(t,x)\,.
\end{align}
As boundary conditions, require that 
\begin{equation}
\Psi(\omega,L)=0,\, \Psi(x\sim 0)=\sin\omega x-\epsilon e^{-i\omega x}\,. 
\end{equation}
These conditions maintain the mirror-like boundary at one extreme $x=L$, while providing an absorption
of energy at $x=0$. The flux of absorbed energy scales like $\epsilon^2\ll1$.
The solution of Eq.~\eqref{eq_inh} subjected to the above boundary conditions is
\begin{equation}
\Psi=i\frac{\cos^2\omega x\sin\pi x/L+\sin^2\omega x\sin \pi x/L}{\omega-\pi/L}+\epsilon\frac{\pi\sin\omega(L-x)}{\omega(\pi-L\omega)(i\epsilon\cos\omega L+(\epsilon-i)\sin\omega L)}\,.\label{sol_nonh_string}
\end{equation}

The original time-domain field is given by the inverse
\begin{equation}
\Phi(t,x)=\frac{1}{2\pi}\int d\omega e^{-i\omega t}\Psi(\omega, x)\,.
\end{equation}
The integral can be done with the help of the residue theorem. We separate the response $\Phi=\Phi_1+\Phi_2$. The first term in Eq.~\eqref{sol_nonh_string} has a simple, {\it real} pole
at $\omega=\omega_0=\pi/L$, and it evaluates to
\begin{equation}
\Phi_1(t,x)=\sin(\pi x/L)e^{-i\pi t/L}\,,
\end{equation}
i.e., it corresponds to the initial data.

The second term has poles at complex values of the frequency, which are also the QNMs of the dissipative system,
\begin{equation}
\omega\approx \frac{n\pi+\epsilon -i\epsilon^2}{L}\,,
\end{equation}
These poles lie close to the normal modes of the system, including those not present in the initial data.
They dictate an exponential decay $\sim e^{-\epsilon^2 t}$, and a consequent lifetime $\tau \sim \epsilon^{-2}$.
Note that this simple exercise shows that all modes are excited when new boundary conditions are turned on.
For NBSs, all the modes cluster around $\omega \sim \mu$, thus we expect to always be in the low-frequency regime
used to estimate the lifetime.

\subsection*{A black hole in a scalar-filled sphere} \label{app:bh_bomb}
%
\begin{figure*}
\centering
\begin{tabular}{ccc}
\includegraphics[width=0.3\textwidth,keepaspectratio]{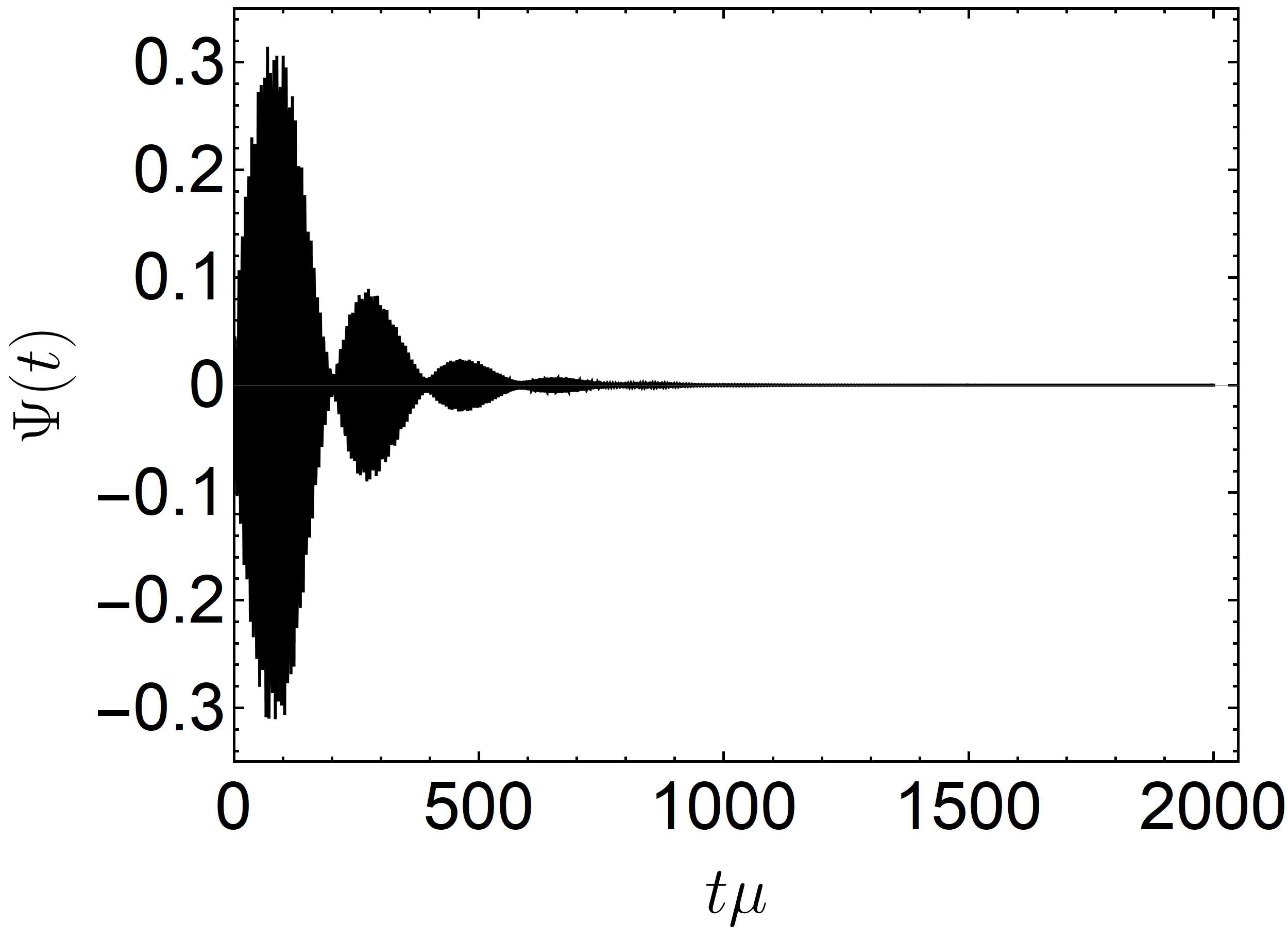} &
\includegraphics[width=0.3\textwidth,keepaspectratio]{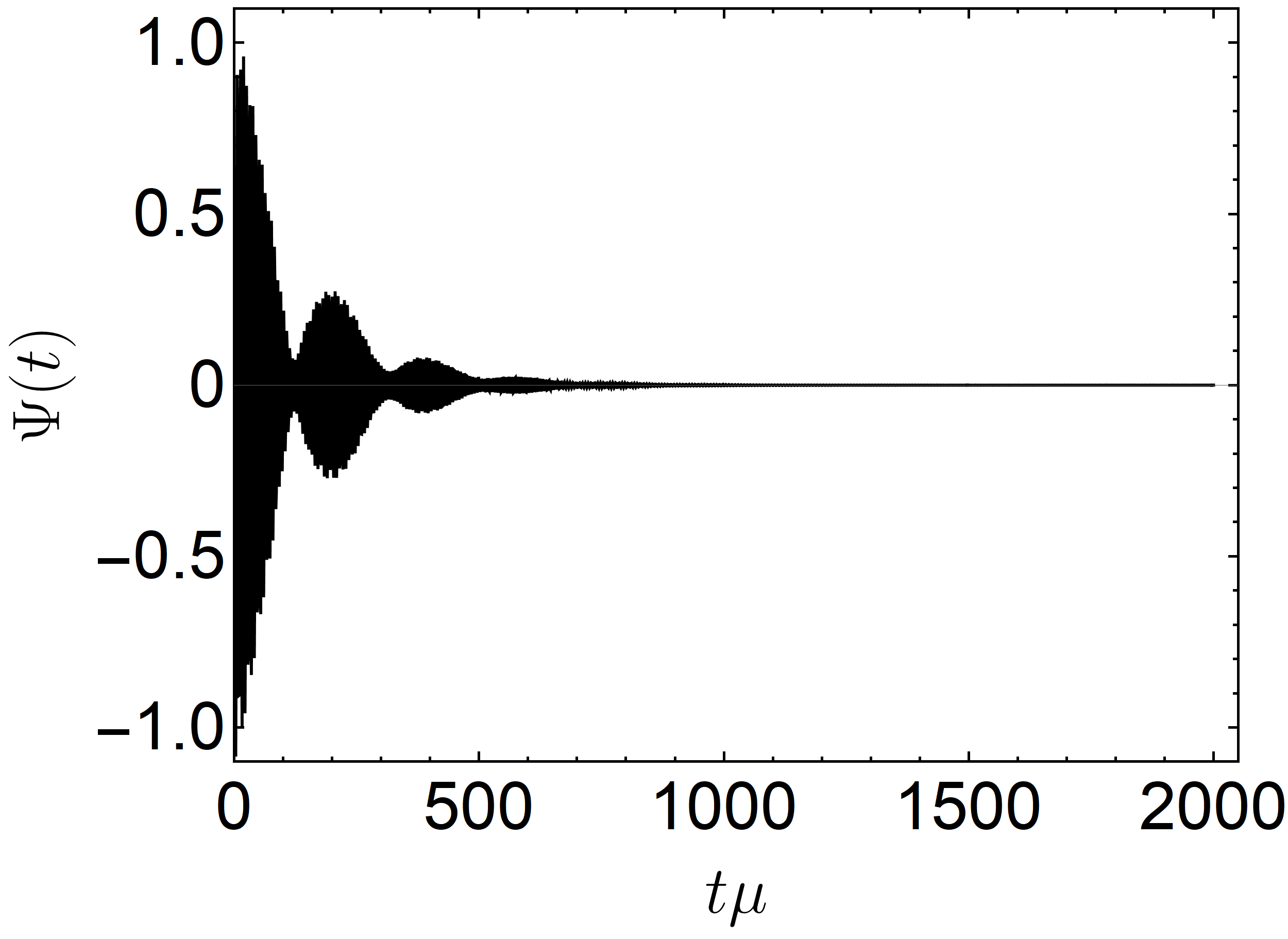}&
\includegraphics[width=0.3\textwidth,keepaspectratio]{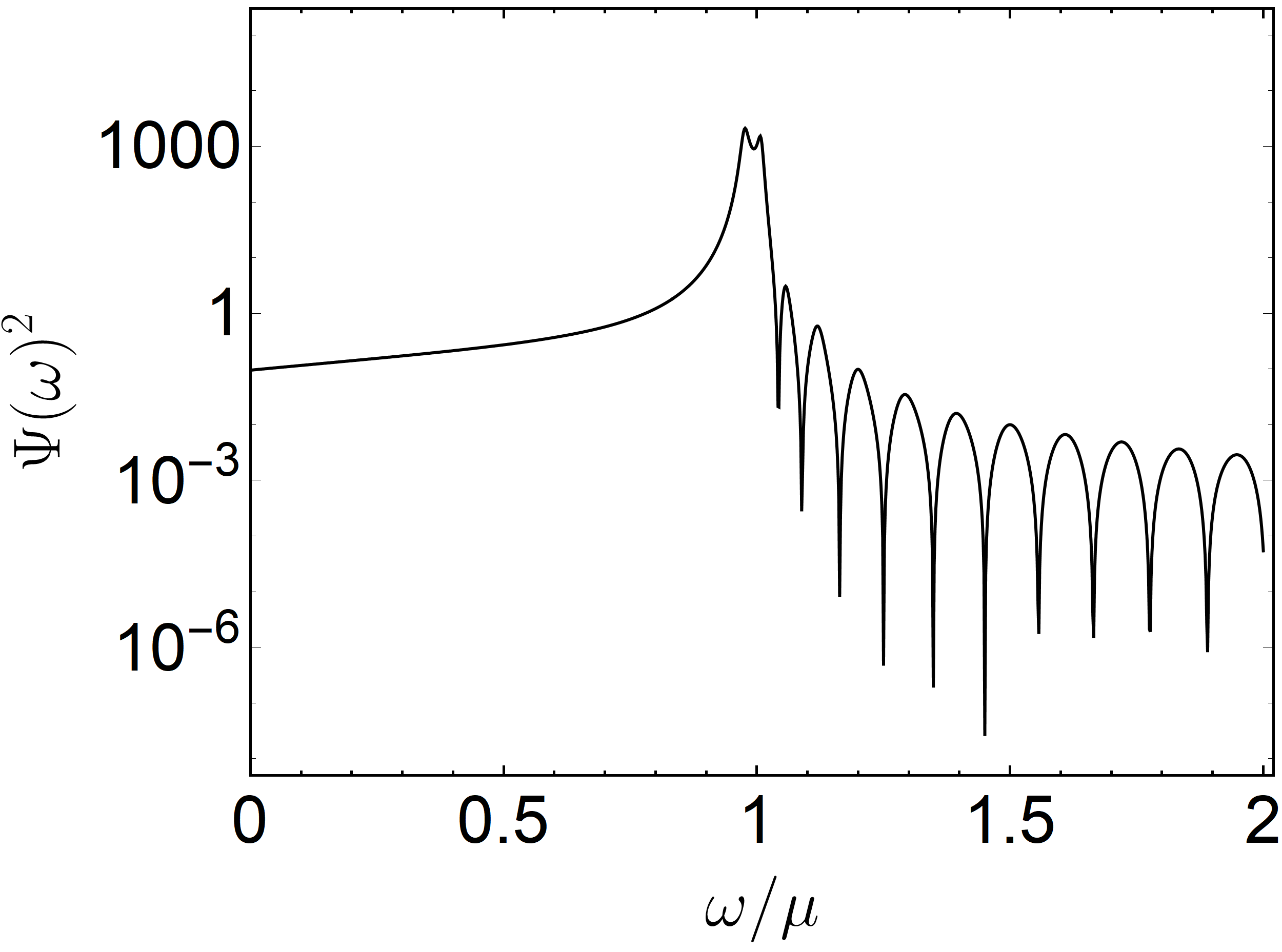}\\
\includegraphics[width=0.3\textwidth,keepaspectratio]{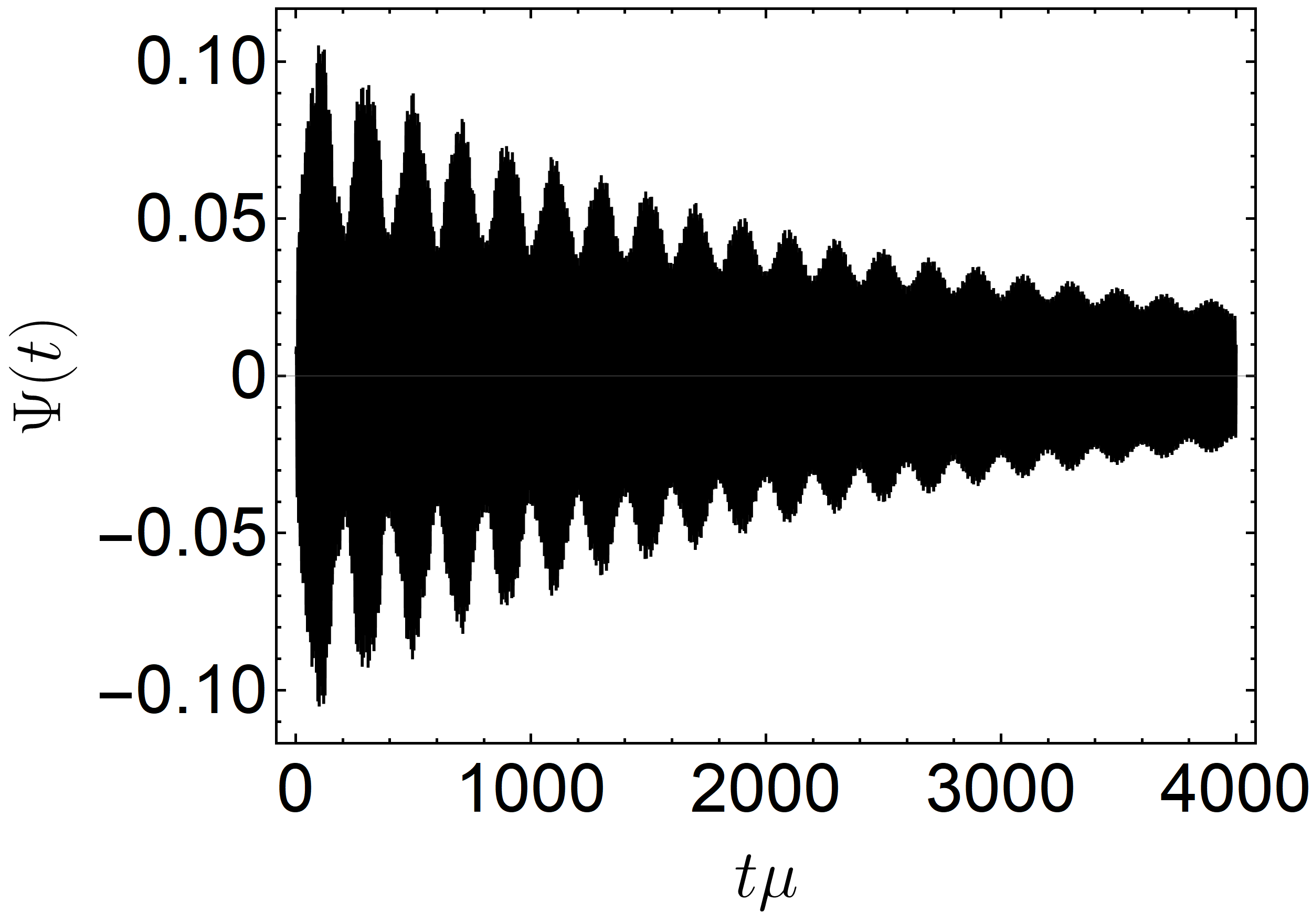} &
\includegraphics[width=0.3\textwidth,keepaspectratio]{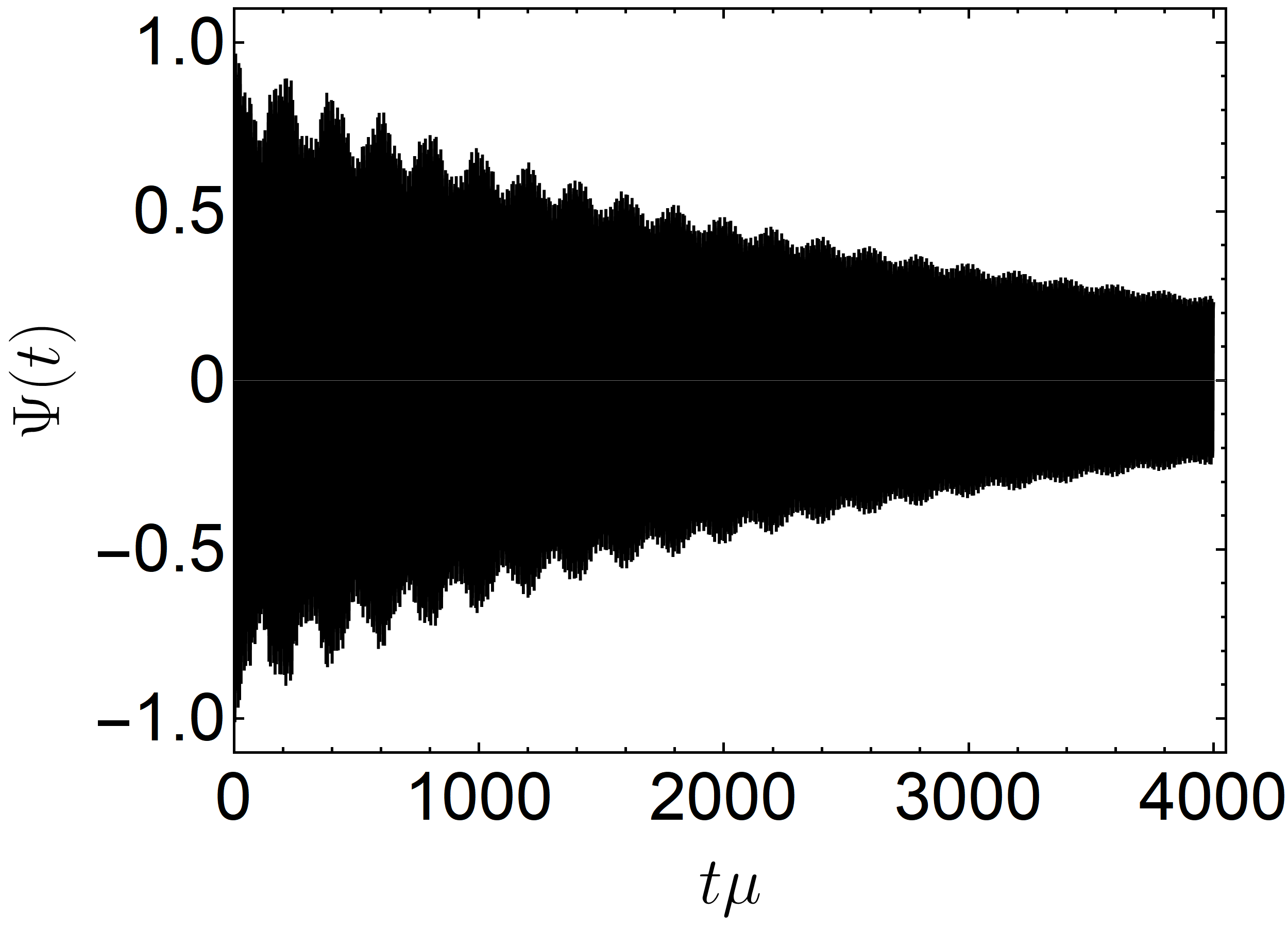}&
\includegraphics[width=0.3\textwidth,keepaspectratio]{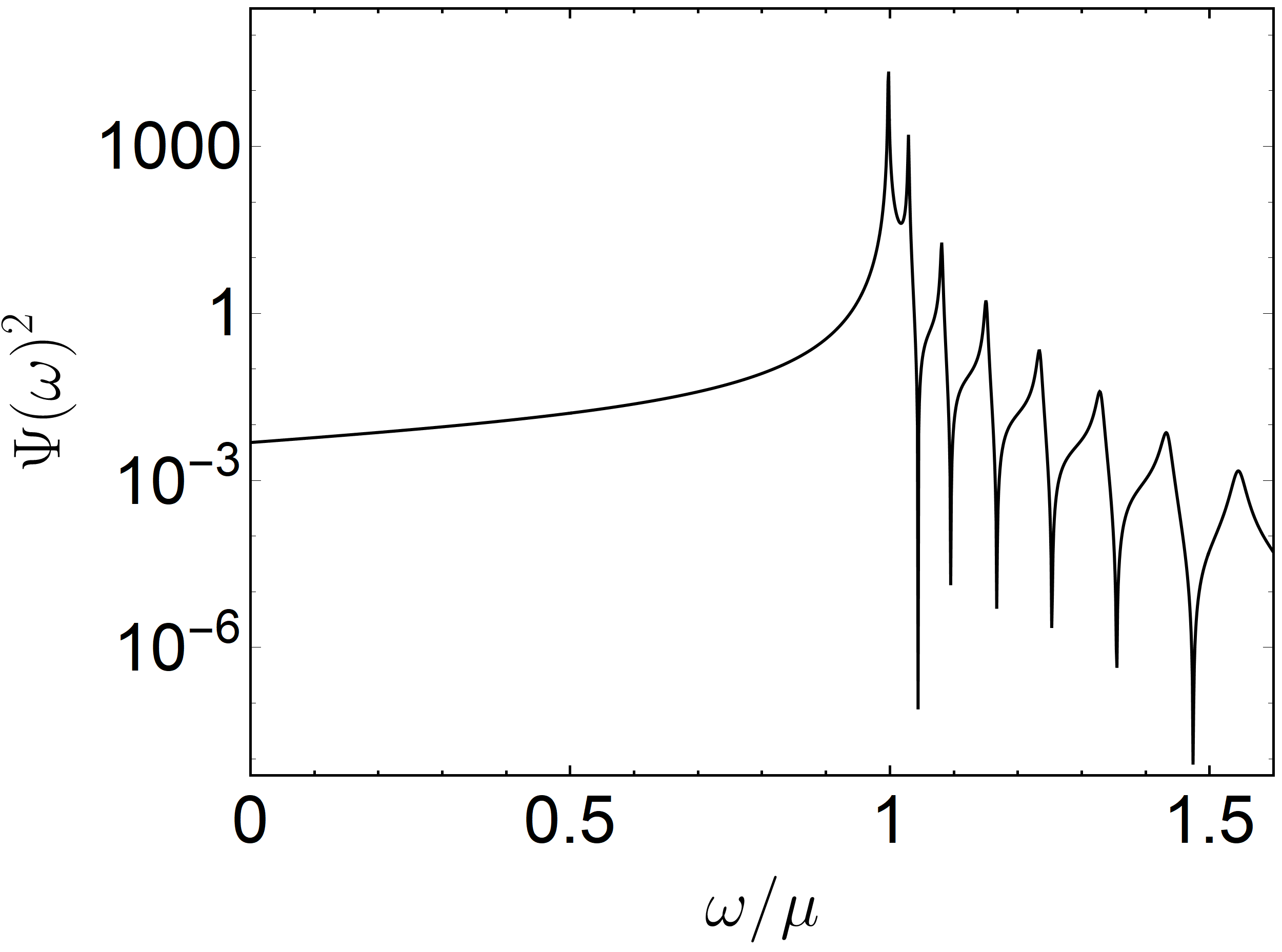}
\end{tabular}
\caption[Massive field in a sphere.]{The evolution of a massive scalar field inside a perfectly reflecting spherical surface of radius $R\mu=20$. In the center of such a sphere, there sits a BH of mass $M_{\rm BH}\mu=0.2$ (upper panels) and $M_{\rm BH}\mu=0.1$ (lower panels). 
{\bf Left:} Scalar field measured on the horizon. 
{\bf Center:} Scalar field measured at $r\mu=10$.
{\bf Right:} Flux measured at the horizon.
\label{fig:BHBomb_evolution}
}
\end{figure*}
A toy model more similar to the problem we wish to study is that of a BH, of mass $M_{\rm BH}$, at the center of a sphere of radius $R$ which was filled with a massive scalar field.
The profile for the scalar is, initially, that of a normal mode (the Klein-Gordon field $\Phi=\Psi/r$),
\begin{equation}
\Psi=\sin \omega_0r\,,
\end{equation}
with $\omega_0=\sqrt{\mu^2+\pi^2/R^2}$. The problem simplifies enormously when the scalar is non-self-gravitating
and is a small disturbance in the background of the BH spacetime. This we assume from now onwards. In such a case all one has to do is evolve the Klein-Gordon equation
in a Schwarzschild geometry, subjected to Dirichlet conditions at the surface of the sphere. The results are summarized in Fig.~\ref{fig:BHBomb_evolution}. While they do not mimic entirely the process of accretion of a self-gravitating NBS by a central BH, these results illustrate some of the possible physics in the more realistic setup.

The figures show the scalar extracted at the horizon (left panel), at a midpoint inside the sphere (middle panel) and the flux per frequency bin (right panel).
The scalar, measured either at the horizon or somewhere within the sphere, decays exponentially. 
The first noteworthy aspect is the sensitive dependence of the decay rate on the size of the BH. Our results are consistent with a decay timescale $\tau \sim (M_{\rm BH}\mu)^{-\beta}$,
with $\beta \sim 4-5$, in agreement with our analysis in Section~\ref{section:A perturber_sitting_at_the_center} and also with a quasinormal mode ringdown of such fields~\cite{Brito:2015oca}.
Note that such suppressed decay for small $M\mu$ couplings happens due to the filtering properties of small BHs, keeping out most of the low-frequency field.
This also explains why the ratio between the field measured at $r \mu=10$ and at the horizon increases when the BH size decreases.
Note also that, in accordance with the simple toy model above, overtones are also excited. This is clearly seen in the Fourier analysis
(rightmost panels in Figure~\ref{fig:BHBomb_evolution}), showing local peaks at all the subsequent overtones, which were absent in the initial data. These correspond to frequencies
$\omega=\sqrt{\mu^2+\pi^2n^2/R^2}\,,n=0,\,1,...$. This is one important difference between this system and NBSs, for which overtones are all bounded in frequency.

\chapter{Gravitational drag by a uniform scalar field} \label{app:drag}

In this Appendix we present the toy model considered in Ref.~\cite{Hui:2016ltb} to compute the gravitational drag acting on a point-like particle travelling through an infinite homogeneous scalar field with a constant non-relativistic velocity~$v\ll 1$. Then, we use (what we believe to be) a more realistic toy model to compute the energy and momentum lost by the infalling body through the plunge into a uniform sphere of scalar field.

Neglecting the self-gravity of the scalar field, the equations describing this process are
\begin{align} \label{SchrodingerAppend}
	i \partial_t \widetilde{\Phi}&=-\frac{1}{2 \mu} \nabla^2\widetilde{\Phi}+\mu U \widetilde{\Phi}\,, \nonumber  \\
	\nabla^2 U&=4 \pi m_p\, \delta(x)\delta(y) \delta(z+v t)\,,
\end{align}
where $m_p$ is the particle's mass.
Now, we change to the frame where the particle is stationary at the origin and the scalar field propagates with momentum $\boldsymbol{k}=\mu v \boldsymbol{e}_z$; so, the gravitational potential is simply $U=-\frac{m_p}{r}$. We consider that the scalar has a
uniform particle density $\rho_0$ in the far past -- before the interaction.  
This is the classical Coulomb scattering problem, and it is known to have the analytic solution
\begin{align} \label{CoulombScat}
	\widetilde{\Phi}&= \sqrt{\rho_0}\, e^{\frac{\pi}{2} \beta} \left|\Gamma(1-i \beta)\right|  e^{-i \left(\frac{k^2}{2 \mu}t-k z\right)} M\left[i \beta, 1, ik(r-z)\right]\,,
\end{align}
where $M$ is the confluent hypergeometric function of the first kind, $r$ is the radial distance from the particle, and the parameter~$\beta$ is
\begin{equation}
	\beta \equiv \frac{m_p \mu}{v}  \,.
\end{equation}
The Klein-Gordon scalar field~$\Phi$ can be obtained from the Schrödinger field~$\widetilde{\Phi}$ through
\begin{equation} \label{KG_Schr}
	\Phi= \frac{1}{\sqrt{\mu}}e^{-i\mu t} \widetilde{\Phi} ,\, \mu \gg |\partial_t \widetilde{\Phi}|\,.
\end{equation}
As expected this solution gives
\begin{equation}
	T_{tt}^S(r\to \infty) = \mu \rho_0\,,
\end{equation}
where we used the non-relativistic limit~$k^2\ll \mu^2$.

A scalar field in a sphere of radius~$R$ centred at the particle exerts a gravitational drag $F_z$ on it,
\begin{equation}
	F_z=- \dot{P}_z^S-\dot{P}_z^{\rm rad}\,,
\end{equation}
using a similar reasoning to the one behind Eq.~\eqref{LossRad}. Here,~$\dot{P}_z^S$ and~$\dot{P}_z^{\rm rad}$ are, respectively, the rate of change of the momentum in the scalar inside the sphere, and the (outgoing) flux of momentum through a surface of radius~$R$; these are calculated through
\begin{align}
	&\dot{P}_z^S=-\int_{r<R} d^3\boldsymbol{r}\,  \partial_t T_{t z}^S\,, \\
	&\dot{P}_z^{\rm rad}=R^2\int_{r=R} d\theta d\varphi \sin \theta\, T_{r z}^S\,.
\end{align}
We take the radius $R$ to be the maximum characteristic length of the problem; to compare with the treatment in Section~\ref{section:Massive_objects_plunging_into_boson_stars} we take it to be the boson star radius. The introduction of this maximum length is necessary and serves as cutoff to the integration, since the Coulomb scattering is known to have an infrared divergence caused by the $1/r$ nature of the gravitational potential. In other words, the gravitational drag is known to diverge in the limit~$R \to \infty$.
Using the divergence theorem, we can rewrite the drag force as
\begin{equation}
	F_z=- \int_{r<R} d^3\boldsymbol{r} \, \left(\partial^t T_ {t z}^S+ \partial^i T^S_ {i z}\right) \,.
\end{equation}
Since we are considering a stationary regime in~\eqref{CoulombScat}, it is easy to check that~$\partial^t T_ {t z}^S$ vanishes. 
Now, using~\eqref{StressEnergy} while keeping only the leading order (Newtonian and non-relativistic) terms,
\begin{equation}
	T_ {i z}^S=\frac{1}{\mu}\,{\rm Re}\left(\partial_i \widetilde{\Phi}^*\partial_z \widetilde{\Phi}\right)- \frac{1}{2}g_ {i z} \left(\frac{1}{\mu}\partial^j\widetilde{\Phi}^* \partial_j \widetilde{\Phi}+i  \widetilde{\Phi}^* \partial_t \widetilde{\Phi}+2 \mu U |\widetilde{\Phi}|^2\right)\,.
\end{equation}
Using Eq~\eqref{SchrodingerAppend} it is straightforward to show
\begin{equation}
	\partial^i T_{i z}^S=-\mu \left(\partial_z U\right)|\widetilde{\Phi}|^2\,,
\end{equation}
which implies that
\begin{equation}
	F_z=\mu\int_{r<R} d^3 \boldsymbol{r}  \left(\partial_z U\right)|\widetilde{\Phi}|^2\,.
\end{equation}
This result is symmetrical to the gravitational force that the particle exerts on the scalar and coincides with Ref.~\cite{Hui:2016ltb}. In the same reference, Hui \textit{et. al.} found that in the limit~$\beta \ll1$ the last integral can be put in the form~\footnote{In Ref.~\cite{Hui:2016ltb} the authors also obtained expressions out of the regime~$\beta\ll1$.}
\begin{align}
	&F_z= \frac{4\pi m_p^2 \,\rho_0 \mu}{v^2} C(v,R\mu)\,, \\
	&C\equiv{\rm Cin}(2 v R \mu)+\frac{\sin (2 v R \mu)}{2v R \mu}-1 \nn \,,
\end{align}
where ${\rm Cin}(x)=\int_0^x(1-\cos x') dx'/x'$ is the cosine integral. For small velocities~$v\ll1/(R\mu)$, the gravitational drag is
\begin{align}
	F_z\simeq\frac{4 \pi }{3}m_p^2 \, \rho_0 \mu^3 R^2\,.
\end{align}
This amounts to a loss of momentum by the particle~$P^{\rm lost}$ of the order 
\begin{equation}
	P^{\rm lost}\simeq F_z \frac{2 R}{v} \simeq  \frac{2}{v} m_p^2\mu^2 M \,,
\end{equation}
where~$2R/v$ is the crossing time and~$M$ is the mass of the scalar contained in the sphere of radius~$R$. Surprisingly, this expression has the same dependence on the physical quantities than the obtained for a more realistic scenario in Section~\ref{section:Massive_objects_plunging_into_boson_stars}; however, this result is a factor of ten larger than the one of that section.

Alternatively, one can consider a toy model closer to the treatment done in Section~\ref{section:Massive_objects_plunging_into_boson_stars}; this consists in linearizing the SP system with respect to an homogeneous sphere of radius $R$ made of scalar field with (particle) density $\rho_0$ and constant gravitational potential $\overline{U}_0<0$. In fact, the assumption of a non-trivial uniform density sphere of scalar field is inconsistent with an homogeneous gravitational potential, due to the Poisson equation. Here, we assume that the Poisson equation only applies to the fluctuations of this medium. We consider that there is no scalar field and the gravitational potential vanishes outside the sphere.
The scalar particles in this medium have energy $\Omega= \mu+\mu \overline{U}_0$, with $- \mu\ll \mu \overline{U}_0<0$. This can be readily verified by plugging the ansatz $\widetilde{\Phi}_0= e^{i\gamma t} \sqrt{\rho_0}/\mu$ in the Schrödinger equation, which inside the sphere reads
\begin{equation}
i \partial_t \widetilde{\Phi}_0=-\frac{1}{2 \mu} \nabla^2\widetilde{\Phi}_0+\mu \overline{U}_0 \widetilde{\Phi}_0\,.	
\end{equation}
That gives $\gamma=-\mu \overline{U}_0$. Then, since the KG scalar field is obtained from the Schrödinger one through~\eqref{KG_Schr}, one gets that, inside the sphere of radius~$R$, the background scalar field is $\Phi_0 = e^{-i \Omega t} \sqrt{\rho_0/\mu}$, with the energy $\Omega= \mu+\mu \overline{U}_0$ satisfying $- \mu\ll \mu \overline{U}_0<0$.

Now, we want to obtain the fluctuations caused by a point-like perturber travelling through the medium at constant (non-relativistic) velocity $v\ll 1$ along the $-\boldsymbol{e}_z$ direction. 
%
These fluctuations are described by the linearized SP system
\begin{align}
&i \partial_t \delta \widetilde{\Phi} =- \frac{1}{2 \mu} \nabla^2\delta \widetilde{\Phi} +\mu U_0 \delta \widetilde{\Phi}+ \mu \widetilde{\Phi}_ 0 \delta U\,, \label{Spert}\\
&\nabla^2\delta U= 4 \pi P \label{Ppert}\,,
\end{align}
with
\begin{equation}
	U_0=\overline{U}_0 \Theta(R-r)\,, \quad \widetilde{\Phi}_0=\frac{\sqrt{\rho_0}}{\mu} \Theta(R-r) e^{-i\mu \overline{U}_0 t}\,,
\end{equation} 
and where the source is given by
\begin{equation}
	P= m_p \frac{\delta(\varphi)}{r^2 \sin \theta} \left[\delta(r+v t) \delta(\theta) \Theta(-t)+ \delta(r-vt) \delta(\theta- \pi)\Theta(t)\right]\,.\nonumber
\end{equation}
Note that the fluctuations $\delta \widetilde{\Phi}$ in the Schrödinger field are related with the fluctuation in the KG field through~$\delta \Phi= e^{-i\mu t}\delta \widetilde{\Phi}$. For simplicity, we are neglecting the self-gravity of the scalar field in the right-hand side of~\eqref{Ppert}. This is a good approximation in the region close to the particle, where the Coulombian potential is dominant.
Using the axially symmetric decompositions
\begin{align}
	P&=\sum_{l=0}^{\infty} \int \frac{d\omega}{\sqrt{2 \pi} r}e^{-i \omega t} Y_l^0(\theta) p(r)\,, \\
	\delta U&=\sum_{l=0}^{\infty} \int \frac{d\omega}{\sqrt{2 \pi} r}e^{-i \omega t} Y_l^0(\theta) u(r)\,, 
\end{align}
where
\begin{equation}\label{p_source}
	p=\sqrt{\frac{2}{\pi}} m_p \frac{Y_l^0(0)}{r v} \delta_m^0  \left[\cos\left(\frac{\omega}{v} r\right)\delta_l^{\rm even}-i \sin \left(\frac{\omega}{v}r\right)\delta_l^{\rm odd}\right]\,,
\end{equation}
the Poisson equation becomes
\begin{align}
	\partial_r^2u-\frac{l(l+1)}{r^2}u=4 \pi p\,.
\end{align}
This admits the homogeneous solutions
\begin{align*}
	u^I=r^{-l},\, u^{II}=r^{l+1}\,, 
\end{align*}
which are regular, respectively, at infinity and at the origin. Using the method of variation of parameters, one gets the inhomogeneous solution
\begin{equation}
	u=-\frac{4\pi}{2l+1} \left(r^{-l} \int_{0}^{r}dr' r'^{l+1}p+r^{l+1}\int_r^{\infty}dr' \frac{p}{r'^l} \right)\,,
\end{equation}
This results in the analytical expression
\begin{align} \label{u_final}
	u&=-i\,\frac{2\sqrt{2\pi}}{2l+1}\frac{m_p}{\omega} Y_l^0(0)\delta_m^0\left(i \frac{r}{v} \omega\right)^{l+1}\Bigg\{\left(i \frac{\omega}{v}r\right)^{-2l-1} \left[\Gamma\left(l+1,i \frac{\omega}{v} r\right)-\Gamma\left(l+1,-i \frac{\omega}{v} r\right)\right]\nonumber\\
	&-\Gamma\left(-l,i \frac{\omega}{v} r\right)-\Gamma\left(-l,-i \frac{\omega}{v} r\right)\Bigg\}\,,
\end{align}
where $\Gamma(a,x)$ is the incomplete gamma function. 
Now, decomposing the scalar fluctuation as
\begin{equation}
	\delta \widetilde{\Phi}=\sum_{l=0}^{\infty} \int \frac{d\omega}{\sqrt{2 \pi} r}e^{-i \left(\omega+\mu \overline{U}_0\right) t}\, Y_l^0(\theta) Z(r)\,,
\end{equation}
equation~\eqref{Spert} becomes
\begin{equation} \label{Zeq}
\partial_{r}^2 Z+\left[2\mu \left(\omega+\mu \overline{U}_0 \Theta(r-R) \right) - \frac{l(l+1)}{r^2}\right]Z=2\mu \sqrt{\rho_0}\, \Theta(R-r) u\,.	
\end{equation}
Outside the sphere o radius~$R$, the solution satisfying the Sommerfeld radiation condition at infinity is simply given by
\begin{equation}
	Z(r)=A \sqrt{r}\, H_{l+\frac{1}{2}}^{(1)}\left(\sqrt{2 \mu\left(\omega+ \mu \overline{U}_ 0\right)}\,r\right)\,,
\end{equation} 
where~$A$ is a complex-constant to be determined through the matching with the interior solution.
Using equation~\eqref{u_final} it is possible to see that the highest frequencies that the perturber excites (efficiently) are~$\omega \sim v/(2 R)$. This has the important consequence that for velocities $v \ll 2 R\mu |\overline{U}_0|$ the emission is strongly suppressed, because the perturber cannot excite (efficiently) waves that travel to infinity. Additionally, in the limit of small velocities~$v\ll 1/(R \mu)$, we have
\begin{align}
	Z(r \sim R)\simeq -\frac{i A}{\pi} \frac{2^{l+\frac{1}{2}}\,\Gamma\left(l+\frac{1}{2}\right)}{\left[2 \mu \left(\omega+\mu \overline{U}_0\right)\right]^{\frac{l}{2}+\frac{1}{4}}}  r^{-l}\,,
\end{align}
where we used the small argument expansion of $H^{(1)}_{l+\frac{1}{2}}$.
Inside the sphere of radius $R$, equation~\eqref{Zeq} has the independent homogeneous solutions
\begin{align}
	Z^I&=\sqrt{r}\,H^{(1)}_{l+\frac{1}{2}}(\sqrt{2 \mu \omega} \,r) \simeq -\frac{i}{\pi} \frac{2^{l+\frac{1}{2}}\,\Gamma\left(l+\frac{1}{2}\right)}{\left(2 \mu \omega\right)^{\frac{l}{2}+\frac{1}{4}}}  r^{-l}\,, \nonumber  \\
	Z^{II}&=\sqrt{r}\,J_{l+\frac{1}{2}}(\sqrt{2 \mu \omega} \,r) \simeq \frac{\left(2 \mu \omega\right)^{\frac{l}{2}+\frac{1}{4}}}{2^{l+\frac{1}{2}}\, \Gamma\left(l+\frac{3}{2}\right)} r^{l+1}\,.
\end{align}
The solution $Z^{II}$ is regular at the origin, and the solution $Z^I$ is (approximately) proportional to $r^{-l}$ everywhere inside the sphere, making it appropriate to match with the exterior solution at $r=R$.  
Using the method of variation of parameters, one obtains that the radial function $Z$ at $r=R$ is
\begin{equation}
	Z(R)=-i \pi \mu \sqrt{\rho_ 0}\,  Z^I(R)\int_0^R dr' Z^{II} \,u(r')\,.
\end{equation}
Then, the constant $A$ can be determined through matching between the interior and exterior solutions,
\begin{align} \label{integr}
	A=-i \pi \mu \sqrt{\rho_ 0}\left(1+\frac{\mu \overline{U}_ 0}{\omega}\right)^{\frac{l}{2}+\frac{1}{4}}\int_0^R dr' Z^{II} \,u(r')\,.
\end{align}
%
Using the large argument expansion of $H^{(1)}_{l+\frac{1}{2}}$ one gets the radial function $Z$ at infinity,
\begin{equation}
	Z_\infty \equiv Z(r\to \infty)=-\frac{2^\frac{1}{4}(-i)^{l-1}}{\sqrt{\pi}} \frac{A\,\sqrt{R}\,e^{i\sqrt{2\mu \left(\omega+\mu \overline{U}_0\right)}\, r}}{(v R \mu)^\frac{1}{4}\alpha^\frac{1}{4}\left(1+\frac{\overline{U}_0 R \mu}{v \alpha}\right)^\frac{1}{4}}\,,
\end{equation}
with the dimensionless parameter
\begin{equation}
\alpha \equiv \frac{\omega R}{v}\,.
\end{equation}
Evaluating the integral in~\eqref{integr} we obtain the analytical expression
\begin{align}
	Z_\infty&=i \pi \delta_m^0 (R \mu)^4 \frac{m_p \sqrt{\rho_0}}{\mu^2} e^{i\sqrt{2\mu \left(\omega+\mu \overline{U}_0\right)}\, r} \frac{(-1)^l Y_l^0(0)(v R \mu)^{\frac{l}{2}-1}}{2^{\frac{l}{2}-\frac{1}{2}}(2l+1) \Gamma\left(l+\frac{3}{2}\right)} \frac{\left(1+\frac{\overline{U}_ 0R \mu}{v \alpha}\right)^{\frac{l}{2}}}{\alpha^{\frac{l}{2}+3}} \nonumber \\
	&\times \Bigg\{\frac{2l+1}{2l+3} \left[\Gamma\left(l+3,i \alpha\right)-\Gamma\left(l+3,-i \alpha\right)\right] +2\frac{(i \alpha)^{2l+3}}{2l+3}\left[\Gamma\left(-l,i \alpha\right)+\Gamma\left(-l,-i \alpha\right)\right]\nn\\
	& +\alpha^2\left[\Gamma\left(l+1, i\alpha\right)-\Gamma\left(l+1, -i\alpha\right)\right]\Bigg\}\,.
\end{align}
We have also solved this problem numerically (without any approximation). This analytical expression describes perfectly the exact results for the first multipoles (essentially~$l\leq3$); these account for most of the radiation.
The energy radiated with frequency between~$\omega$ and $\omega+d\omega$ is 
\begin{align}
	\frac{d E^{\rm rad}}{d \omega}&=\frac{\sqrt{2} }{R}\left(v R \mu \right)^{\frac{1}{2}}\left[\mu+\frac{\alpha v}{R }\left(1+\frac{\overline{U}_0R \mu}{v \alpha}\right)\right] \alpha^\frac{1}{2}\,{\rm Re}\left[ \left(1+\frac{\overline{U}_0R \mu}{v \alpha}\right)^\frac{1}{2}\right]\sum_{l=0}^{\infty} \left|Z_ \infty\right|^2 \nn \\
	&\simeq \frac{\sqrt{2} \mu}{R}\left(v R \mu \right)^{\frac{1}{2}}  \alpha^\frac{1}{2}\,{\rm Re}\left[ \left(1+\frac{\overline{U}_0R \mu}{v \alpha}\right)^\frac{1}{2}\right]\sum_{l=0}^{\infty} \left|Z_ \infty\right|^2\,,
\end{align}  
where in the last equality we used that the scalar fluctuations are non-relativistic.
This results in the total radiated energy
\begin{equation}
	E^{\rm rad}=\frac{\sqrt{2}}{R^3}\,  \left(v R \mu\right)^{\frac{3}{2}} \sum_{l=0}^{\infty} \int_{\frac{\left|\overline{U}_0\right| R \mu}{v}}^\infty d \alpha \alpha^{\frac{1}{2}} {\rm Re}\left[ \left(1+\frac{\overline{U}_0 R \mu}{v \alpha}\right)^\frac{1}{2}\right] \left|Z_ \infty\right|^2\,.
\end{equation}
The energy lost by the perturber in this process is
\begin{equation}
E^{\rm lost}=\frac{\sqrt{2}}{R^5 \mu^2}\,  \left(v R \mu\right)^{\frac{5}{2}} \sum_{l=0}^{\infty} \int_{\frac{\left|\overline{U}_0\right| R \mu}{v}}^\infty d \alpha \alpha^{\frac{3}{2}} {\rm Re}\left[ \left(1+\frac{\overline{U}_0 R \mu}{v \alpha}\right)^\frac{3}{2}\right] \left|Z_ \infty\right|^2\,.
\end{equation}
In the case of a vanishing gravitational potential~$\overline{U}_0=0$, we see that for small velocities the radiated energy goes with $\sim v^{-\frac{1}{2}}$ and the energy lost by the perturber with~$v^{\frac{1}{2}}$; note that~$Z_\infty \sim v^{\frac{l}{2}-1}$. In the case of a non-trivial gravitational potential, the radiated energy is highly suppressed for small velocities; this is because smaller velocities excite lower frequencies -- these may not be capable of escaping the gravitational influence of the scalar configuration.

The spectral flux of linear momentum radiated along~$z$ is given by
\begin{equation}
	\frac{d P_z^{\rm rad}}{d \omega}=\frac{4}{R^2} \left(v R \mu\right)\alpha\, \Theta\left(1+\frac{\overline{U}_0 R \mu}{v\alpha}\right)\left(1+\frac{ \overline{U}_0 R \mu}{v \alpha}\right) \sum_{l=0}^{\infty} \frac{ (l+1)\,{\rm Re}\left(Z_\infty^l\left(Z_\infty^{l+1}\right)^*\right)}{\sqrt{\left(2l+1\right)\left(2l+3\right)}} \,.
\end{equation}
So, the total linear momentum radiated during this process is
\begin{equation}
	P_z^{\rm rad}=\frac{4}{R^4 \mu} \left(v R \mu\right)^2 \sum_{l=0}^{\infty} \frac{l+1}{\sqrt{(2l+1)(2l+3)}} \int_{\frac{\left|\overline{U}_0\right| R \mu}{v}}^\infty d \alpha\, \alpha \left(1+\frac{ \overline{U}_0 R \mu}{v \alpha}\right) \,{\rm Re}\left(Z_\infty^l\left(Z_\infty^{l+1}\right)^*\right)\,.
\end{equation}
The loss in momentum for a small perturber~$m_p \mu \ll v$ is simply
\begin{align}
	P_z^{\rm lost}=\frac{E^{\rm lost}}{v}\,.
\end{align}
In the case of a vanishing gravitational potential, for small velocities the radiated momentum goes with~$v^\frac{1}{2}$ and perturber's loss in momentum with~$\sim v^{-\frac{1}{2}}$. Again, with a non-trivial gravitational potential both quantities are suppressed in the limit of small velocities.

Our toy model shows that: (i) the gravitational potential of a scalar configuration tends to suppress both the radiation and the loss in momentum for plunging perturbers, specially in the small velocity limit;~\footnote{Actually, although we do not present it in this work, we solved the full problem -- including the self-gravity of the scalar -- in a way similar to Section~\ref{section:Massive_objects_plunging_into_boson_stars} but with constant velocity. We found qualitative agreement with the toy model considered here; however, including the self-gravity of the scalar yields to a larger suppression of radiation and loss of momentum.}
(ii) when neglecting the gravity of the scalar, the loss in momentum for a perturber plunging in a uniform sphere of scalar field at a constant small velocity follows~$P^{\rm lost}\sim v^{-\frac{1}{2}}$. This behavior is different than the one found in Ref.~\cite{Hui:2016ltb}; in that reference besides neglecting the gravity of the scalar, the authors study a stationary regime in an infinite scalar field medium (introducing a cut-off length~$R$ \textit{a posteriori}).

For a full realistic plunge into an NBS -- including the self-gravity of the scalar and the accelerated free fall of the perturber  -- see Section~\ref{section:Massive_objects_plunging_into_boson_stars}.
\chapter{Angle-action variables}
\label{angleaction}

In this Appendix we show how to develop and use the angle-action formalism, useful to describe in a compact way the perturbations acting on a binary in Keplerian orbit.

\section*{Newtonian dynamics in the Delaunay variables}
\label{angleaction1}
From the reduced Newtonian Lagrangian in the CM coordinates, using spherical coordinates ($r,\theta,\varphi$),
\begin{equation}
\tilde{L}\equiv\frac{L}{m\nu} = \frac{G m}{r} + \frac{1}{2}\dot{r}^{2} + \frac{1}{2}r^{2}\left(\dot{\theta}^{2}+\sin^{2}\theta\,\dot{\varphi}^{2}\right) \,,
\end{equation}
we determine the conjugate momenta $p_{x}=\partial\tilde{L}/\partial\dot{x}$,
\begin{equation}
p_{r}=\dot{r},\, p_{\theta}=r^{2}\dot{\theta},\, p_{\varphi}=r^{2}\sin^2\theta\,\dot{\varphi}\,,
\end{equation}
and, performing a Legendre transformation, the reduced Hamiltonian,
\begin{align}
\mathcal{\tilde{H}}_{0} &\equiv p_{r}\dot{r}+p_{\theta}\dot{\theta}+p_{\varphi}\dot{\varphi}-\tilde{L}\nonumber\\
& = -\frac{G m}{r} + \frac{1}{2}p_{r}^{2} + \frac{1}{2r^{2}}p_{\theta}^{2} + \frac{1}{2r^{2}\sin^2\theta}p_{\varphi}^{2} \,.
\end{align}
The angular momentum $\mathbf{L} = \mathbf{r}\wedge\mathbf{v}$ is then given by
\begin{equation}
L_{r}=0,\, L_{\theta}= -\frac{p_{\varphi}}{\sin\theta},\, L_{\varphi} = p_{\theta}\,.
\end{equation}
We want to go from the canonical set of variables $\left(r,\theta,\varphi,\,p_{r},p_{\theta},p_{\varphi}\right)$ to a set of canonical angle-action variables, by taking into account the symmetries of the system. We use the modified Delaunay variables that are well suited to described the Keplerian two-body problem. The actions are given by,
\begin{align}
\label{JitoOrbitElem}
& J_{3}=\frac{Gm}{\sqrt{-2E}} ,\, J_{2}=\frac{Gm}{\sqrt{-2E}}-L ,\, J_{1}=L-L_{z} \,.
\end{align}
The Hamiltonian is then simply
\begin{equation}
\mathcal{\tilde{H}}_{0} = -\frac{G^{2}m^{2}}{2J_{3}^2}\,.
\end{equation}
We can then derive the frequencies $\Omega_{i}=\partial\tilde{H}/\partial J_{i}$,
\begin{equation}
\Omega_{3}= \frac{G^{2}m^{2}}{J_{3}^{3}},\, \Omega_{2} = 0,\, \Omega_{1} = 0\,.
\end{equation}
The angles $\theta_{i}$, conjugate variables of the action $J_{i}$, are then linear in time,
\begin{equation}
\theta_{3}=\Omega_{3}(t-t_{0})+(\theta_{3})_{0} \,,\ \ \theta_{2}=(\theta_{2})_{0}\,,\ \ \theta_{1}=(\theta_{1})_{0}\,,
\end{equation}
with $(\theta_{i})_{0}$ the values of the angle variables at time $t_{0}$.
We can also relate the modified Delaunay variables to the orbital elements $a,e,l,\iota,\omega,\psi$, we get 
\begin{align}
J_{3}&=\sqrt{Gma}, \, J_{2}=\sqrt{Gma}\left(1-\sqrt{1-e^{2}}\right) \,,\nonumber\\
J_{1}&=\sqrt{Gma(1-e^{2})}\left(1-\cos\iota\right) \,,\\
\theta_{3}&=l+\omega+\psi \,,\ \, \theta_{2}=-\left(\omega+\psi\right) \,,\ \, \theta_{1}=-\psi \,.\label{JitoOrbitElem2}
\end{align}
Note in particular that these variables are well-defined when $e=0$ and $\iota=0$, which will allow us to perform an expansion for small eccentricity.
\section*{Perturbation theory}
\label{angleaction2}
When the total Hamiltonian is no longer integrable, it is not possible to write it as a function of the actions only.
The perturbed Hamiltonian in the modified Delaunay variables can be written as
\begin{equation}
\mathcal{\tilde{H}} = \mathcal{\tilde{H}}_{0}\left(\mathbf{J}\right) + \mathcal{\tilde{H}}_{1}\left(\boldsymbol{\theta},\mathbf{J},t\right)\,,
\end{equation}
where $\mathcal{\tilde{H}}_{0}=-G^{2} m^{2}/(2 J_{3}^{2})$ is the Newtonian Hamiltonian previously studied, and $\tilde{\mathcal{H}}_{1}$ is the perturbation, assumed to be small, $\mathcal{O}(H_{ij})\equiv\mathcal{O}(\varepsilon)\ll 1$. We see that the total Hamiltonian now depends on the actions and angles, but also on time. The dependence on time can be removed by introducing a new coordinate $\tau$ and its conjugate variable $\mathcal{T}$, and transforming the time-dependent Hamiltonian $\mathcal{\tilde{H}}$ into a time-independent Hamiltonian $\mathcal{\tilde{\tilde{H}}}$ in the following way,
\begin{equation}
\mathcal{\tilde{\tilde{H}}} = \Omega\mathcal{T} + \mathcal{\tilde{H}}\left(\boldsymbol{\theta},\mathbf{J},\tau\right)\,. \label{Ham_tau}
\end{equation}
We see that the Hamilton equations for $\tau$ are
\begin{equation}
\dot{\tau}=\frac{\partial\mathcal{\tilde{\tilde{H}}}}{\partial\mathcal{T}}=\Omega\,, \qquad \dot{\mathcal{T}} = -\frac{\partial\mathcal{\tilde{H}}_{1}}{\partial\tau}\,,
\end{equation}
such that $\tau = \Omega t$. The other equations for the angle-action variables are unchanged, and are now given by,
\begin{align}
\dot{J_{3}} &= -\frac{\partial\mathcal{\tilde{H}}_{1}}{\partial\theta_{3}} ,\, \dot{J_{2}} = -\frac{\partial\mathcal{\tilde{H}}_{1}}{\partial\theta_{2}} ,\, \dot{J_{1}} = -\frac{\partial\mathcal{\tilde{H}}_{1}}{\partial\theta_{1}} \,,\nonumber\\
\dot{\theta_{3}}&=\frac{G^{2}m^{2}}{J_{3}^{3}} +\frac{\partial\mathcal{\tilde{H}}_{1}}{\partial J_{3}},\, \dot{\theta_{2}} =\frac{\partial\mathcal{\tilde{H}}_{1}}{\partial J_{2}} ,\, \dot{\theta_{1}}= \frac{\partial\mathcal{\tilde{H}}_{1}}{\partial J_{1}} \,.\label{HamiltonEqs}
\end{align}
Using the relations~\eqref{JitoOrbitElem} linking the angle-action coordinates $\left(\boldsymbol{\theta},\mathbf{J}\right)$, and the Hamilton equations~\eqref{HamiltonEqs}, we can see that we directly have the variation of the orbit elements $a,e,l,\iota,\omega,\psi$. As the perturbation is small, we can use the unperturbed (Newtonian) results to evaluate $\tilde{\mathcal{H}}_{1}$ and its derivatives. Then by averaging over one (Newtonian) orbit, we get the secular evolution of the binary. However the Hamiltonian $\mathcal{\tilde{\tilde{H}}}$ is quite complicated and so are the equations~\eqref{HamiltonEqs}.

In order to circumvent these technical difficulties we use Hamiltonian perturbation theory to define a new set of canonical angle-action coordinates $\left(\boldsymbol{\theta}_{\mathrm{new}},\,\boldsymbol{J}_{\mathrm{new}}\right)$ such that the Hamiltonian will only depend on the action variables. We call $\left(\boldsymbol{\theta}^{0},\,\boldsymbol{J}^{0}\right)$ the old variables (including $\tau$ and $\mathcal{T}$). We have
\begin{equation}
\mathcal{\tilde{\tilde{H}}}\left(\boldsymbol{J}_{\mathrm{new}}\right) = \mathcal{H}_{0}\left(\boldsymbol{J}^{0}\right) + \mathcal{\tilde{H}}_{1}\left(\boldsymbol{\theta}^{0},\mathbf{J}^{0}\right)\,.
\end{equation}
We now consider the generating function
\begin{equation}
\tilde{S}\left(\boldsymbol{\theta}_{\mathrm{new}},\,\boldsymbol{J}^{0}\right) = \boldsymbol{\theta}_{\mathrm{new}}\cdot\boldsymbol{J}^{0} +\tilde{s}\left(\boldsymbol{\theta}_{\mathrm{new}},\,\boldsymbol{J}^{0}\right)\,,
\end{equation}
where $\tilde{s}=\mathcal{O}\left(\varepsilon\right)$. Then we can rewrite the Hamiltonian, up to $\mathcal{O}\left(\varepsilon^{2}\right)$, as
\begin{equation}
\mathcal{\tilde{\tilde{H}}}\left(\boldsymbol{J}_{\mathrm{new}}\right) = \mathcal{H}_{0}\left(\boldsymbol{J}_{\mathrm{new}}\right) -\boldsymbol{\Omega}^{0}\cdot\frac{\partial\tilde{s}}{\partial\boldsymbol{\theta}_{\mathrm{new}}} + \mathcal{\tilde{H}}_{1}\left(\boldsymbol{\theta}_{\mathrm{new}},\mathbf{J}_{\mathrm{new}}\right)\,.\nonumber
\end{equation}
where $\boldsymbol{\Omega}^{0}\equiv\frac{\partial\mathcal{H}_{0}}{\partial\mathbf{J}^{0}}$. Now we expand both the perturbed Hamiltonian $\mathcal{\tilde{H}}_{1}$ and $\tilde{s}$ in Fourier series,
\begin{align}
\mathcal{\tilde{H}}_{1}\left(\boldsymbol{\theta}_{\mathrm{new}},\boldsymbol{J}_{\mathrm{new}}\right) & = \sum_{\mathbf{k}}h_{\mathbf{k}}\left(\mathbf{J}_{\mathrm{new}}\right)\mathrm{e}^{i\mathbf{k}\cdot\boldsymbol{\theta}_{\mathrm{new}}} \,,\\
\tilde{s}\left(\boldsymbol{\theta}_{\mathrm{new}},\boldsymbol{J}_{\mathrm{new}}\right) & = i\sum_{\mathbf{k}}s_{\mathbf{k}}\left(\mathbf{J}_{\mathrm{new}}\right)\mathrm{e}^{i\mathbf{k}\cdot\boldsymbol{\theta}_{\mathrm{new}}} \,.
\end{align}
Then the Hamiltonian becomes, up to $\mathcal{O}\left(\varepsilon^{2}\right)$,
\begin{equation}
\label{Htt}
\mathcal{\tilde{\tilde{H}}}\left(\boldsymbol{J}_{\mathrm{new}}\right) = \mathcal{H}_{0}\left(\boldsymbol{J}_{\mathrm{new}}\right) +h_{\mathbf{0}}\left(\boldsymbol{J}_{\mathrm{new}}\right)+\sum_{\mathbf{k}\neq\mathbf{0}}\left[h_{\mathbf{k}}\left(\mathbf{J}_{\mathrm{new}}\right) +\mathbf{k}\cdot\boldsymbol{\Omega}^{0}\left(\mathbf{J}_{\mathrm{new}}\right)s_{\mathbf{k}}\left(\mathbf{J}\right)\right]\mathrm{e}^{i\mathbf{k}\cdot\boldsymbol{\theta}_{\mathrm{new}}} \,.\nonumber
\end{equation}
As the l.h.s. of Eq.~\eqref{Htt} depends only on the action variables $\mathbf{J}_{\mathrm{new}}$, the r.h.s. should also depends only on this variables. This gives the Fourier coefficients of the generating functions,
\begin{equation}\label{GeneratingFunction}
s_{\mathbf{k}}\left(\mathbf{J}\right) = - \frac{h_{\mathbf{k}}\left(\mathbf{J}\right)}{\mathbf{k}\cdot\boldsymbol{\Omega}^{0}\left(\mathbf{J}\right)} \qquad \text{for k}\neq 0\,.
\end{equation}
This transformation is valid only when $\mathbf{k}\cdot\boldsymbol{\Omega}^{0}\left(\mathbf{J}\right) \neq 0$. The case
\begin{equation}
\mathbf{k}\cdot\boldsymbol{\Omega}^{0}\left(\mathbf{J}\right) = 0 \,,
\end{equation}
is called the \textit{problem of small divisors} and it describes the appearance of a resonance at the corresponding frequency. In that case the formalism we are using to describe the binary dynamics is no more valid.

We now consider the coordinate transformation, defined by~\eqref{GeneratingFunction}, to a new set of canonical variables $\left(\boldsymbol{\theta}',\,\boldsymbol{J}'\right)$. The Hamiltonian obtained after this transformation is
\begin{equation}
\tilde{\mathcal{H}}'\left(\mathbf{J}'\right) \equiv \tilde{\mathcal{H}}_{0}\left(\mathbf{J}'\right) +h_{0}\left(\mathbf{J}'\right) \,.
\end{equation}
It describes the dynamics of the system up to first order included. The new variables are related to the old ones by the relations,
\begin{align}
\mathbf{J}' &= \mathbf{J} +\sum_{\mathbf{k}}\frac{h_{\mathbf{k}}\left(\mathbf{J}\right)}{\mathbf{k}\cdot\boldsymbol{\Omega}^{0}\left(\mathbf{J}\right)}\mathbf{k}\mathrm{e}^{i\mathbf{k}\cdot\boldsymbol{\theta}'} \,,\\
\boldsymbol{\theta}' &= \boldsymbol{\theta} + i\sum_{\mathbf{k}}\frac{\partial}{\partial\mathbf{J}}\left(\frac{h_{\mathbf{k}}\left(\mathbf{J}\right)}{\mathbf{k}\cdot\boldsymbol{\Omega}^{0}\left(\mathbf{J}\right)}\right)\mathrm{e}^{i\mathbf{k}\cdot\boldsymbol{\theta}'} \,.
\end{align}
Finally the dynamics of the system is governed by Hamilton's equation
\begin{align}
\dot{\mathbf{J}'} & = -\frac{\partial\tilde{\mathcal{H}}'}{\partial\boldsymbol{\theta}'} = 0 \,,\\
\dot{\boldsymbol{\theta}'} & = \frac{\partial\tilde{\mathcal{H}}'}{\partial\boldsymbol{J}'} = \frac{\partial\tilde{\mathcal{H}}_{0}}{\partial\boldsymbol{J}'} + \frac{\partial h_{0}}{\partial\boldsymbol{J}'} \,.
\end{align}

\chapter{Initial data for extreme compact objects collision}\label{app:ECO_initialdata}
In this Appendix we define in detail a possible choice for the initial data describing an ECO binary.

We start modelling the ECO interior, finding (in the framework of a $3+1$ decomposition, see Sec.~\ref{sec:3+1}) a
conformally flat three-metric
\begin{equation}
\prescript{3}{}{ds}^2_{\rm ECO}=\varphi^4 \prescript{3}{}{d\eta}^2=\varphi^4 \left( dR^2+R^2d\Omega^2 \right)\,,
\label{eq:ds2ECO}
\end{equation}
solution of the Hamiltonian constraint equation~\eqref{eq:H_ECO}, which, with the ansatz~\eqref{eq:ds2ECO}, can
be written as
\begin{equation} \label{eq:ECO_Hamiltonian_constraint}
\nabla^2_{\eta}\varphi=-2\pi \rho \varphi^{5}\,.
\end{equation}
We first consider a single, static ECO; then, we shall consider two close static ECOs at rest. Our solution can,
in principle, be matched with the BL solution in the exterior discussed in Sec.~\ref{subsec:BL-ECO}, where $R$ is the
isotropic coordinate defined in Eq.~\eqref{eq:R_isotropic_to_ECO}. The interior solution is defined for $R\le R_0$,
where $R=R_0$ is the ECO surface. 

We consider, for a single, static ECO, a density profile of the form
\begin{equation}
\rho=\frac{\varphi^{-5}}{2\pi}\left[
\frac{3 M}{2R_0^3}\Theta\left( R_0-R\right)  \right]\,.\label{eq:ECOrho}
\end{equation}
Therefore, Eq.~\eqref{eq:ECO_Hamiltonian_constraint} gives, 
\begin{equation}
  \nabla^2_{\eta}\varphi=
  \frac{3M}{2R_0^3}  \Theta\left( R_0-R\right)\,.\label{eq:phi_ECO}
\end{equation}
This choice can be considered either as a toy model, or as a specific example of an ECO interior. We stress that the
interior structure of the ECO does not enter in the GW signal computed in this thesis, since we assume that --~due to
the small lapse~-- the internal degrees of freedom do not significantly affect the GW emission and thus can safely be
neglected (see Sec.~\ref{subsec:BL-ECO}).

The exterior solution (i.e. Schwarzschild's metric) for $R\ge R_0$ is $\varphi=1+M/(2R)$. 
The solution of Eq.~\eqref{eq:phi_ECO} matching with the exterior is
\begin{equation}
\varphi=1+\frac{M}{2}\left( \frac{3}{2R_0}-\frac{R^2}{2R_0^3}\right)+\frac{M}{2}\left( \frac{1}{R}+\frac{R^2}{2R_0^3}-\frac{3}{2R_0}\right) \Theta\left( R-R_0\right)\,.
\end{equation}
Let us now consider two ECOs at rest, with the same positions as the BHs in the BL initial data (see
Sec.~\ref{subsec:BL}), $(0,0,\pm Z_0)$, having radii $\widetilde{R}_1$, $\widetilde{R}_2$, and ``bare'' masses $m_1$,
$m_2$, corresponding to the ADM masses (see Sec.~\ref{subsec:BL-ECO}) $M_1=m_1(1+m_2/(4Z_0))$, $M_2=m_2(1+m_1/(4Z_0))$;
the ADM mass of the final object is $M=m_1+m_2$.

\begin{figure}
\centering
\includegraphics[width=0.5\textwidth,keepaspectratio]{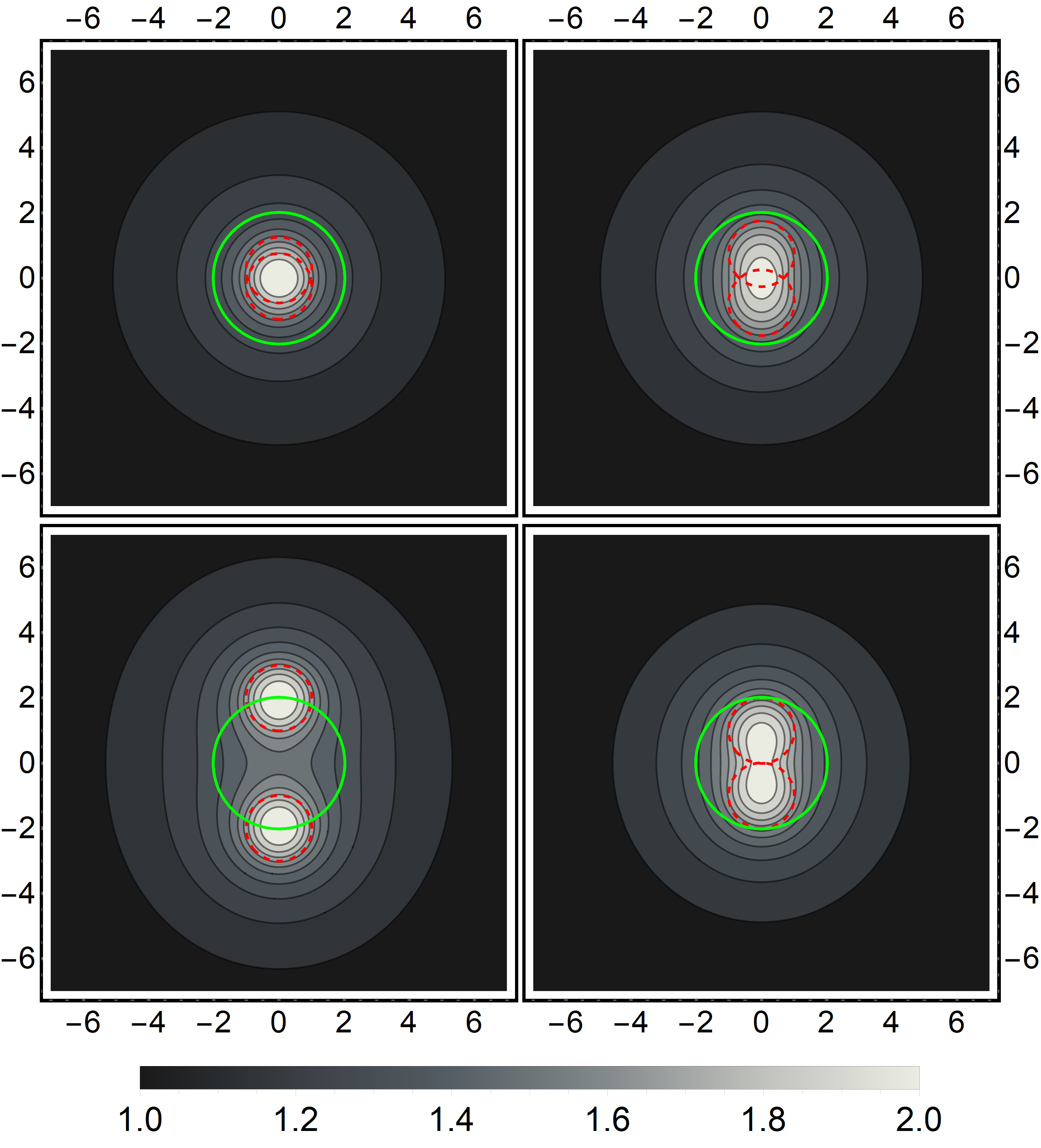}
\caption[Binary ECOs conformal factor.]{Curves with the same value of $\varphi$ for an equal mass binary ECO spacetime ($m_1=m_2=M/2$,
  $R_1=R_2=R_0$). The axes are normalized to the mass of the final ECO ($Y/M, Z/M$), and each frame represents a $Y-Z$
  spatial slice of the conformal factor. Starting from the upper-left panel and moving clockwise in the panels, ECOs are
  at a distance $Z_0=M\{1/8,3/8,1/2,1\}$ from the origin of the reference frame. The dashed red circles corresponds to the
  colliding ECOs surfaces and the solid green one to the final static ECO.  }
\label{fig:BECO_initialvalue}
\end{figure}

 
The constraint equation has the form 
\begin{equation}
 \nabla^2_{\eta}\varphi=
  -\frac{3 m_1}{2R_1^3}\Theta(\widetilde{R}_1-\sqrt{\Delta-2Z_0R\cos\theta})-\frac{3 m_2}{2R_2^3}\Theta(\widetilde{R}_2-\sqrt{\Delta+2Z_0R\cos\theta})\,,\label{eq:equationBECO}
\end{equation}
where $\Delta=R^2+Z_0^2$ and $\cos\theta=Z/R$.
Matching with the BL solution~\eqref{eq:BL_id_newtonian} gives:
\begin{equation}
\varphi=1-\frac{m_1}{2}\Upsilon\left( \widetilde{R}_1\right)-\frac{m_2}{2}\Upsilon\left( \widetilde{R}_2\right)+\frac{R Z_0 \cos\theta}{2}\left[\frac{m_1}{\widetilde{R}_1^3}-\frac{m_2}{\widetilde{R}_2^3} \right]
  +\frac{m_1 g_{\left( \widetilde{R}_1,Z_0\right)}+m_2 g_{\left( \widetilde{R}_2, -Z_0\right)}}{2}\,,
\end{equation}
where
\begin{align}
\Upsilon\left( \widetilde{R}_i\right)&=\frac{R^2+Z_0^2-3\widetilde{R}_i^2}{2\widetilde{R}_i^3}\nn\\
\label{eq:upsilon_and_g}
g_{\left( \widetilde{R}_i,Z_0\right)}&=\Bigg[\frac{1}{\sqrt{\Delta+2 Z_0 R \cos\theta}} -\frac{3}{2\widetilde{R}_i} +
  \frac{\Delta+2 Z_0 R \cos\theta}{2\widetilde{R}_i^3} \Bigg]\Theta\left[ \sqrt{\Delta+2 Z_0 R \cos\theta}-\widetilde{R}_i \right]\,.
\end{align}
For an equal mass binary ECO ($m_1=m_2=M/2$, $R_1=R_2=R_0$), the conformal factor reduces to  
\begin{equation} \label{eq:id_ECOs_equalmasses}
\varphi= 1-\frac{M}{2}\Upsilon\left( R_0\right) +\frac{M}{4} \left[ g_{\left( R_0,Z_0\right)}+ g_{\left( R_0, -Z_0\right)}\right]\,,
\end{equation}
with $\Upsilon$ and $g$ defined in Eq.~\eqref{eq:upsilon_and_g}. An example of this configuration is represented in
Fig.~\ref{fig:BECO_initialvalue} for various ECOs distances.

We remark that this solution has been determined under the assumption that the two ECOs do not overlap, i.e. that
$Z_0>R_0$. Since (for $\epsilon\ll1$) $R_0\simeq M/2$, this implies that $Z_0/M\gtrsim0.5$. When the ECOs overlap, the
solution~\eqref{eq:id_ECOs_equalmasses} should be considered as a rough approximation. We remark, however, that the
model of the interior does not affect the GW signal derived in Sec.~\ref{subsec:ECO-collision}, because the internal
degrees of freedom are, with good approximation, decoupled from the evolution of the exterior spacetime.

  \cleardoublepage

\chapter{Klein Gordon equation in binary black holes spacetimes}\label{app:separateKG}

In this Appendix we show how to separate the KG equation in axisymmetric spacetimes. We perform this separation in the context of EsGB (see Chapter\til\ref{chapter:EsGB}), where, thanks to non-trivial couplings between the scalar field and the curvature, the r.h.s of the KG equation is non zero. Setting the coupling constant of EsGB to zero $\left(\eta=0\right)$ instead, this procedure refers to separating the KG equation in GR (see Chapter\til\ref{chapter:CLAP}).

Let us consider the Klein-Gordon equation for a massless scalar field in EsGB with the quadratic coupling function in Eq.\til\eqref{eq:coupling_function}, 
\be \label{eq:KG_equation_appendix}
\square\Phi=-\frac{\eta}{4} \Phi\rgb\,.
\ee
In the equation above, the d'Alembertian operator is defined as $\frac{1}{\sqrt{-g}}\partial_\mu\left(g^{\mu\nu}\sqrt{-g}\partial_\nu\right)$, and the spacetime metric represents the perturbative geometry of a BBH: $g_{\mu\nu}=g^{(0)}_{\mu\nu}+h_{\mu\nu}$. Here, $g^{(0)}_{\mu\nu}$ is given by the usual Schwarzschild metric (Eq.\til\eqref{eq:Schwarzschild_metric}) and $h_{\mu\nu}$ by Eq.\til\eqref{eq:BLrecast}. Expanding the D'Alambertian operator in powers of the BH separation $Z_0$, and neglecting higher order terms $O(Z_0^3)$, Eq.\til\eqref{eq:KG_equation_appendix} takes the form:
\be \label{eq:KG_pert}
\oo \square^{(0)}+ Z_0^2 \square^{(1)} \cc \Phi = -\frac{\eta}{4}  \oo \rgb^{(0)}+ Z_0^2 \rgb^{(1)} \cc \Phi\,,
\ee
where each contribution on the right hand side can be computed through the BBH spacetime in Eq.\til\eqref{eq:pert_Schw_CLAP_GR_BL},
\begin{align}\label{eq:R0andR1}
\rgb^{(0)}&=\frac{48 M^2}{r^6}\,,\nn\\
\rgb^{(1)}&=-\frac{\alpha\left(\theta\right)}{M r^6} \Bigg(r \left(r (2 M-r) \frac{d^2 g}{dr^2}+(r-5 M) \frac{d g}{dr}\right)+3 g (4 M-r)\Bigg)\,,
\end{align}
with 
\begin{equation}\label{eq:alpha_definition}
\alpha\left(\theta\right)=1+3 \cos (2 \theta )\,.
\end{equation}

Thus, we expand the scalar field in scalar spherical harmonics $Y^{\ell m}(\theta,\phi)$, as in
Eq.~\eqref{eq:Phi_ansatz}, where the harmonic functions 
normalization reads as $ \int d\Omega \oo Y^{\ell m}\cc^* Y^{\ell'm'}=\delta_{\ell \ell'}\delta_{m m'}$.
As shown in Ref.\til\cite{Cano:2020cao}, since Schwarzschild's spacetime ($Z_0=0$) is spherically symmetric, the spherical harmonics are eigenfunctions of the KG operator on $g^{(0)}_{\mu\nu}$, while the first order KG operator ($\square^{(1)}$) couples harmonics with different $\ell$. This means that each solutions of the zero-th order problem contains only one definite value of the index $\ell$. The index $m$, instead, always factors out from the equation since the background is axisymmetric. Since the first order KG operator is proportional to $Z_0^2$, we assume,
\begin{align} \label{eq:Phi_ansatz_2}
\Phi=\frac{\psi_{\ell\,m}\left( t,r\right) Y^{\ell\,m}\left(\theta,\phi\right)}{r}+Z_0^2 \sum_{\ell'\neq\ell}\frac{\psi_{\ell'\, m}\left( t,r\right) Y^{\ell'\, m}\left(\theta,\phi\right)}{r}\,.
\end{align}
It is important to remark that with the ansatz in Eq.\til\eqref{eq:Phi_ansatz_2} we restrict to excitations with a single value of $\ell$. Perturbation mixing multiple values of $\ell$s simultaneously may lower even more the threshold of instability of BBH spacetimes in EsGB (see Chapter \ref{chapter:EsGB}).

Inserting Eq.\til\eqref{eq:Phi_ansatz_2} in Eq.\til\eqref{eq:KG_pert} we find (dropping the $\left(\theta,\phi\right)$ dependence in the spherical harmonics),
\begin{align} \label{eq:KG_pert_2}
& \square^{(0)} \left[\frac{\psi_{\ell\, m} Y^{\ell\,m} }{r}\right]+ Z_0^2  \square^{(1)} \left[ \frac{\psi_{\ell\,m} Y^{\ell\,m} }{r} \right]+ Z_0^2 \sum_{\ell'\neq \ell} \square^{(0)} \left[ \frac{\psi_{\ell'\,m} Y^{\ell'\,m} }{r}\right] =\nn\\
&-\frac{\eta}{4} \Bigg[ \frac{\psi_{\ell\,m} Y^{\ell\,m} }{r} \rgb^{(0)} + Z_0^2 \frac{\psi_{\ell m}  Y^{\ell\,m} }{r}  \rgb^{(1)} +Z_0^2 \sum_{\ell'\neq \ell}\frac{\psi_{\ell' m}  Y^{\ell'\,m}}{r}\rgb^{(0)}\Bigg]\,.
\end{align}
Projecting the above equation on the complete basis of spherical harmonics, the components $\psi_{\ell'\,m}$ with $\ell'\neq\ell$ vanish, because the spherical harmonics are eigenfunctions of $\square^{(0)}$. Thus, the only remaining $O(Z_0^2)$ term on the l.h.s of Eq.~\eqref{eq:KG_pert_2} can be written explicitly as,
\begin{align}
  & \square^{(1)} \left[ \frac{\psi_{\ell\,m} Y^{\ell\,m} \left(\theta,\phi\right)}{r} \right]=-\frac{\partial Y^{\ell\, m}}{\partial \theta}\frac{3 g \sin (2 \theta ) }{8 M^2 r^3}\psi_{\ell\,m}\nn\\
&  +Y^{\ell\, m}\Bigg(- r\left(-r^2 f dg/dr+4 M g\right)\frac{\partial \psi_{\ell\,m}}{\partial r}-2r^3 f g\frac{\partial^2 \psi_{\ell\,m}}{\partial r^2}\nn\\
&+\left(-r^2 f dg/dr+2 g (\ell (\ell+1) r+2 M)\right)\psi_{\ell\,m}\Bigg)\frac{\alpha\left(\theta\right)\,}{16 M^2 r^4} \,,
\label{eq:KGZ02}
\end{align}
and we remind that $f=1-2M/r$.
 
As expected, the projection on $Y^{\ell\,m}$ of the $O(0)$ term on the l.h.s in Eq.~\eqref{eq:KG_pert_2} provides the standard form of the KG equation in
Schwarzschild's spacetime, 
\begin{equation} \label{eq:order0KGproj}
-\frac{1}{r f}\Bigg(\frac{\partial^2\psi_{\ell\,m}}{\partial t^2}-f^2\frac{\partial^2\psi_{\ell\, m}}{\partial r^2}  -f\frac{df }{dr}\frac{\partial\psi_{\ell\, m}}{\partial r}+f\frac{\ell (\ell+1) r+2 M }{r^3}\psi_{\ell\, m}\Bigg)\,.
\end{equation}
Finally, projecting the full Eq.\til\eqref{eq:KG_pert} on $Y^{\ell\, m}$ using Eq.\til\eqref{eq:R0andR1} and the results in Eqs.\til\eqref{eq:order0KGproj}-\eqref{eq:KGZ02}, we obtain the desired
{\it decoupled equation}

\begin{align}
\label{eq:KG_notortoise_app}
&\frac{\partial^2 \psi_{lm}}{\partial t^2}+\frac{\partial^2 \psi_{lm}}{\partial r^2} \oo U_0+Z_0^2\tilde{U}_0\cc+\frac{\partial \psi_{lm}}{\partial r}  \oo U_1+Z_0^2\tilde{U}_1\cc+\nn\\
&\psi_{lm} \oo \oo W_0+\frac{\eta}{4}\tilde{W}_0\cc +Z_0^2 \oo W_1+ \frac{\eta}{4} \tilde{W}_1\cc\cc=0\,,
\end{align}
with radial potentials given by,
\begin{align}
\label{eq:potentials_notortoise}
&U_0(r)=-f^2\,,\nn\\
&\tilde{U}_0(r)=\frac{\left( r-2 M \right) f q^{(1)}_{\ell\,m} g}{8M^2r}\,,\nn\\
&U_1(r)= -f \frac{df}{dr} \,,\nn\\
& \tilde{U}_1(r)= -\frac{q^{(1)}_{\ell\,m} (2 M-r) \left(4 M g(r)-f r^2 dg/dr\right)}{16 M^2 r^3}\,,\nn\\
&W_0(r)=f\frac{\ell (\ell+1) r+2 M }{r^3}\,,\nn\\
&\tilde{W}_0(r)=\ooq \frac{48   M^2 (2 M-r)}{r^7}\ccq\,,\nn\\
&W_1(r)=\frac{q^{(1)}_{\ell\,m}}{16 M^2 r^4} (f r^2 (r-2 M) dg/dr+2 g (2 M-r) (l (l+1) r+2 M))+3q^{(2)}_{\ell\,m}\frac{(r-2M)g}{4M^2r^3}\,,\nn\\
&\tilde{W}_1(r)= -\frac{2  M q^{(1)}_{lm} \oo 2M-r\cc}{r^7}\Bigg(3 \oo 4M-r\cc \Delta_1 + r\oo \oo r-5M \cc\frac{d \Delta_1}{dr}  + \oo 2M-r\cc r \frac{d^2 \Delta_1}{dr^2} \cc \Bigg) \,.
\end{align}
The coefficients in Eqs.\til\eqref{eq:potentials_notortoise} are defined as
\begin{align} \label{eq:q1lm}
q^{(1)}_{\ell\,m}  \equiv&\int d\Omega \left( Y^{\ell\,m}\right)^* Y^{\ell\,m} \alpha\left(\theta\right)
  \,,\\ \label{eq:q2lm}
  q^{(2)}_{\ell\,m}\equiv &\int d\Omega \sin \theta \cos \theta \left( Y^{\ell\,m}\right)^*
  \frac{d Y^{\ell\,m}}{d\theta}  \,.
\end{align}
Since
\begin{equation}
\label{eq:q12lm}
q^{(1)}_{00}=q^{(2)}_{00}=0\,,
\end{equation}
the $\ell=0$ equation is not affected by the $O(Z_0^2)$ corrections. Instead, for $0<\ell\le2$ one gets,
\begin{align} 
&  q^{(1)}_{1-1}=q^{(1)}_{11}=-\frac{4}{5},\,q^{(1)}_{10}=\frac{8}{5}\,,\nonumber\\
&  q^{(1)}_{2-2}=q^{(1)}_{22}=-\frac{8}{7},\,q^{(1)}_{2-1}=q^{(1)}_{21}=\frac{4}{7},\,q^{(1)}_{20}=\frac{8}{7}\,,\nonumber\\
& q^{(2)}_{1-1}=q^{(2)}_{11}=\frac{1}{5},\,q^{(2)}_{10}=-\frac{2}{5}\,,\nonumber\\
& q^{(2)}_{2-2}=q^{(2)}_{22}=\frac{2}{7},\,q^{(2)}_{2-1}=q^{(2)}_{21}=-\frac{1}{7},\,q^{(2)}_{20}=-\frac{2}{7}\,.
\end{align}

  \cleardoublepage
  
\SgIncludeBib{references2}

\end{document}